\documentclass[preprint]{elsarticle}

\usepackage{lineno,hyperref}
\modulolinenumbers[5]
\usepackage{tabularx}

\newcommand\clearrow{\global\let\rowmac\relax}
\clearrow

\usepackage[T1]{fontenc}
\usepackage[utf8]{inputenc}
\usepackage{amsmath}
\usepackage{natbib}
\usepackage{graphicx}
\usepackage{gensymb}
\usepackage{url}
\usepackage{mathtools}
\usepackage[varg]{txfonts}
\usepackage{textcomp}
\usepackage[super]{nth}
\usepackage{caption}
\usepackage{subcaption}

\usepackage{wasysym}
\usepackage{esvect}

\usepackage{siunitx}
\usepackage{lscape}
\usepackage{longtable}
\usepackage{threeparttable}

\usepackage{adjustbox}
\usepackage{booktabs}
\usepackage{float}
\usepackage{wrapfig}

\biboptions{authoryear}

\begin{document}

\begin{frontmatter}

\title{A Physical Survey of Meteoroid Streams: Comparing Cometary Reservoirs}

\author[uwopa,wiese]{N. Buccongello}
\ead{nbuccong@uwo.ca}

\author[uwopa,wiese]{P. G. Brown}
\author[uwopa,wiese]{D. Vida}
\author[jacobs]{A. Pinhas}

\address[uwopa]{Department of Physics and Astronomy, University of Western Ontario, London, Ontario, N6A 3K7, Canada}
\address[wiese]{Western Institute for Earth and Space Exploration, University of Western Ontario, London, Ontario, N6A 5B7, Canada}
\address[jacobs]{Columbus Technologies and Services Inc. \& Jacobs Space Exploration Group, NASA Marshall Space Flight Center, Huntsville, Alabama, 35812, USA}

\begin{abstract}
In this work, we present an optical survey of mm-sized meteoroids using the Canadian Automated Meteor Observatory's (CAMO) mirror tracking system. The system tracks meteors to magnitude +7.5 through an image-intensified telescopic system which has a spatial accuracy of $\sim$\SI{1}{\metre} and a temporal resolution of \SI{10}{\milli \second}.
We analyze 41 meteors from 13 showers with known parent bodies, recorded between 2016 and 2022. We fit a numerical ablation and fragmentation model to our data which models meteoroid fragmentation as erosion into \SIrange{10}{500}{\micro \metre} constituent grains and uses the observed wake as a hard constraint on the model parameters. We measure average bulk meteoroid densities which are consistent with in situ measurements: 602 $\pm$ \SI{155}{\kilo \gram \per \cubic \metre} for Jupiter-family and 345 $\pm$ \SI{48}{\kilo \gram \per \cubic \metre} for Halley-type showers. The Geminids had the highest measured bulk density of 1387 $\pm$ \SI{240}{\kilo \gram \per \cubic \metre}, consistent with carbonaceous material. We fail to reproduce the high bulk density (>\SI{3000}{\kilo \gram \per \cubic \metre}) for Jupiter-family meteoroids previously reported in the literature derived using fragmentation models on data sets with fewer observational constraints. We also provide estimates of the meteoroid grain sizes, grain mass distributions, and energy necessary to trigger the erosion for meteoroids in the analyzed showers.
\end{abstract}

\end{frontmatter}


\section{Introduction}

Comets are among the most pristine bodies in our solar system and their well-preserved primordial nuclei are direct windows as to how the earliest solar system solids formed \citep{Blum_2022}. They also may have played a role in water delivery and the subsequent rise of life on Earth \citep{Thomas2007}. Meteoroids, dust particles released from comets, are also proxies for the physical properties of their parent bodies. Thus, inferring the properties of cometary meteoroids permits a deeper understanding of cometary nuclei, and indirectly, the agglomeration processes that operated early in solar system history within the protoplanetary disk. From early observations of meteors, a physical picture of cometary meteoroids as ``dust-balls'' has emerged \citep{Hawkes1975}, with aggregates of small silicate grains glued together similar to the structure of interstellar dust \citep{Greenberg_Li_1999}. This picture of cometary meteoroids has been largely confirmed by the findings of the Rosetta spacecraft \citep{Hornung_2016}.

Our understanding of the formation and evolutionary pathways of comets at various length scales remains incomplete \citep{Blum_2022}. While in situ comet observations have provided substantial information about a few nuclei \citep[e.g. 1P/Halley, 81P/Wild, 67P/Churyumov-Gerasimenko; ][]{cochran2015composition}, such missions are few, expensive and time-consuming. There also remains a question as to how representative these few nuclei are of their respective cometary populations. For example, do Jupiter-family comets (from the scattered disk) have similar properties to Halley-type or nearly isotropic comets which formed at different locations in the solar nebula \citep{Nesvorny2017a, Vokrouhlicky2019}?

Meteoroids of cometary origin whose orbits pass close to Earth provide a unique opportunity to survey cometary material with relative ease. For a number of meteor showers, parent bodies are known \citep{Jenniskens_2006}. Therefore, if one could determine the physical properties associated with meteoroids from a specific shower, those properties will also be diagnostic of that shower's parent body and lead to better constraints on the wider cometary population.

Here we report a survey of meteoroid physical properties that samples a dozen meteor showers, each with a known parent body. For each shower, we have identified several mm-sized shower meteoroids which were recorded by the mirror-tracking system of the Canadian Automated Meteor Observatory (CAMO). The mirror tracking system captures videos of meteors at high temporal (\SI{10}{\milli \second}) and spatial (a few meters) resolutions from two stations. 

By fitting a meteoroid ablation and fragmentation model to the observed meteor characteristics as a function of time (lightcurve, velocity, wake) we can forward model the data to determine the meteoroid's physical properties, such as its bulk density, grain size distribution and other parameters defined in the model \citep{vida2023orionids}. Uniquely to CAMO, the wake (brightness distribution behind a meteor due to grain release and subsequent ablation) is used as a model constraint, permitting a more robust forward model of the parameters. CAMO routinely records wake for more than 90\% of all tracked meteors \citep{Subasinghe_2016}. 

In Section \ref{sec:priorwork} we provide a review of the known physical properties of meteor showers and cometary parents. We focus on meteoroid bulk densities and grain mass distributions and review the works which determined these values in the literature. In Section \ref{sec:equipment} we describe the equipment used to observe and collect data, as well as our data selection and reduction process. Section \ref{sec:modelling} explains our modelling process and how we used the ablation model to determine the physical properties of each meteoroid. Section \ref{sec:resultsdiscussion} presents our key results, compares them to the previous literature and discusses the implications. Section \ref{sec:conclusions} summarizes this work and highlights our main findings.

\section{Meteoroid Physical properties: Previous Work}
\label{sec:priorwork}
Our knowledge of the properties intrinsic to cometary nuclei comes from direct measurements as well as estimates of meteoroid properties from associated meteor showers. For the latter, a major limitation in the past has been the temporal and spatial resolution of observations. While meteor lightcurves are typically well measured, position measurements are limited to resolutions of tens of meters at best and wake measurements are virtually impossible due to limited pixel resolution and the movement of meteors across a static image plane \citet{Shadbolt1995, Borovicka_2007, Kikwaya_2011, Vojacek_2017}. This means that past attempts to infer meteoroid properties via ablation modelling are likely inaccurate and/or ambiguous, as imprecise measurement of meteor deceleration or lack of resolution to resolve wake make it difficult to constrain one unique set of model parameters. 

For example, even when \citet{Campbell-Brown2013} attempted to model ten CAMO recorded meteors, each with very precise lightcurves and dynamics, two different models (the erosion model from \citet{Borovicka_2007} and the thermal disruption model from \cite{Campbell-Brown2004}) were able to accurately predict meteoroid lightcurves and deceleration with significantly different fragmentation processes. However, both models were unable to simultaneously reproduce the wake. This work hopes to improve on these past modelling attempts by also fitting the wake in order to sufficiently constrain the model results. 

In this section, a brief summary of the known physical properties of mm to cm-sized cometary meteoroids and their parent comets is presented for both general context and comparison. A full list of showers and their parent bodies examined in this paper are shown in Table \ref{tab:dataset}, while a full list of their properties are listed in \ref{ap:bulkdensity} and \ref{ap:massindex}.

\subsection{Meteoroid physical properties: Meteor Population Groups}

Bulk density is arguably the most important meteoroid physical property in relation to parent bodies.  Additionally, potential damage to satellites and spacecraft caused by impacts is strongly affected by whether the particle is a porous dust aggregate or a monolithic body \citep{Drolshagen2019}. For meteors, bulk density is difficult to determine due to complicating effects such as fragmentation, variations in ablation behaviour and shape, as well as differences in structure and chemical composition \citep{Kikwaya_2009}. With these caveats in mind, attempts have been made in the past to estimate bulk densities of various meteor populations including meteor showers and comet nuclei.

Some of the earliest work to categorize small meteoroid populations by physical characteristics was that of \citet{Ceplecha_1967a} and \citet{Ceplecha_1968}. They estimated the resistance to ablation of small meteoroids from Super-Schmidt observations and found that most could be classified into 4 categories (A, B, C and D) based on a one-dimensional strength/ablation parameter $K_{B}$. This parameter depends on the beginning height of the meteor, its initial velocity, and its entry angle. In general, the average meteor begin height increases and the strength decreases from classes A to D. From this trend, \citet{Ceplecha_1988} associated dense ordinary chondrite material with group A. Intermediate (carbonaceous chondrite) material with smaller perihelion and aphelion close to Jupiter (such as the Geminids) belonged to group B \citep{Ceplecha_1977a}.

Most relevant to this work, classical meteor showers with known parent comets are contained in group C, which is further subdivided into three populations: $C_{1}$ with ecliptic short period orbits, $C_{2}$ with randomly inclined long period orbits and $C_{3}$ with both short-period orbits and random inclinations \citep{Ceplecha_1988}. Finally, group D contains extremely soft cometary material (such as the Draconids) \citep{Ceplecha_1967a}. \citet{Ceplecha_1977a} later updated their findings to include data from multiple mass ranges and to correct bulk densities that were overestimated in previous work. \citet{Ceplecha_1988} further extended these strength populations with supplementary data from TV observations and small camera data by introducing small changes to $K_{B}$ accounting for the different system sensitivities. Note that this one-dimensional approach only identified relative strength differences; the absolute properties (like bulk density) are not explicitly estimated.

\citet{BellotRubio_2002} later re-analysed the Super-Schmidt data and determined quantitative estimates of bulk density for groups A, B and C using ablation model comparisons to data. It is important to note however, that \citet{BellotRubio_2002} used single-body ablation theory as opposed to quasi-continuous fragmentation, so the measured densities are likely underestimated \citep{Borovicka_2006}. Finally, \citet{Kikwaya_2011} was able to estimate densities for group A and C meteors using data from a number of TV and camera systems. This work used the thermal ablation model of \citet{Campbell-Brown2004} which assumed meteoroids disintegrated into constituent grains upon being heated to a threshold temperature. Results from each study are summarized in Table 5.10 in \citet{Kikwaya_2011}, and we provide an updated version in Table \ref{tab:bulkdensities}.

\begin{table}
\begin{center}

\caption{Average bulk densities (\SI{}{\kilo \gram \per \metre \cubed}) in Ceplecha’s four meteor classes from different studies. Adapted from \citet{Kikwaya_2011}.}
\label{tab:bulkdensities}

\begin{tabular}{|p{1.8cm}|p{2.2cm}|p{1.6cm}|p{1.4cm}|p{1.3cm}|p{1.3cm}|}
\hline
Work & Instrument & Group A & Group B & Group C & Group D\\
\hline
Ceplecha (1967) & Photographic Super-Schmidt & 4000 & 2200 & 1400 & - \\
\hline
Ceplecha (1977) & Photographic Super-Schmidt & 2100 & 1000 & 600 & 200\\
\hline
Ceplecha (1988) & Photographic and TV & 1400-2700 & 650-1700 & 550-910 & 180-380\\
\hline
Bellot Rubio et al. (2002) & Photographic Super-Schmidt & 2400 & 1400 & 400 & -\\
\hline
Kikwaya et al. (2011) & TV Systems & 3800 $\pm{800}$ & - & 800 & -\\
\hline
\end{tabular}

\end{center}
\end{table}

\newpage

\subsection{Meteoroid Physical Properties: Inferences from Observations of Meteor Showers}

Several works have previously attempted to directly estimate the bulk density of meteoroids in specific meteor showers. As part of the Harvard Meteor Program, \citet{Verniani_1967} analyzed sporadic and shower meteors captured by Super-Schmidt cameras. They used the luminous efficiency value from \citet{Verniani1965} to establish a mass scale from which meteoroid densities were determined \citep{Verniani_1967}. However, \citet{Ceplecha_1988} noted that this work underestimated densities in a statistical sense due to the use of single-body theory and a low luminous efficiency value \citep[several times lower than what is currently accepted; ][]{Popova_2019}. As a result, the estimated meteoroid mass was significantly increased, biasing the selection of the dataset by removing most asteroidal-type objects. 

\citet{Verniani_1969} summarized the individual estimates of mass and deceleration for Super-Schmidt meteors by \citet{Jacchia1967} and through this summarization, postulated that ``most meteors are of cometary origin, and are porous, crumbly objects composed of loosely conglomerate sponge-like material''. He also noted that the dataset was not a random sample and that the result only applies to fairly large (gram-sized) meteoroids, so the generality of this conclusion as applied to all meteoroids has been challenged \citep{Ceplecha_1998}. Nevertheless, it establishes the highly friable and low-density nature of some meteoroids, particularly at smaller masses. 

\citet{Babadzhanov_2002} used a model incorporating quasi-continuous fragmentation to fit the lightcurves of photographic meteors and estimate the bulk density of meteoroids from various showers. This work was among the first attempts at directly measuring accurate bulk densities but failed to include the meteor deceleration \citep{Borovicka_2006}. Neglecting deceleration leaves ambiguity in bulk density results, as multiple combinations of parameters can be found to match the same lightcurve by simply changing the ablation coefficient. \citet{Babadzhanov_2009} later expanded on this work with double station photographic observations but again neglected to include deceleration in the model fits. 

\citet{BellotRubio_2002} used single-body theory to re-examine the bulk densities of Super-Schmidt meteors and was able to estimate densities for some shower meteors while also categorizing them into Ceplecha's meteor groups. This work was limited again due to the small number of data points per Super Schmidt meteor and neglect of meteoroid fragmentation \citep{Borovicka_2006}.  

\citet{Vojacek_2017} reported what is likely the most accurate and complete measurement of mm to cm-sized meteoroid densities to date using the \citep{Borovicka_2007} erosion model to simultaneously fit meteor light curves and deceleration. However, only a small number of shower meteors were reduced; sporadics composed the majority of the analyzed events. 

Results from each study are summarized in Table \ref{tab:showerdensities}. Few studies have estimated the densities of individual shower meteoroids. \citet{hughes1995perseid} found a density of 320~\SI{}{\kilo \gram \per \metre \cubed} for the Perseids. \citet{Kikwaya_2011} focused on sporadic meteors, but 10 of the analyzed meteors were Perseids. It was found that they had an average density of \SIrange{420}{820}{\kilo \gram \per \metre \cubed}. However, it should be noted that this work measured densities much higher than what would be expected for certain meteor populations (e.g. meteors on Jupiter-family orbits). It is suspected that the data used in this study was not of a high enough resolution to observe the true meteor deceleration for such JFC meteors, so the model used was biased to higher-density meteors that did not decelerate as significantly.

\hspace{1cm}
\begin{landscape}
\begin{center}
\begin{table}

\caption{Average bulk densities (\SI{}{\kilo \gram \per \metre \cubed}) of meteor showers which are the focus of the current work, as determined from various sources. Showers are labelled by their official three-letter IAU Meteor Data Center code.}
\label{tab:showerdensities}

\begin{tabular}{ |p{2.1cm}|p{1cm}|p{1.9cm}|p{1.7cm}|p{1cm}|p{1.9cm}|p{1.7cm}|p{1.9cm}|p{1.9cm}| }
 \hline
 Work & CAP & GEM & LEO & LYR & NTA & ORI & PER & STA \\
 \hline
 Verniani (1967) & 140 & 1060 & 600 & 390 & 260 & 250 & 290 & 280 \\
 \hline
 Babadzhanov (2002) & - & 2900 ($\pm$ 600) & 400 ($\pm$ 100) & - & 1500 ($\pm$ 200) & - & 1300 ($\pm$ 200) & 1500 ($\pm$ 200) \\
 \hline
 Babadzhanov and Kokhirova (2009) & 2100 & 2900 ($\pm$ 600) & 400 ($\pm$ 100) & - & 1600 ($\pm$ 400) & 900 ($\pm$ 500) & 1200 ($\pm$ 200) & 1600 ($\pm$ 400) \\
 \hline
 Bellot Rubio et al. (2002) & 450 & 1940 & - & - & 420 & - & 600 & -\\
 \hline
 Vojacek (2017) & - & 1788 & - & 230 & 1967 & - & 167 & 1967\\
 \hline
 
\end{tabular}
\end{table}
\end{center}
\end{landscape}

\subsection{Meteoroid Physical Properties: Inferences from Interplanetary Dust Particles}
 
The closest analogs to shower meteoroids that are available for laboratory examination are interplanetary dust particles (IDPs) \citep{Bradley_2007}. These particles originate from a broad range of dust-producing bodies extending from the inner asteroid belt to the Kuiper belt and in some cases are among the most primitive astromaterials studied in a laboratory \citep{Greenberg_2000}. While the specific parent bodies of IDPs are indeterminable, certain IDPs known as chondritic porous (CP) show characteristics strongly suggestive of links to cometary nuclei. They therefore represent the best material for comparison to cometary meteors \citep{Bradley2007} and provide a powerful comparison for our work.

\begin{figure}[H]
    \centering
    \includegraphics[width=\linewidth]{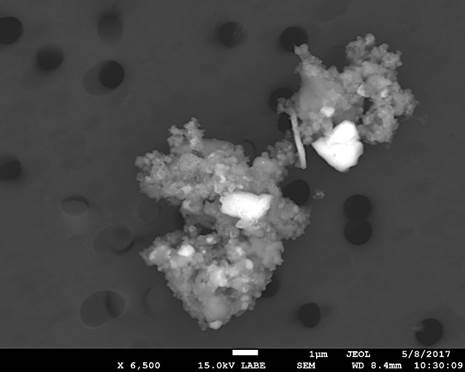}
    \caption{An IDP from the Aircraft Collected Particle (ACP) collection (per NASA Cosmic Dust Collection).}
    \label{fig:IDP}
\end{figure}

Figure \ref{fig:IDP} shows an IDP a few $\mu$m in size, collected in the atmosphere by NASA aircraft. IDPs such as this one are very reminiscent in structure to cometary meteors as both particles have been released from a (likely) cometary parent body, are made of smaller grains and have undergone some space weathering before eventually reaching Earth's atmosphere. 

CP IDPs usually have extremely high porosities (70\%-80\%) and low bulk densities of around \SI{500}{\kilo \gram \per \metre \cubed} for 10-50~$\mu$m-sized grains \citep{Greenberg_2000}, though they range anywhere from 300 to \SI{6000}{\kilo \gram \per \metre \cubed} \citep{Bradley_2007}. CP IDPs with densities above \SI{3500}{\kilo \gram \per \metre \cubed} usually contain a large FeNi sulphide grain \citep{Fraundorf_1982} \citep{Love_1994}. 

\citet{Joswiak2007} used He release combined with an entry model to distinguish high and low entry speed IDPs - the faster particles were associated with cometary parents. This IDP population had bulk densities ranging from 700 to \SI{1700}{\kilo \gram \per \metre \cubed}, providing the best laboratory measure of what might be expected for cometary meteoroid bulk densities. Note that this cometary IDP data set, derived from the work of \citet{Love_1994}, is likely biased toward higher densities as highly porous IDPs (or cluster IDPs) are underrepresented due to their fragile nature \citep{Rietmeijer98}.
 
While CP IDPs are likely of cometary origin, they have some characteristics which suggest they are not perfectly identical to cometary grains that have been freshly released from a comet's nucleus. IDPs have been released from their parent body 100,000 years ago or longer, while material identified with cometary meteor showers is likely to have only been released on shorter timescales \citep[1000s of years; ][]{Greenberg_2000} due to the times required for dynamical orbit evolution which merges showers into the sporadic background population \citep{Pauls2005}. This means that IDPs have undergone considerably more processing and weathering when compared to cometary dust, leading to physical differences in both structure and porosity \citep{Greenberg_2000}. It has also been suggested that meteoroids could represent an intermediate step between mm to cm-sized cometary dust particles and \SIrange{10}{100}{\micro \metre} IDPs \citep{Greenberg_2000}. The main limitation with IDPs is that they cannot be reliably tied back to a single parent comet \citep{Trigo-rodriguez2021}, though their heating history during atmospheric entry does provide some clues into their dynamical origin \citep{Bradley2007}.

\subsection{Parent Comet Physical Properties}

For cometary nuclei, density measurements represent a global estimate of the bulk density and macroporosity of these kilometre-sized bodies. While such values will differ from the bulk density and microporosity inherent to mm to cm-sized meteoroids ejected from the parent, these results are still insightful. 

There are several techniques for measuring nuclear bulk density as summarized in \citet{Groussin19}, most of them indirect. Such techniques include density inferences from non-gravitational effects (e.g. outgassing), radar observations, tidal disruption and inversion of shape models based on lightcurves. 

Among these techniques, the most accurate is a spacecraft flyby. Volume measurements can be made through optical imaging, though these rely on derived shape models which are usually incomplete as not all sides of the comet nucleus can be observed at once \citep{Weissman_2008}. Mass measurements have also been attempted by measuring the comet's gravitational effects on spacecraft, but this is also challenging as the small cometary mass, and therefore weak gravitational force, does not have an appreciable effect on the spacecraft's high-velocity orbit \citep{Weissman_2008}. 

In situ measurements of this type have only been accomplished for 6 comets: 1P/Halley, 9P/Tempel, 19P/Borelly, 67P/Churyumov-Gerasamenko, 81P/Wild, and 103P/Hartley. Note there have also been ``blind'' flybys of comet 21P/Giacobini-Zinner and 26P/Grigg-Skjellerup where cameras were not present on the spacecraft \citep{Groussin19}.
Only one of these comets has a meteor shower directly probed in this study, 1P/Halley, which was found to have a nucleus bulk density likely between 260 and \SI{1200}{\kilo \gram \per \metre \cubed} from various studies \citep{Greenberg_Li_1999, Skorov_1999}. By far the most accurate nuclear density to date was made by Rosetta during its rendezvous with 67P/Churyumov-Gerasamenko, producing an estimated global bulk density of 532 $\pm$ \SI{7}{\kilo \gram \per \metre \cubed} \citep{Jorda_2016}. 

Any comet that has not been investigated with an in situ mission requires an indirect way of measuring bulk density. This usually involves using telescopic observations to estimate the non-gravitational forces produced from a comet's surface, which when combined with the comet's orbital information, can constrain the mass range \citep{Weissman_2008}. Volume can be constrained from telescopic observations of dust production which give an estimate of the free-sublimating areas \citep{rickman1987estimates}, however irregular shapes are commonplace for comets, so this process can be quite difficult. Regardless, these measurements do provide reference comparisons for our work. Cometary nuclei with reliable bulk density measurements for parent bodies germane to our study are summarized in \ref{ap:bulkdensity}. 

Some studies that required cometary nuclei bulk densities for their models assume values based on average comet densities \citep[e.g. ][ for C/1861 G1 (Thatcher)]{Molina_2013}. We chose to not include these values as they are not actual measurements. One notable exception is \citet{Masiero_2021} and references therein, who assumed a density of \SI{3000}{\kilo \gram \per \metre \cubed} for (3200) Phaethon. This was an informed assumption, as it was suggested in \citet{Clark_2010} that CK4 carbonaceous chondrites are material analogs for Phaethon, so this value does have a physical basis.

From a summary of all available density measurements, \citet{Groussin19} found that cometary nuclei have an average bulk density of 480 $\pm$ \SI{220}{\kilo \gram \per \metre \cubed} and high porosity (70-80\%). They also concluded that the nuclei are very weakly bound on global scales with tensile strengths less than \SI{100}{\pascal}.

\subsection{Comet Dust and Grain Implications}

The dust particles measured in situ shortly after release from comets are also of particular interest, as these are the same material (albeit with less evolutionary processing) that eventually become meteor showers on Earth. This means their bulk density, grain size range and distribution are especially relevant for comparison to this work.

The size frequency distribution of constituent grains is a fundamental property of meteoroids. A closely associated measure, mass index, represents the particle mass distribution either in a meteoroid stream or the sporadic background \citep{Pokorny2016, Janches_2019}. This index can be represented in a number of ways, however we will specifically focus on the differential mass distribution index $s$, where the difference in the number of particles $dN$ between masses $m$ and $m + dm$ scales with $dN \propto m^{-s}dm$ \citep{mckinley1961meteor}.

When the value of $s$ nears 2, the masses are relatively equally distributed between all particle sizes. Values above 2 signify a population where the total mass is dominated by smaller particles, while values below 2 signify a population with the total mass dominated by larger particles.

From the three-channel spectrometer on Vega-2, it was found that comet 1P/Halley dust has a bulk density of \SI{300}{\kilo \gram \per \metre \cubed} \citep{Krasnopolsky_1988}. \citet{Fulle_2000} used a coma model applied to 1P/Halley that incorporated a grain size distribution and data from the Optical Probe Experiment (OPE) and the Dust Impact Detection System (DIDSY) on Giotto to estimate a dust bulk density of \SIrange{50}{500}{\kilo \gram \per \metre \cubed} (with a favoured value of around \SI{100}{\kilo \gram \per \metre \cubed}). This study also indicated grain masses ranging from \SIrange{1e-12}{1e-3}{\kilogram}, with a differential mass distribution $s$ = 1.53 $\pm$ 0.1 \citep{Fulle_2000}.

More recently, Rosetta used a number of instruments to measure dust near comet 67P. Firstly, the Grain Impact Analyser and Dust Accumulator (GIADA) was used to reveal two unique dust populations which make up the vast majority of the collected particles \citep{Choukroun2020}. The first population being ``fluffy'' particles with extremely low densities and the second being ``compact'' particles with higher densities. These values were later updated after further data acquisition, resulting in an overall dust bulk density of $785^{+520}_{-115}$~\SI{}{\kilo \gram \per \metre \cubed} \citep{Fulle_2000}. GIADA also detected a small number of particles which belong to a third population. These particles are thought to be aggregates of minerals, with high densities similar to Fe sulphides (\SI{>4000}{\kilo \gram \per \metre \cubed}), though they make up a small percentage of the total particles detected \citep{Fulle_2017}. 

The Cometary Secondary Ion Mass Analyzer (COSIMA) was also used in the Rosetta mission in order to collect cometary dust particles from 67P's coma \citep{Hornung_2016}. It was found that the particles collected ranged in size from \SIrange{15}{300}{\micro \metre}, though they were mostly comprised of the larger mass grains with an equivalent mass distribution $s$ = 1.77 $\pm$ 0.1. This work also analyzed 10 example particles that were collected by COSIMA, their sizes ranging from \SIrange{40}{309}{\micro \metre}, and directly estimated their bulk densities, with results ranging from \SIrange{135}{302}{\kilo \gram \per \metre}. While the mean density of these example particles was \SI{211.5}{\kilo \gram \per \metre \cubed}, this work suggests a value of \SI{160}{\kilo \gram \per \metre \cubed} for smaller \SI{100}{\micro \metre}-sized particles. 

Additionally, the Micro-Imaging Dust Analysis System (MIDAS) experiment on Rosetta used atomic force microscopy to observe the shape of nm to $\mu$m-sized dust \citep{Mannel2016}. MIDAS recorded a few thousand dust particles whose sizes ranged from \SIrange{0.04}{8}{\micro \metre} \citep{Kim_2023}. These particles were dominated by the larger mass grains with equivalent mass distribution $s$ = 1.22 - 1.29, though the authors note these smaller $\mu$m size particles were likely produced by fragmentation as opposed to released by gas pressure, meaning they may not represent the larger particles found in the coma \citep{Kim_2023}. 

Using atmospheric estimates from meteor observations, \citet{Blaauw_2019} provides a detailed literature review containing measurements of the mass indices of meteor showers and their relevant mass/magnitude ranges. See \ref{ap:massindex} for the full table. \citet{Vojacek_2017} found similar results to that of 67P when analyzing JFC-related meteoroids with the erosion ablation model. They found grains \SI{5}{\micro \metre} to \SI{2.5}{\milli \metre} in size had a mass index $s$ = 1.8. We extracted the grain ranges and mass indices from this work and added the values to \ref{ap:massindex}. In general for mm-sized sporadic meteors, \citet{Vojacek_2017} found a grain mass range approximately from \SIrange{5e-12}{3e-4}{\kilo \gram} with a mean mass index $s$ = 2.0. 

Mass indices have also been measured for the meteoroid population from some of the parent comets relevant to this work. These are also summarized in \ref{ap:massindex}. In general, the mass distribution of the dust in cometary trails is dominated by larger mass particles, having a mass index $s$ = 1.67 - 2.00 \citep{Engrand2023}. 

Grain masses and their distributions can help constrain or confirm models of comet formation and subsequently planet/solar system formation. For example, it has been shown that the peak mass of dust ejected from 67P is between 1~cm and 1~m, and cm-sized particles are abundant on the surface of the comet \citep{Blum_2017}. The presence of these primordial ``pebble'' sized particles, their measured mass distribution, and the predicted/measured gas emission from 67P are all consistent with comet formation expected in the pebble accretion solar system formation model due to streaming instability followed by gravitational collapse \citep{Trigo-rodriguez2021}.

\section{Equipment and Data Reduction}
\label{sec:equipment}

All data used in this work was collected by the Canadian Automated Meteor Observatory (CAMO) mirror tracking system. In the following section, we provide a brief overview of this system. Further hardware, software and analysis details can be found in \citet{Weryk_2013} and \citet{Vida_2021}. 

\subsection{Canadian Automated Meteor Observatory (CAMO)}

CAMO is a fully-automated video meteor detection facility originally designed for simultaneous radar and video observations. CAMO now has several optical instrument suites, including influx cameras, EMCCD systems, spectral cameras, and a mirror tracking system. Collectively these instruments are designed to address specific goals, such as measuring meteoroid mass influx at Earth \citep{Campbell-Brown2016}, providing constraints for meteor ablation modelling \citep{Campbell-Brown2013} and collecting simultaneous optical meteor data to compare with radar measurements by the Canadian Meteor Orbit Radar \citep{Brown2020}. 

CAMO is composed of two sites located in Ontario, approximately \SI{45}{\kilo \metre} apart in the townships of Tavistock (43.26420$^{\circ}$N, 80.77209$^{\circ}$W, \SI{329}{\metre}) and Elginfield (43.19279$^{\circ}$N, 81.31565$^{\circ}$W, \SI{324}{\metre}). Each site has identical systems whose pointings overlap in the atmosphere, ensuring triangulation for a large sample of meteors.

The CAMO mirror tracking system has been operational since 2009. It is designed to record mm-sized meteors with high-temporal and spatial precision. When sky conditions are dark and clear at both CAMO stations, a high-speed image-intensified wide-field camera (34$^{\circ}$ $\times$ 34$^{\circ}$ field of view, operated at 80 frames per second) is used to quickly detect meteors (within \SI{0.1}{\second}). Once detected, a meteor track is formed based on real-time detection over eight frames. Using positional information from this initial track, dual-axis galvanometers position mirrors to follow the meteor in flight. This optical scanner directs the meteor image into a telescope equipped with an image intensifier and high-speed camera (1.5$^{\circ}$ $\times$ 1.5$^{\circ}$) with a limiting sensitivity near +7 magnitude. This telescopic narrow-field camera has an effective astrometric precision of 1 arc second and is operated at 100 frames per second. This provides high-precision deceleration measurements limited only by meteor morphology and not pixel scale \citep{Vida2021}. A schematic of the CAMO wide-field and narrow-field mirror tracking system is shown below in Figure \ref{fig:CAMO} (adapted from \citet{Vida_2021}).

\begin{figure}[H]
    \centering
    \includegraphics[width=\linewidth]{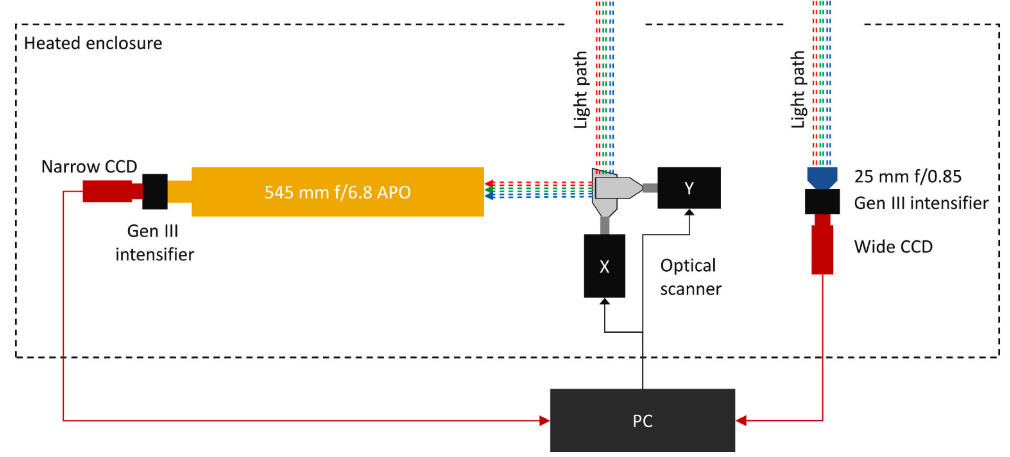}
    \caption{Block Diagram of the CAMO Mirror tracking system. Reproduced with permission from \citep{Vida_2021}.}
    \label{fig:CAMO}
\end{figure}

Due to its tracking capability and a higher resolution than traditional optical systems \citep{Koten2019}, CAMO allows routine measurements of meteor wake. More than 90\% of all CAMO-tracked meteors show significant wake \citep{Subasinghe_2016}. Therefore, the wake can be used as an additional constraint in ablation modelling, allowing more robust estimates of model parameters for CAMO-recorded meteors \citep{Armitage_2020}. This makes the CAMO mirror system a unique instrument as it simultaneously measures a high-precision trajectory, dynamics, photometry, and wake. The details of the methodology, calibration, and measurement approach used in this study are given in \ref{ap:methodology}.

\section{Modelling Methodology}
\label{sec:modelling}
\subsection{Selection of Data}

The dataset examined in this paper is comprised of 1912 CAMO-tracked meteor events, collected between 2010-2022, from 13 different showers associated with 12 unique parent comets. Only the best 41 meteors were chosen and modelled after being manually inspected to determine suitability for detailed data reduction. 

Suitability was based on each event's fulfillment of multiple criteria. Only those meteors were chosen which:
\begin{enumerate}
    \item Were associated with a shower where the parent body is known, shower association was done using the data from the Global Meteor Network \citet{Vida2021}.
    \item Were recorded entirely in the field of view for all four cameras (both the wide and narrow-field cameras at both locations).
    \item The entirety of the meteor's lightcurve was observed. Note that the narrow-field camera is about 2 magnitudes more sensitive than the wide-field camera but tracking only begins after a $\sim0.1$~s delay \citep{Vida_2021}.
    \item Were well tracked by the system and the meteor stayed near the centre of the frame for the majority of the event. This minimizes any geometric or calibration errors which might occur near the edge of the frame.
    \item Did not show extreme amounts of gross fragmentation in order to allow valid application of the erosion model. Rarely, meteors showed very complex fragmentation and disruption into multiple large fragments. Such meteors are hard to measure due to their morphology (e.g. overlapping fragments) which severely reduces the astrometric accuracy \citep{Vida_2021} and were ignored.
    \item For showers with more than 5 well-suited events, preference was given to events that occurred on the same day in order to save time making astrometric plates which require manual work.
\end{enumerate}

Once all of the foregoing filters were applied, 54 events were chosen for reduction, distributed between various showers as shown in Table \ref{tab:dataset}. Thirteen of these were removed from the data set post reduction, as the improved accuracy of their radiants/speeds precluded their original cometary shower association, leaving 41 events to be modelled. Given the time-consuming nature of the reduction process, we generally restricted the number of meteors analysed from any one shower to five when an excess was available. Individual trajectory, orbits and physical modelling results for all of these events are summarized in \ref{ap:metmeasurements}.

\begin{center}
\begin{table}[H]
\adjustbox{max width=\textwidth}{
\begin{tabularx}{\textwidth}{>{\centering\arraybackslash}p{1.8cm}p{4.7cm}>{\centering\arraybackslash}p{2cm}>{\centering\arraybackslash}p{2cm}}
\toprule
\textbf{Shower} & \textbf{Parent Body} & \textbf{Total CAMO Recorded Events} & \textbf{Events Modelled in this Work} \\
\midrule
AUR & C/1911 (Kiess) & 14 & 1 \\
CAP & 169P/NEAT & 108 & 3 \\
ELY & C/1983 H1 (IRAS-Araki-Alcock) & 13 & 2 \\
ETA & 1P/Halley & 87 & 5 \\
GEM & (3200) Phaethon & 166 & 5 \\
JPE & C/1979 Y1 (Bradfield) & 21 & 1 \\
LEO & 55P/Tempel-Tuttle & 32 & 1 \\
LYR & C/1861 G1 (Thatcher) & 22 & 2 \\
MON & C/1971 F1 (Mellish) & 1 & 1 \\
NTA/STA & 2P/Encke & 517 & 9 \\
PER & 109P/Swift-Tuttle & 407 & 5 \\
ORI & 1P/Halley & 522 & 4 \\
TAH & 73P/Schwassmann–Wachmann 3 & 2 & 2 \\
\midrule
Total: 13 & 12 & 1912 & 41 \\
\bottomrule
\end{tabularx}}

\caption{Data summary. Meteor showers are represented by their 3 letter IAU code along with their respective parent bodies. Numbers of both observed and modelled events are shown.}
\label{tab:dataset}

\end{table}
\end{center}

\subsection{Meteor Ablation Model}

The ablation model used in this work is an implementation of the \citet{Borovicka_2007} erosion model. This model has successfully been used in the past to interpret a larger set of mm and cm-sized meteors detected with video instruments in \citet{Vojacek_2017} and \citet{vojacek2019}.

The model is based on the classical single-body equations  \citep{Ceplecha_1998},

\begin{equation} \label{eq:drag}
    \frac{dv}{dt} = -Km^{-1/3} \rho_a v^2
\end{equation}

\begin{equation} \label{eq:mass_loss}
    \left(\frac{dm}{dt}\right)_{ablation} = -K\sigma m^{2/3} \rho_a v^3
\end{equation}

\begin{equation} \label{eq:luminosity} 
    I = -\tau \frac{v^2}{2} \left(\frac{dm}{dt}\right)_{ablation}  + m v \frac{dv}{dt}
\end{equation}

\noindent where $m$ is the meteoroid mass, $v$ is the velocity, $t$ is the time, $\rho_a$ is the air mass density, $\sigma$ is the ablation coefficient \citep{Borovicka_2007}, $I$ is the luminous intensity, and $\tau$ is the dimensionless luminous efficiency. We use the empirical $\tau$ model derived by \citet{vida2023orionids}, which is based on measurements of faint meteors using CAMO. The equations are rewritten in terms of the shape density coefficient $K$, where $K = \Gamma A \rho_m^{-2/3}$, with $\Gamma$ being the drag coefficient (assumed to be unity), $A$ being the shape factor (assumed 1.21 for a sphere) and $\rho_m$ being the meteoroid density.

The erosion model assumes that at a certain altitude, the meteoroid begins quasi-continuous fragmentation where constituent \SI{}{\micro \metre} grains are gradually released, each ablating independently \citep{Borovicka_2007}. This process is called erosion and continues until all of the meteoroid mass has been consumed. Treating ablation as an atom-by-atom mass loss and the erosion as mass loss through grain release, we can rewrite the total mass loss equation as a combination of both ablation and erosion \citep{vida2023direct}: 

\begin{equation}
    \frac{dm}{dt} = \left(\frac{dm}{dt}\right)_{ablation} + \left(\frac{dm}{dt}\right)_{erosion}
\end{equation}

\noindent where mass loss due to erosion is described by the erosion coefficient $\eta$, analogous to the ablation coefficient such that:

\begin{equation}
    \left(\frac{dm}{dt}\right)_{erosion} = -K\eta m^{2/3}\rho_a v^3
\end{equation}

The masses of grains released in erosion are distributed according to a power-law distribution with a mass index $s$ between the minimum and maximum masses $m_l$ and $m_u$. The grains have a fixed density at $\rho_{\text{grain}} = $\SI{3500}{\kilo \gram \per \cubic \metre}, implying that no meteoroid can have a bulk density larger than this grain density, as the porosity cannot be higher than 100\%. We wanted to ensure the meteoroids from our highest density shower (the Geminids) were well within this upper limit to not bias results. We note that the grain density cannot be measured from the model as identical fits can be obtained with different values of grain density \cite{Borovicka_2007}. A grain density of \SI{3000}{\kilo \gram \per \cubic \metre} was assumed in \citep{Borovicka_2007} when modelling Draconid meteors, however we set this value to \SI{3500}{\kilo \gram \per \cubic \metre}, appropriate for refractory silicate grains and as this is closer to the grain density measured in most meteorites \citep{Flynn2017}. In this model, we also implicitly assume that the porosity of the meteoroid is simply the ratio of the bulk and grain densities. 

The outputs of the model are the dynamics (velocity) and light curve over time and height. The model provides two reference points for the dynamics: the point of maximum brightness on the trail and the position of the leading fragment at each time step \citep{Borovicka_2007}. A reference point is chosen depending on the meteor's morphology and the way observations were reduced (e.g. if the leading fragment was resolved and picked in the video, then the leading fragment position provided by the model is used for fitting). The light curve at each time step is computed by summing the total luminous intensity (Eq. \ref{eq:luminosity}) of all ablating fragments from which the magnitude is computed as:

\begin{equation}
    M = -2.5 \log_{10} \frac{\sum I}{P_{0M}} \,,
\end{equation}

\noindent where $P_{0M}$ is the power of a zero-magnitude meteor. A value of $P_{0M} = \SI{840}{\watt}$ was used, as appropriate for meteors in CAMO's spectral bandpass \citep{weryk2013simultaneous}.

The erosion model provides improved fits for both a meteor's measured deceleration and lightcurve when compared to previous models which instead use sudden disruption of the simulated meteoroid \citep{Borovicka_2007}. For a more detailed explanation of all equations and parameter definitions used in the erosion model, the reader is referred to \citet{Popova_2019} and \citet{vida2023direct}. 

For our modelling, a number of parameters are defined from the observed trajectory and are well known, while others are assumed. The remaining variables are the parameters we are trying to estimate based on the model fit to observations. Table \ref{tab:modelparams} summarizes all model quantities and their adopted values where appropriate. While the uniqueness of model fits is not necessarily implied, it was found that constraining the model with meteor photometry, dynamics and wake almost always produced physical parameters limited to a narrow range.

\begin{table}
\begin{tabular}{lllll}
\textbf{Parameter} & \textbf{Description} & \textbf{Type} & \textbf{Value/Range} & \textbf{Units} \\
\hline
A & Shape Factor & Constant & 1.21 & - \\
$\Gamma$ & Drag Coefficient & Constant & 1 & - \\
$v_0$ & Initial Velocity & Constant & Measured per Event & \SI{}{\kilo \metre \per \second} \\
$Z_{\text{c}}$ & Zenith Angle & Constant & Measured per Event & deg \\
$\tau$ & Luminous Efficiency & Model & \citet{vida2023orionids} & - \\
PSF $\sigma$ & Point Spread Function & Constant & 3 & m \\
$\rho_{\text{grain}}$ & Grain Density & Constant & 3500 & \SI{}{\kilo \gram \per \metre \cubed} \\
$\rho$ & Bulk Density & Tuned & $\leq$ 3500 & \SI{}{\kilo \gram \per \metre \cubed} \\
$m_0$ & Initial Mass & Tuned & - & kg \\
$\sigma$ & Ablation Coefficient & Tuned & - & \SI{}{\kilogram \per \mega \joule} \\
$h_e$ & Erosion Start Height & Tuned & - & km \\
$\eta$ & Erosion Coefficient & Tuned & - & \SI{}{\kilogram \per \mega \joule} \\
$m_u$ & Max. Grain Mass & Tuned & - & kg \\
$m_l$ & Min. Grain Mass & Tuned & - & kg \\
$s$ & Grain Mass Index & Tuned & 1.3-2.8 & - \\
\end{tabular}

\caption{Parameters used in our ablation model. The constants were not changed during the modelling process as these were either measured with high precision or assumed. The tuned parameters were adjusted in order to fit the simulated meteor to the observed data points. The luminous efficiency $\tau$ was sampled from the empirical model of \citet{vida2023orionids}.}
\label{tab:modelparams}

\end{table}

To reconstruct the wake we assumed a Gaussian point spread function (PSF). For most events, we used an instrument-measured value of the standard deviation of the PSF of \SI{3}{\metre}. However, we discovered that during the week in 2021 when the 5 Eta-Aquariid events were recorded, the narrow field camera was slightly out of focus. As a result, a PSF of \SI{10}{\metre} was used when modelling only these events as this value was required in order to properly fit the observed wakes and have them match at both sites.

Finally, while initial velocity is measured for each event based on the initial trajectory solution, in practice we do slightly adjust this value in order to fit the observed lag. We define the lag as the distance the meteoroid falls behind an object of constant velocity equal to the meteoroid's initial velocity \citep{subasinghe2017luminous}, effectively only showing the deceleration of the meteoroid. The velocity usually must be increased slightly as the narrow-field camera tracking is delayed and the meteor starts to decelerate by the time the camera locks on. An improved orbit is computed once an initial velocity is fit in the model.

\subsection{Model Fitting Process}

\begin{figure}[H]
    \centering
    \includegraphics[width=\linewidth]{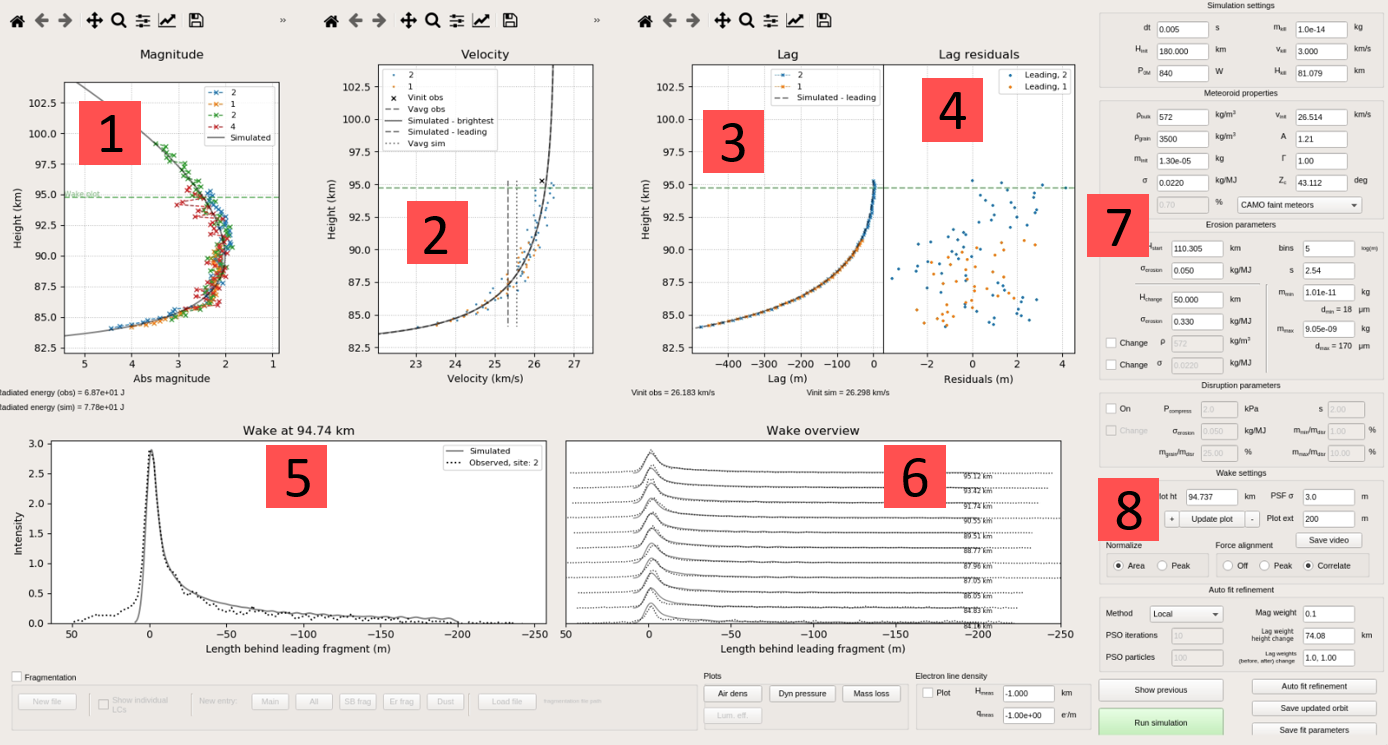}
    \caption{An example event showing the usage of the ablation model GUI in comparing observations (colored points) to the simulated meteor fit (solid black line). Individual numbered subplots represent:
        1) Lightcurve plot
        2) Velocity plot 
        3) Lag (Deceleration) plot 
        4) Lag residuals
        5) Wake plot at a specific altitude
        6) Wake overview, providing a broad summary of the wake at all heights
        7) Meteoroid and erosion parameters
        8) Wake settings, allowing for height navigation of the wake plot
        }
    \label{fig:modelfit}
\end{figure}

The model fitting process was performed manually and iteratively, using a graphical user interface (GUI) that displays plots of the observed lightcurve, deceleration and wake overlaid with simulated meteor values. The GUI allows for a more intuitive user experience during the fitting process as all plots can be visualized at once and in real time. The GUI also contains adjustable meteoroid and erosion parameters for the simulated meteoroid and is shown in Figure \ref{fig:modelfit}. The full details of the fitting procedure are given in \citet{vida2023orionids}. Manual fitting proceeds in a trial-and-error fashion until a set of parameters is reached that simultaneously fits all of the observed data, as best as the analyst feels is possible, across all observed heights. At this point, these physical parameters are adopted as being representative of the observed meteoroid within the model assumptions.

As a check, multiple human analysts performed independent fits. In general, most of our modelled events were fit by at least three analysts and in some cases as many as five.

\begin{figure}[H]
\centering
    \includegraphics[width=\linewidth,height=6cm]{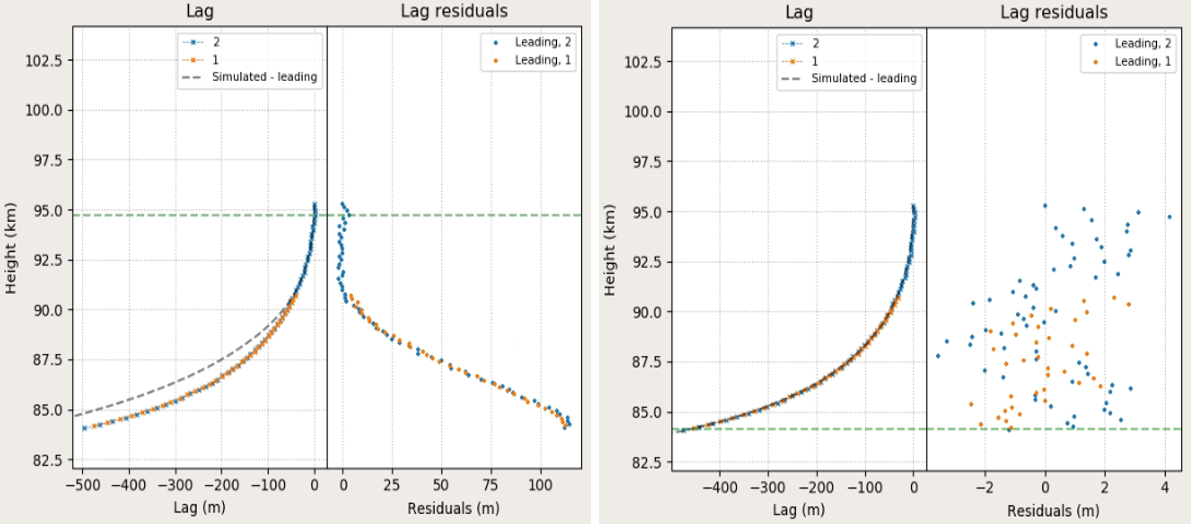}
    \caption{Left: An example of a poor lag fit. The model fit begins to deviate from observations below \SI{90}{\kilo \metre} height, with the model meteoroid decelerating much faster than the observed meteoroid. The lag residuals are large, increasing up to of order \SI{100}{\metre} and show a systematic trend.
    Right: An example of a good lag fit. Here deceleration is modelled well, agreeing with the observed meteor for the entire event. Residuals are small, on the order of \SIrange{2}{3}{\metre} and show random scatter about the vertical (0) line.}
    \label{fig:lagfits}
\end{figure}

As the initial fits are done subjectively by analysts, the question remains, what exactly constitutes an ``accurate'' fit? For the lag, the goodness of fit is usually easy to define as the residuals are shown in the GUI itself (label 4 in Figure \ref{fig:modelfit}). From experience, we found that the upper limit for acceptable lag residuals is around \SI{20}{\metre} for any individual data point, indicating that the along-the-track residuals are on that order while position picks transverse to the trajectory are usually below a meter. The lag for an example event is shown in Figure \ref{fig:lagfits}, with both a good and bad fit shown for comparison.

\begin{figure}[H]
\centering
    \includegraphics[width=\linewidth, height=7.5cm]{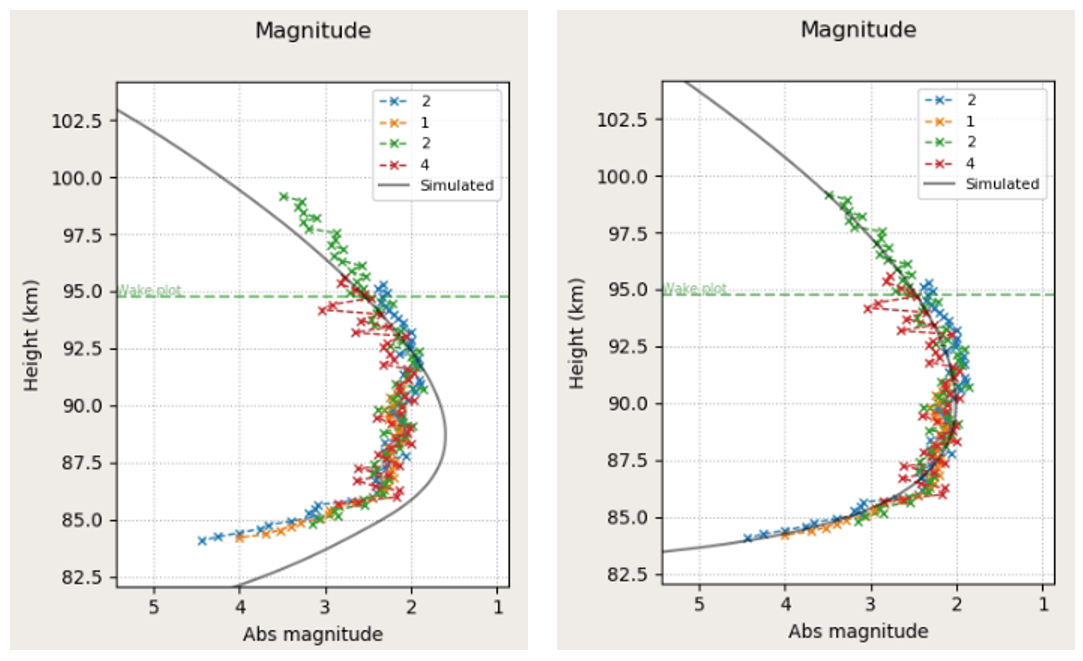}
    \caption{Left: An example of a poor lightcurve fit. The magnitude is underestimated at the start of the event and overestimated for the peak magnitude until the end of the event. The lightcurve shape is clearly inaccurate.
    Right: An example of a good lightcurve fit. The magnitude is modelled well throughout the entire event and the lightcurve shape agrees within the spread of observed data.}
    \label{fig:LCfits}
\end{figure}

The corresponding lightcurve for this example event is shown in Figure \ref{fig:LCfits}. The lightcurve contains data from four separate instruments, so photometric data from each station did not always perfectly agree at all heights due to differences in local observing conditions, thus showing the magnitude residuals is less useful. The variance between lightcurves from the four cameras was normally less than 0.3 magnitudes. Therefore, the simulated meteor was fit to the lightcurve(s) with the best quality data, where observing conditions at each station were taken into account by the analyst to decide which lightcurve was the most physical. Analysts preferred stations with clear skies that also accurately tracked the meteor for the entire duration of the event.

Deciding on the goodness of fit for the wake is more challenging as both wake intensity and shape are simulated via the release of hundreds of grains in many mass bins for all heights. The wake is computed as the integrated brightness of all grains (with the PSF applied) perpendicular to the trajectory in predefined steps, resulting in a curve of brightness as a function of distance behind the meteoroid. The uncalibrated observed wake is then scaled to the model wake by normalizing the total area under the model wake curve to that of the observed curve. 

\begin{figure}
\centering
    \includegraphics[width=\linewidth]{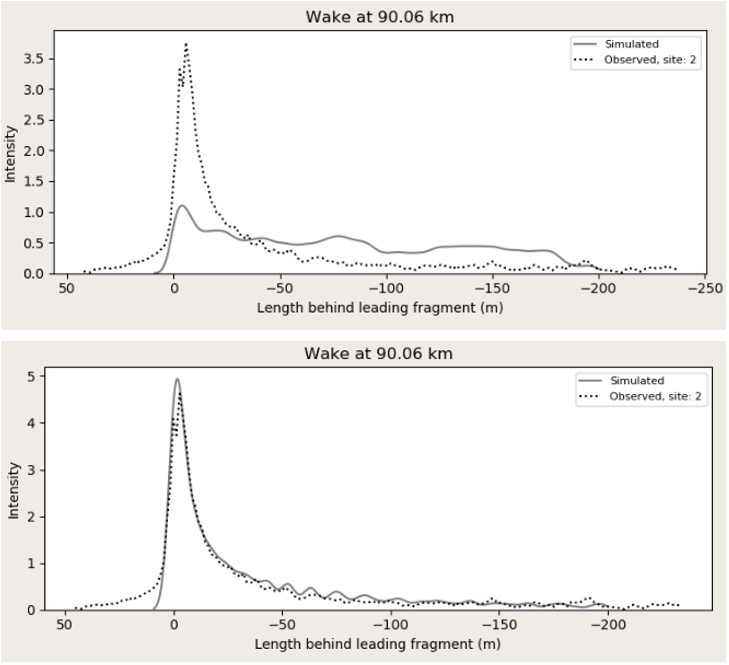}
    \caption{Top: An example of a poor wake fit (at \SI{90.06}{\kilo \metre} height). The modelled peak intensity of the meteoroid is greatly underestimated and the intensity of the wake behind the meteoroid is overestimated. The overall wake shape does not agree with the observed meteor.
    Bottom: An example of a good wake fit. Here the modelled peak intensity of the meteoroid matches observations and the intensity of the wake is accurate at all distances behind the meteoroid.}
    \label{fig:wakefits}
\end{figure}

Attempts to quantify the quality of the wake fit were found to be too computationally intensive as computing the wake adds significant computational overhead. Thus, each fit was first only performed on the light curve and the lag, and after an initial fit was obtained, the wake was consulted to improve the fit. Consequently, the goodness of fit of the wake was more subjective than for the lightcurve or the dynamics. The wake at a single height for an example event is shown in Figure \ref{fig:wakefits}, with both a good and bad fit shown for comparison.

As the minimum/maximum grain sizes and the gain mass distribution have a large impact on the wake but a much smaller impact on the lightcurve and lag, our approach was to first get a rough estimate for the key meteoroid parameters (e.g. bulk density, initial mass, ablation coefficient) using the lightcurve and lag as guides, before introducing changes to the grain properties and fitting the wake. Preference was given to fits which agreed with the shape/distribution of the entire observed wake from both sites, as opposed to only agreeing with the intensity of the light coming from near the head of the meteor, as the grains were the main parameters of focus in this final stage of fitting.

While model fits were always preferred if they matched observations well at all heights, for some events this was not possible. For these more difficult events, preference was given to fits that matched the observed data earlier in the event, but did not match so well near the end of the event. This is because we are mainly interested in the bulk of the initial meteoroid, which by the end of an event, has usually been ablated, eroded and/or fragmented leaving an unrepresentative refractory portion of the actual meteoroid remaining. Such partial fits also reflect the relative simplicity of our model in comparison to the complexity of some meteoroid fragmentation. We postulate that in those cases where the wake could only be fit at one portion of the trajectory the grain distribution changed over time, which was not modelled.

The one added complexity we allowed in the model was to permit an analyst to choose one change in the erosion coefficient, ablation coefficient and/or bulk density at one certain height during the meteor flight in order to improve the model fit. This decision was informed by the observed changes in the lightcurve of some meteors (e.g. those with double-peaked lightcurves). However, as this introduces more complication into the model, fits were preferred which contained as few changes as possible; we only resorted to such a change when absolutely necessary. If two fits were similar, the first having a change in bulk density and erosion coefficient, while the second only had an erosion change, the second fit was preferred. 

For each event, the best manual fit among those completed independently by human analysts was chosen as a starting point for an automated refinement step described in detail in \citet{vida2023orionids}. This algorithm concentrates on the local space surrounding the hyper dimensional point defined by the analyst-determined parameters and minimizes a cost function determined by the light curve and lag. In most cases this refinement step produced only a modest change from the initial parameters; and was adopted only if the resulting overall fit, and especially the wake, was fit as good or better than the initial solution.

\section{Results and Discussion}
\label{sec:resultsdiscussion}

Our final dataset is comprised of 41 modelled meteors, three of which are presented as example fits in Figure \ref{fig:examplefits}. All meteoroid orbits are given in Table \ref{tab:orbits}, and their physical properties and erosion parameters are given in \ref{ap:metmeasurements}. We summarize all of our average parameter results found for each shower in Table \ref{tab:resultsummary}. Key findings are then shown in the graphs that follow, where solid black error bars represent the estimated uncertainty that we have assigned, namely the standard error of the mean. This is meant to give a sense of the dispersion about the mean for each shower. Graphs showing the comparison between observations and the model for each event, including the wake across a range of heights, are given in Supplementary Materials.

\begin{figure}[H]
    \centering
    \includegraphics[width=\linewidth]{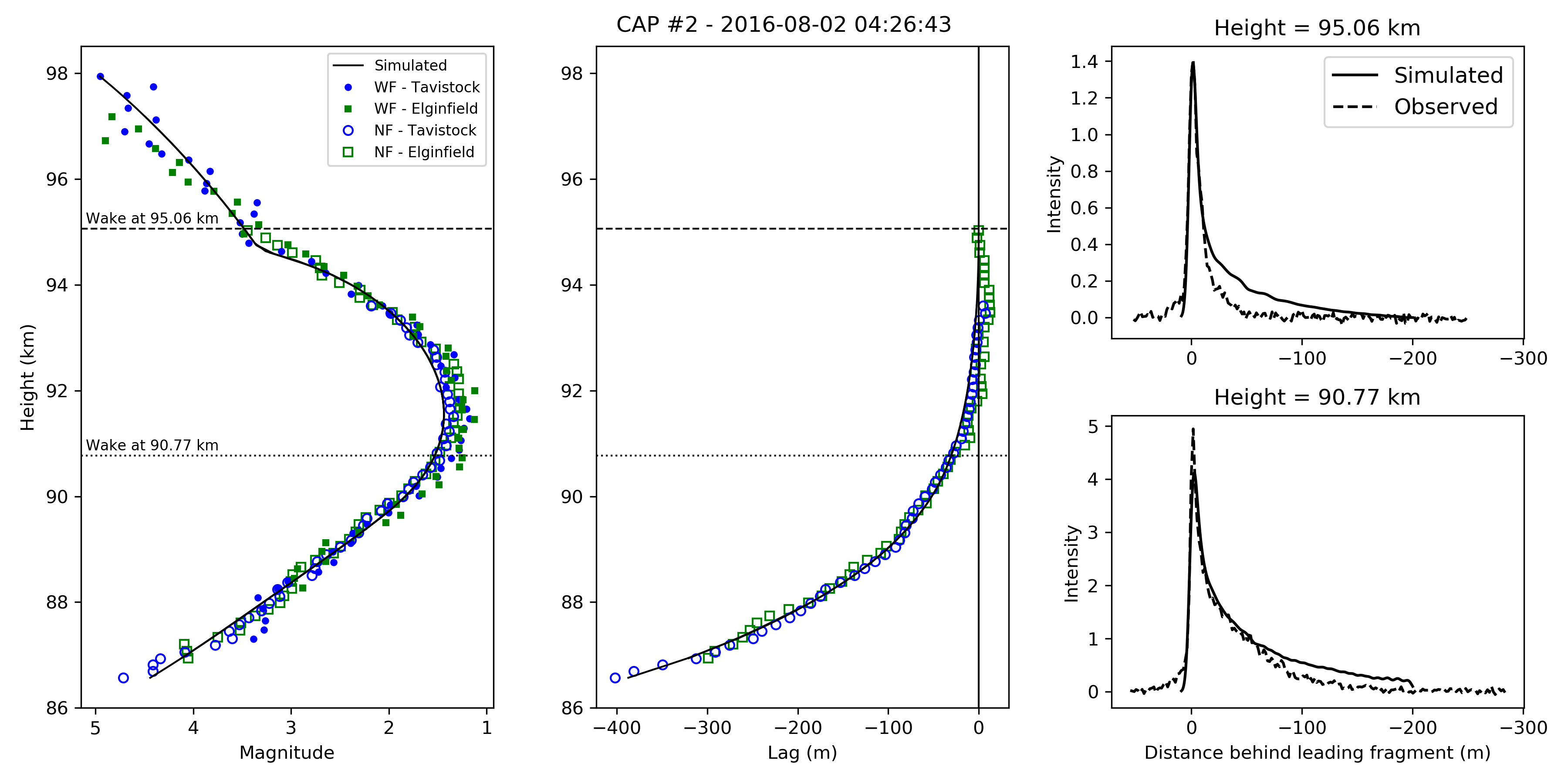}
\end{figure}
\begin{figure}[H]
    \centering
    \includegraphics[width=\linewidth]{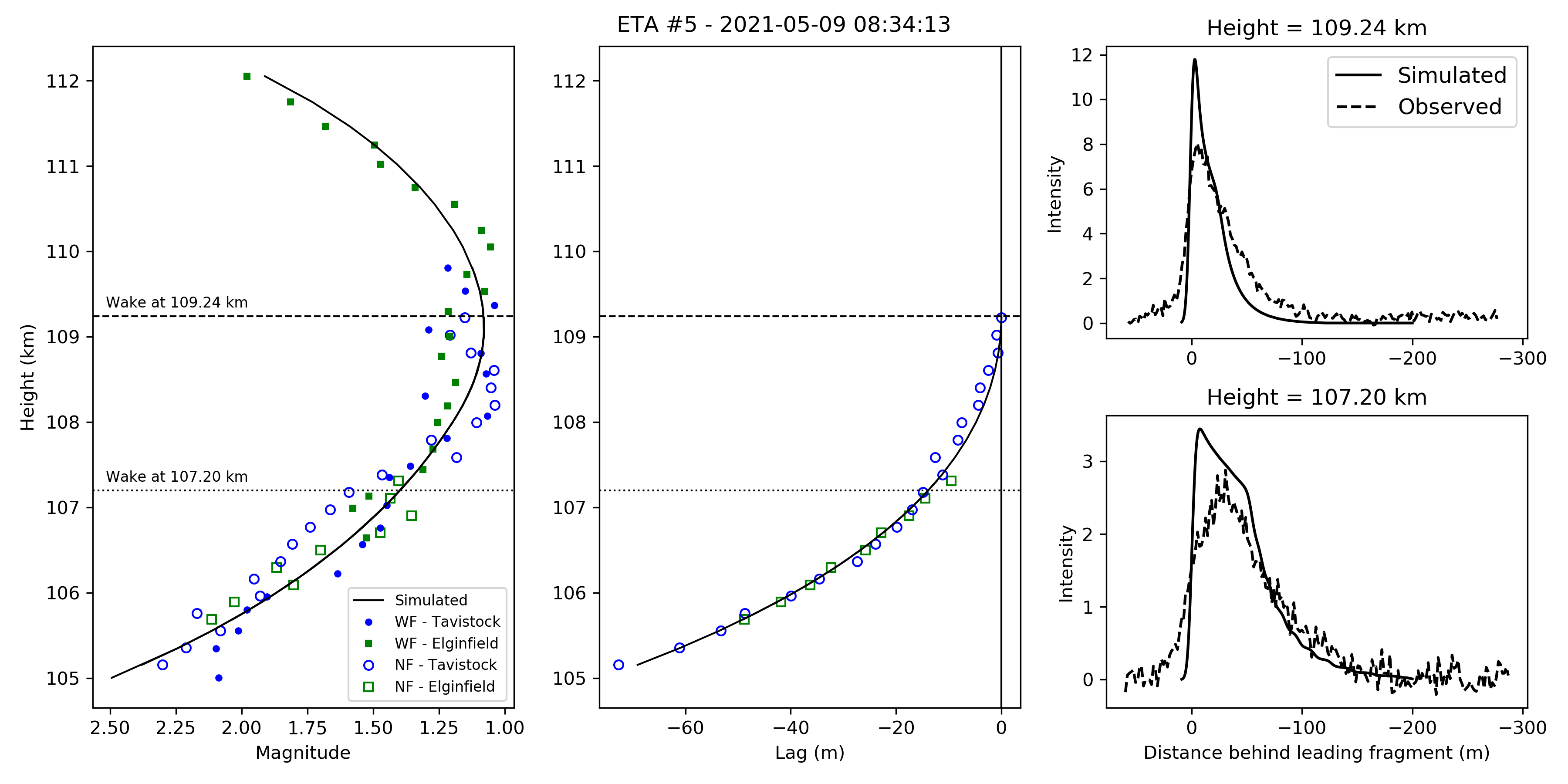}
\caption{An Alpha-Capricornid (top), Eta-Aquariid (middle) and Geminid (bottom) are shown as examples of the final model fits. Lightcurves, lags and two snapshots of the wake are shown for each event. The solid black line represents the simulated meteor while the colour data points are measured values from the observed meteor.}
\label{fig:examplefits}
\end{figure}

\begin{figure}[H]
    \centering
    \includegraphics[width=\linewidth]{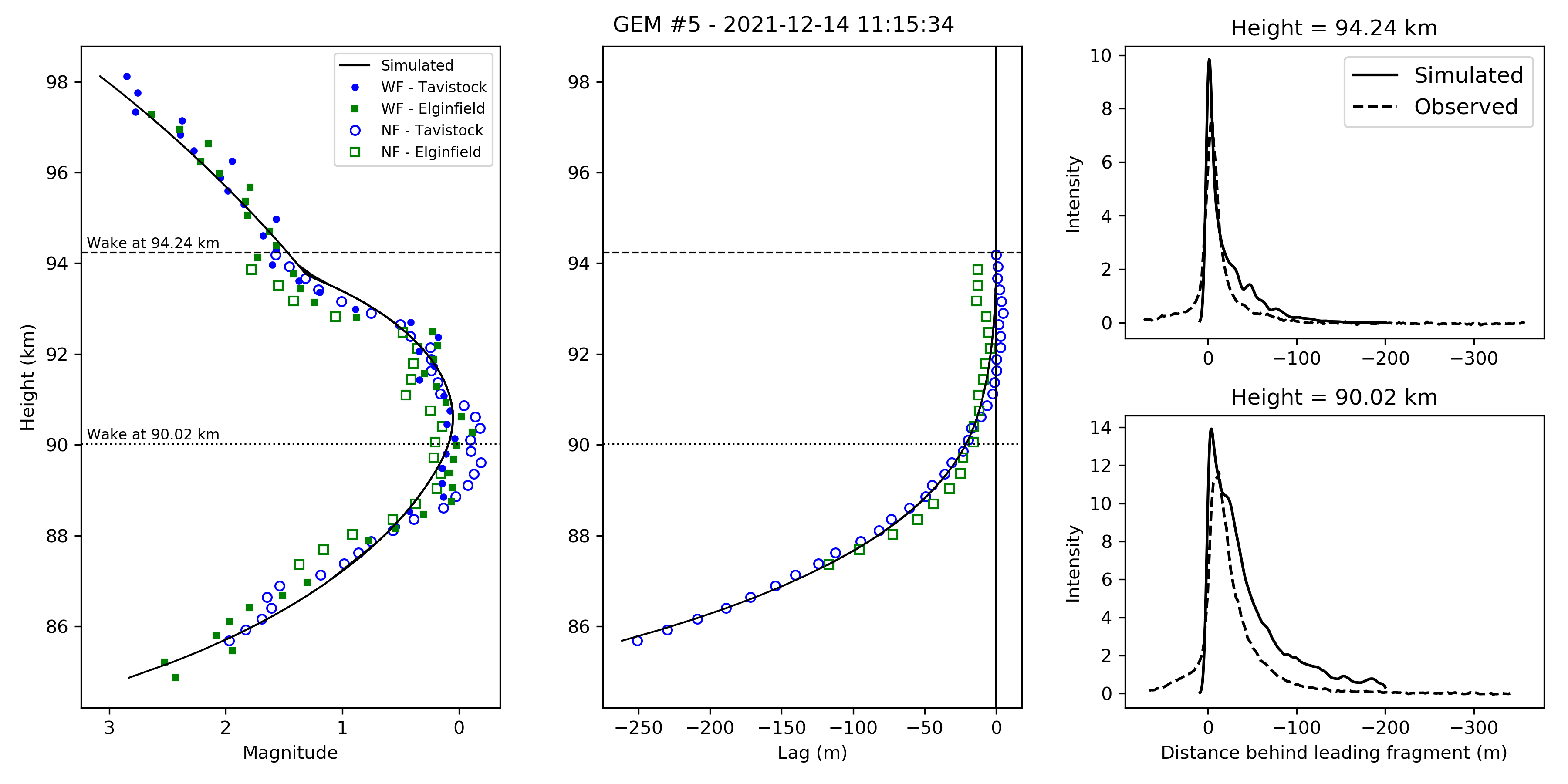}
    \caption{Continued.}
\end{figure}

\begin{landscape}

\begin{table}[h]
\centering

\begin{tabular}{|l|l|l|l|l|l|l|l|l|l|l|l|l|l|}
\hline
Shower & $\rho$ & $m_0$ & $V_0$ & $\sigma$ & $h_e$ & $\eta$ & $h_{e2}$ & $\eta_2$ & $s$ & $m_l$ & $m_u$ & $E_S$ & $E_V$ \\
 & (\SI{}{\kilo \gram \per \metre \cubed}) & (kg) &  (\SI{}{\kilo \metre \per \second}) & (\SI{}{\kilogram \per \mega \joule}) & (km) & (\SI{}{\kilogram \per \mega \joule}) & (km) & (\SI{}{\kilogram \per \mega \joule}) & & (kg) & (kg) &  \SI{}{\mega \joule \per \metre \squared} & \SI{}{\mega \joule \per \kilo \gram} \\
\hline
CAP & 830 & 9.33E-06 & 24.755 & 0.017 & 99.505 & 0.197 & 93.147 & 0.232 & 2.46 & 7.53E-11 & 3.05E-08 & 1.50 & 1.09 \\
 & $\pm$187 & $\pm$2.69E-06 & $\pm$0.376 & $\pm$0.004 & $\pm$0.313 & $\pm$0.047 & $\pm$1.553 & $\pm$0.168 & $\pm$0.12 & $\pm$2.27E-11 & $\pm$3.01E-08 & $\pm$0.21 & $\pm$0.26 \\
 \hline
TAH & 260 & 7.20E-05 & 16.355 & 0.020 & 92.463 & 0.091 & 88.150 & 0.410 & 2.39 & 3.47E-11 & 5.37E-10 & 1.60 & 1.22 \\
 & $\pm$33 & $\pm$3.20E-05 & $\pm$0.068 & $\pm$0.004 & $\pm$1.581 & $\pm$0.033 & $\pm$0.350 & $\pm$0.290 & $\pm$0.32 & $\pm$2.26E-11 & $\pm$3.84E-10 & $\pm$0.46 & $\pm$0.44 \\
 \hline
ETA & 305 & 1.20E-05 & 66.740 & 0.025 & 114.738 & 0.326 & 107.151 & 0.244 & 1.69 & 6.49E-11 & 4.68E-10 & 2.78 & 3.56 \\
 & $\pm$55 & $\pm$1.75E-06 & $\pm$0.275 & $\pm$0.001 & $\pm$0.617 & $\pm$0.081 & - & - & $\pm$0.27 & $\pm$2.63E-11 & $\pm$8.57E-11 & $\pm$0.16 & $\pm$0.48 \\
 \hline
LEO & 379 & 4.76E-06 & 72.106 & 0.029 & 115.739 & 1.035 & - & - & 2.21 & 8.89E-12 & 9.00E-09 & 1.00 & 1.37 \\
 & - & - & - & - & - & - & - & - & - & - & - & - & - \\
 \hline
MON & 668 & 1.07E-05 & 42.941 & 0.019 & 113.460 & 0.039 & 90.162 & 0.006 & 2.01 & 1.03E-11 & 4.74E-09 & 0.47 & 0.34 \\
 & - & - & - & - & - & - & - & - & - & - & - & - & - \\
 \hline
ORI & 282 & 1.18E-05 & 67.671 & 0.027 & 116.151 & 0.229 & 108.081 & 0.362 & 2.14 & 8.73E-12 & 1.70E-08 & 1.03 & 1.48 \\
 & $\pm$52 & $\pm$5.26E-06 & $\pm$0.509 & $\pm$0.003 & $\pm$1.141 & $\pm$0.035 & $\pm$2.719 & $\pm$0.205 & $\pm$0.12 & $\pm$1.24E-12 & $\pm$1.10E-08 & $\pm$0.15 & $\pm$0.34 \\
 \hline
PER & 365 & 5.55E-06 & 59.875 & 0.027 & 111.205 & 0.340 & 104.469 & 0.079 & 2.16 & 5.24E-11 & 5.88E-08 & 1.33 & 2.22 \\
 & $\pm$134 & $\pm$8.58E-07 & $\pm$0.622 & $\pm$0.001 & $\pm$0.512 & $\pm$0.070 & $\pm$0.334 & $\pm$0.043 & $\pm$0.21 & $\pm$4.39E-11 & $\pm$4.10E-08 & $\pm$0.14 & $\pm$0.39 \\
 \hline
AUR & 412 & 1.07E-05 & 65.490 & 0.021 & 111.757 & 0.144 & - & - & 2.35 & 2.87E-12 & 3.31E-07 & 1.51 & 1.50 \\
 & - & - & - & - & - & - & - & - & - & - & - & - & - \\
 \hline
ELY & 860 & 2.30E-06 & 46.229 & 0.028 & 107.016 & 0.385 & - & - & 2.38 & 9.93E-12 & 7.02E-08 & 1.52 & 1.55 \\
 & $\pm$30 & $\pm$1.75E-07 & $\pm$0.065 & $\pm$0.002 & $\pm$0.486 & $\pm$0.045 & - & - & $\pm$0.02 & $\pm$6.50E-14 & $\pm$2.98E-08 & $\pm$0.32 & $\pm$0.33 \\
 \hline
JPE & 284 & 1.21E-05 & 65.684 & 0.021 & 114.102 & 0.159 & 99.690 & 0.010 & 2.28 & 9.49E-12 & 2.13E-09 & 1.12 & 1.37 \\
 & - & - & - & - & - & - & - & - & - & - & - & - & - \\
 \hline
LYR & 438 & 1.02E-05 & 48.547 & 0.035 & 110.716 & 0.146 & 100.091 & 0.015 & 2.48 & 1.02E-10 & 5.08E-09 & 1.03 & 1.41 \\
 & $\pm$36 & $\pm$7.73E-06 & $\pm$0.045 & $\pm$0.024 & $\pm$4.699 & $\pm$0.052 & - & - & $\pm$0.06 & $\pm$1.00E-10 & $\pm$5.02E-09 & $\pm$0.38 & $\pm$0.91\\
 \hline
TAU & 647 & 8.72E-06 & 29.773 & 0.027 & 104.797 & 0.062 & 90.490 & 0.223 & 2.11 & 2.39E-11 & 7.54E-08 & 1.00 & 0.87 \\
 & $\pm$58 & $\pm$2.24E-06 & $\pm$0.625 & $\pm$0.002 & $\pm$1.141 & $\pm$0.002 & $\pm$2.131 & $\pm$0.066 & $\pm$0.13 & $\pm$1.15E-11 & $\pm$2.30E-08 & $\pm$0.15 & $\pm$0.13 \\
 \hline
GEM & 1387 & 6.82E-06 & 35.971 & 0.028 & 101.372 & 0.171 & 93.452 & 0.437 & 1.85 & 3.57E-10 & 1.72E-07 & 2.08 & 1.38 \\
 & $\pm$240 & $\pm$3.35E-06 & $\pm$0.172 & $\pm$0.002 & $\pm$1.914 & $\pm$0.037 & $\pm$0.773 & $\pm$0.205 & $\pm$0.20 & $\pm$1.86E-10 & $\pm$9.39E-08 & $\pm$0.45 & $\pm$0.38 \\
 \hline
 \end{tabular}

\caption{Average model results for all analyzed showers. Average values are shown with the standard error of the mean as uncertainty. $\eta_2$ represents the second erosion coefficient that was found after an erosion change at height $h_{e2}$.  $E_V$ represents the energy per unit mass needed to begin erosion and $E_S$ is the energy per unit cross section needed to begin erosion. A dash - is shown for parameters where there was only one value or none, so the standard error of the mean could not be calculated.}
\label{tab:resultsummary}

\end{table}
\end{landscape}

\subsection{General characteristics of observed meteors}

\begin{figure}[H]
    \centering
      \includegraphics[width=\linewidth]{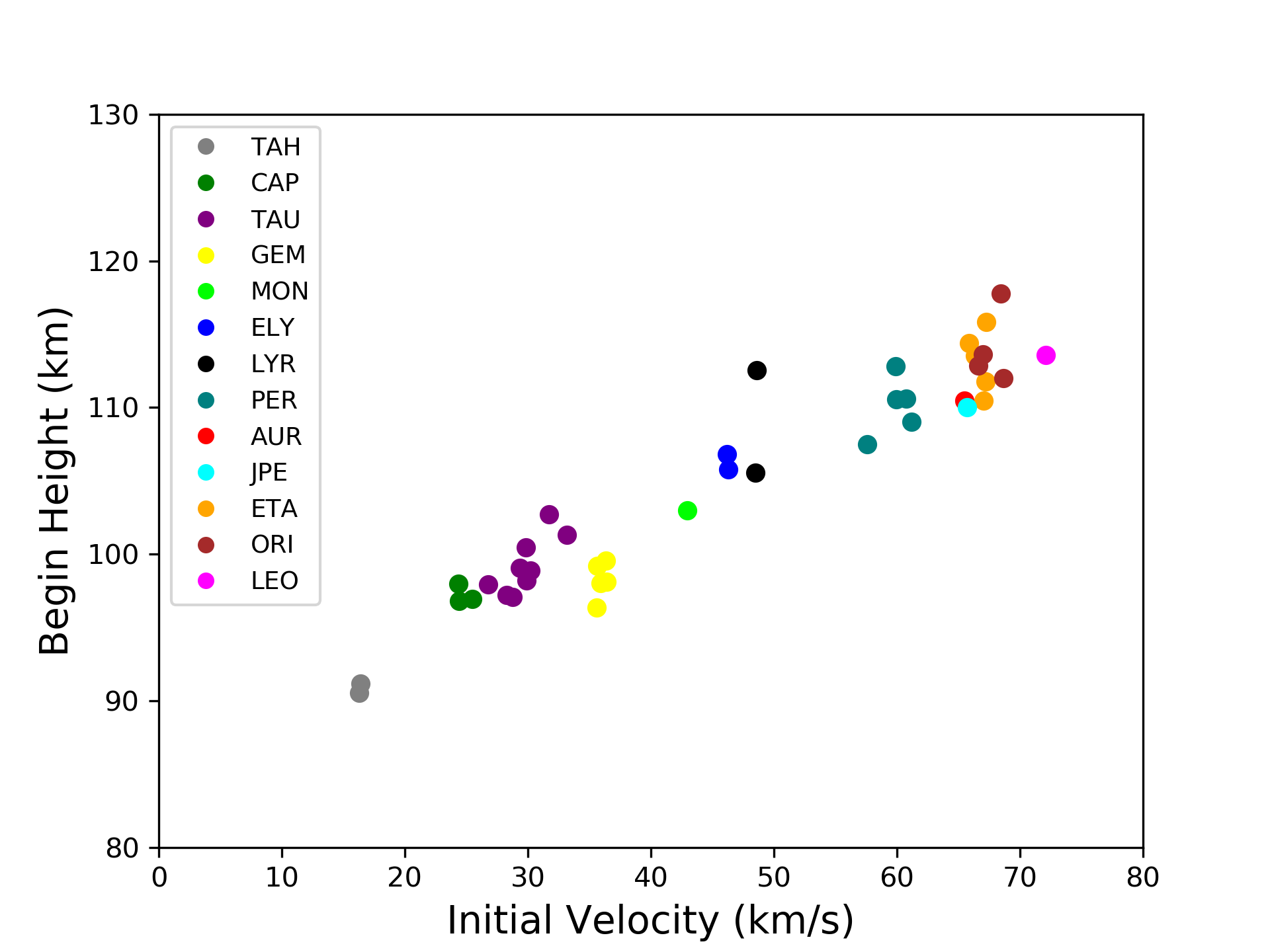}
       \caption{The observed begin height (km) as a function of initial velocity (\SI{}{\kilo \metre \per \second}). Events are colour coded by shower.}
       \label{fig:beginheight}
\end{figure}

Figure \ref{fig:beginheight} shows the expected trend of higher velocity events appearing at higher altitudes, a well-known result \citep{Vida_2018, Borovicka2019a}. Variations in begin height by shower may reflect material, structural or strength differences \citep{Koten2004}. Our studied showers cover a large range of regular meteor velocities and begin heights. For our shower meteors, the one notable deviation from the general pattern is for the Geminids which penetrate deeper, consistent with their stronger nature. 

The Tau-Herculids were our slowest ($\sim$16~\SI{}{\kilo \metre \per \second}) and lowest beginning shower ($\sim\SI{91}{\kilo \metre}$) followed by the Alpha-Capricornids, the Taurids and then the Geminids with increasing velocities and begin heights as expected. The single Leonid, the Orionids, the Eta Aquarids and the single July Pegasid were our fastest ($>\SI{65}{\kilo \metre \per \second}$) and highest beginning showers ($>\SI{109}{\kilo \metre}$). 

\begin{figure}[H]
    \centering
    \includegraphics[width=\linewidth]{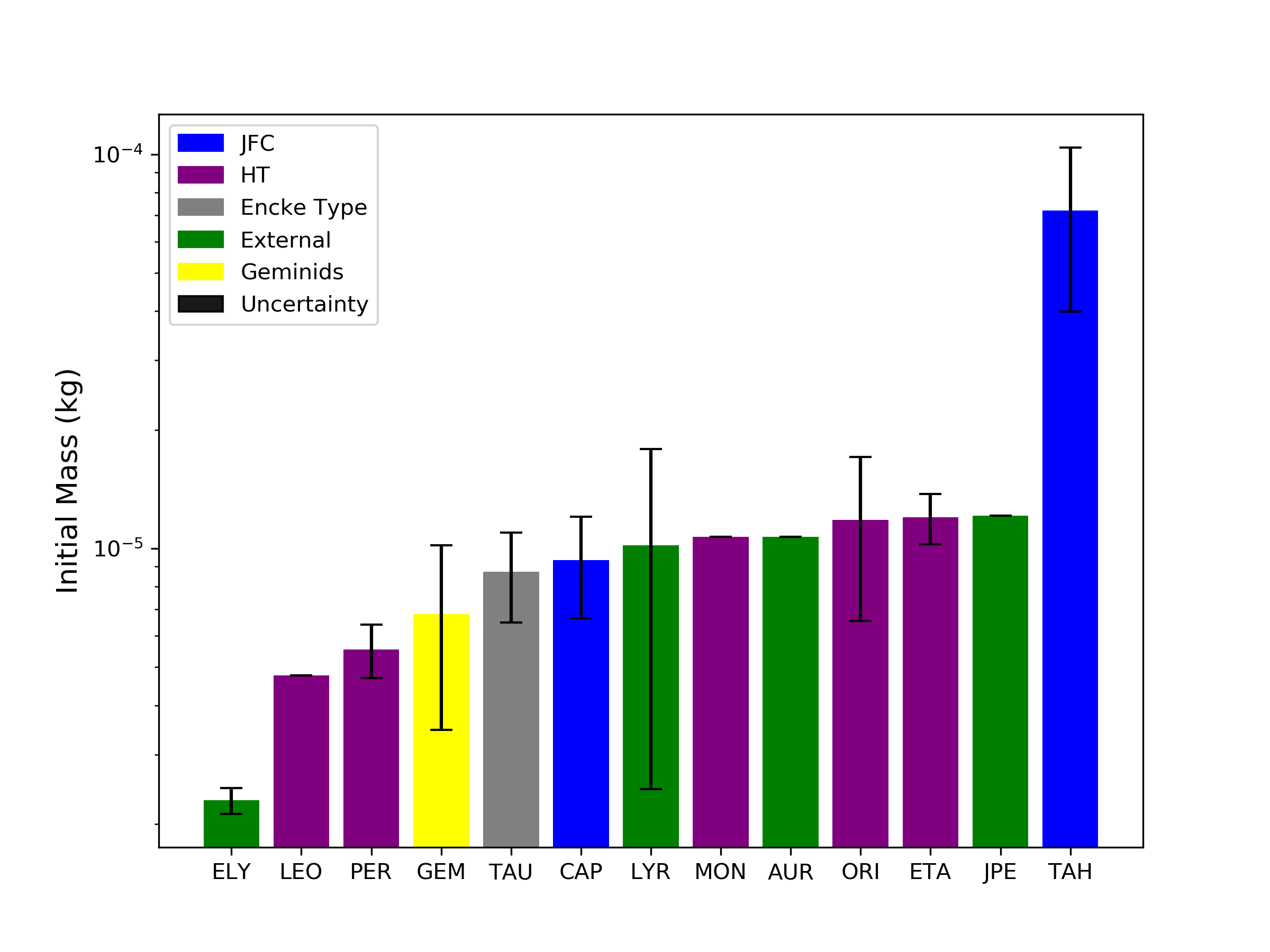}
    \caption{Average meteoroid initial mass (kg) for each analyzed shower. The showers are colour coded according to their parent comet orbital population following the classification of \citet{Levison1996}.}
\label{fig:metmass}
\end{figure}

To provide a sense of the meteoroid mass range covered in our reductions, Figure \ref{fig:metmass} presents the average meteoroid mass and standard error of the mean per shower. The Eta Lyrids were the lowest mass events, averaging $\sim\SI{2.3e-6}{\kilo \gram}$ while on the other end of the spectrum, the Tau-Herculids were the highest mass events (simply due to their low speed), averaging $\sim\SI{7e-5}{\kilo \gram}$. Most showers had fairly similar average masses in the range of 5 - 11 $\times$ 10$^{-6}$~kg. Our survey covers a little more than one order of magnitude in mass, with a representative average being near $\sim\SI{1e-5}{\kilo \gram}$. Meteoroids in this size range provide a unique area of study, as particles of this size are too large for dust detectors or infrared emission analysis, yet too small to be observed optically from Earth while they are still in space \citep{Campbell-Brown2004}.

\begin{figure}[H]
    \centering
    \includegraphics[width=\linewidth]{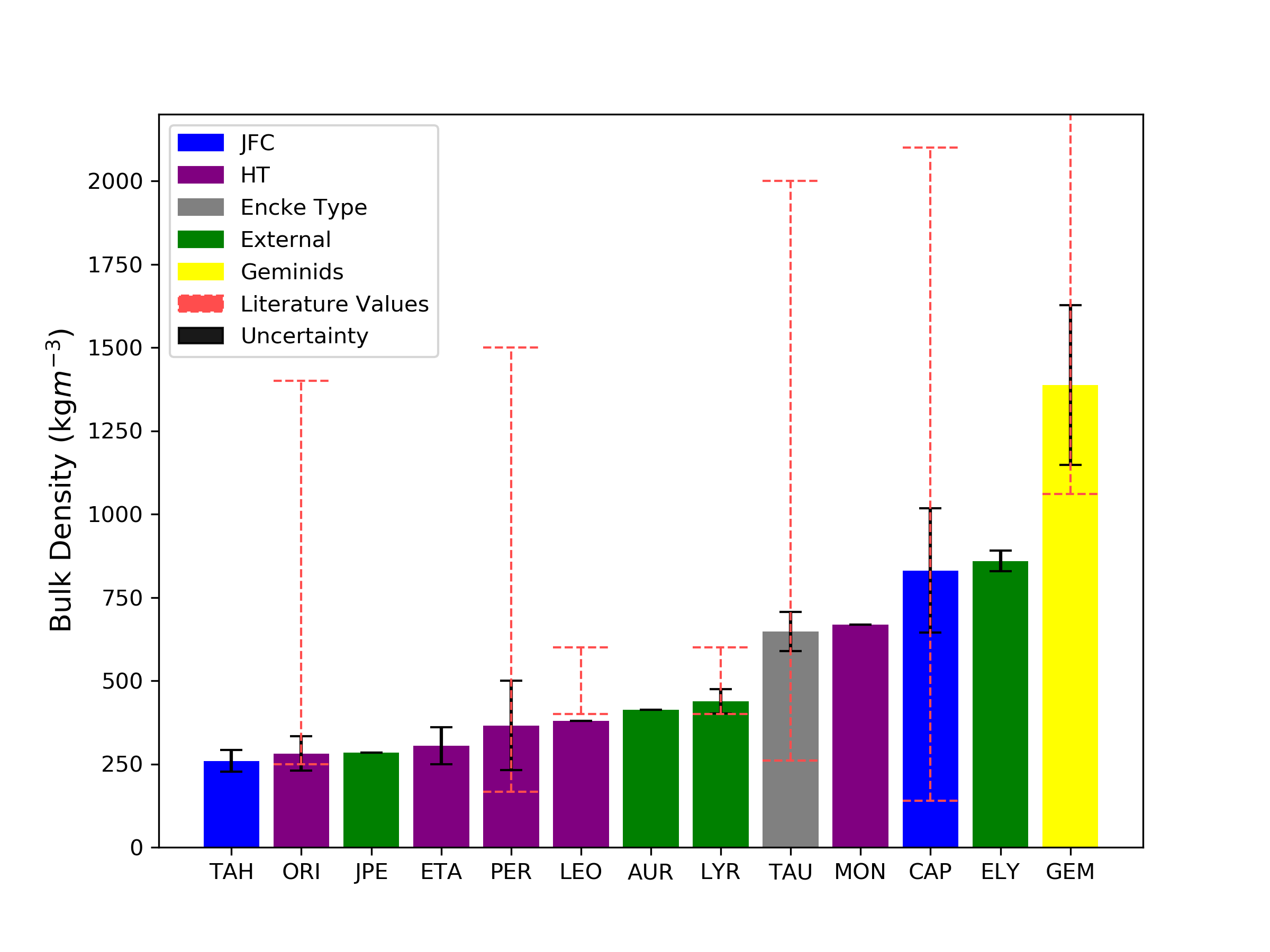}
    \caption{Average meteoroid bulk density (\SI{}{\kilo \gram \per \metre \cubed}) for each observed shower. The showers are colour coded according to their parent comet orbital population following the classification of \citet{Levison1996}. The red dashed bars represent the range of previously measured literature values - see Table \ref{tab:showerdensities}. (GEM maximum literature value is 3500~\SI{}{\kilo \gram \per \metre \cubed}).}
    \label{fig:metdens}
\end{figure}

The average meteoroid bulk density for each shower is shown in Figure \ref{fig:metdens}. These values are compared to previously measured shower bulk densities from the literature in Table \ref{tab:showerdensities}. 

Our typical values for meteoroids from cometary meteoroid streams show densities in the \SIrange{300}{1000}{\kilo \gram \per \metre \cubed} range, consistent with Rosetta measurements of bulk density as well as the bulk density of cometary nuclei in general. These densities are also consistent with group C and D material from \citet{Ceplecha_1988} meteor classes, as one would expect. Overall the Tau-Herculids were the lowest average density shower (\SI{259}{\kilo \gram \per \metre \cubed}) and the Geminids were the highest average density shower (\SI{1387}{\kilo \gram \per \metre \cubed}) and our only shower consistent with group B material from \citet{Ceplecha_1988} meteor classes. 

Of the showers that have previously published densities in the literature (the Orionids, Perseids, Leonids, Lyrids, Taurids, Alpha-Capricornids and Geminids) our densities are in agreement within uncertainty, though usually on the lower end of the literature range.

\subsection{Grain mass distribution}

\begin{figure}[H]
    \centering
    \includegraphics[width=\linewidth]{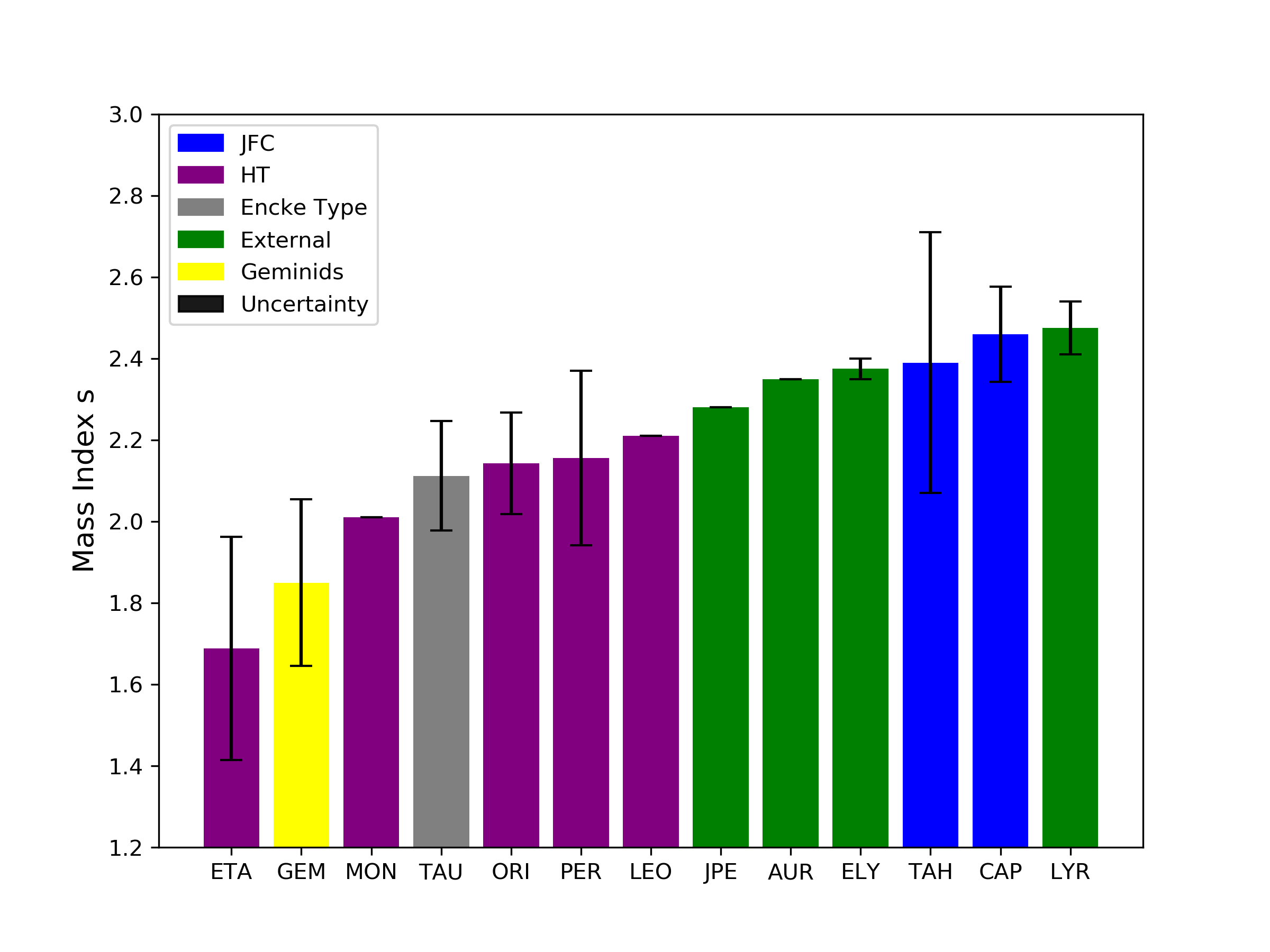}
    \caption{The average differential grain mass index $s$ for each analyzed shower. The showers are colour coded according to their parent comet orbital population following the classification of \citet{Levison1996}.}
\label{fig:massindex}
\end{figure}

One outcome of our model fits is that we also determine the most likely differential mass distribution of constituent grains. Figure \ref{fig:massindex} shows the average differential grain mass distribution index $s$ for each shower. The vast majority of our analyzed showers were dominated by small grains ($s$ > 2), the extreme being the Lyrids ($s$ = 2.48 $\pm$ 0.06), though the number statistics are small. This is the opposite trend to what is seen in looking at the larger individual meteoroid mass distribution within streams (see \ref{ap:massindex}) which are almost entirely large particle dominated with $s$ < 2.

We found Halley-type showers have on average more large mass grains compared to the Jupiter-family and external showers which tend to have more small mass grains. However, we caution that uncertainties are quite large for some showers due to outlier events with very high $s$ values. Furthermore, for some showers, only one event was modelled so the standard error of the mean was indeterminable. A larger sample size would be needed to confirm the veracity of this trend.

We found only two showers whose meteoroids are preferably composed of larger grains: the Eta-Aquariids ($s$ = 1.69~$\pm$0.27) and the Geminids ($s$ = 1.85~$\pm$0.20). Both of these shower's parent bodies, (3200) Phaethon and 1P/Halley, also have mass index values for larger meteoroids with an $s$ < 2. The other daughter shower of 1P/Halley, the Orionids, was also found in past literature to have $s$ < 2. However, the Orionids were found in this work to have an average $s$ = 2.14 $\pm$ 0.10. Given the large uncertainties and small number statistics, these differences may not be significant. 
 
Looking at in situ $s$ measurements for the mass distribution of smaller grains, we see mixed results when compared to the literature. The dust from 67P (a comet on a JFC orbit) seems to certainly be dominated by larger size grains ($s$ between 1.22 - 1.77 for grains between 0.04 and \SI{300}{\micro \metre}), depending on the instrument and the population of dust studied. As noted by \citet{Ellerbroek2019}, more than half of the actual dust that impacted any of the sensor instruments was actually lost on impact, but the general trend of larger grains seems robust. 

\citet{Vojacek_2017} found an $s$ similar to 67P for JFC meteoroid grain fits. However, our work finds the two JFC showers we looked at do not agree with the results from 67P, as these showers had two of the highest $s$ values of all examined showers. Specifically, we found the Tau-Herculids had an average $s$ = 2.39 $\pm$ 0.32 and the Alpha-Capricornids had an $s$ = 2.46 $\pm$ 0.12. As summarized in \ref{ap:massindex}, these showers' parent comets 73P/Schwassmann-Wachmann and 169P/NEAT have been found to produce meteoroids which at Earth have $s \lesssim$ 2 \citep{dubietis2004observational, egal2023modeling}. 

Overall, it seems clear the Geminids and likely the Eta-Aquariids are large grain dominated, while the Lyrids are probably small grain dominated. For the other showers as well as comet populations as a whole, the mass index $s$ may not be simple to constrain. The disagreement between our shower results and the literature shows that grain mass index and the differential mass index $s$, governing the distribution of shower meteoroids at masses orders of magnitudes larger, are likely distinct.

\subsection{Energy needed to trigger erosion}

\begin{figure}[H]
    \centering
    \includegraphics[width=\linewidth,height=8.4cm]{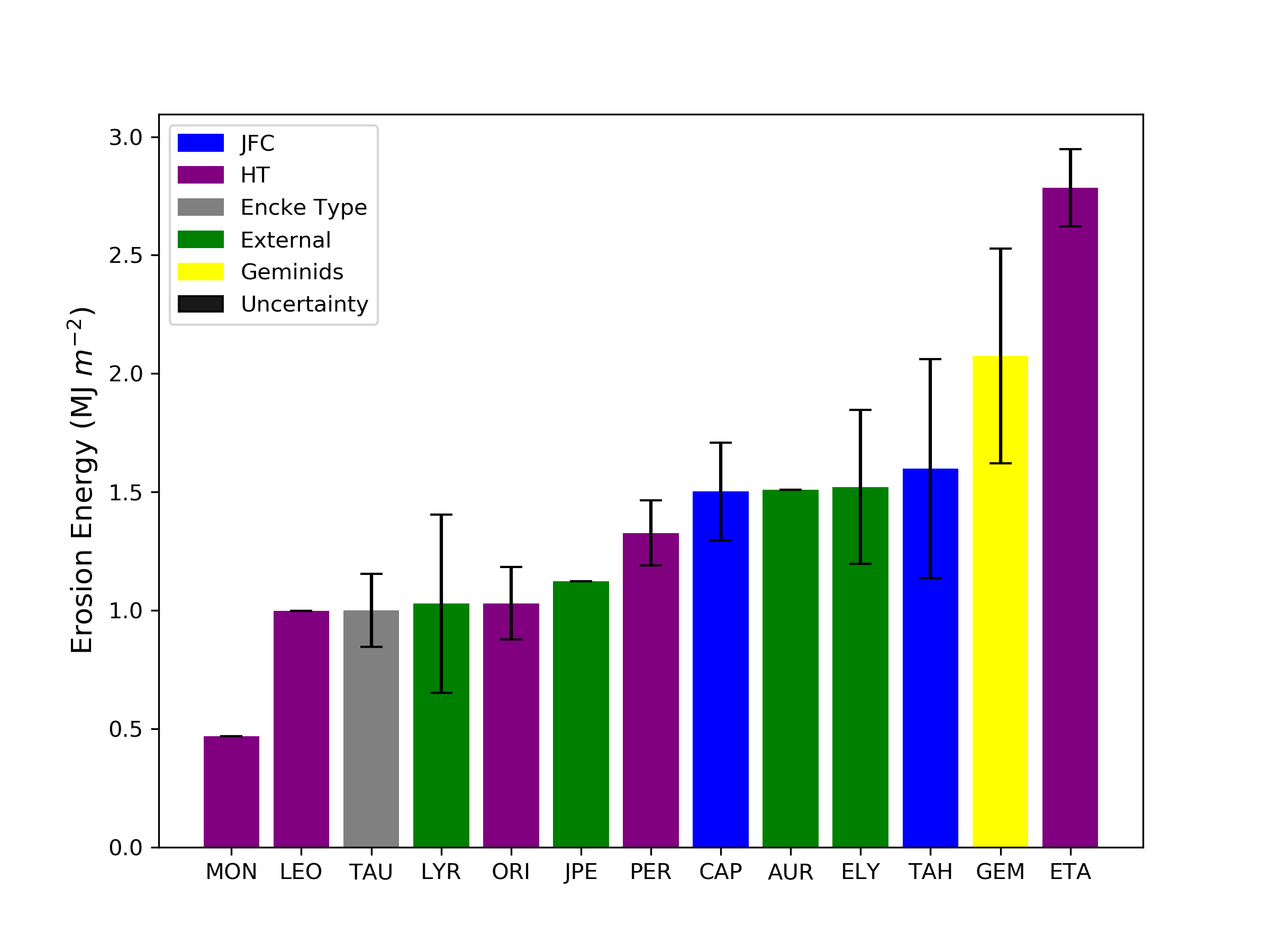}
\end{figure}

\begin{figure}[H]
    \centering
    \includegraphics[width=\linewidth,height=8.4cm]{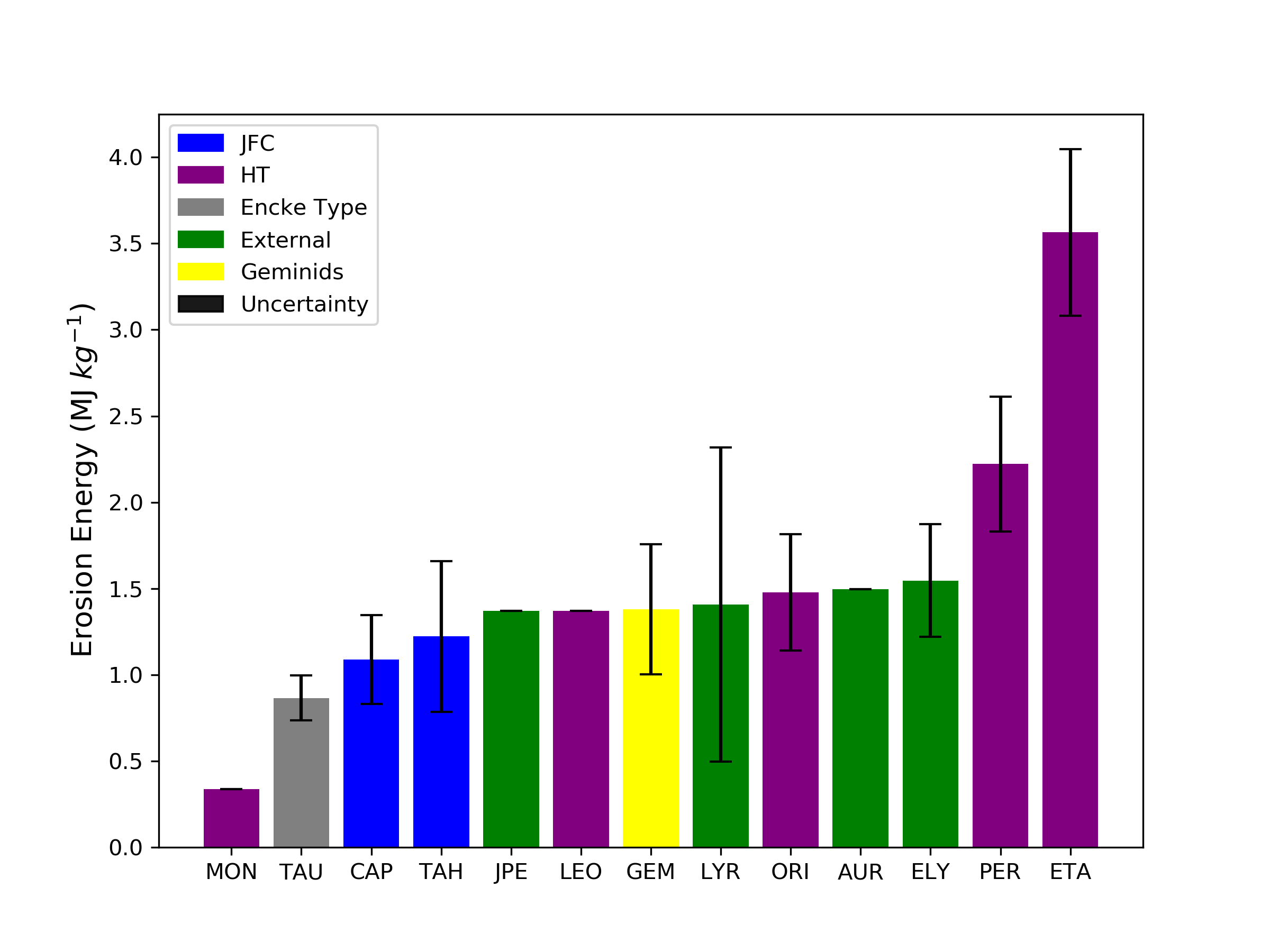}
    \caption{Top: The average energy per unit cross section (\SI{}{\mega \joule \per \metre \squared}) required to begin erosion for each shower ($E_S$). Bottom: The average energy per unit mass (\SI{}{\mega \joule \per \kilogram})  required to begin erosion for each shower ($E_V$). For both insets, the showers are colour coded according to their parent comet orbital population following the classification of \citet{Levison1996}.}
    \label{fig:erosionenergy}
\end{figure}

Figure \ref{fig:erosionenergy} shows a comparison of the average energy per unit mass and per unit cross section needed to trigger erosion \citep{Borovicka_2007}. Most showers seem to require similar amounts of energy within uncertainty, between 1 and \SI{2}{\mega \joule} per unit mass or cross section, to begin the erosion process. This is interesting as the bulk density of the shower appears uncorrelated with erosion energy. For example, the Tau-Herculids, our lowest density shower with weak material, was found to have near identical erosion energies to the Geminids, our highest density shower. 

Our shower with the lowest erosion energy was the single Monocerotid event, which required 2 to 3 times less energy compared to the other showers. This is likely because of the large difference in erosion start height and observed begin height ($\Delta h$) of $\sim$10~km for this event, leading to an underestimation of erosion energy. We thus do not consider this value robust. The $\Delta h$ of this event was the highest of all modelled events, leading to an unusually high erosion start height, decreased atmospheric density and therefore decreased calculated erosion energy. 

The Eta-Aquariids were found to have the highest erosion energy in both insets of Figure \ref{fig:erosionenergy}, despite being one of the lowest density showers. They also require about 2.5$\times$ more energy to start erosion than the Orionids, despite originating from the same parent body. A likely reason for this behaviour is that the Eta-Aquariids have the shallowest entry angle of all our analyzed showers ($\sim20^{\circ}$), as the shower radiant is at a relatively southern latitude. Thus the Eta-Aquariids observed by CAMO in Canada had to pass through a lot of atmosphere before becoming visible. If the Orionids and Eta-Aquariids really do have similar physical properties, then this finding indicates that a better figure of merit when investigating the erosion threshold would be the instantaneous energy, instead of the total energy received by the meteoroid before the start of erosion.

This difference with the Orionids could mean that the concept of erosion energy is less significant to erosion onset than the dynamic pressure which scales with erosion start height; both the Orionids and Eta Aquriids have similar erosion start heights.  Alternatively, it is possible that the younger age of the shower \citep{Egal2020} compared to the Orionids could result in higher binding energy; however without reliable age estimates of each shower meteoroid, it is unclear if this is plausible.  

The exact mechanism that triggers erosion is still unclear. It could be due to the penetration of heat through the surface of the meteoroid, with volatiles causing refractory grains to be ejected and/or ablation of grain bonds. In this case, we expect grain release to occur at nearly constant erosion energies. Alternatively, if it is due to a continuous mechanical failure of the surface layer as the dynamic pressure on the meteoroid increases, grain release would occur at near-constant dynamic pressures. The fact that most of our showers show similar average erosion energies argues for the former interpretation, while the disparity between the Orinids and the Eta-Aquariids argues for the latter. 

\citet{hulfeld2021three} used direct simulation Monte Carlo (DSMC) to model the fragmentation and ablation of a Draconid meteoroid and used parameters which were determined based on the erosion model to kick-start the simulation. They found that they can explain the fragmentation purely mechanically (i.e. driven by the dynamic pressure), however more spectral observations are needed to determine the role of volatiles during erosion.

\newpage

\subsection{Ablation and erosion in relation to grain size}

\begin{figure}[H]
    \centering
    \includegraphics[width=\linewidth,height=8.2cm]{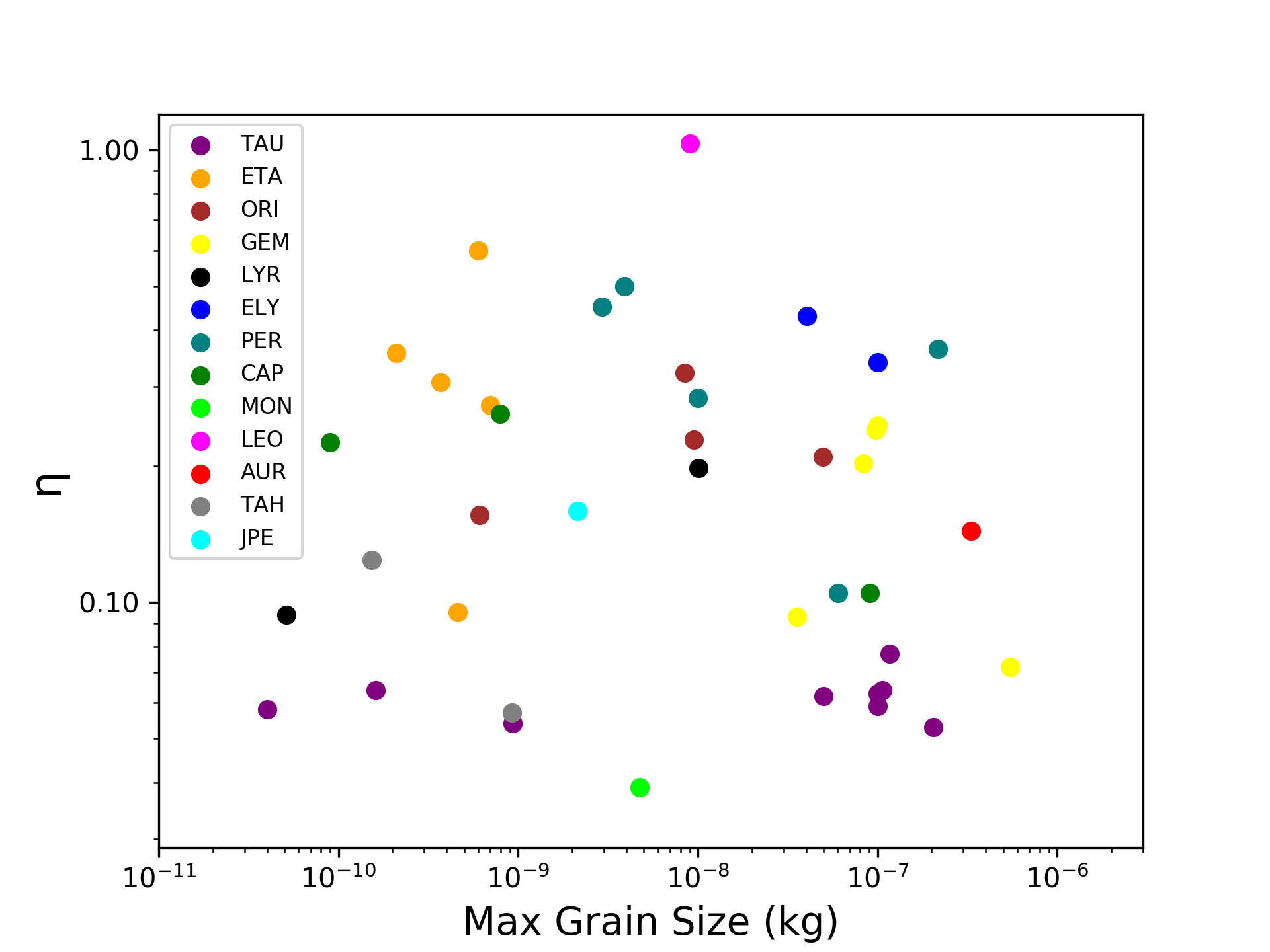}
\end{figure}
\begin{figure}[H]
    \centering
    \includegraphics[width=\linewidth,height=8.2cm]{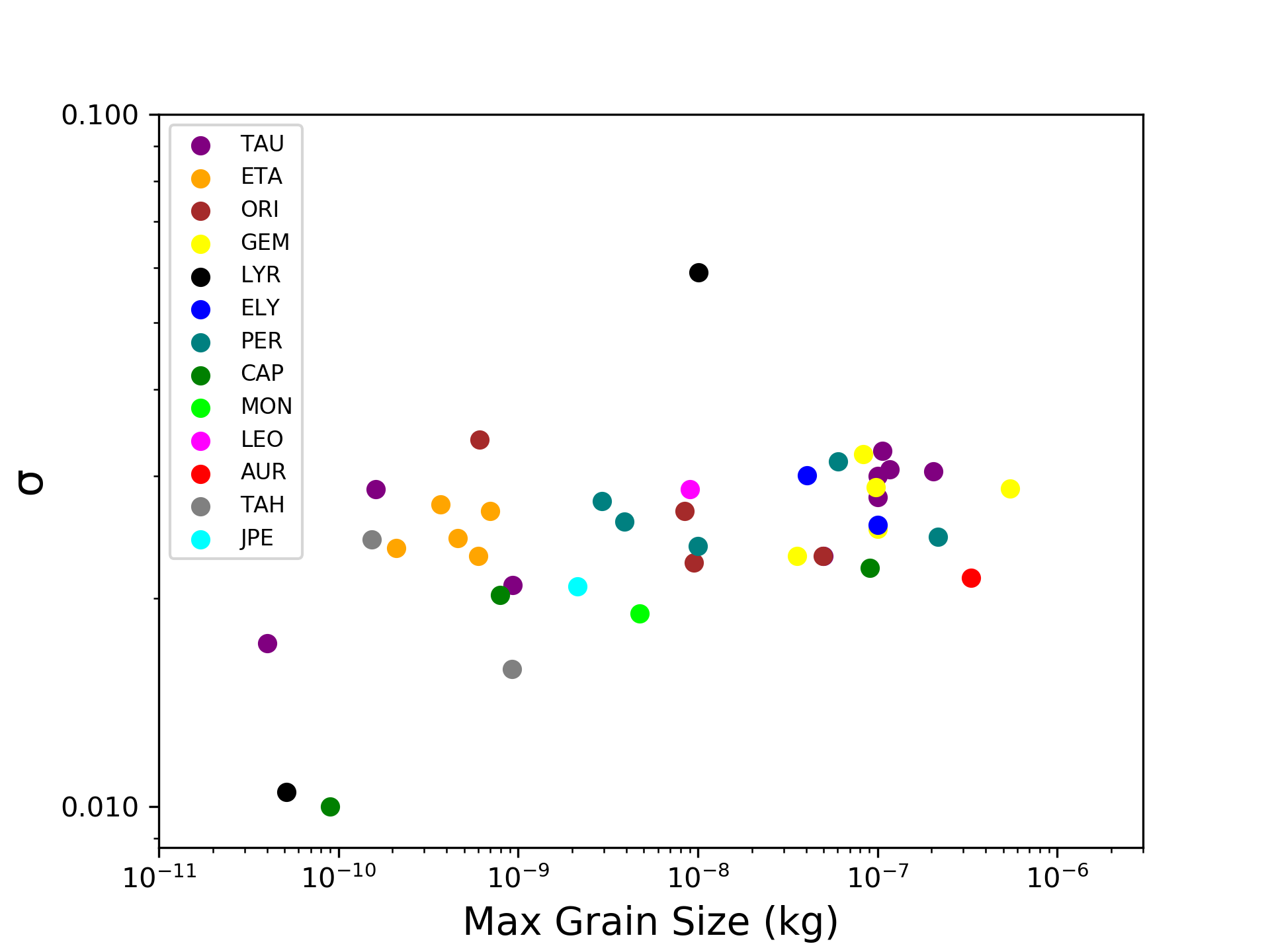}
    \caption{Top: Erosion coefficient $\eta$ (\SI{}{\kilogram \per \mega \joule}) for each event as a function of maximum grain size (kg).
    Bottom: Ablation coefficient $\sigma$ (\SI{}{\kilogram \per \mega \joule}) for each event as a function of maximum grain size (kg).
    For both plots, events are colour coded by shower.}
    \label{fig:etasigma}
\end{figure}

\begin{figure}[H]
    \centering
    \includegraphics[width=\linewidth]{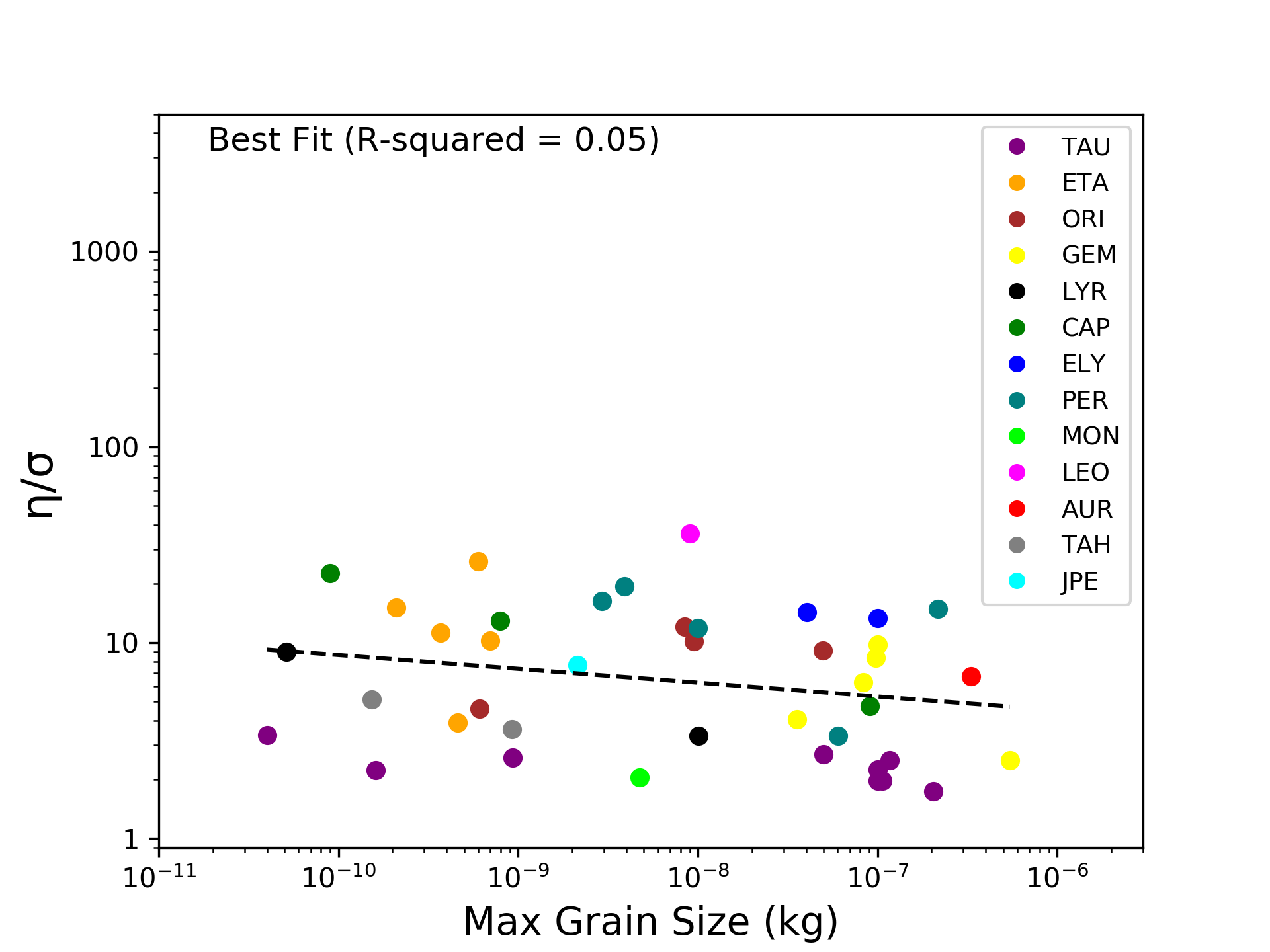}
    \caption{The ratio of the erosion coefficient $\eta$ to the ablation coefficient $\sigma$ as a function of maximum grain size (kg). Events are colour coded by shower.}
    \label{fig:coeffratios}
\end{figure}

Removal of grains during ablation is characterized by the erosion coefficient which indicates how easily grains can be removed from the main meteoroid and is therefore a structural indicator. The ablation coefficient is an indicator of the ease of removal of atoms and is linked to the material properties of the meteoroid.

As shown in the top inset of Figure \ref{fig:etasigma} the erosion coefficient $\eta$ is independent of the grain upper mass limit; the same is true for the minimum grain mass. It also is independent of the differential grain mass index. This suggests that the constituent grain size plays little role in erosion.

The bottom inset of Figure \ref{fig:etasigma} shows that ablation coefficient $\sigma$ is at most very weakly positively correlated with the grain upper mass limit as suggested by \citet{Vojacek_2017}, though our results are consistent with no trend as well. 

The ratio of $\eta$/$\sigma$ is given in Figure \ref{fig:coeffratios}, and shows a very slight downward trend, driven by the increasing ablation coefficient. However, we do not find this trend to be statistically significant, as linear regression results in R$^2$ = 0.05. This means that only 5$\%$ of the variance in the ratio of $\eta$/$\sigma$ can be predicted or explained by the maximum grain size; much less significant than what was originally found by \citet{Vojacek_2017}. 

These results suggest that it does not appear to be more difficult to release large grains from a meteoroid, but it might be slightly harder to vaporize or ablate large grains. This makes sense physically as the release of grains is determined more by the breakdown of the meteoroid matrix (which we presume remains homogenous and independent of grain size) and not the breakdown of grains themselves. The ablation of a meteoroid consists of the vaporisation of the entire body and its material, including the grains. Thus it is sensible to imagine that larger grains are more difficult to vaporise.

\subsection{Changes in \texorpdfstring{$\eta$}{} and \texorpdfstring{$\sigma$}{}}

Overall we find that a majority of events (56\%) required a change in erosion coefficient while (10\%) required a change in ablation coefficient in order to be accurately modelled. It was found that most showers contained meteoroids which were physically concordant in terms of whether their meteoroids required an erosion change. For example, the Alpha-Capricornids, Tau-Herculids, Geminids, Orionids, and Perseids required erosion changes for between 67\% - 100\% of shower meteors, and all events which required ablation changes were contained within these showers. Material belonging to these stated showers appears to be somewhat heterogeneous as they seem more complex than one would likely expect for mm-sized meteoroids, a similar conclusion as was reached by \citet{vojacek2019}. 

In contrast, the Southern Taurids, Eta-Lyrids and Eta-Aquarids required erosion changes for only 0\% - 29\% of analyzed events, suggestive of less complex, homogeneous meteoroids. This consistency in ablation behaviour within a shower was also seen by \citet{Vojacek_2017} where the majority of the Draconids and the Taurids they analyzed required a second stage of fragmentation. While they find an opposite trend for the Taurids than we do, it is likely explained by the fact that they observed Taurids orders of magnitude more massive than in our study, and as a result are more likely to contain complex structures.

For the remaining showers, the Northern Taurids, Lyrids, Leonids, Aurigids, Monocerotids and July Pegasids, only one or two events were modelled, so we are unable to say whether they have a more complex structure at mm-sizes due to small number statistics.

Our results suggest that for some showers, meteoroids are fairly homogeneous, made up of very similar material. However, other showers consist of meteoroids that are complex, likely containing multiple materials of varying mechanical strengths, such as was found for the Taurids by \citet{Borovicka2020}. It is possible that showers which usually require a change in erosion or ablation could have some physical differences between their outer and the inner meteoroid material as suggested by \citet{Vojacek_2017}. This result is similar to what \citet{Blum_2022} suggests for comets themselves at larger scales.

\subsection{Dispersion of Physical Parameters}

Table \ref{tab:resultsummary} shows the average parameter values we found for each analyzed shower, as well as the standard error of the mean as an indicator of the level of heterogeneity of the shower material properties. 

\citet{Vojacek_2017} found that the dispersion of some parameters, namely the ablation and erosion coefficients increased as meteoroid mass increased. We do not see this same trend as both our $\sigma$ and $\eta$ values did not vary much for events within the same shower, regardless of the analyzed meteoroid masses. The most probable reason we do not see this trend is that our measurements cover a very narrow range of masses between \SIrange{1e-6}{1e-4}{\kilo \gram}, while \citet{Vojacek_2017} observed meteoroids with masses spanning over 5 orders of magnitude. 

The erosion start height and erosion change height seem to be the parameters with the least dispersion, showing little variability between shower events (< 2\%). This points to similar structural properties between meteoroids of the same shower. Erosion start height is directly related to erosion energy as the atmosphere becomes more dense with decreasing altitude, causing more particles of air to collide with the meteoroid and transfer their energy \citep{Popova_2019}. Hence if all meteoroids from a shower begin erosion at the same height, they likely require very similar erosion energies as well, and therefore have similar bulk strengths. This is confirmed in looking at Figure \ref{fig:erosionenergy} where the spread in values about the mean is relatively small for the majority of showers. This result further supports the concept of relative homogeneity between meteoroids in the same shower at mm-sizes.

In absolute terms, the erosion energy is not significantly different for most showers, also suggesting commonality in the underlying meteoroid structure. The main exception are the Geminids, known to be structurally strong and likely distinct from cometary streams. 

The bulk density, ablation coefficient, first and second erosion coefficient, and mass index vary from the average by about 35\% for our showers. These are the key parameters that define the physical makeup of a meteoroid, again reflecting the relative homogeneity of the material in the same shower. The absolute value of $\sigma$ was relatively similar for all showers, ranging between \SIrange{0.020}{0.031}{\kilogram \per \mega \joule}, a value consistent with cometary material \citep{Ceplecha_1998} and suggesting that the material properties of shower meteoroids are even more similar than their structural strength. 

The minimum and maximum grain size were the parameters we found with the highest scatter, showing differences of 95\% relative to the mean. This is an interesting finding as the mass distribution of the grains ($s$) remained fairly consistent between events of the same shower. The range of the minimum grain masses was between \SIrange{9e-12}{3e-10}{\kilo \gram} and for the maximum grain masses between \SIrange{9e-9}{3e-7}{\kilo \gram}, although this range only varied one order of magnitude within the same shower. This may reflect the manual modelling approach as we found the grain masses were the hardest parameter to constrain as similar solutions can be made if other ablation properties are varied in concert with the grain mass limits. Thus we do not view the intrashower grain mass limit differences as significant.

\subsection{Key Shower Comparison}

A summary of our model results per shower is given in Table \ref{tab:resultsummary}. As \citet{Vojacek_2017} is the only other study performing comparable model fits to our work, we compare our findings in detail to theirs. For the Geminids we found an average bulk density of 1387 $\pm$ \SI{240}{\kilo \gram \per \metre \cubed} and a grain mass index $s$ = 1.85 $\pm$ 0.20, compared to an average of \SI{1788}{\kilo \gram \per \metre \cubed} and $s$ = 1.90 from \citet{Vojacek_2017}. These are similar results confirming the well-known result that Geminid meteoroid material is intrinsically stronger than typical low-density cometary material. 

In contrast to our work, however, \citet{Vojacek_2017} found that mm-sized Taurids are denser than the Geminids, with an average bulk density of \SI{1967}{\kilo \gram \per \metre \cubed}, while we find significantly lower bulk density averaging 647 $\pm$ \SI{58}{\kilo \gram \per \metre \cubed}, more consistent with larger Taurids which were found to have cometary densities \citep{Borovicka2020}. We also find a slightly higher average $s$ value for the Taurids of $s$ = 2.11 $\pm$ 0.13 compared to $s$ = 1.95 found by \citet{Vojacek_2017}, however our value is within the statistical error. These disparities may not be entirely surprising as both the Northern and Southern Taurids can easily be contaminated with sporadic meteoroids \citep{Borovicka2020}, suggesting that some of our Taurid meteors may be sporadics. 

In order to confirm our analyzed meteors were in fact Taurids, we compared the radiants for each of our Taurid events with the observations of the Global Meteor Network \citep{Vida2021}. We find that of our 9 Taurid events, 2 are NTAs (2021-11-08 04:53:19 and 2021-11-08 04:13:48), and the remaining 7 are STAs. One event is close to the border for STA radiants, but even if this one event was a sporadic it still would not explain the characteristic differences we find across the rest of our Taurid events. It is possible that mm-sized Taurids contain a variety of materials with different bulk densities and strengths. Such heterogeneity in the Taurids (and also several other streams) is also reported among the fireball population \citep{Borovicka2022physical}

This is also seen in Table \ref{tab:showerdensities} where both the NTAs and STAs have one of the widest ranges in densities of previously measured literature values for any shower (\SIrange{260}{1967}{\kilo \gram \per \metre \cubed}). Despite this wide range of bulk densities, $s$ values from the literature for both Taurid meteoroids and dust near the parent comet 2P/Encke point toward material dominated by large mass grains ($s$ < 2). This differs from our results and again suggests different processes and evolution for meteoroids/young dust as compared to the constituent grains making up the building blocks of meteoroids. 

For the Perseids, \citet{Vojacek_2017} found extremely low densities (\SIrange{100}{250}{\kilo \gram \per \metre \cubed}), consistent with very weak, fresh cometary material. We find an average bulk density $\sim$ 2 times higher (365 $\pm$ \SI{134}{\kilo \gram \per \metre \cubed}), but generally in agreement within systematic uncertainties. Note our average density for the shower is weighted upwards significantly because of one event (2021-08-14 07:06:31, $\rho$ = \SI{900}{\kilo \gram \per \metre \cubed}), which if removed as an outlier, would result in an average density of \SI{231}{\kilo \gram \per \metre \cubed} for the shower, a nearly identical result to \citet{Vojacek_2017}. The Perseid grain mass index of $s$ = 2.16 $\pm$ 0.21 also agrees within uncertainty with the value found by \citet{Vojacek_2017} ($s$ = 2.07). We suggest a special focus should be put on the Perseids as they are one of the showers that pose the most significant risk to spacecraft due to their high activity and speed, making measuring an accurate bulk density particularly important. So while we only found one higher density Perseid, this is 20\% of our data set and possibly indicative of a range of physical properties within the stream. This is consistent with the findings of \citet{Matlovic2019a, Borovicka2022physical}.

The Lyrids produced an average bulk density of 438 $\pm$ \SI{36}{\kilo \gram \per \metre \cubed}, which is about $\sim$2 times higher than the results of \citet{Vojacek_2017}, but again similar given systematic uncertainties. The only other study that estimated Lyrid bulk density, \citet{Verniani_1967}, found an intermediate value of \SI{390}{\kilo \gram \per \metre \cubed}. There are no measurements of the bulk density of the Lyrid's parent comet, C/1861 G1 (Thatcher). Our conservative estimate is that the average bulk density of the Lyrids is almost certainly <\SI{1000}{\kilo \gram \per \metre \cubed}. Likely it is even much lower, as the highest density meteoroid we found was \SI{474}{\kilo \gram \per \metre \cubed} and both \citet{Vojacek_2017} and \citet{Verniani_1967} point to lower densities than ours. Interestingly, our measured grain differential mass index $s$ agrees well with \citet{Vojacek_2017} such that the meteoroids in the shower are primarily composed of small grains. 

The Lyrids also contain the one event (2020-04-22 09:00:11) which required a bulk density change, increasing from an initial value of \SI{401}{\kilo \gram \per \metre \cubed} to \SI{2966}{\kilo \gram \per \metre \cubed} about a third of the way through the trajectory. This increase is likely the sign of either a refractory inclusion such as a Calcium and Aluminum-rich inclusion (CAI), refractory metal elements/beads \citep[e.g.][]{Rudraswami2014} or possibly an Fe sulphide refractory inclusion such as the ones found in the dust of comet 67P \citep{Fulle_2017}, 81P/Wild2 \citep{Brownlee2012} and common in micrometeorites \citep[e.g.][]{Taylor2012}. Looking qualitatively at the latter half of the event, the meteor appears as a small single body or leading fragment consistent with such an inclusion. Finally, this higher density inclusion would also explain the prolonged "hump" seen at the end of the lightcurve (see the plot in Supplementary Materials). It also further supports the notion that high-temperature condensates were widely mixed across large radial distances in the early solar nebula as suggested by returned samples from comet 81P/Wild2 \citep{Brownlee2014}.

\subsection{Bulk Density and Orbital Population}

The Tisserand parameter with respect to Jupiter (T$_{\mathrm{J}}$) is a simple (and approximate) means of associating an orbit with its likely small body population \citep{Kresak1972}. Orbits with T$_{\mathrm{J}}$ > 3 are more likely of asteroidal origin, those with 2 < T$_{\mathrm{J}}$ < 3 originate from ecliptic comets, though mixing between cometary and asteroidal populations occurs in 3 < T$_{\mathrm{J}}$ < 3.08 \citep{Tancredi2014}. Objects with T$_{\mathrm{J}}$ < 2 are classified as nearly-isotropic comets (NIC).

We can also further categorize these orbits into several sub-types. Following the taxonomy proposed by \citet{Levison1996}, we subdivide NIC orbits into external ($a > 40$~AU) and Halley-type (HT) ($a < 40$~AU). We divide ecliptic comets into Encke-type orbits having T$_{\mathrm{J}}$ > 3 with a semi-major axis larger than that of Jupiter ($a > a_{\mathrm{J}}$), and into Jupiter-family comet (JFC) orbits with T$_{\mathrm{J}}$ < 3.

One exception to this classification are the Geminids. Though their parent body (3200) Phaethon, if it were an active comet, would technically be on an Encke-type orbit, we separate them into a special class as Phaethon is compositionally unique among meteor shower parent bodies \citep{halliday1988geminid}. The asteroid is spectrally close to carbonaceous chondrites \citep{maclennan2023thermal} and possibly made of anhydrous carbonaceous material \citep{geem20223200}.

\begin{figure}[H]
    \centering
    \includegraphics[width=\linewidth]{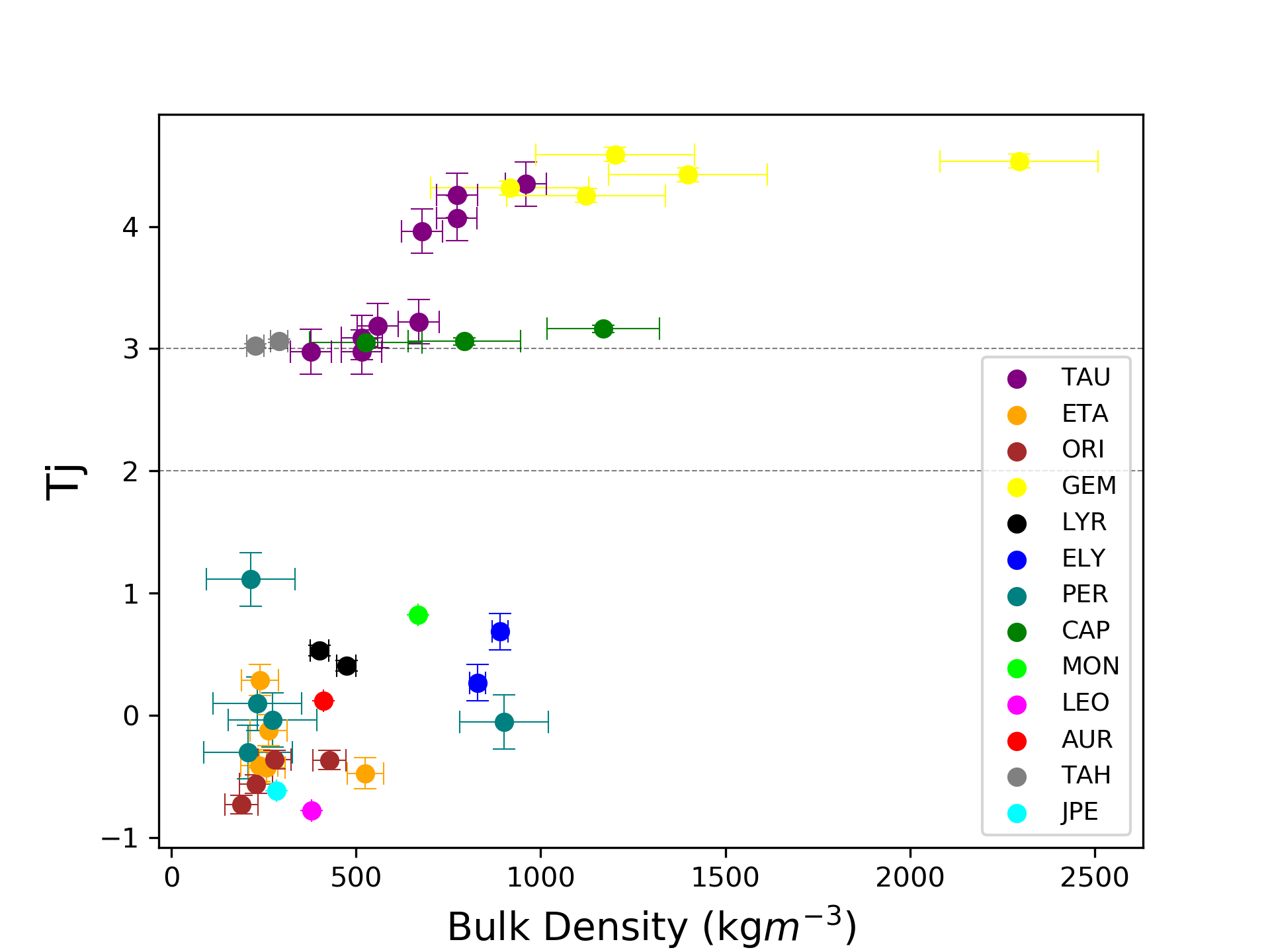}
    \caption{The Tisserand parameter with respect to Jupiter as a function of meteoroid bulk density (\SI{}{\kilo \gram \per \metre \cubed}). Events are colour coded by shower.}
    \label{fig:tis}
\end{figure}

Generally, we may expect a relationship between T$_{\mathrm{J}}$ and bulk density. \citet{Moorhead_2017} proposed that bulk densities of meteoroids are strongly correlated with T$_{\mathrm{J}}$, perhaps even more so than with the K$_B$ parameter from \citet{Ceplecha_1967a}. Figure \ref{fig:tis} shows this relationship for our data and our measurements of meteoroid bulk density as a function of the orbital population are compared to \citet{Kikwaya_2011} and \citet{Vojacek_2017} in Table \ref{tab:bulkvstiss}. Overall, our bulk densities were lower than either study for both Halley-type and Jupiter-family meteoroids. The authors note that Poynting-Robertson circulation of JFC material may systematically increase T$_{\mathrm{J}}$ over time above T$_{\mathrm{J}}$ = 3. Thus the Taurid and Alpha-Capricornid meteoroids are actually JFC-derived, despite having meteoroid orbits just over T$_{\mathrm{J}}$ = 3. 

\begin{table}[ht]
\centering
\renewcommand{\arraystretch}{1.7}
\begin{tabular}{|c|c|c|c|}
\hline
Work &  \multicolumn{2}{c|}{Bulk Densities (\SI{}{\kilo \gram \per \metre \cubed})} \\
\cline{2-3}
 & Halley-type & Jupiter-family \\
\hline
\citet{Kikwaya_2011} & 360$^{+400}_{-100}$ - 1510$^{+400}_{-900}$  & 3190 $^{+490}_{-480}$ \\
\hline
\citet{Vojacek_2017} & 960 $\pm$ 220 & 850 $\pm110$ \\
\hline
This work & 345 $\pm$ 48  & 602 $\pm$ 155 \\
\hline
\end{tabular}
\caption{The average bulk densities for meteoroids of various $T_J$ classes from three studies. A range of values is shown instead where the average value is not available.}
\label{tab:bulkvstiss}
\end{table}

We see a slight positive correlation between bulk density and T$_{\mathrm{J}}$, but this is mainly due to the Geminids. There are clearly higher density streams that have NIC/HTC-type orbits, though all have averages below \SI{1000}{\kilo \gram \per \cubic \metre}. We find this correlation is significantly weaker than was found by \citet{Kikwaya_2011}, though that study selected a wide range of meteoroids, mostly sporadics, not just cometary streams with known parents. Our showers with JFC parents (e.g. Taurids, Alpha-Capricornids) do not have unusually high bulk density, at variance with the results of \citet{Kikwaya_2011}. For example, our lowest density shower, the Tau-Herculids is from a JFC parent. The difference we believe may be that the increased accuracy of CAMO data and the wake it observes helped significantly constrain the bulk densities, while the observations of many meteors in \citet{Kikwaya_2011} lacked wake and were less accurate. 

High densities above \SI{3000}{\kilo \gram \per \metre \cubed} for meteoroids on JFC orbits was the most surprising result of \citet{Kikwaya_2011}, as cometary densities <\SI{1000}{\kilo \gram \per \metre \cubed} would be expected. We found an average density of 602 $\pm$ \SI{155}{\kilo \gram \per \metre \cubed} for such meteoroids, close to the value of 850 $\pm$ \SI{110}{\kilo \gram \per \metre \cubed} found by \citet{Vojacek_2017}. Our model uses realistic densities of silicate grains of \SI{3500}{\kilo \gram \per \metre \cubed} and sets that as a hard limit on the bulk density, however, the model can still ascertain bulk densities up to that point and would be able to reproduce the high densities found by \citet{Kikwaya_2011}. One reason that could explain the difference in densities between our study and the other two is that our study only looked at two JFC showers, the Alpha-Capricornids and the Tau-Herculids. Both of these showers are relatively dynamically young \citep{Jenniskens2010}, whereas the other studies looked at sporadic meteors which are dynamically older \citep{Campbell-Brown2005}. The increased processing and weathering of these older meteoroids might mean a decrease in porosity, however it is unlikely this process could push bulk densities >\SI{3000}{\kilo \gram \per \metre \cubed}, reminiscent of monolithic silicates. For example, IDPs (which are smaller than the mm to cm-sized meteoroids represented in Table \ref{tab:bulkvstiss}) have dynamical ages in excess  of 10$^5$ years, much longer than sporadic meteoroids, and their densities only range from \SIrange{700}{1700}{\kilo \gram \per \metre \cubed} \citep{Love_1994}. Moreover, the building process for larger cometary dust suggests that meteoroid bulk densities should decrease with size \citep{Divine1986}. Given all these considerations, it is unlikely weathering alone can explain the difference between our results and the results of \citet{Kikwaya_2011}.

For HTC meteoroids, we find lower densities when compared to the other two works, both of which agree fairly well with each other. This difference could again be a bias in our sample as we are including only shower meteoroids while the other two studies also included sporadics. \citet{Vojacek_2017} analyzed three Perseids which had an average bulk density of only \SI{167}{\kilo \gram \per \metre \cubed}. \citet{Kikwaya_2011} analyzed 10 Perseids which had an average bulk density of 620 $\pm$ \SI{200}{\kilo \gram \per \metre \cubed}. From these results, and our results in Figure \ref{fig:metdens} it is fairly clear that the HTC shower meteors have extremely low density. Other than the \SI{900}{\kilo \gram \per \metre \cubed} Perseid and the \SI{668}{\kilo \gram \per \metre \cubed} Monocerotid, all of our modelled HT meteoroids had densities less than \SI{500}{\kilo \gram \per \metre \cubed}.

For meteoroids on asteroidal orbits \citet{Kikwaya_2011} found densities between \SIrange{3000}{5000}{\kilo \gram \per \metre \cubed}. The only ``asteroidal'' shower in our analysis was the Geminids, where we found densities between \SIrange{900}{2300}{\kilo \gram \per \metre \cubed}, with an average value of $1387 \pm$ \SI{240}{\kilo \gram \per \metre \cubed}. Despite their asteroidal orbits, we exclude the Taurids from this group as their parent body (2P/Encke) is clearly a comet. 


\section{Conclusions}
\label{sec:conclusions}

We have applied a numerical implementation of the erosion model from \citet{Borovicka_2007} to a total of 41 mm-sized faint meteors from various meteor showers with known cometary parent bodies. The model was fit to the lightcurve, atmospheric deceleration, and wake observed by the Canadian Automated Meteor Observatory's mirror tracking system. The data were forward modelled to measure meteoroid physical properties, which were compared to in situ measurements of comet dust and used to characterize general properties of cometary material from different populations (HTC, JFC, asteroidal).

Our main conclusions are:
\begin{enumerate}
   \item{Our measured shower meteoroid bulk densities agree with both the in situ parent comet nucleus/dust densities and the lower end of previous in-atmosphere estimates from the literature.}
   \item{We confirm that bulk density slightly increases as a function of $T_J$ \citep{Moorhead_2017} up to $\sim$\SI{1000}{\kilo \gram \per \metre \cubed}, but we do not reproduce high bulk densities of meteoroids on JFC orbits ($\sim$\SI{3000}{\kilo \gram \per \metre \cubed}) measured by \citet{Kikwaya_2011}.}
   \item{Grain mass indices $s$ < 2 were confirmed for the Geminids and the Eta-Aquariids, while $s$ > 2 was confirmed for the Lyrids.}
   \item{For most showers, meteoroids within the same shower appear to be materially homogeneous, showing similar material properties with little variance as reflected by very similar ablation and erosion model parameters.}
   \item{Individual meteoroids from some showers (CAP, TAH, GEM, ORI, and PER) appear to have complex structural morphologies, reflecting a change in erosion or ablation properties during flight. Some show a leading fragment morphology while wake observations show that the distribution of eroded grains changes during the flight. Such events were harder to model as we assumed the grain distribution is constant as to keep the number of free parameters reasonable.}
   \item{Most showers had similar values of the average energy required to trigger erosion, suggesting that grain bonds require fixed energy to be released across showers. However, a large difference between ORI and ETA erosion energies hints at dynamic pressure as a significant factor in the onset of erosion.}
   \item{A Lyrid meteor was found which required an increase in bulk density from \SI{401}{\kilo \gram \per \metre \cubed} to $\sim$\SI{3000}{\kilo \gram \per \metre \cubed}. This is consistent with a refractory inclusion.}
\end{enumerate}

\section{Acknowledgements}
This work was supported in part by the NASA Meteoroid Environment Office under cooperative agreement 80NSSC21M0073. PGB also acknowledges funding support from the Natural Sciences and Engineering Research council of Canada and the Canada Research Chairs program. We thank Z. Krzeminski for help in optical data reduction, J. Gill, and M. Mazur for software support and camera operations and V. Voj{\'{a}}{\v{c}}ek for data access. 

\bibliography{references}

\newpage

\appendix

\begin{landscape}

\section{Summary of Bulk Density Estimates of Cometary Nuclei From Literature}
\label{ap:bulkdensity}
\begin{center}

\begin{table}[H]

\caption{Estimated bulk densities (\SI{}{\kilo \gram \per \metre \cubed}) of comet nuclei previously investigated by in situ spacecraft missions. (1P/Halley is moved to the next table as it is a direct comparison for this work).}

\begin{tabular}{ |p{4.4cm}|p{4.8cm}|p{4.4cm}|p{3.1cm}| }
 \hline
 Comet & Work & Method & Bulk Density (\SI{}{\kilo \gram \per \metre \cubed}) \\
 \hline
 9P/Tempel & Richardson et al. (2007) & Analysis of Ejecta Plume & 400 (+600, -200) \\
 & Davidsson et al. (2007) & Nongravitational Forces & 450 ($\pm$250) \\
 & Sosa $\&$ Fernández (2009)  & Nongravitational Forces & 200 ($\pm$100) \\
 & Thomas et al. (2013) & Deep Impact Ejecta & 470 (+780, -240) \\
 \hline
 19P/Borelly & Davidsson $\&$ Gutiérrez (2004) & Nongravitational Forces & 180-300 \\
 & Farnham $\&$ Cochran (2002) & Nongravitational Forces & 490 (+340, -200) \\
 \hline
 21P/Giacobini-Zinner & Trigo-Rodriguez $\&$ Llorca (2006) & Strength/Density Interpolation & $<$200 \\
 \hline
 67P/Churyumov-Gerasimenko & Preusker et al. (2015) & OSIRIS Image Analysis & 535 ($\pm$35) \\
 & Jorda et al. (2016) & OSIRIS Image Analysis & 532 ($\pm$7) \\
 & Pätzold et al. (2016) & Gravity Field Analysis & 533 ($\pm$6) \\
 \hline
 81P/Wild & Davidsson et al. (2006) & Nongravitational Forces & $\le$600-800 \\
 & Sosa $\&$ Fernández (2009) & Nongravitational Forces & 300 ($\le$800) \\
 \hline
 103P/Hartley 2 & Thomas et al. (2013) & MRI Image Analysis & 200-400 \\
 & Richardson $\&$ Bowling (2014) & Erosion Rate & 220 (+300, -80) \\
 \hline
\end{tabular}
\end{table}

\hspace{1cm}

\begin{table}[H]
\caption{Estimated bulk densities (\SI{}{\kilo \gram \per \metre \cubed}) of parent comet nuclei for showers analyzed in this work. $^{(a)}$ Preferred value, the possible range is 30 - 4900 \SI{}{\kilo \gram \per \metre \cubed}. \\}

\begin{tabular}{ |p{3.6cm}|p{3.5cm}|p{4.6cm}|p{3.5cm}| }
 \hline
 Comet & Relevant Showers & Work & Bulk Density (\SI{}{\kilo \gram \per \metre \cubed}) \\
 \hline
 1P/Halley & ORI and ETA & Peale (1989) & $1000^{(a)}$ \\
 & & Sagdeev et al. (1988) & 600 (+900, -400) \\
 & & Skorov $\&$ Rickman (1999) & 500-1200 \\
 & & Greenberg $\&$ Li (1999) & 260-510 \\
 & & Sosa $\&$ Fernández (2009) & 500 ($\pm$300) \\
 \hline
 2P/Encke & STA and NTA & Harmon $\&$ Nolan (2005) & 500-1000 \\
 & & Sosa $\&$ Fernández (2009) & 800 ($\pm$800) \\
 \hline
 169P/NEAT & CAP & Kasuga et al. (2010) & 200-500 \\
 \hline
 (3200) Phaethon & GEM & Hanuš et al. (2018) & 1670 ($\pm$470) \\
 & & Masiero et al. (2021) & 3000 \\
 \hline
\end{tabular}

\end{table}
\end{center}

\hspace{1cm}

\section{Differential Mass Index \texorpdfstring{$s$}{} of Shower and Comet Dust From Literature}
\label{ap:massindex}
\begin{center}

\begin{table}[H]
\caption{Differential mass indices ($s$) for the relevant major showers (part 1). Adapted from \citep{Blaauw_2019}, references are provided therein. $^{(a)}$ Note: average grain distribution mass indices were added from \cite{Vojacek_2017} for the GEM, PER, LYR, and Taurids.}

\begin{tabular}{|p{1cm}|p{5cm}|p{6cm}|p{5cm}|}
\hline
Shower & $s$ & Work & Magnitude/Mass Range \\
\hline
ETA & 1.40 & Wiess (1961) & +1.75 to +0.5 \\
 & 2.17 ($\pm$ 0.24) & Cevolani et al. (1998) & limit of $10^{-8}$ to $10^{-9}$ kg \\
 & 1.85 ($\pm$ 0.07) & Blaauw et al. (2011a) & $10^{-8}$ kg \\
\hline
ORI & 1.95 & Schult et al. (2018) & $10^{-8}$ kg \\
 & 2.06 ($\pm$ 0.17) & Cevolani et al. (1998) & limit of $10^{-8}$ to $10^{-9}$ kg \\
 & 1.65 ($\pm$0.04) to 0.77 ($\pm$0.06) & Blaauw et al. (2011a) & $10^{-8}$ kg \\
\hline
 GEM & 2.3 & Browne et al. (1957) & +2.0 \\
 & 1.62 & Browne et al. (1957) & +7.0 to +5.0 \\
 & 2.7 & Kresakova (1966) & +2.5 to -3.5 \\
 & 1.5 ($\pm$ 0.3) to 1.0 ($\pm$ 0.05) & Webster et al. (1966) & +4.5 \\
 & 1.5 & Belkovich et al. (2001) & +3.0 \\
 & 1.63 ($\pm$ 0.05) & Chenna Reddy et al. (2008) & +3.5 \\
 & 1.64 & Schult et al. (2018) & $10^{-8}$ kg \\
 & 1.71 ($\pm$ 0.02) to 0.78 ($\pm$ 0.04) & Hughes (1973) & $10^{-6}$ to $10^{-8}$ kg \\
 & 2.03 ($\pm$ 0.04) & Cevolani et al. (1998) & limit of $10^{-8}$ to $10^{-9}$ kg \\
 & 1.58 to 1.63 ($\pm$ 0.04) & Blaauw et al. (2011a) & $10^{-8}$ kg \\
 & $1.9^{(a)}$ & Vojacek (2017) & $10^{-5}$ to $10^{-10}$ kg \\
\hline
LEO & 1.35 ($\pm$ 0.02) to 1.48 ($\pm$ 0.05) & Plavcoa 1968 & +2.5 to 0 \\
 & 1.5 ($\pm$ 0.6) & Brown et al. 1998 & +4.5 \\
 & 1.64 ($\pm$ 0.17) & Brown et al. 1998 & +5.0 \\
 & 1.9 & Koten et al. 2008 & +5.0 \\
 & 1.91 ($\pm$ 0.08) & Cevolani et al. 1998 & limit of $10^{-8}$ to $10^{-9}$ kg \\
\hline
\end{tabular}

\end{table}

\hspace{1cm}

\begin{table}[H]
\caption{Differential mass indices ($s$) for the relevant major showers (part 2). Adapted from \citep{Blaauw_2019}, references are provided therein. $^{(a)}$ Note: average grain mass indices were added from \cite{Vojacek_2017} for the GEM, PER, LYR, and Taurids.}

\begin{tabular}{|p{2cm}|p{4cm}|p{6cm}|p{5cm}|}
\hline
Shower & $s$ & Work & Magnitude/Mass Range \\
\hline 
LYR & 2.1 $\pm$ 0.12 & Cevolani et al. (1998) & $10^{-8}$ to $10^{-9}$ kg \\
& 2.32$^{(a)}$ & Vojacek (2017) & $10^{-5}$ to $10^{-13}$ kg \\
\hline
TAURIDS & 1.37 $\pm$ 0.1 & Sokolova $\&$ Sergienko (2016) & +3 \\
& 1.95$^{(a)}$ & Vojacek (2017) & $10^{-8}$ to $10^{-12}$ kg \\
\hline
 PER & 1.62 & Bel'kovich $\&$ Ishmukhametova (2006) & +3 \\
 & 2 & Browne et al. (1957) & +2 \\
 & 1.59 & Browne et al. (1957) & +7 to +5 \\
 & 1.57 & Belkovich et al. (1995) & +3 \\
 & 1.45 & Schult et al. (2018) & $10^{-8}$ kg \\
 & 1.74 $\pm$ 0.01 & Cevolani et al. (1998) & $10^{-8}$ to $10^{-9}$ kg \\
 & 2.07$^{(a)}$ & Vojacek (2017) & $10^{-6}$ to $10^{-9}$ kg \\
\hline
\end{tabular}

\end{table}

\hspace{1cm}

\begin{table}[H]
\caption{Differential mass indices ($s$) for the dust ejected by parent comets related to major showers in our study.}

\begin{tabular}{|p{4.5cm}|p{2.5cm}|p{2cm}|p{4cm}|p{3.5cm}|}
\hline
Parent Comet & Relevant Shower & $s$ & Work & Mass/Size Range \\
\hline
1P/Halley & ORI $\&$ ETA & 1.85 & McDonnell et al. (1987) & $10^{-13}$ to $10^{-8}$ kg \\
 & ORI and ETA & 1.75 ($\pm$ 0.07) & McDonnell et al. (1991) & $10^{-13}$ to $10^{-9}$ kg \\
 & ORI and ETA & 1.53 ($\pm$ 0.1) & Fulle et al. (2000) & $10^{-12}$ to $10^{-3}$ kg \\
\hline
2P/Encke & STA and NTA & 1.73 ($\pm$ 0.067) & Epifani et al. (2001) & 11.5 to 15 $\mu$m \\
 & STA and NTA & 1.87 ($\pm$ 0.1) & Fulle (1989) & 40 $\mu$ to 20 cm \\
\hline
109P/Swift-Tuttle & PER & 1.5 ($\pm$ 0.17) & Sarmecanic et al. (1998) & 0.6 to 10 $\mu$m \\
\hline
73P/Schwassmann-Wachmann & TAH & 1.85$-$2.0 & Vaubaillon $\&$ Reach (2010) & $10^{-15}$ to $10^{-8}$ kg \\
\hline
\end{tabular}

\end{table}
\end{center}
\end{landscape}

\section{Reduction Methodology}
\label{ap:methodology}

\begin{wrapfigure}{r}{0.6\textwidth}
    \centering
    \includegraphics[width=0.6\textwidth]{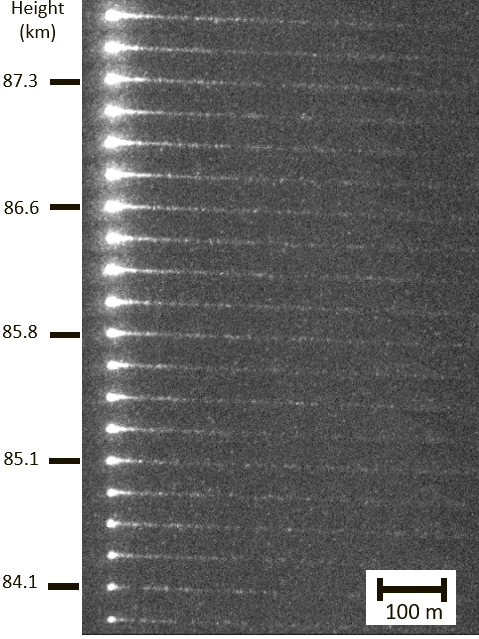}
    \caption{Vertical stack of a meteor recorded using the CAMO narrow-field system with time increasing from top to bottom. Each frame has been rotated until the horizontal and leading edges are aligned to display morphology and visibility of wake. The event was a Southern Taurid recorded October 10, 2019 at 08:49:12 UTC.}
\end{wrapfigure}

\subsection{Data Reduction: CAMO Weblog}

All events investigated in this work were reduced according to the CAMO data reduction pipeline presented in \citep{Vida_2021}. Nevertheless, a brief explanation of the process is given here. All events detected by CAMO on a given night are sent to a central server where their orbits are computed with up-to-date astrometric plates. Every morning, a new web page on the internal CAMO weblog is generated with that night's events, including each event's wide- and narrow-field videos and their initially estimated orbits. After a manual inspection, suitable events are chosen for data reduction.

\subsection{Data Reduction: Wide-Field Data}

\begin{figure}[H]
    \centering
    \includegraphics[width=\linewidth]{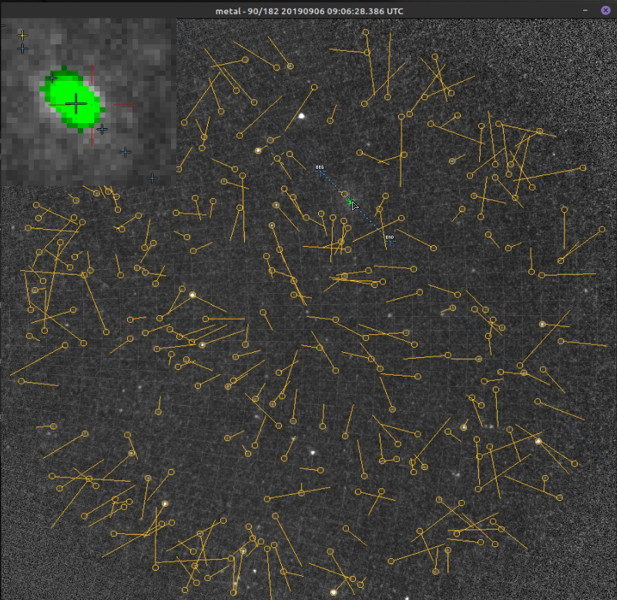}
    \caption{Example wide-field video frame. Astrometric picks are seen as small crosses spanning the meteor track, while the extent of the photometry mask on the frame is represented as green highlighted pixels. Yellow circles and lines are star positions and their residuals used to create the astrometric plate.}
    \label{fig:metal}
\end{figure}

Flat fields for the wide-field data are created by co-adding many video frames from a given night which are separated in time, eliminating star trails. Astrometric plates and photometry are calibrated manually using the meteor video data to ensure a high level of quality. Position picks on the meteor are also done manually \citep{Weryk_2013}. The centroiding region's position and radius are manually chosen for each frame of the wide-field video and the extent of the meteor is manually selected for accurate photometry, as shown in Figure \ref{fig:metal}. Astrometric picks are run through a trajectory solver based on \citep{Borovicka_1990}, and a preliminary heliocentric meteoroid orbit is computed.

\subsection{Data Reduction: Narrow-Field Data}

\begin{figure}[H]
    \centering
    \includegraphics[width=\linewidth]{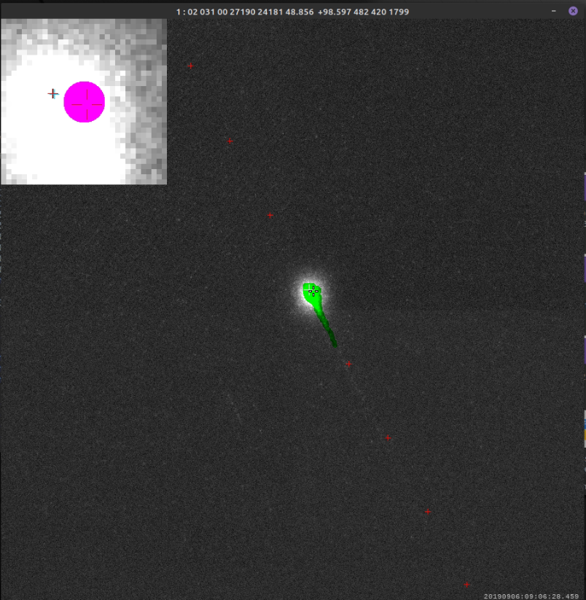}
    \caption{Example narrow field video of an event recorded in 2019, recorded by  the wide field camera and reduced using the software \emph{Mirfit}. Astrometric picks are seen as red plus signs spanning the meteor track, while photometric picks are seen as green highlighted pixels.}
    \label{fig:mirfit}
\end{figure}

Reduction of narrow-field data is done using the \texttt{Mirfit} software, shown in Figure \ref{fig:mirfit}. The astrometric plate is computed during a dedicated calibration procedure which runs every hour during which the mirrors are pointed at up to 50 bright stars in the mirror field of regard \citep{Vida2021}. Astrometric solution quality is checked by reverse mapping star catalog positions onto each video frame, and ensuring they match the true position of stars. Position picks on the meteor are done manually as with the wide-field data, however accuracy can become difficult and even subjective depending on the meteors specific morphology and fragmentation behaviour. As a consequence, the precision of a meteor's trajectory is commonly limited by its morphology \citep{Vida2021}, despite the high resolution of the CAMO mirror tracking system. To keep astrometric picks as consistent as possible, and to ensure common features are tracked from both sites, an approach is usually taken to centroid on the most consistent leading fragment that exists throughout the event. For an event with unresolvable features and no consistent leading fragment, picks are set manually to the leading edge of the meteor. Photometry is again performed by selecting and summing the pixels in each frame that belong to the meteor.

After narrow-field astrometric picks are finished for both sites, the meteor's trajectory is more accurately determined using the \citep{Vida_2020_I} and \citep{Vida_2020_II} trajectory solver. The reference trajectory solution going forward only uses narrow-field data, while the wide-field data is only used for photometry due to its limited accuracy compared to the narrow-field.

\begin{landscape}
\section{Event Results}
\label{ap:metmeasurements}

\begin{longtable}{r l r r r r r r r r r r}
\caption{Radiants and orbits of the observed meteors. Rows below every entry list $1\sigma$ uncertainties. The uncertainties are measurement precision and not the total accuracy. The geocentric velocity was computed using the model-corrected speed.} \label{tab:orbits} \\
	\hline\hline 
	
Name & 	Date and time (UTC) & $\lambda_{\astrosun}$ & $\alpha_g$ & $\delta_g$ & $v_g$ & $a$ & $e$ & $q$ & $\omega$ & $i$ & $\pi$\\
 & & (deg) & (deg) & (deg) & (\SI{}{\kilo \metre \per \second}) & (AU) &  & (AU) & (deg) & (deg) & (deg) \\
\hline
\endfirsthead
\caption{continued.} \\
	\hline\hline 
	
Name &	Date and time (UTC) & $\lambda_{\astrosun}$ & $\alpha_g$ & $\delta_g$ & $v_g$ & $a$ & $e$ & $q$ & $\omega$ & $i$ & $\pi$ \\
& & (deg) & (deg) & (deg) & (\SI{}{\kilo \metre \per \second}) & (AU) &  & (AU) & (deg) & (deg) & (deg) \\
\hline
\endhead 
\hline
\endfoot
AUR \#1 & 2019-09-06 09:06:28 & 163.2328 & 96.735 & 39.084 & 64.543 & 6.340 & 0.8961 & 0.6587 & 105.46 & 147.90 & 268.69 \\
  &   &   & 0.018 & 0.061 & 0.078 & 0.214 & 0.0043 & 0.00230 & 0.41 & 0.07 & 0.41 \\
CAP \#1 & 2019-08-01 02:44:24 & 128.3338 & 305.719 & -8.669 & 21.747 & 2.523 & 0.7561 & 0.6154 & 265.06 & 7.41 & 33.43 \\
  &   &   & 0.018 & 0.063 & 0.009 & 0.003 & 0.0004 & 0.00022 & 0.03 & 0.04 & 0.03 \\
CAP \#2 & 2016-08-02 04:26:43 & 130.0875 & 307.519 & -8.715 & 21.660 & 2.500 & 0.7539 & 0.6152 & 265.16 & 7.05 & 35.27 \\
  &   &   & 0.008 & 0.034 & 0.001 & 0.001 & 0.0001 & 0.00008 & 0.01 & 0.02 & 0.01 \\
CAP \#3 & 2019-07-26 02:41:50 & 122.5958 & 302.485 & -10.116 & 22.960 & 2.541 & 0.7732 & 0.5761 & 269.54 & 7.53 & 32.17 \\
  &   &   & 0.016 & 0.045 & 0.003 & 0.001 & 0.0001 & 0.00020 & 0.02 & 0.03 & 0.02 \\
ELY \#1 & 2021-05-11 03:56:57 & 50.4038 & 291.663 & 44.747 & 44.810 & -24.690 & 1.0407 & 1.0038 & 188.92 & 74.69 & 239.33 \\
  &   &   & 0.083 & 0.520 & 0.313 & 473.693 & 0.0447 & 0.00085 & 0.70 & 0.12 & 0.70 \\
ELY \#2 & 2021-05-12 07:16:55 & 51.5041 & 294.700 & 43.976 & 44.943 & 33.253 & 0.9697 & 1.0063 & 187.24 & 76.26 & 238.75 \\
  &   &   & 0.022 & 0.067 & 0.047 & 1.051 & 0.0063 & 0.00007 & 0.08 & 0.01 & 0.08 \\
ETA \#1 & 2021-05-14 08:15:57 & 53.4747 & 343.532 & 1.471 & 66.324 & 14.168 & 0.9554 & 0.6317 & 103.43 & 163.98 & 156.89 \\
  &   &   & 0.005 & 0.025 & 0.007 & 0.034 & 0.0002 & 0.00013 & 0.02 & 0.05 & 0.02 \\
ETA \#2 & 2021-05-09 08:22:50 & 48.6487 & 340.040 & 0.065 & 66.141 & 17.178 & 0.9649 & 0.6036 & 100.41 & 163.64 & 149.05 \\
  &   &   & 0.007 & 0.058 & 0.031 & 0.174 & 0.0009 & 0.00068 & 0.10 & 0.12 & 0.10 \\
ETA \#3 & 2021-05-06 08:25:03 & 45.7467 & 338.722 & 0.011 & 64.903 & 8.905 & 0.9388 & 0.5454 & 92.92 & 162.00 & 138.65 \\
  &   &   & 0.012 & 0.042 & 0.015 & 0.037 & 0.0004 & 0.00017 & 0.03 & 0.10 & 0.03 \\
ETA \#4 & 2021-05-06 08:35:50 & 45.7540 & 338.138 & -0.238 & 65.421 & 11.389 & 0.9500 & 0.5694 & 96.06 & 162.33 & 141.80 \\
  &   &   & 0.006 & 0.023 & 0.081 & 0.605 & 0.0037 & 0.00153 & 0.29 & 0.07 & 0.29 \\
ETA \#5 & 2021-05-09 08:34:13 & 48.6563 & 339.922 & -0.329 & 66.270 & 17.254 & 0.9646 & 0.6102 & 101.19 & 164.37 & 149.83 \\
  &   &   & 0.027 & 0.099 & 0.017 & 0.105 & 0.0006 & 0.00026 & 0.04 & 0.21 & 0.04 \\
GEM \#1 & 2021-12-14 10:59:13 & 262.3692 & 113.361 & 32.232 & 34.680 & 1.411 & 0.8988 & 0.1428 & 323.72 & 23.41 & 226.09 \\
  &   &   & 0.013 & 0.042 & 0.026 & 0.003 & 0.0002 & 0.00025 & 0.05 & 0.09 & 0.05 \\
GEM \#2 & 2021-12-14 09:03:00 & 262.2871 & 114.065 & 32.347 & 33.869 & 1.301 & 0.8892 & 0.1441 & 324.46 & 23.27 & 226.74 \\
  &   &   & 0.018 & 0.056 & 0.020 & 0.003 & 0.0002 & 0.00040 & 0.07 & 0.10 & 0.07 \\
GEM \#3 & 2021-12-14 10:45:53 & 262.3598 & 113.619 & 31.757 & 34.579 & 1.378 & 0.8994 & 0.1386 & 324.55 & 22.68 & 226.90 \\
  &   &   & 0.020 & 0.042 & 0.053 & 0.005 & 0.0005 & 0.00032 & 0.06 & 0.12 & 0.06 \\
GEM \#4 & 2021-12-14 08:16:13 & 262.2541 & 113.755 & 32.245 & 33.790 & 1.306 & 0.8884 & 0.1458 & 324.18 & 22.75 & 226.44 \\
  &   &   & 0.027 & 0.062 & 0.028 & 0.004 & 0.0001 & 0.00036 & 0.09 & 0.12 & 0.09 \\
GEM \#5 & 2021-12-14 11:15:34 & 262.3807 & 113.945 & 31.342 & 34.145 & 1.318 & 0.8960 & 0.1371 & 325.23 & 21.67 & 227.61 \\
  &   &   & 0.010 & 0.057 & 0.066 & 0.006 & 0.0006 & 0.00015 & 0.05 & 0.17 & 0.05 \\
JPE \#1 & 2019-07-05 07:36:03 & 102.7611 & 343.534 & 9.883 & 64.740 & 26.681 & 0.9764 & 0.6294 & 256.76 & 147.85 & 359.51 \\
  &   &   & 0.005 & 0.032 & 0.047 & 1.381 & 0.0026 & 0.00100 & 0.18 & 0.06 & 0.18 \\
LEO \#1 & 2019-11-26 10:02:58 & 243.5703 & 158.636 & 19.971 & 71.247 & 20.855 & 0.9527 & 0.9861 & 183.57 & 162.45 & 67.14 \\
  &   &   & 0.020 & 0.127 & 0.035 & 1.076 & 0.0049 & 0.00011 & 0.20 & 0.19 & 0.20 \\
LYR \#1 & 2020-04-22 09:00:11 & 32.4214 & 272.376 & 33.009 & 47.214 & 35.071 & 0.9738 & 0.9193 & 214.25 & 80.26 & 246.67 \\
  &   &   & 0.009 & 0.029 & 0.019 & 0.704 & 0.0025 & 0.00026 & 0.08 & 0.01 & 0.08 \\
LYR \#2 & 2020-04-22 05:31:10 & 32.2798 & 272.471 & 32.732 & 47.308 & 27.141 & 0.9662 & 0.9185 & 214.49 & 80.63 & 246.77 \\
  &   &   & 0.045 & 0.130 & 0.014 & 4.239 & 0.0065 & 0.00085 & 0.24 & 0.10 & 0.24 \\
MON \#1 & 2021-12-12 07:52:16 & 260.2039 & 101.629 & 6.899 & 41.481 & 18.673 & 0.9891 & 0.2031 & 126.59 & 36.62 & 206.80 \\
  &   &   & 0.004 & 0.020 & 0.004 & 0.080 & 0.0001 & 0.00017 & 0.02 & 0.03 & 0.02 \\
NTA \#1 & 2021-11-08 04:53:19 & 225.7053 & 54.445 & 20.959 & 28.094 & 2.178 & 0.8332 & 0.3632 & 293.21 & 1.71 & 158.88 \\
  &   &   & 0.016 & 0.041 & 0.008 & 0.002 & 0.0002 & 0.00008 & 0.02 & 0.05 & 0.02 \\
NTA \#2 & 2021-11-08 04:13:47 & 225.6777 & 55.931 & 22.426 & 29.704 & 2.311 & 0.8592 & 0.3254 & 296.97 & 3.31 & 162.63 \\
  &   &   & 0.009 & 0.020 & 0.007 & 0.003 & 0.0002 & 0.00003 & 0.01 & 0.03 & 0.01 \\
ORI \#1 & 2019-10-29 09:55:41 & 215.4110 & 100.228 & 15.626 & 65.684 & 12.246 & 0.9581 & 0.5126 & 89.38 & 163.51 & 124.79 \\
  &   &   & 0.028 & 0.072 & 0.040 & 0.482 & 0.0020 & 0.00053 & 0.07 & 0.15 & 0.07 \\
ORI \#2 & 2019-10-29 08:21:22 & 215.3456 & 100.468 & 15.624 & 66.063 & 17.141 & 0.9693 & 0.5260 & 87.47 & 163.71 & 122.81 \\
  &   &   & 0.008 & 0.021 & 0.022 & 0.323 & 0.0010 & 0.00050 & 0.08 & 0.04 & 0.08 \\
ORI \#3 & 2020-10-12 06:58:50 & 199.1292 & 88.489 & 14.968 & 67.525 & 24.397 & 0.9728 & 0.6635 & 71.32 & 163.08 & 90.45 \\
  &   &   & 0.011 & 0.030 & 0.016 & 0.464 & 0.0009 & 0.00013 & 0.01 & 0.06 & 0.01 \\
ORI \#4 & 2020-10-12 08:17:16 & 199.1831 & 89.018 & 15.195 & 67.747 & 27.291 & 0.9752 & 0.6762 & 69.70 & 163.65 & 88.88 \\
  &   &   & 0.070 & 0.166 & 0.065 & 0.896 & 0.0040 & 0.00175 & 0.21 & 0.32 & 0.21 \\
PER \#1 & 2021-08-14 08:19:44 & 141.4821 & 50.780 & 58.120 & 58.890 & 11.331 & 0.9167 & 0.9439 & 149.03 & 113.40 & 290.51 \\
  &   &   & 0.136 & 0.119 & 0.083 & 0.721 & 0.0094 & 0.00126 & 0.38 & 0.15 & 0.38 \\
PER \#2 & 2021-08-14 07:06:41 & 141.4334 & 50.847 & 58.517 & 58.863 & 13.808 & 0.9317 & 0.9433 & 149.01 & 112.85 & 290.44 \\
  &   &   & 0.165 & 0.192 & 0.251 & 3.231 & 0.0263 & 0.00223 & 0.75 & 0.13 & 0.75 \\
PER \#3 & 2021-08-13 06:14:53 & 140.4385 & 49.408 & 57.993 & 60.166 & -40.033 & 1.0237 & 0.9478 & 150.73 & 114.05 & 291.17 \\
  &   &   & 0.035 & 0.159 & 0.104 & 126.582 & 0.0147 & 0.00065 & 0.26 & 0.15 & 0.26 \\
PER \#4 & 2021-08-14 06:57:36 & 141.4274 & 51.937 & 58.638 & 59.753 & -122.420 & 1.0077 & 0.9391 & 148.72 & 113.32 & 290.15 \\
  &   &   & 0.070 & 0.102 & 0.048 & 10.432 & 0.0077 & 0.00078 & 0.24 & 0.11 & 0.24 \\
PER \#5 & 2021-08-16 09:06:33 & 143.4350 & 58.536 & 59.488 & 56.510 & 4.376 & 0.7952 & 0.8962 & 137.78 & 110.56 & 281.21 \\
  &   &   & 0.116 & 0.094 & 0.086 & 0.116 & 0.0082 & 0.00210 & 0.55 & 0.06 & 0.55 \\
STA \#1 & 2019-10-09 08:57:21 & 195.5101 & 28.702 & 5.895 & 24.343 & 1.472 & 0.7394 & 0.3836 & 116.68 & 5.49 & 132.19 \\
  &   &   & 0.008 & 0.047 & 0.008 & 0.000 & 0.0002 & 0.00031 & 0.03 & 0.04 & 0.03 \\
STA \#2 & 2019-10-21 04:13:58 & 207.2013 & 39.409 & 9.985 & 25.990 & 1.656 & 0.7776 & 0.3684 & 116.12 & 5.52 & 143.32 \\
  &   &   & 0.011 & 0.033 & 0.007 & 0.001 & 0.0002 & 0.00018 & 0.02 & 0.03 & 0.02 \\
STA \#3 & 2019-09-27 04:15:13 & 183.5076 & 21.507 & 4.695 & 27.724 & 1.470 & 0.8084 & 0.2817 & 127.50 & 5.27 & 130.99 \\
  &   &   & 0.009 & 0.021 & 0.008 & 0.001 & 0.0002 & 0.00012 & 0.01 & 0.03 & 0.01 \\
STA \#4 & 2019-10-09 07:15:23 & 195.4402 & 31.284 & 7.129 & 27.698 & 1.608 & 0.8083 & 0.3083 & 123.15 & 6.51 & 138.59 \\
  &   &   & 0.009 & 0.025 & 0.006 & 0.001 & 0.0001 & 0.00014 & 0.01 & 0.03 & 0.01 \\
STA \#5 & 2021-11-12 09:20:55 & 229.9120 & 64.345 & 17.223 & 31.267 & 2.277 & 0.8771 & 0.2800 & 122.32 & 5.82 & 172.24 \\
  &   &   & 0.006 & 0.029 & 0.008 & 0.002 & 0.0001 & 0.00011 & 0.01 & 0.04 & 0.01 \\
STA \#6 & 2021-11-07 03:42:30 & 224.6520 & 52.930 & 15.191 & 27.176 & 2.222 & 0.8213 & 0.3972 & 109.19 & 4.01 & 153.85 \\
  &   &   & 0.015 & 0.039 & 0.019 & 0.004 & 0.0005 & 0.00020 & 0.01 & 0.04 & 0.01 \\
STA \#7 & 2021-11-12 08:46:07 & 229.8877 & 56.697 & 14.451 & 26.538 & 2.345 & 0.8166 & 0.4301 & 104.93 & 5.32 & 154.84 \\
  &   &   & 0.011 & 0.061 & 0.015 & 0.003 & 0.0004 & 0.00030 & 0.03 & 0.06 & 0.03 \\
TAH \#1 & 2022-05-31 03:27:37 & 69.3868 & 209.361 & 28.253 & 12.092 & 3.117 & 0.6826 & 0.9894 & 199.82 & 11.15 & 269.23 \\
  &   &   & 0.014 & 0.021 & 0.007 & 0.004 & 0.0006 & 0.00003 & 0.02 & 0.01 & 0.02 \\
TAH \#2 & 2022-05-31 03:35:42 & 69.3922 & 209.470 & 28.311 & 11.907 & 2.987 & 0.6687 & 0.9897 & 199.83 & 11.04 & 269.25 \\
  &   &   & 0.053 & 0.108 & 0.031 & 0.016 & 0.0022 & 0.00014 & 0.08 & 0.04 & 0.08 \\
\end{longtable}
\end{landscape}

\newpage 

\begin{table}
    \caption{Modelled physical properties: the initial mass $m_0$, the initial velocity $v_0$, the zenith angle $Z_c$, the ablation coefficient $\sigma$, the bulk density $\rho$, energy per unit cross section needed to begin erosion $E_S$ and the energy per unit mass section needed to begin erosion $E_V$.}
    {
    \begin{tabular}{r r r r r r r r}
    \hline\hline 
Name & $m_0$ & $v_0$ & $Z_c$ & $\sigma$ & $\rho$ & $E_S$ & $E_V$ \\
 & (kg) & (\SI{}{\kilo \metre \per \second}) & (deg) & (\SI{}{\kilogram \per \mega \joule}) & (\SI{}{\kilo \gram \per \metre \cubed}) & (\SI{}{\mega \joule \per \metre \squared}) & (\SI{}{\mega \joule \per \kilogram}) \\
\hline
AUR \#1 & \num{1.1e-05} & 65.490 & 41.042 & 0.021 & 412 & 1.5 & 1.5 \\
CAP \#1 & \num{4.1e-06} & 24.420 & 57.069 & 0.010 & 1170 & 1.4 & 1.0 \\
CAP \#2 & \num{1.3e-05} & 24.341 & 49.423 & 0.022 & 794 & 1.2 & 0.7 \\
CAP \#3 & \num{1.1e-05} & 25.505 & 59.950 & 0.020 & 527 & 1.9 & 1.6 \\
ELY \#1 & \num{2.5e-06} & 46.164 & 55.940 & 0.030 & 829 & 1.8 & 1.9 \\
ELY \#2 & \num{2.1e-06} & 46.294 & 25.540 & 0.025 & 890 & 1.2 & 1.2 \\
ETA \#1 & \num{1.5e-05} & 67.245 & 71.529 & 0.023 & 525 & 2.3 & 1.8 \\
ETA \#2 & \num{1.2e-05} & 67.065 & 71.965 & 0.024 & 258 & 3.3 & 4.4 \\
ETA \#3 & \num{6.9e-06} & 65.844 & 73.017 & 0.027 & 263 & 2.6 & 4.1 \\
ETA \#4 & \num{1.7e-05} & 66.355 & 71.194 & 0.024 & 240 & 2.7 & 3.4 \\
ETA \#5 & \num{9.8e-06} & 67.192 & 70.295 & 0.027 & 238 & 2.9 & 4.3 \\
GEM \#1 & \num{5.9e-06} & 36.415 & 40.804 & 0.032 & 1123 & 3.2 & 2.0 \\
GEM \#2 & \num{3.1e-06} & 35.644 & 20.766 & 0.025 & 2295 & 0.5 & 0.2 \\
GEM \#3 & \num{2.5e-06} & 36.319 & 38.362 & 0.029 & 916 & 2.0 & 1.9 \\
GEM \#4 & \num{2.6e-06} & 35.569 & 14.348 & 0.029 & 1202 & 2.7 & 2.1 \\
GEM \#5 & \num{2.0e-05} & 35.906 & 43.789 & 0.023 & 1399 & 2.0 & 0.7 \\
JPE \#1 & \num{1.2e-05} & 65.684 & 40.920 & 0.021 & 284 & 1.1 & 1.4 \\
LEO \#1 & \num{4.8e-06} & 72.106 & 31.083 & 0.029 & 379 & 1.0 & 1.4 \\
LYR \#1 & \num{2.4e-06} & 48.502 & 12.061 & 0.059 & 401 & 1.4 & 2.3 \\
LYR \#2 & \num{1.8e-05} & 48.592 & 46.660 & 0.011 & 474 & 0.7 & 0.5 \\
MON \#1 & \num{1.1e-05} & 42.941 & 38.336 & 0.019 & 668 & 0.5 & 0.3 \\
NTA \#1 & \num{5.4e-06} & 30.210 & 24.808 & 0.028 & 670 & 0.4 & 0.4 \\
NTA \#2 & \num{2.4e-05} & 31.711 & 29.961 & 0.033 & 378 & 0.9 & 0.7 \\
ORI \#1 & \num{6.0e-06} & 66.615 & 28.309 & 0.023 & 428 & 0.7 & 0.9 \\
ORI \#2 & \num{6.9e-06} & 66.989 & 31.854 & 0.027 & 279 & 0.9 & 1.3 \\
ORI \#3 & \num{2.8e-05} & 68.430 & 46.656 & 0.022 & 190 & 1.1 & 1.3 \\
ORI \#4 & \num{6.7e-06} & 68.650 & 34.867 & 0.034 & 229 & 1.4 & 2.4 \\
PER \#1 & \num{7.5e-06} & 59.927 & 30.263 & 0.024 & 232 & 1.2 & 2.0 \\
PER \#2 & \num{5.7e-06} & 59.900 & 39.738 & 0.028 & 900 & 1.1 & 0.8 \\
PER \#3 & \num{6.5e-06} & 61.181 & 45.743 & 0.026 & 274 & 1.8 & 2.7 \\
PER \#4 & \num{5.7e-06} & 60.775 & 41.341 & 0.025 & 207 & 1.5 & 2.9 \\
PER \#5 & \num{2.4e-06} & 57.590 & 27.559 & 0.032 & 214 & 1.0 & 2.7 \\
STA \#1 & \num{4.1e-06} & 26.757 & 49.030 & 0.029 & 960 & 1.0 & 0.8 \\
STA \#2 & \num{4.4e-06} & 28.265 & 39.157 & 0.021 & 773 & 1.3 & 1.1 \\
STA \#3 & \num{7.0e-06} & 29.866 & 46.149 & 0.030 & 774 & 1.6 & 1.2 \\
STA \#4 & \num{1.3e-05} & 29.841 & 36.278 & 0.017 & 679 & 0.4 & 0.3 \\
STA \#5 & \num{5.0e-06} & 33.180 & 44.523 & 0.023 & 515 & 0.6 & 0.7 \\
STA \#6 & \num{1.1e-05} & 29.357 & 37.752 & 0.031 & 516 & 1.6 & 1.4 \\
STA \#7 & \num{4.0e-06} & 28.769 & 45.329 & 0.031 & 559 & 1.1 & 1.2 \\
TAH \#1 & \num{1.0e-04} & 16.423 & 14.686 & 0.016 & 227 & 1.1 & 0.8 \\
TAH \#2 & \num{4.0e-05} & 16.287 & 15.812 & 0.024 & 292 & 2.1 & 1.7 \\
    \hline 
\end{tabular}
    }
    \label{tab:physical_properties}
\end{table}

\begin{landscape}
\begin{longtable}{r r r r r r r r r r r r r}
    \caption{Modelled erosion properties. The erosion parameters are the height of the erosion beginning $h_e$, the dynamic pressure at erosion beginning $p_{dyn,e}$, the erosion coefficient $\eta$, the upper grain mass limit $m_u$, the lower grain mass limit $m_l$, the grain mass index $s$, the height of the erosion change $h_{e2}$, the changed erosion coefficient $\eta_2$, the changed ablation coefficient $\sigma_2$, the changed bulk density $\rho_2$, the total mass loss prior to erosion change $\Delta m_e$ and the mass of the leading fragment $m_2$.}
    \label{tab:erosion_properties} \\
    \hline\hline 
Name & $h_e$ & $p_{dyn,e}$ & $\eta$ & $m_u$ & $m_l$ & $s$ & $h_{e2}$ & $\eta_2$ & $\sigma_2$ & $\rho_2$ & $\Delta m_e$ & $m_2$ \\
& (km) & (kPa) & (\SI{}{\kilogram \per \mega \joule}) & (kg) & (kg) & & (km) & (\SI{}{\kilogram \per \mega \joule}) & (\SI{}{\kilogram \per \mega \joule}) & (\SI{}{\kilo \gram \per \metre \cubed}) & (\%) & (kg) \\
\hline
\endfirsthead
\caption{continued.} \\
	\hline\hline 
	
Name & $h_e$ & $p_{dyn,e}$ & $\eta$ & $m_u$ & $m_l$ & $s$ & $h_{e2}$ & $\eta_2$ & $\sigma_2$ & $\rho_2$ & $\Delta m_e$ & $m_2$ \\
 & (km) & (kPa) & (\SI{}{\kilogram \per \mega \joule}) & (kg) & (kg) &  & (km) & (\SI{}{\kilogram \per \mega \joule}) & (\SI{}{\kilogram \per \mega \joule}) & (\SI{}{\kilo \gram \per \metre \cubed}) & (\%) & (kg) \\
\endhead 
\hline
\endfoot
AUR \#1 & 111.76 & 0.276216 & 0.144 & \num{3e-07}   & \num{3e-12}   & 2.35 & -      & -     & -     & -    & -    & -           \\
CAP \#1 & 99.61 & 0.312081 & 0.226 & \num{9e-11}   & \num{3e-11}   & 2.31 &  91.59 & 0.064 & 0.010 & 1170 & 100.0 & \num{1e-11} \\
CAP \#2 & 99.99 & 0.367011 & 0.105 & \num{9e-08}   & \num{1e-10}   & 2.38 &  94.70 & 0.400 & 0.037 &  794 & 100.0 & \num{7e-12} \\
CAP \#3 & 98.92 & 0.399987 & 0.261 & \num{8e-10}   & \num{1e-10}   & 2.69 &  90.58 & 0.330 & 0.020 &  527 & 90.1 & \num{1e-06} \\
ELY \#1 & 107.50 & 0.264116 & 0.430 & \num{4e-08}   & \num{1e-11}   & 2.35 & -      & -     & -     & -    & -    & -           \\
ELY \#2 & 106.53 & 0.283041 & 0.340 & \num{1e-07}   & \num{1e-11}   & 2.40 & -      & -     & -     & -    & -    & -           \\
ETA \#1 & 116.50 & 0.146492 & 0.600 & \num{6e-10}   & \num{3e-11}   & 1.50 & -      & -     & -     & -    & -    & -           \\
ETA \#2 & 113.22 & 0.300823 & 0.095 & \num{5e-10}   & \num{1e-10}   & 2.77 & 107.15 & 0.244 & 0.024 &  258 & 94.0 & \num{7e-07} \\
ETA \#3 & 115.77 & 0.162777 & 0.307 & \num{4e-10}   & \num{2e-11}   & 1.53 & -      & -     & -     & -    & -    & -           \\
ETA \#4 & 114.50 & 0.192403 & 0.356 & \num{2e-10}   & \num{2e-10}   & 1.34 & -      & -     & -     & -    & -    & -           \\
ETA \#5 & 113.70 & 0.227164 & 0.273 & \num{7e-10}   & \num{2e-11}   & 1.30 & -      & -     & -     & -    & -    & -           \\
GEM \#1 & 98.54 & 0.625700 & 0.203 & \num{8e-08}   & \num{5e-10}   & 1.56 &  93.61 & 1.043 & 0.013 & 1123 & 100.0 & \num{1e-10} \\
GEM \#2 & 108.45 & 2.674469 & 0.246 & \num{1e-07}   & \num{9e-11}   & 1.73 &  92.47 & 0.005 & 0.025 & 2295 & 100.0 & \num{1e-09} \\
GEM \#3 & 100.96 & 0.625700 & 0.241 & \num{1e-07}   & \num{1e-10}   & 1.75 &  95.96 & 0.764 & 0.029 &  916 & 100.0 & \num{1e-10} \\
GEM \#4 & 97.51 & 0.793490 & 0.072 & \num{5e-07}   & \num{5e-11}   & 2.65 &  91.34 & 0.020 & 0.029 & 1202 & 13.4 & \num{2e-06} \\
GEM \#5 & 101.40 & 1.890779 & 0.093 & \num{4e-08}   & \num{1e-09}   & 1.56 &  93.88 & 0.352 & 0.023 & 1399 & 100.0 & \num{1e-14} \\
JPE \#1 & 114.10 & 0.248769 & 0.159 & \num{2e-09}   & \num{9e-12}   & 2.28 &  99.69 & 0.010 & 0.021 &  284 & 100.0 & \num{9e-12} \\
LEO \#1 & 115.74 & 0.178062 & 1.035 & \num{9e-09}   & \num{9e-12}   & 2.21 & -      & -     & -     & -    & -    & -           \\
LYR \#1 & 106.02 & 0.407315 & 0.198 & \num{1e-08}   & \num{2e-10}   & 2.54 & 100.09 & 0.015 & 0.019 & 2966 & 88.3 & \num{3e-07} \\
LYR \#2 & 115.42 & 0.507783 & 0.094 & \num{5e-11}   & \num{1e-12}   & 2.41 & -      & -     & -     & -    & -    & -           \\
MON \#1 & 113.46 & 0.388371 & 0.039 & \num{5e-09}   & \num{1e-11}   & 2.01 &  90.16 & 0.006 & 0.019 &  668 & 100.0 & \num{1e-10} \\
NTA \#1 & 108.67 & 0.518893 & 0.063 & \num{1e-07}   & \num{1e-11}   & 2.36 & -      & -     & -     & -    & -    & -           \\
NTA \#2 & 104.52 & 0.234843 & 0.064 & \num{1e-07}   & \num{1e-12}   & 1.91 &  98.78 & 0.030 & 0.033 &  378 & 100.0 & \num{5e-10} \\
ORI \#1 & 118.17 & 0.206496 & 0.210 & \num{5e-08}   & \num{1e-11}   & 2.47 & -      & -     & -     & -    & -    & -           \\
ORI \#2 & 116.85 & 0.187465 & 0.322 & \num{8e-09}   & \num{1e-11}   & 1.92 & 104.80 & 0.102 & 0.027 &  279 & 100.0 & \num{3e-13} \\
ORI \#3 & 116.70 & 0.374453 & 0.229 & \num{9e-09}   & \num{9e-12}   & 2.20 & 113.48 & 0.767 & 0.022 &  190 & 100.0 & \num{2e-13} \\
ORI \#4 & 112.88 & 1.244747 & 0.156 & \num{6e-10}   & \num{5e-12}   & 1.98 & 105.96 & 0.216 & 0.010 &  229 & 100.0 & \num{4e-11} \\
PER \#1 & 110.66 & 0.262668 & 0.283 & \num{1e-08}   & \num{5e-12}   & 2.32 & 104.58 & 0.078 & 0.024 &  232 & 100.0 & \num{2e-13} \\
PER \#2 & 113.22 & 0.175846 & 0.450 & \num{3e-09}   & \num{7e-12}   & 2.40 & 103.50 & 0.035 & 0.028 &  900 & 100.0 & \num{3e-13} \\
PER \#3 & 110.37 & 0.447796 & 0.500 & \num{4e-09}   & \num{2e-10}   & 1.30 & 105.00 & 0.200 & 0.026 &  274 & 100.0 & \num{2e-14} \\
PER \#4 & 110.82 & 0.157552 & 0.363 & \num{2e-07}   & \num{1e-11}   & 2.40 & 104.79 & 0.001 & 0.025 &  207 & 97.8 & \num{1e-07} \\
PER \#5 & 110.96 & 0.282627 & 0.105 & \num{6e-08}   & \num{1e-11}   & 2.36 & -      & -     & -     & -    & -    & -           \\
STA \#1 & 103.32 & 0.217121 & 0.064 & \num{2e-10}   & \num{6e-11}   & 1.35 &  86.41 & 0.330 & 0.029 &  960 & 99.9 & \num{5e-09} \\
STA \#2 & 102.05 & 0.446384 & 0.054 & \num{9e-10}   & \num{1e-11}   & 1.72 & -      & -     & -     & -    & -    & -           \\
STA \#3 & 102.32 & 0.501228 & 0.059 & \num{1e-07}   & \num{1e-10}   & 2.64 & -      & -     & -     & -    & -    & -           \\
STA \#4 & 110.00 & 0.292474 & 0.058 & \num{4e-11}   & \num{1e-11}   & 2.50 &  89.58 & 0.097 & 0.017 &  679 & 83.1 & \num{2e-06} \\
STA \#5 & 108.72 & 0.374596 & 0.062 & \num{5e-08}   & \num{1e-12}   & 2.09 &  88.45 & 0.330 & 0.023 &  515 & 99.6 & \num{2e-08} \\
STA \#6 & 100.74 & 0.507980 & 0.077 & \num{1e-07}   & \num{9e-12}   & 2.17 &  89.00 & 0.327 & 0.031 &  516 & 100.0 & \num{2e-14} \\
STA \#7 & 102.83 & 0.394352 & 0.053 & \num{2e-07}   & \num{1e-11}   & 2.27 & -      & -     & -     & -    & -    & -           \\
TAH \#1 & 94.04 & 0.564395 & 0.057 & \num{9e-10}   & \num{6e-11}   & 2.71 &  88.50 & 0.120 & 0.016 &  227 & 96.9 & \num{3e-06} \\
TAH \#2 & 90.88 & 0.496037 & 0.124 & \num{2e-10}   & \num{1e-11}   & 2.07 &  87.80 & 0.700 & 0.024 &  292 & 25.1 & \num{3e-05} \\
\end{longtable}
\end{landscape}

\section{Supplementary Info - Modelled Meteor Lightcurves, Lags and Wakes}

\begin{figure}[H]
    \centering
    \includegraphics[width=\linewidth]{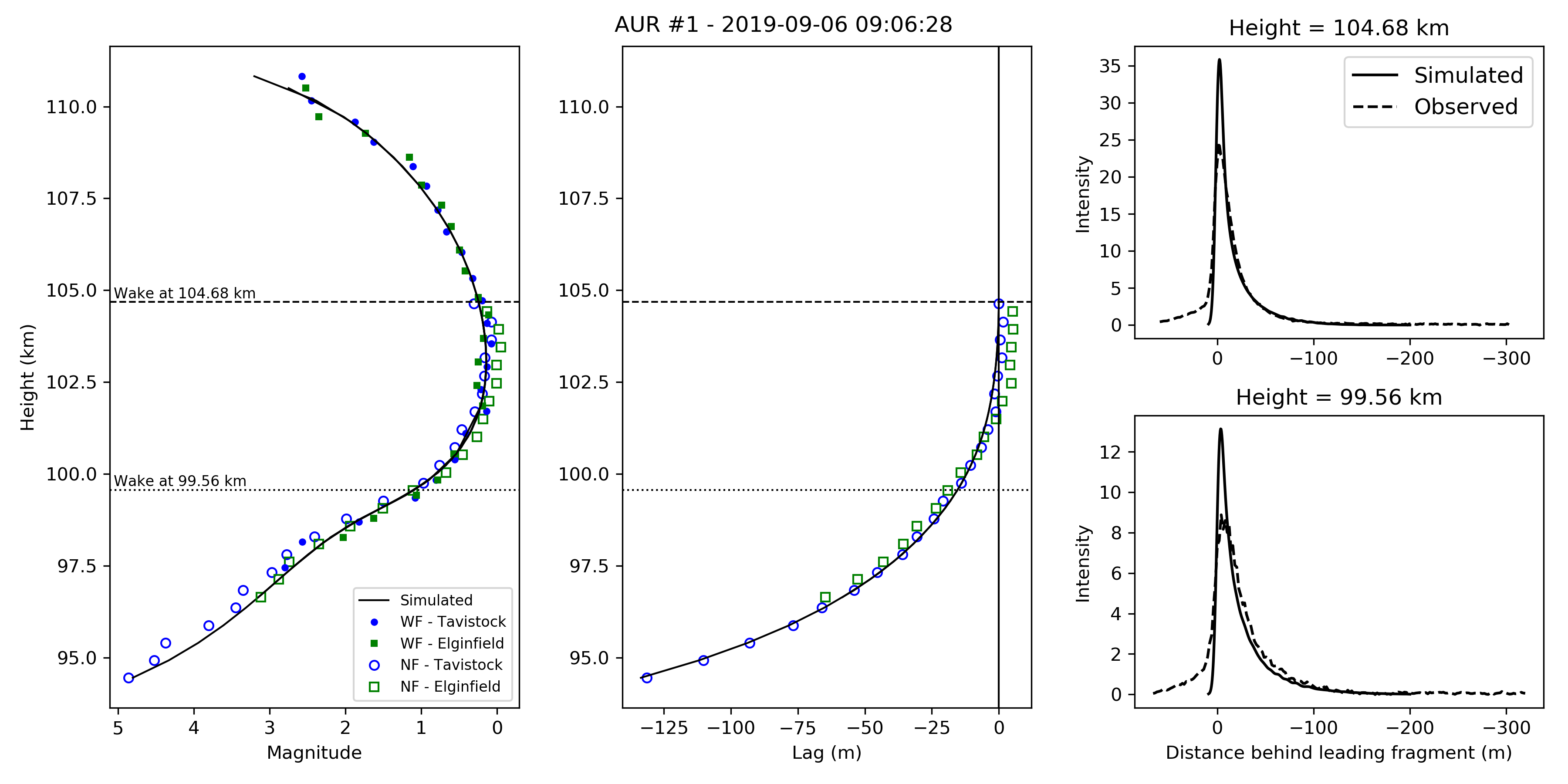}
    \caption{AUR1 2019-09-06 09:06:28}
\end{figure}
\begin{figure}
    \centering
    \includegraphics[width=\linewidth]{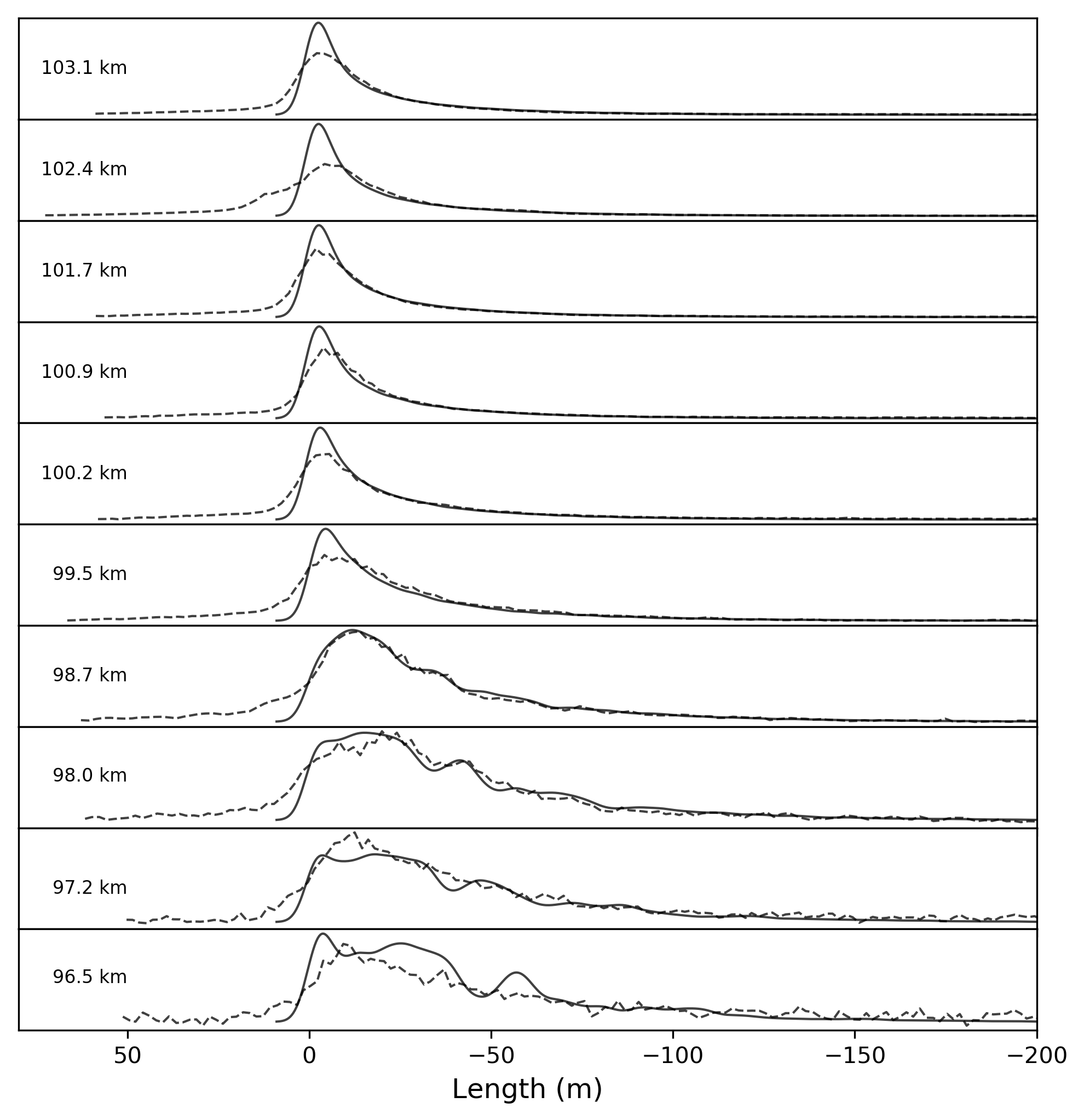}
    \caption{AUR1 2019-09-06 09:06:28}
\end{figure}

\begin{figure}
    \centering
    \includegraphics[width=\linewidth]{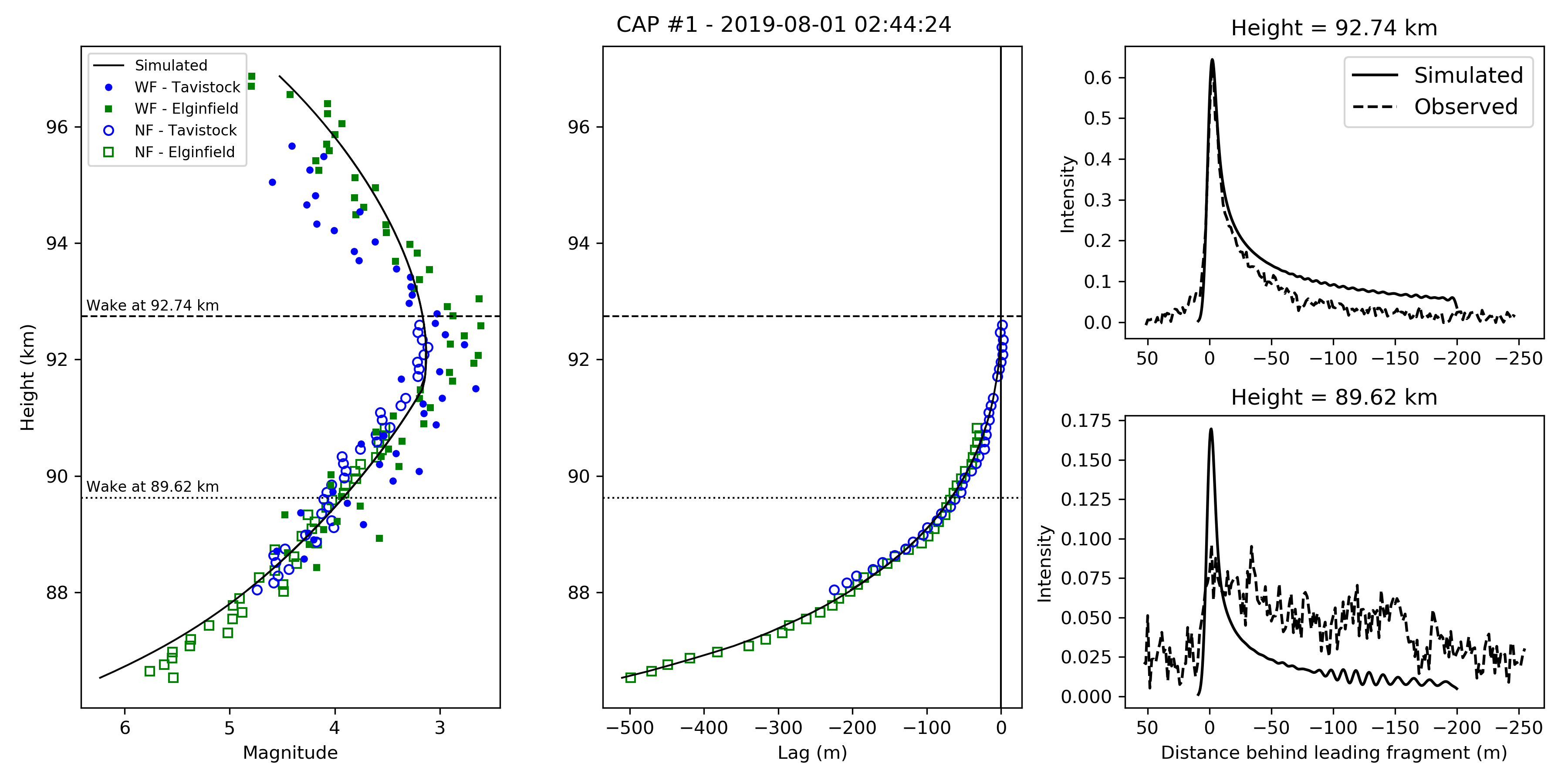}
    \caption{CAP1 2019-08-01 02:44:24}
\end{figure}
\begin{figure}
    \centering
    \includegraphics[width=\linewidth]{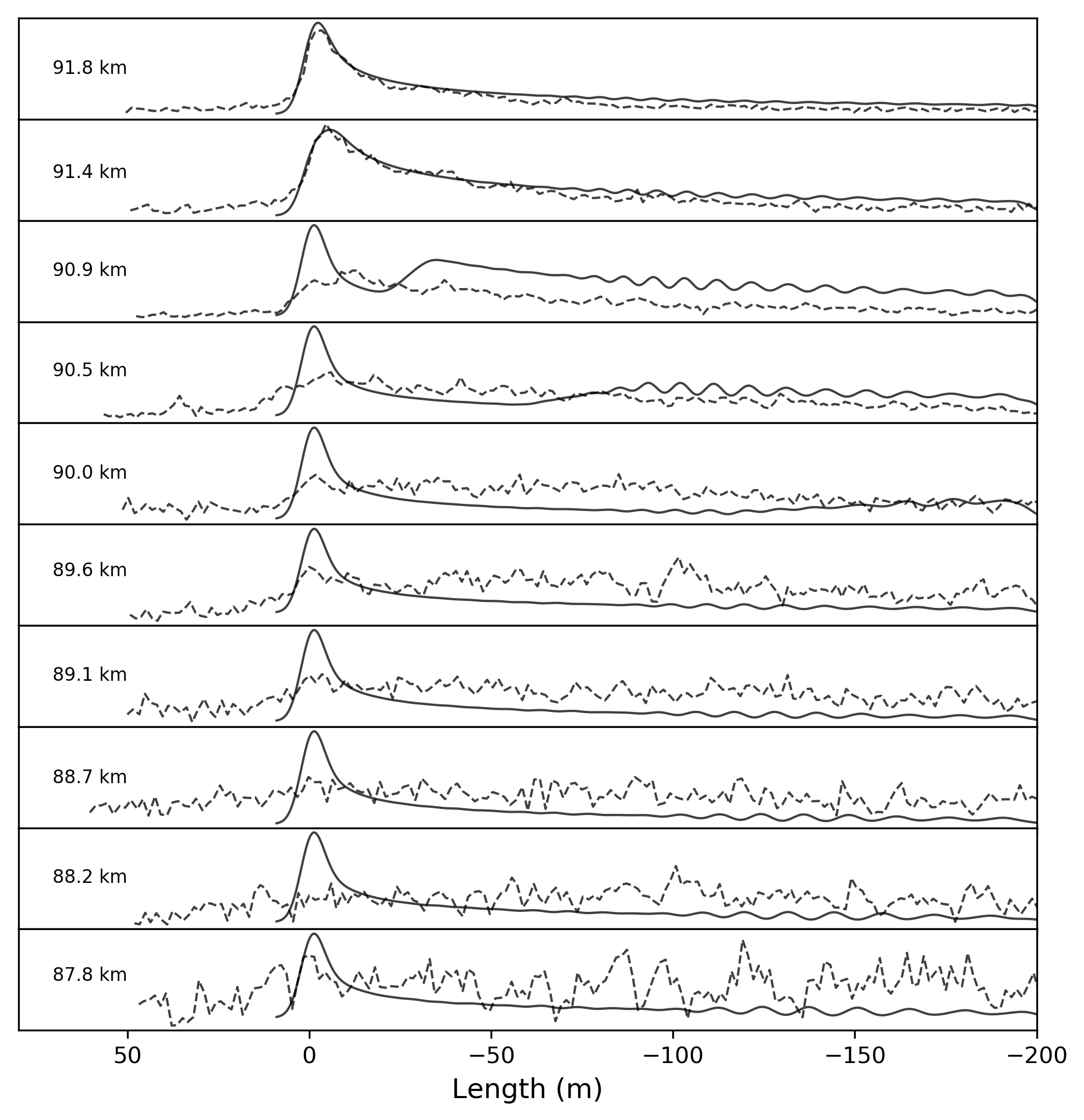}
    \caption{CAP1 2019-08-01 02:44:24}
\end{figure}

\begin{figure}
    \centering
    \includegraphics[width=\linewidth]{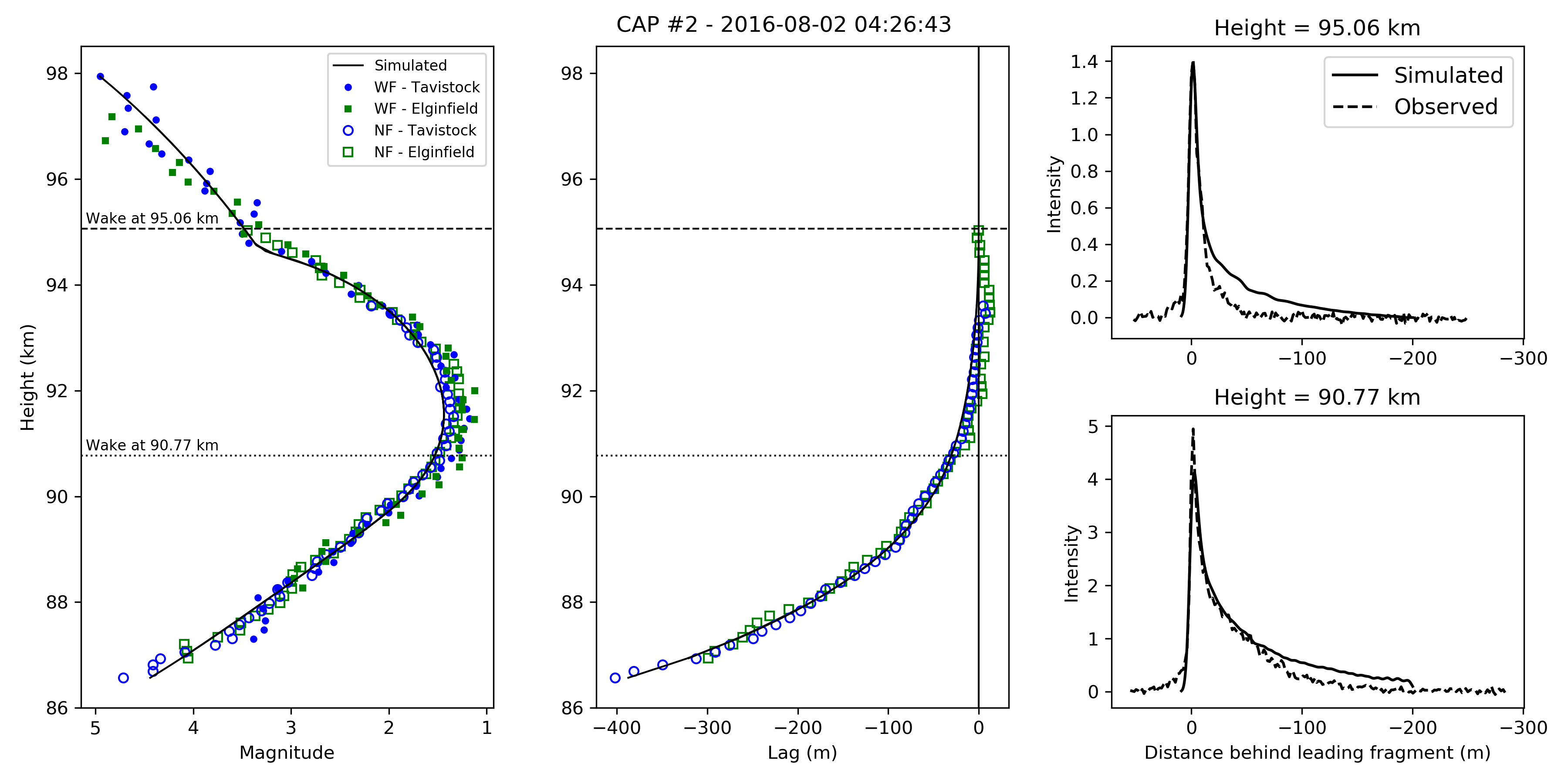}
    \caption{CAP2 2016-08-02 04:26:43}
\end{figure}
\begin{figure}
    \centering
    \includegraphics[width=\linewidth]{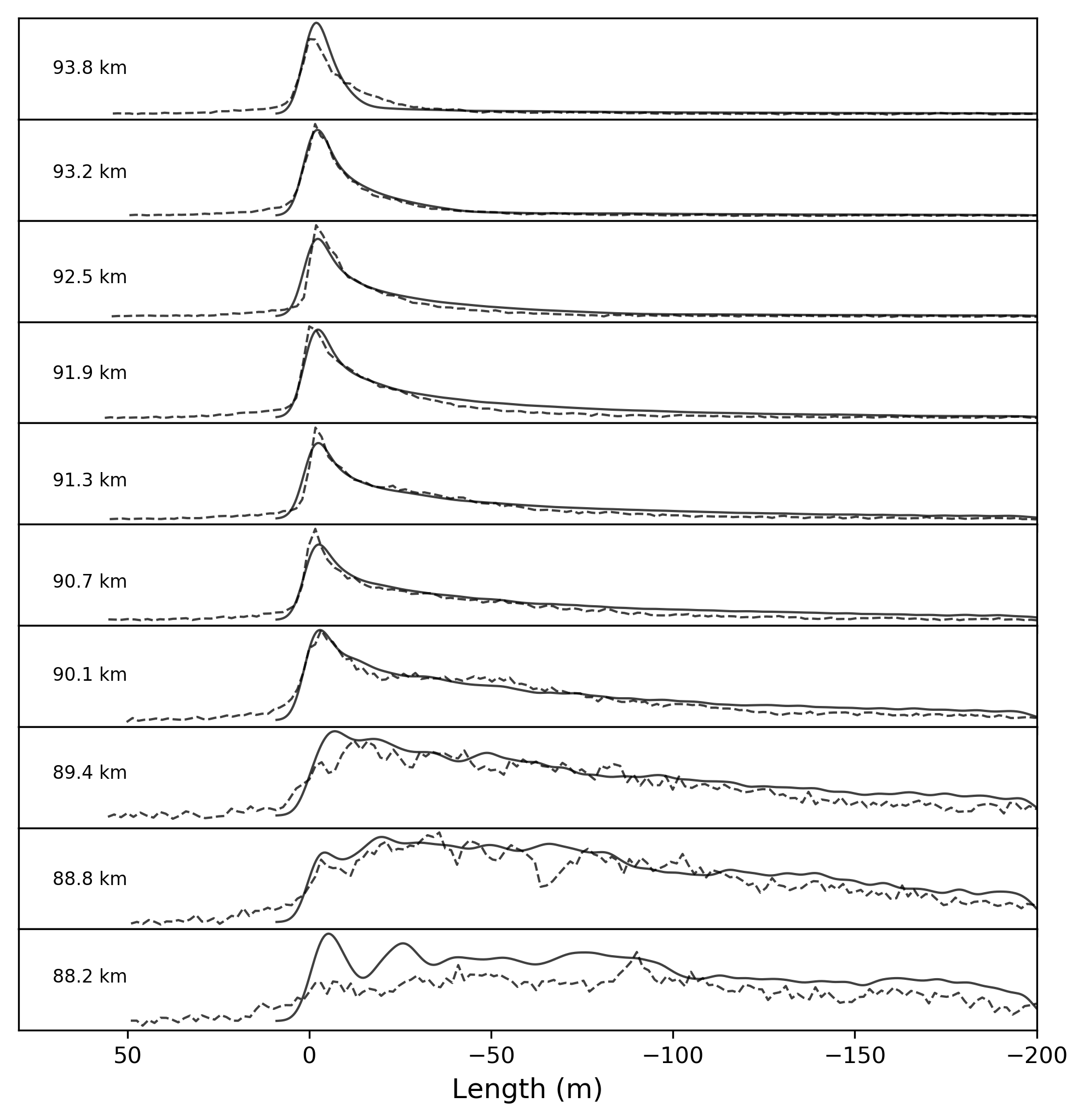}
    \caption{CAP2 2016-08-02 04:26:43}
\end{figure}

\begin{figure}
    \centering
    \includegraphics[width=\linewidth]{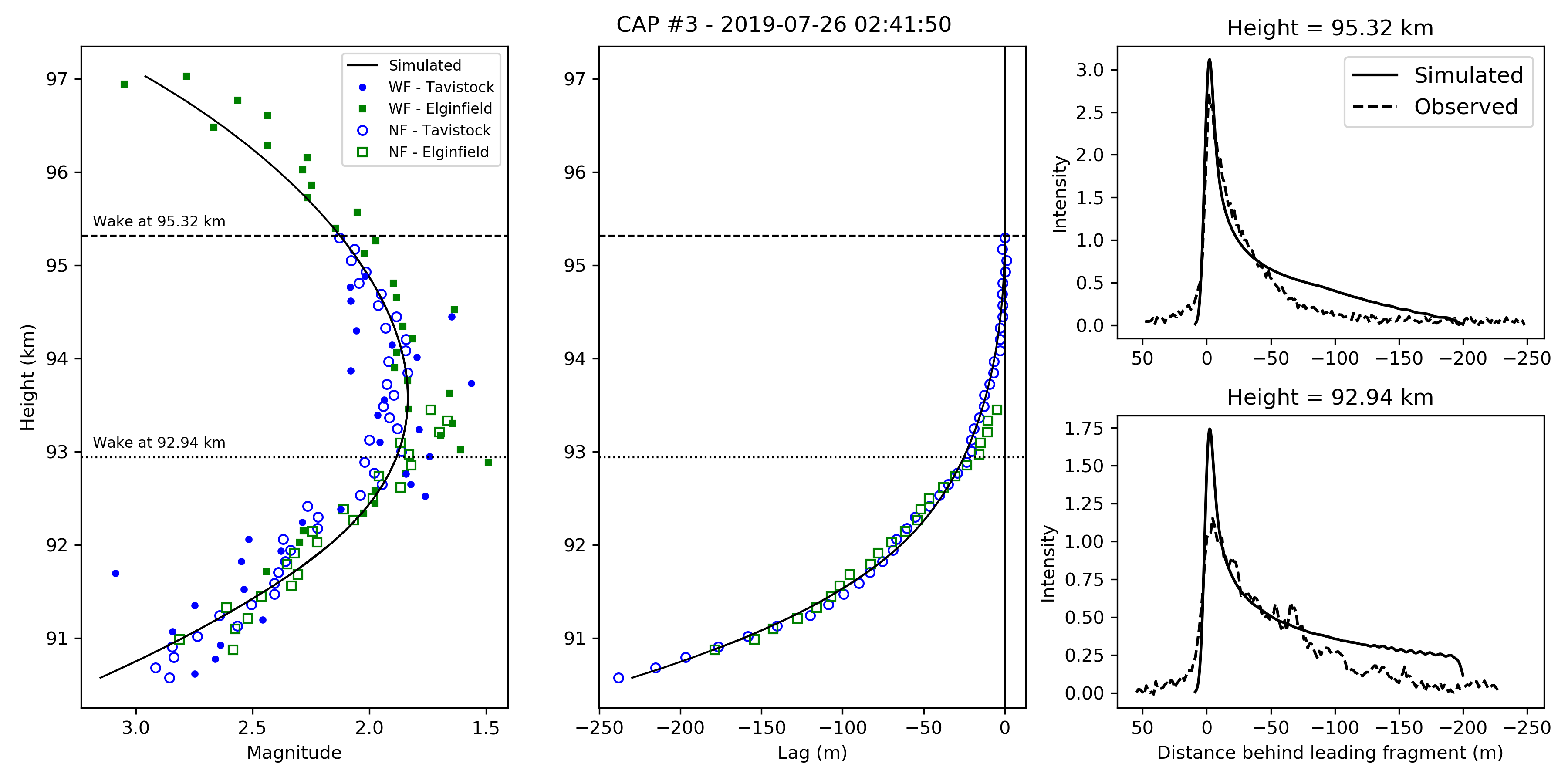}
    \caption{CAP3 2019-07-26 02:41:50}
\end{figure}
\begin{figure}
    \centering
    \includegraphics[width=\linewidth]{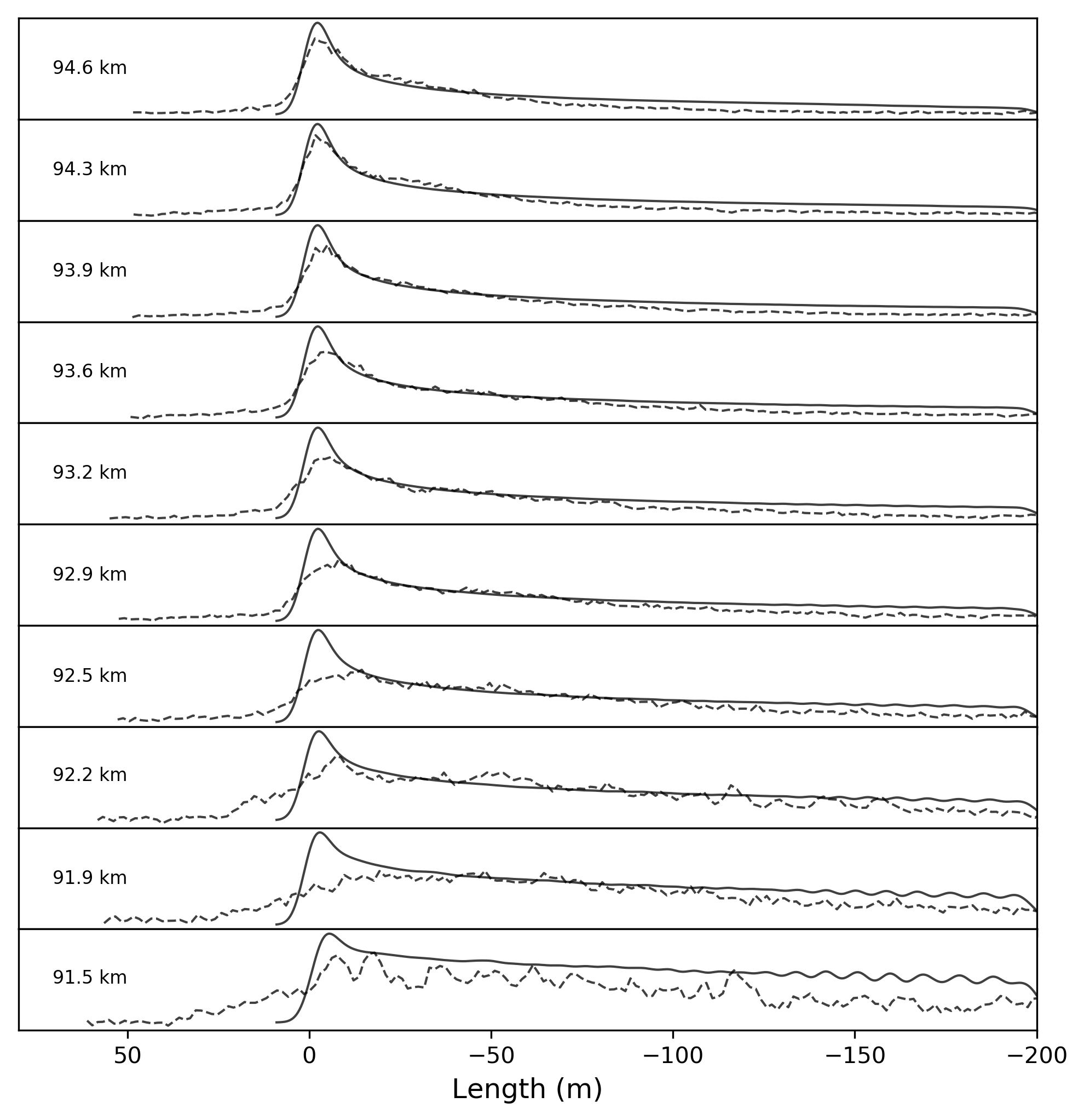}
    \caption{CAP3 2019-07-26 02:41:50}
\end{figure}

\begin{figure}
    \centering
    \includegraphics[width=\linewidth]{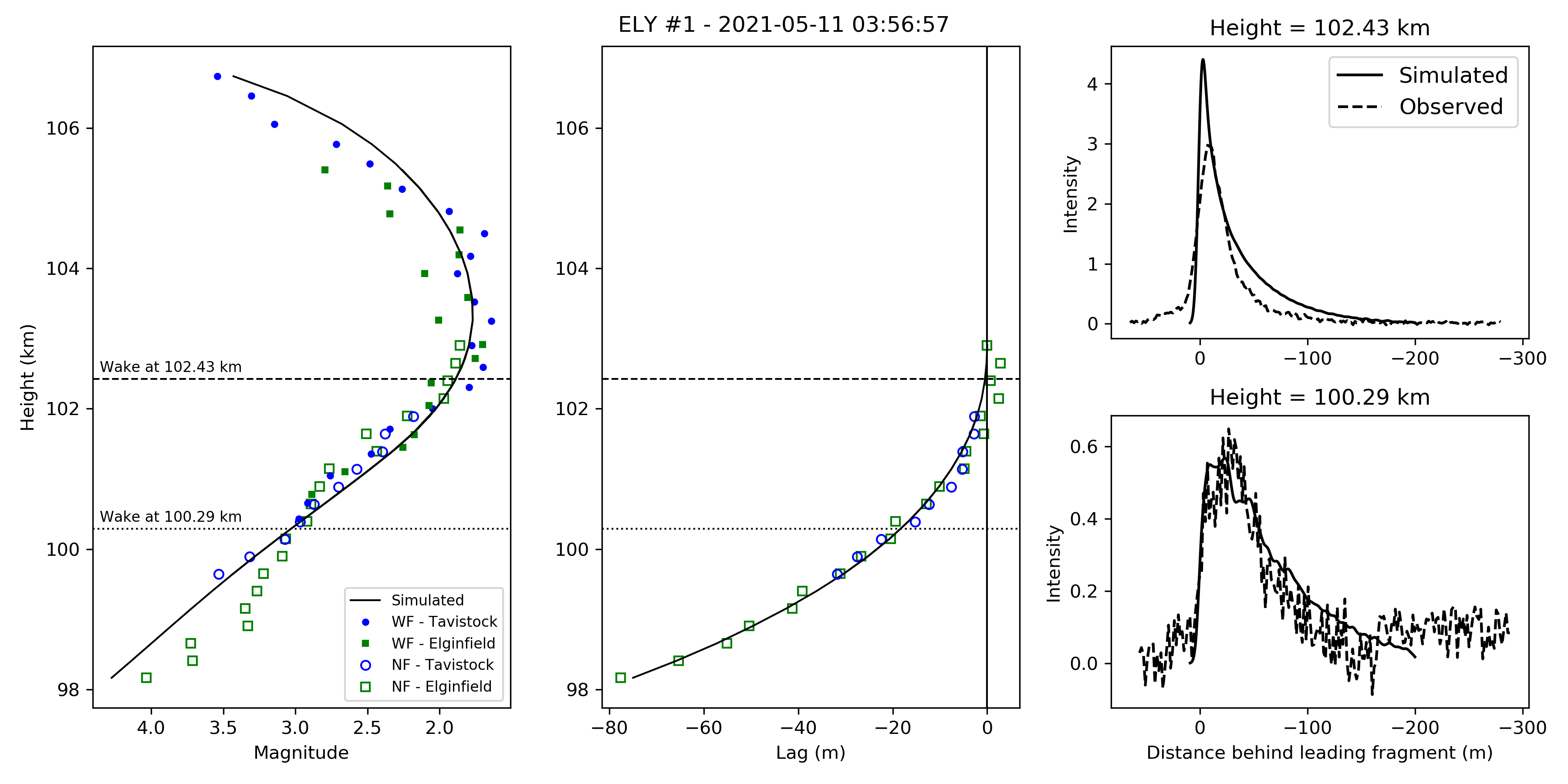}
    \caption{ELY1 2021-05-11 03:56:57}
\end{figure}
\begin{figure}
    \centering
    \includegraphics[width=\linewidth]{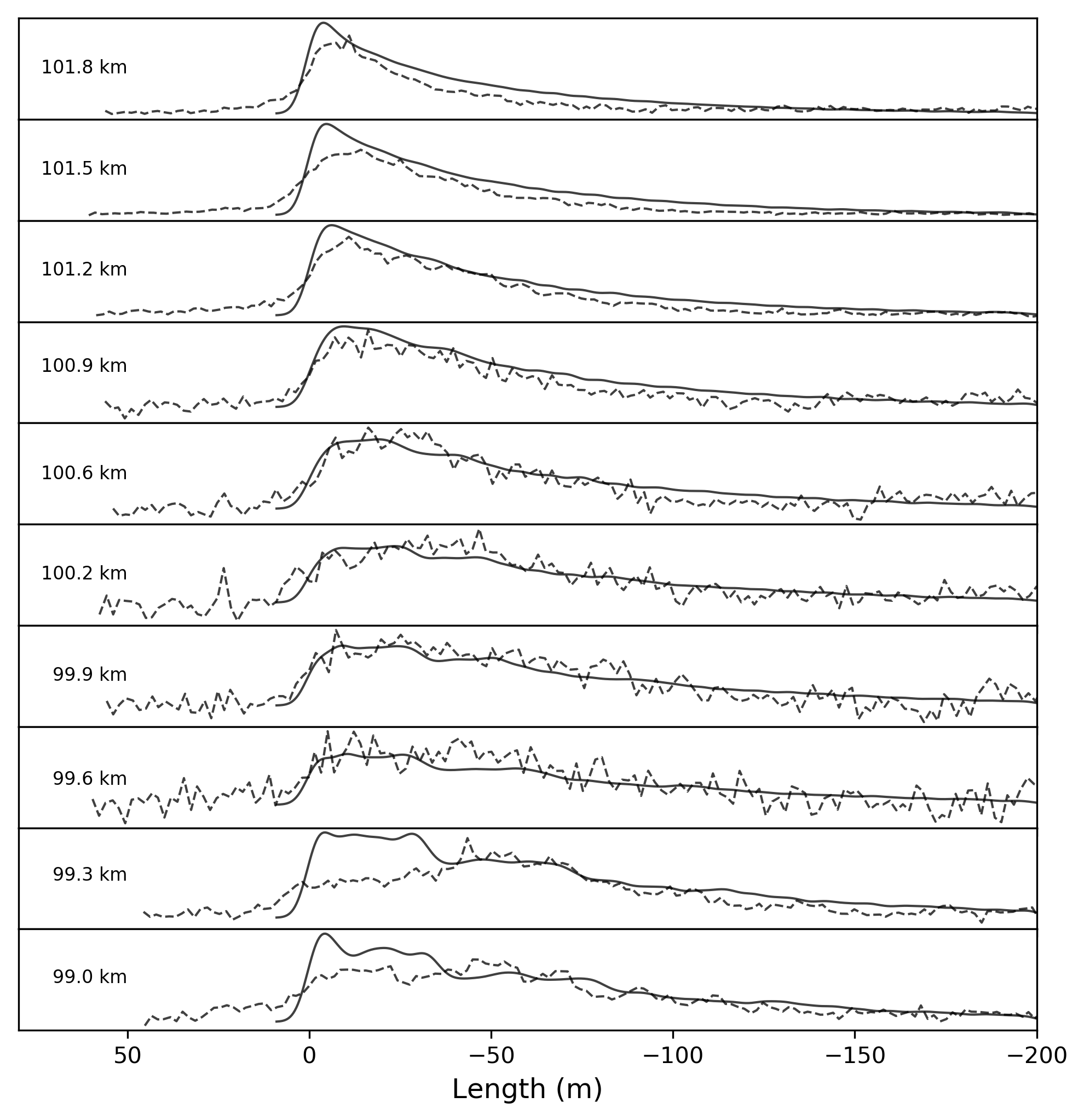}
    \caption{ELY1 2021-05-11 03:56:57}
\end{figure}

\begin{figure}
    \centering
    \includegraphics[width=\linewidth]{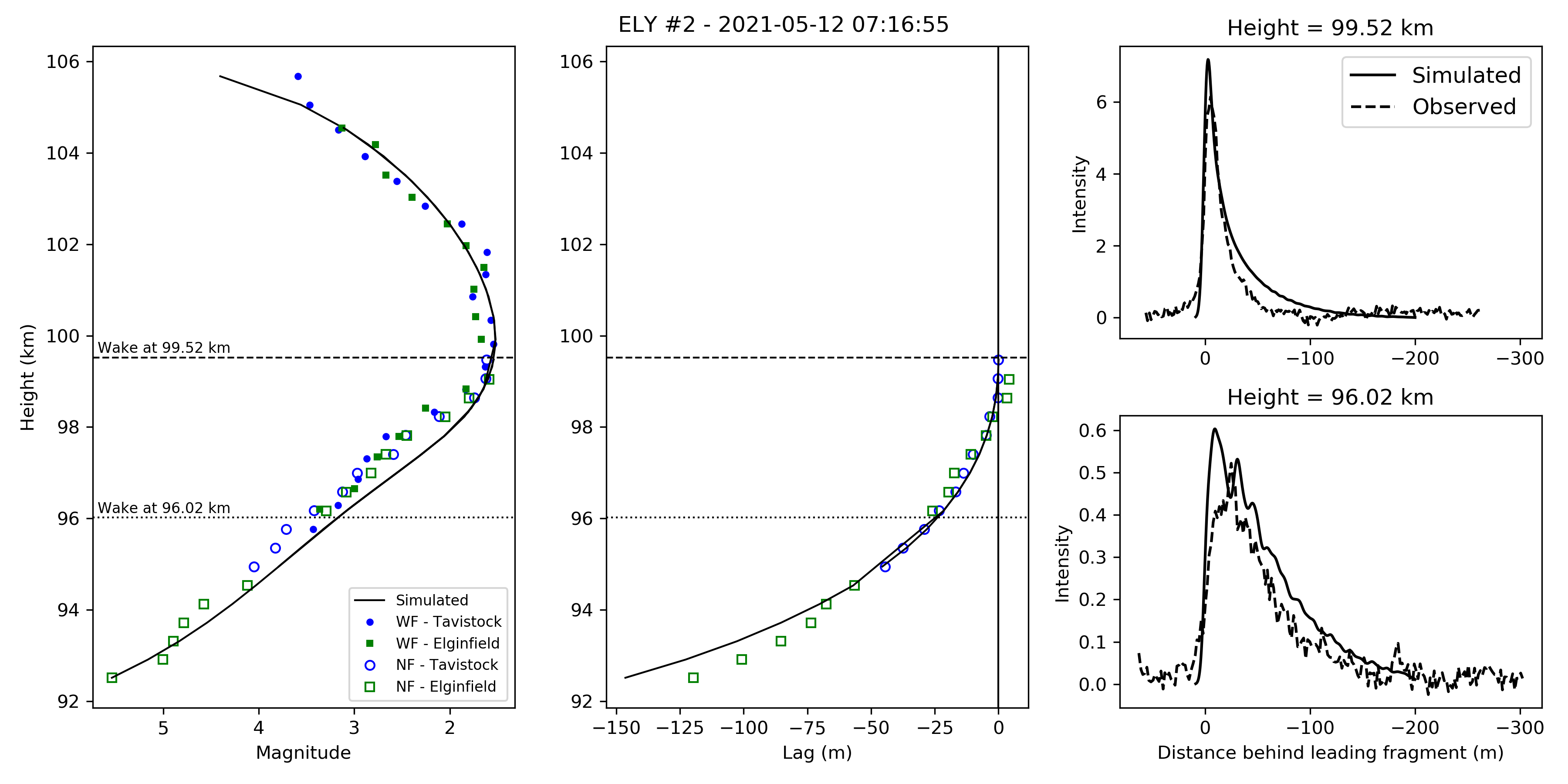}
    \caption{ELY2 2021-05-12 07:16:55}
\end{figure}
\begin{figure}
    \centering
    \includegraphics[width=\linewidth]{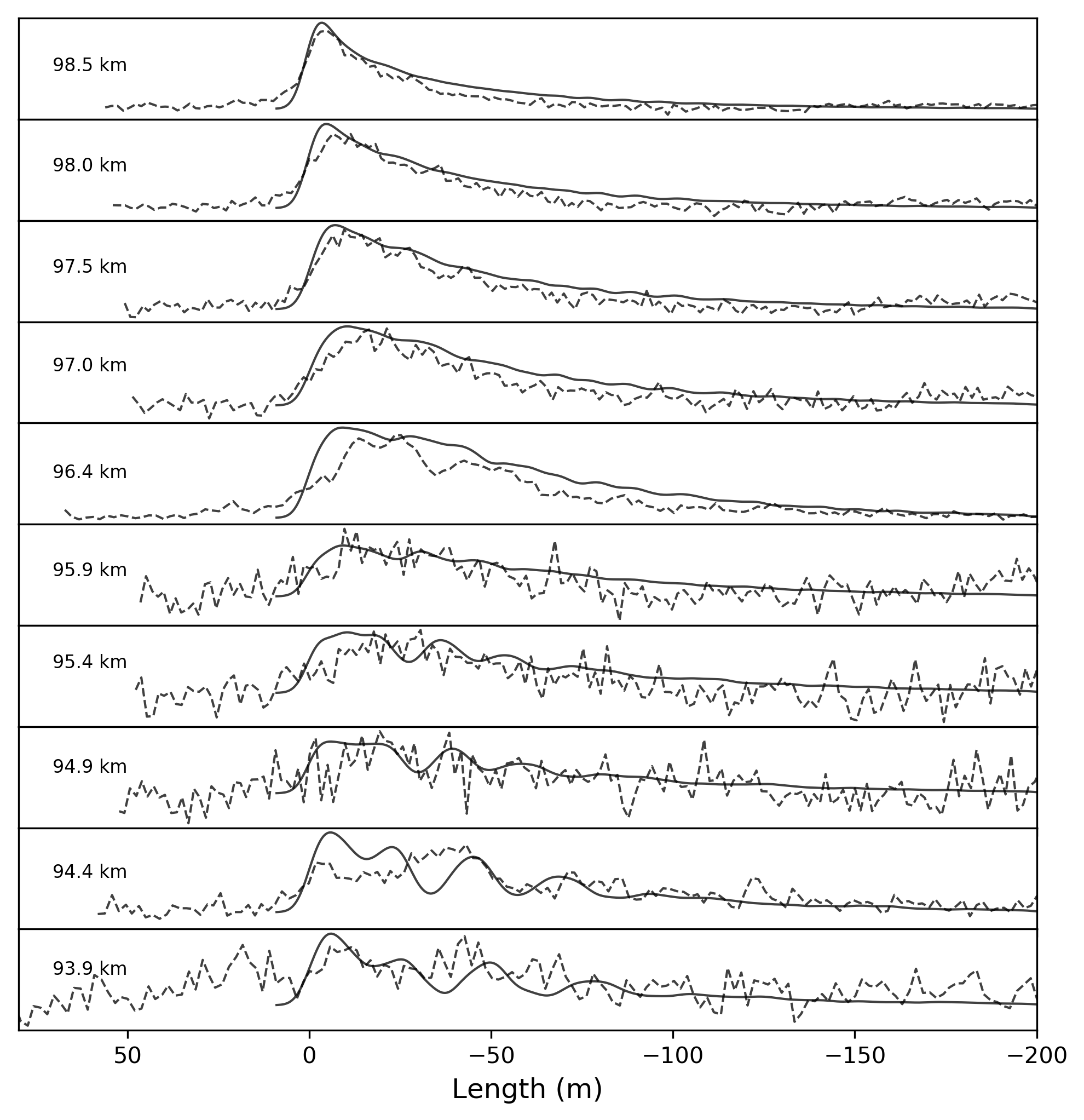}
    \caption{ELY2 2021-05-12 07:16:55}
\end{figure}

\begin{figure}
    \centering
    \includegraphics[width=\linewidth]{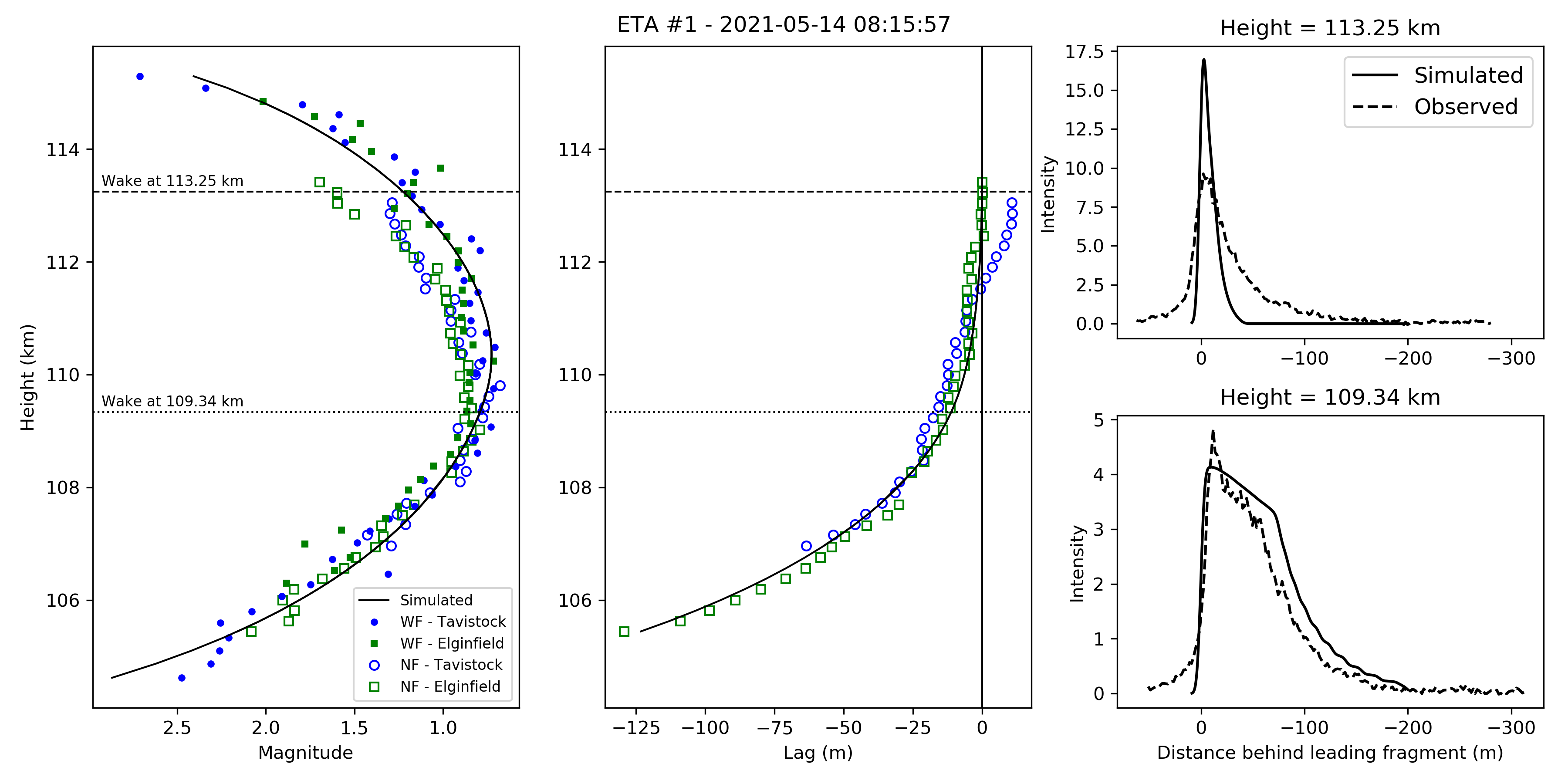}
    \caption{ETA1 2021-05-14 08:15:57}
\end{figure}
\begin{figure}
    \centering
    \includegraphics[width=\linewidth]{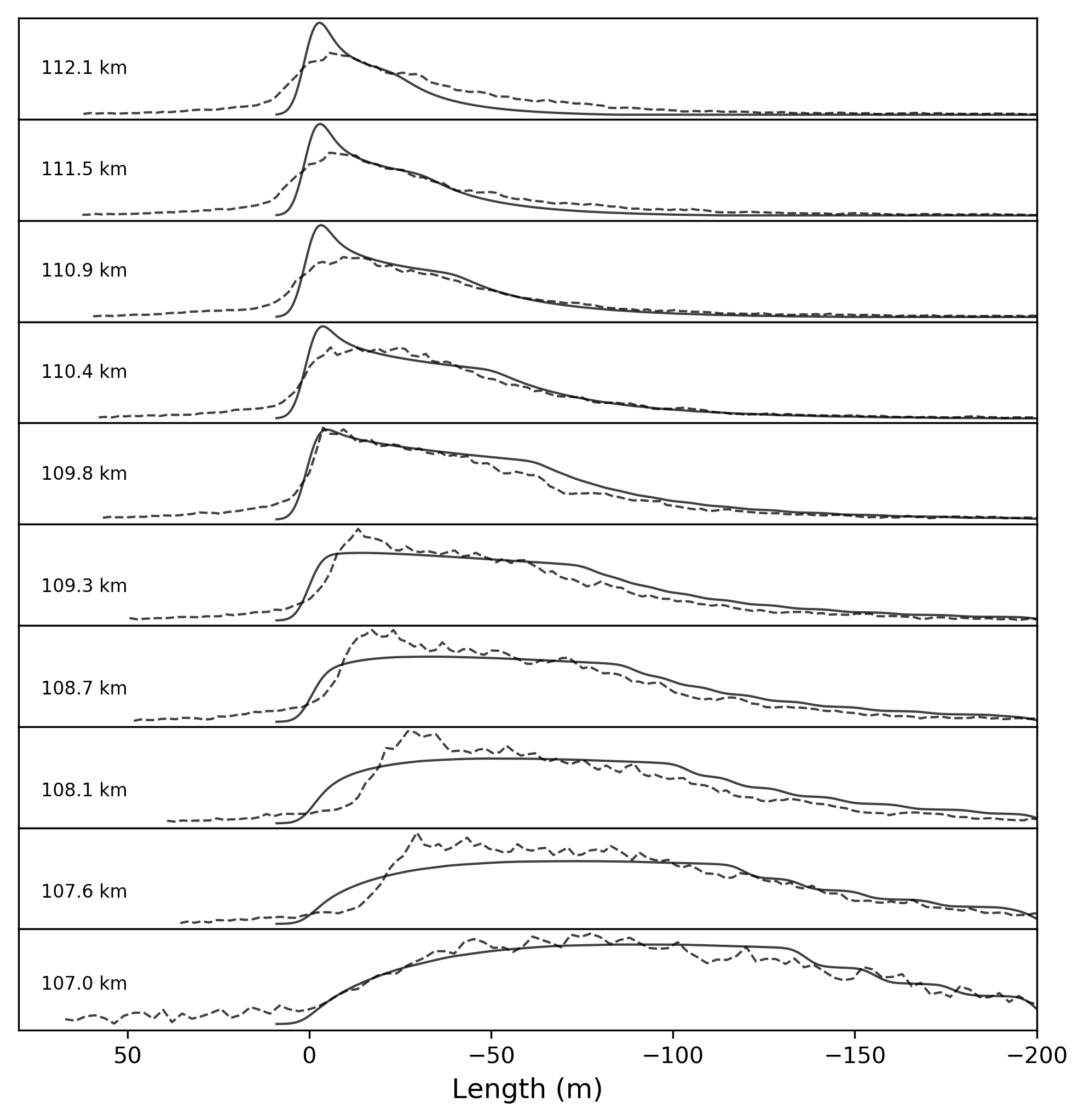}
    \caption{ETA1 2021-05-14 08:15:57}
\end{figure}

\begin{figure}
    \centering
    \includegraphics[width=\linewidth]{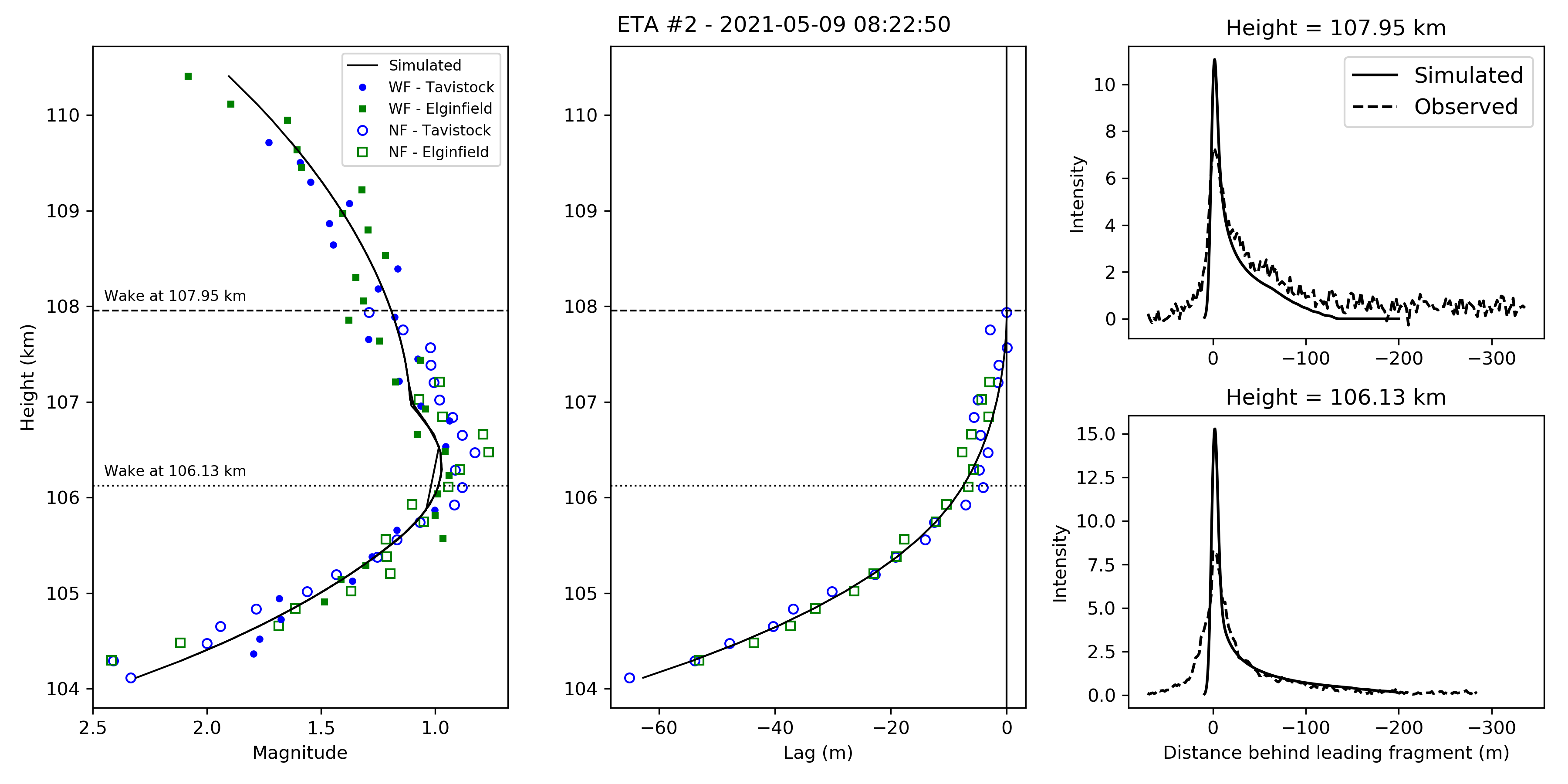}
    \caption{ETA2 2021-05-09 08:22:50}
\end{figure}
\begin{figure}
    \centering
    \includegraphics[width=\linewidth]{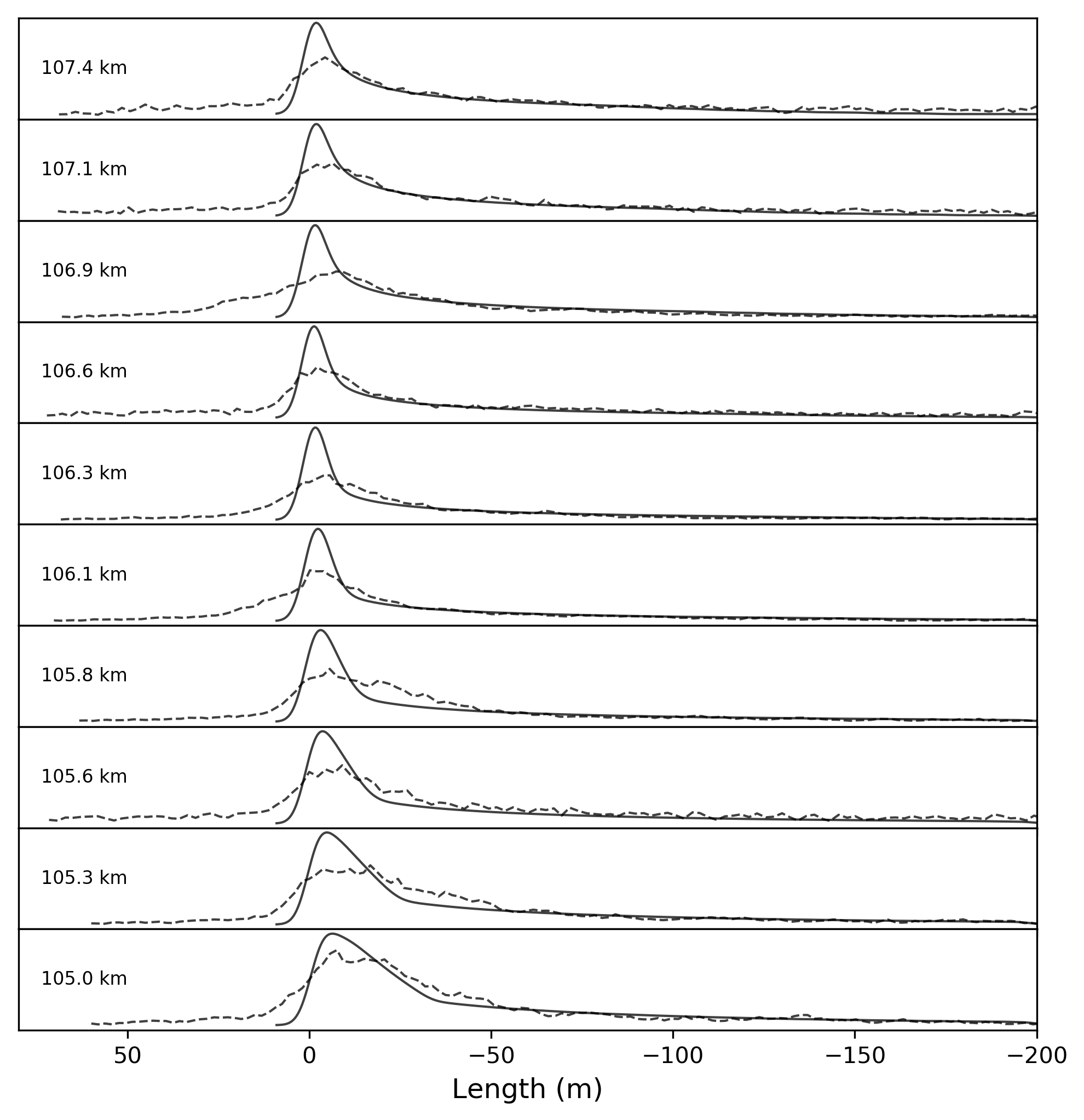}
    \caption{ETA2 2021-05-09 08:22:50}
\end{figure}

\begin{figure}
    \centering
    \includegraphics[width=\linewidth]{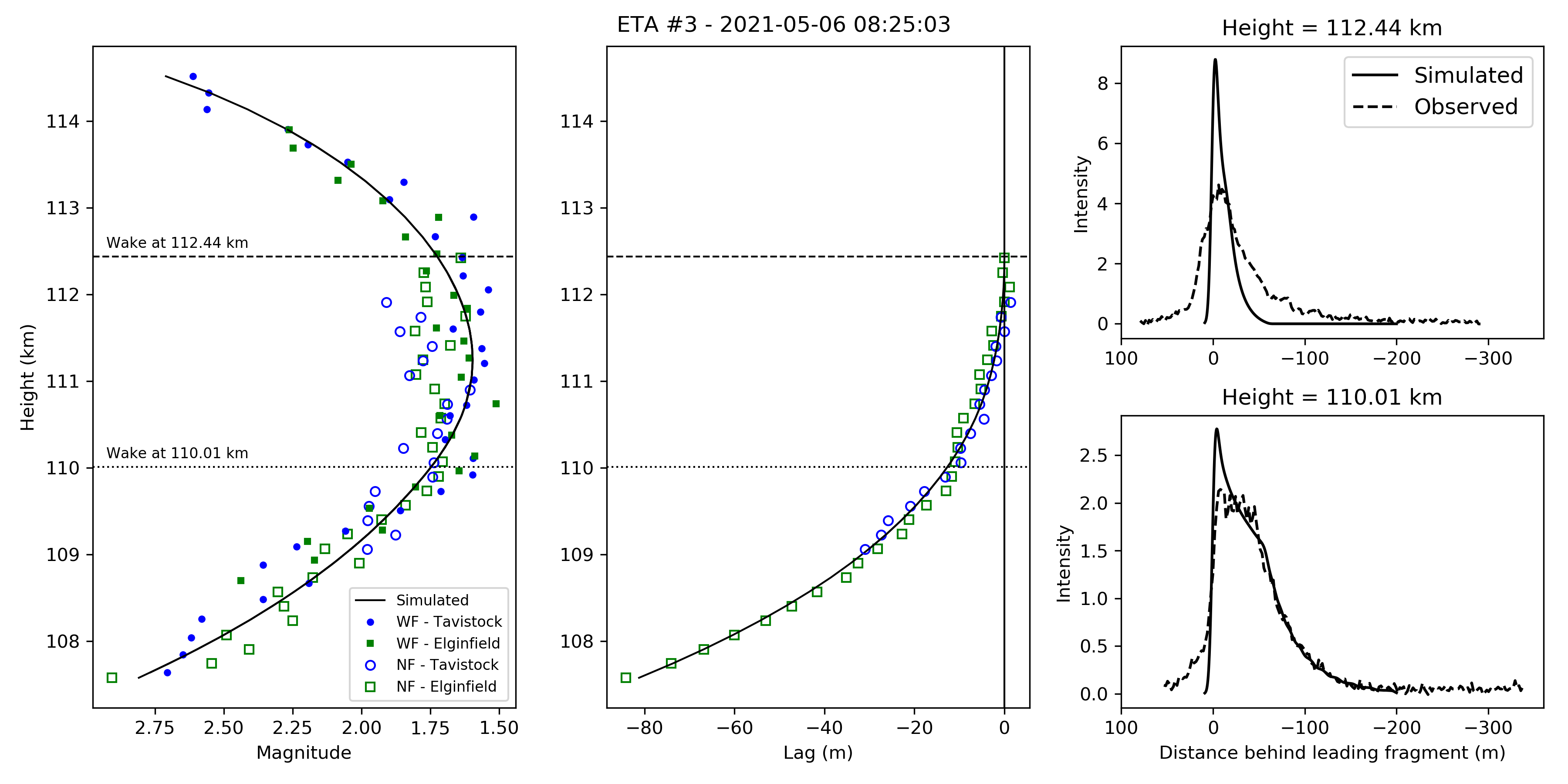}
    \caption{ETA3 2021-05-06 08:25:03}
\end{figure}
\begin{figure}
    \centering
    \includegraphics[width=\linewidth]{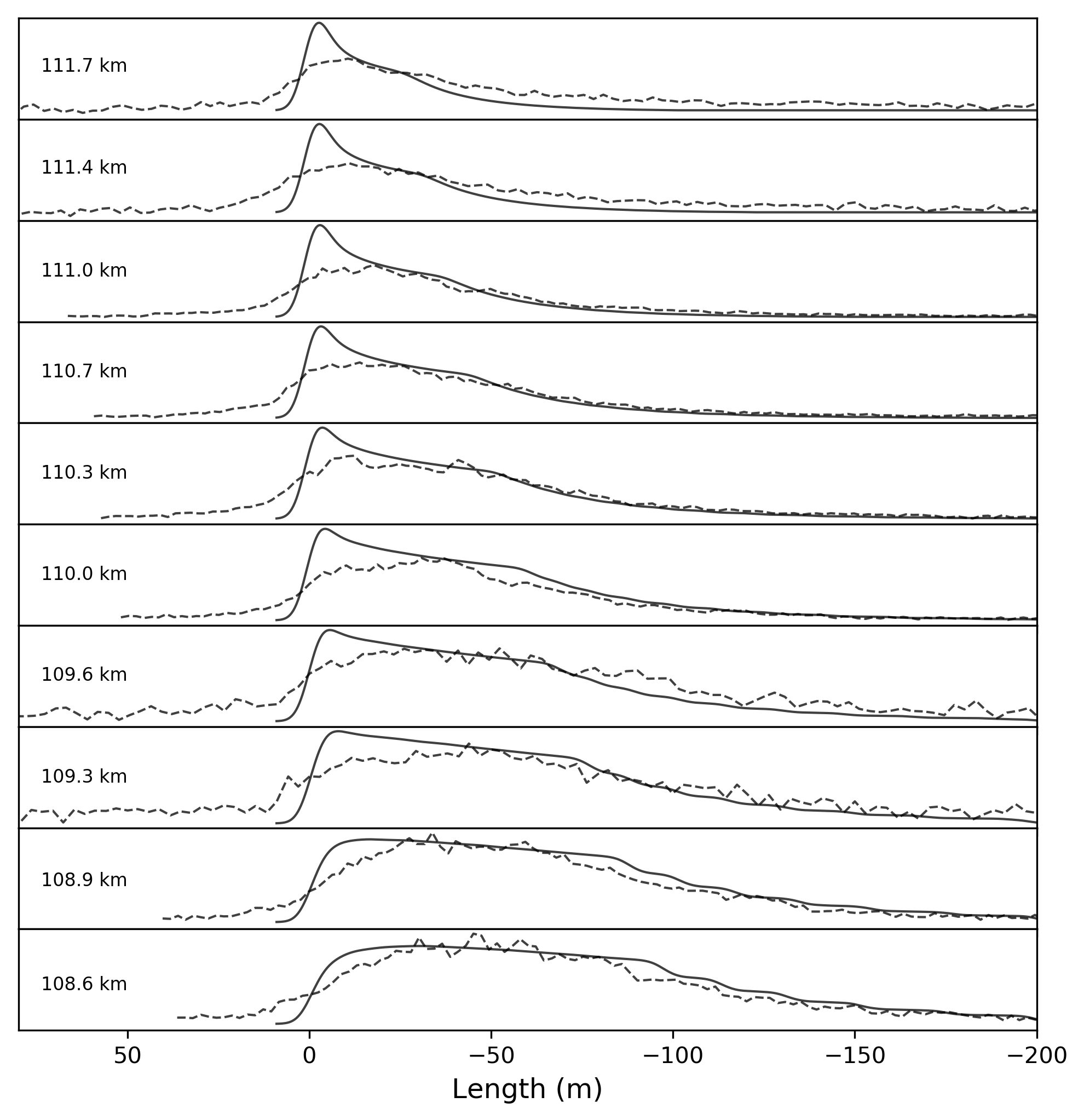}
    \caption{ETA3 2021-05-06 08:25:03}
\end{figure}

\begin{figure}
    \centering
    \includegraphics[width=\linewidth]{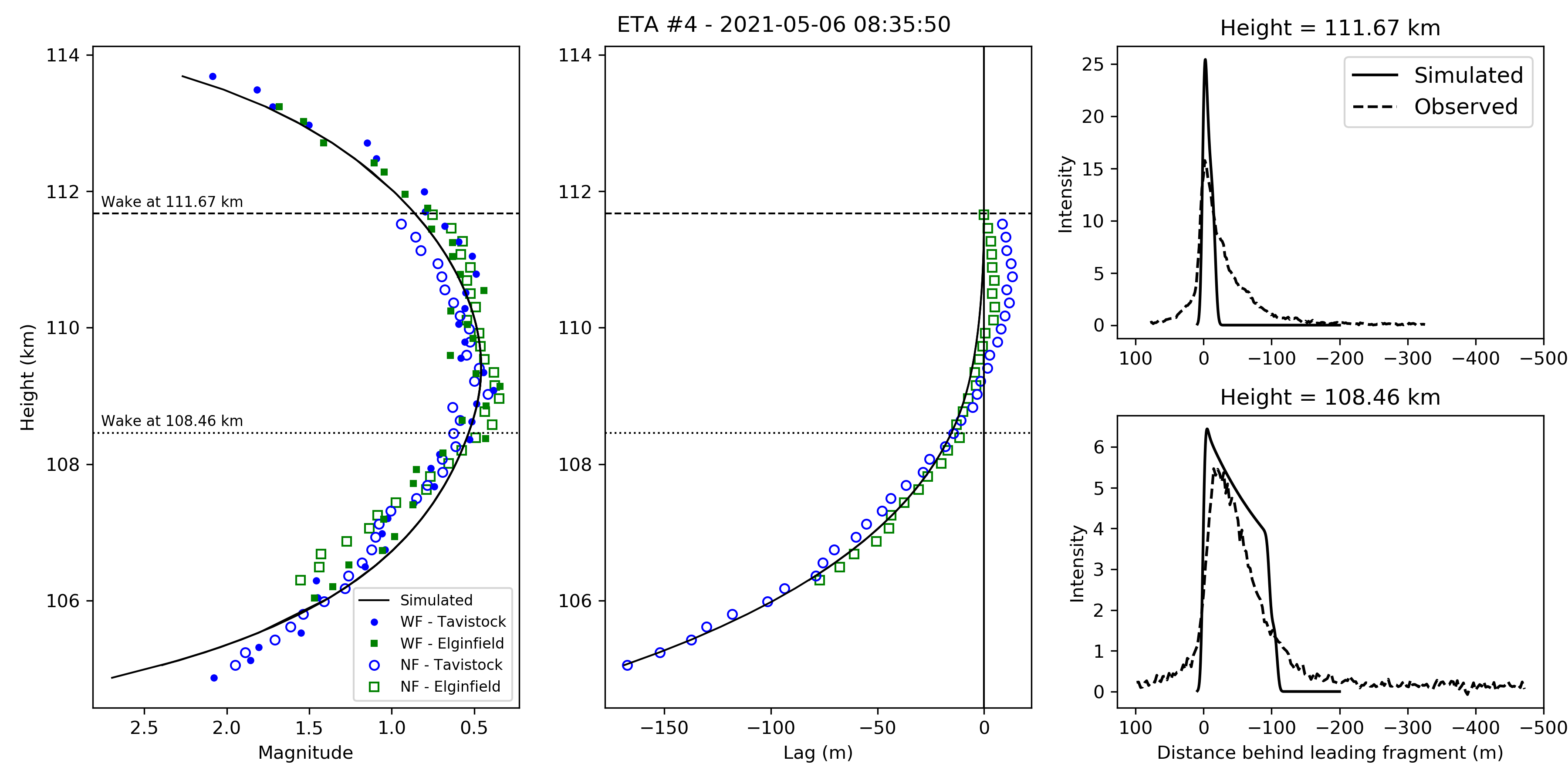}
    \caption{ETA4 2021-05-06 08:35:50}
\end{figure}
\begin{figure}
    \centering
    \includegraphics[width=\linewidth]{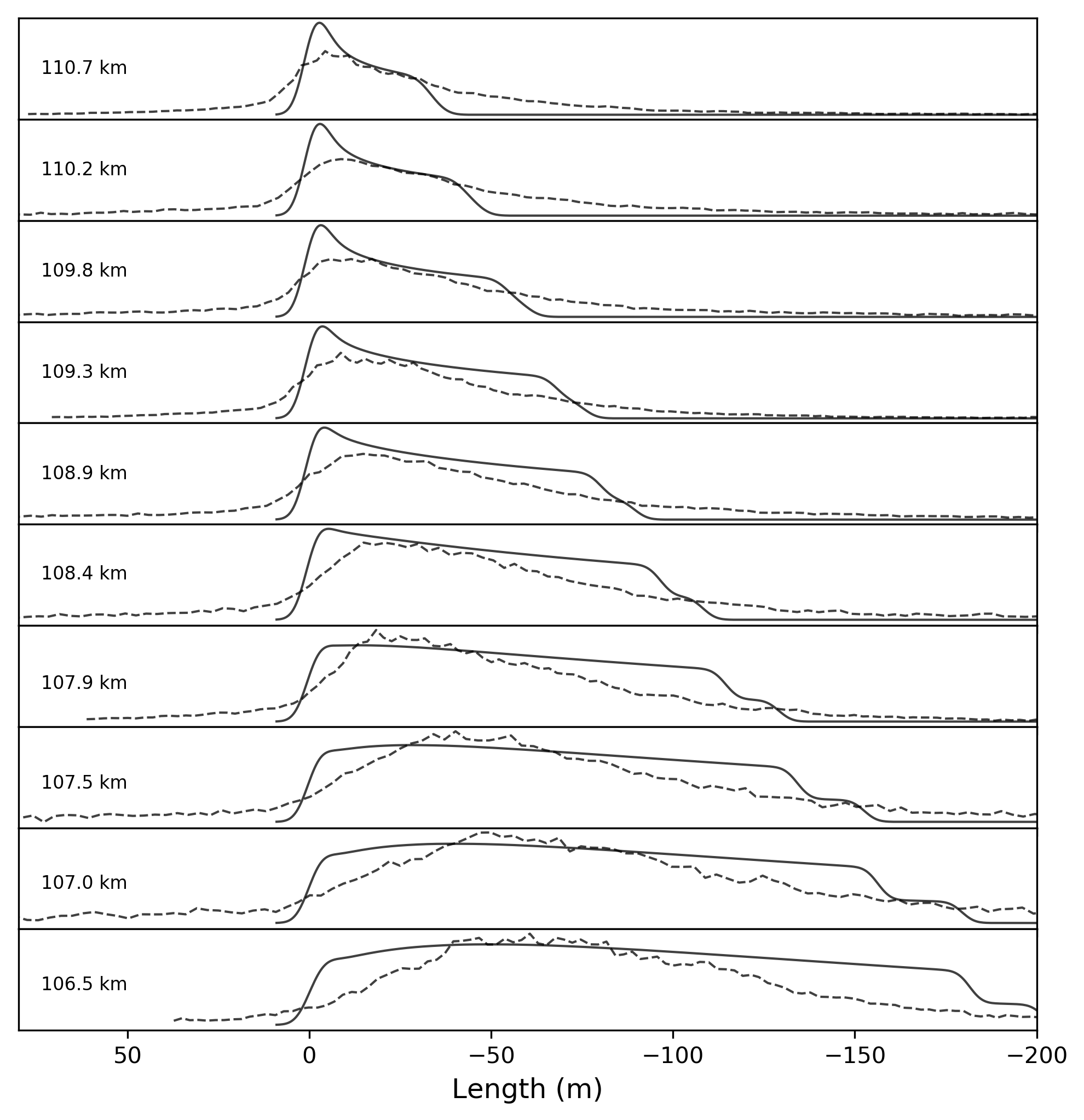}
    \caption{ETA4 2021-05-06 08:35:50}
\end{figure}

\begin{figure}
    \centering
    \includegraphics[width=\linewidth]{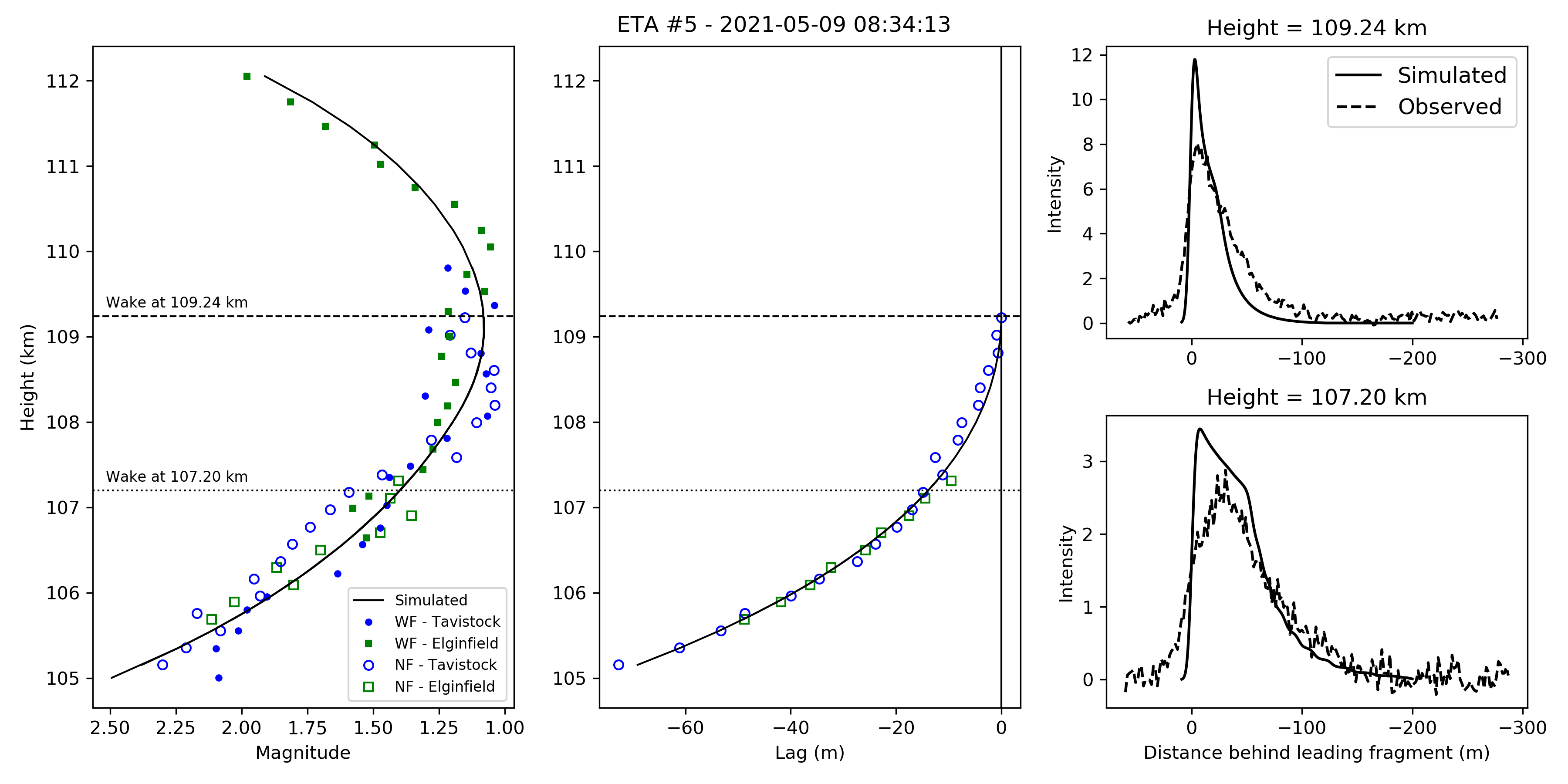}
    \caption{ETA5 2021-05-09 08:34:13}
\end{figure}
\begin{figure}
    \centering
    \includegraphics[width=\linewidth]{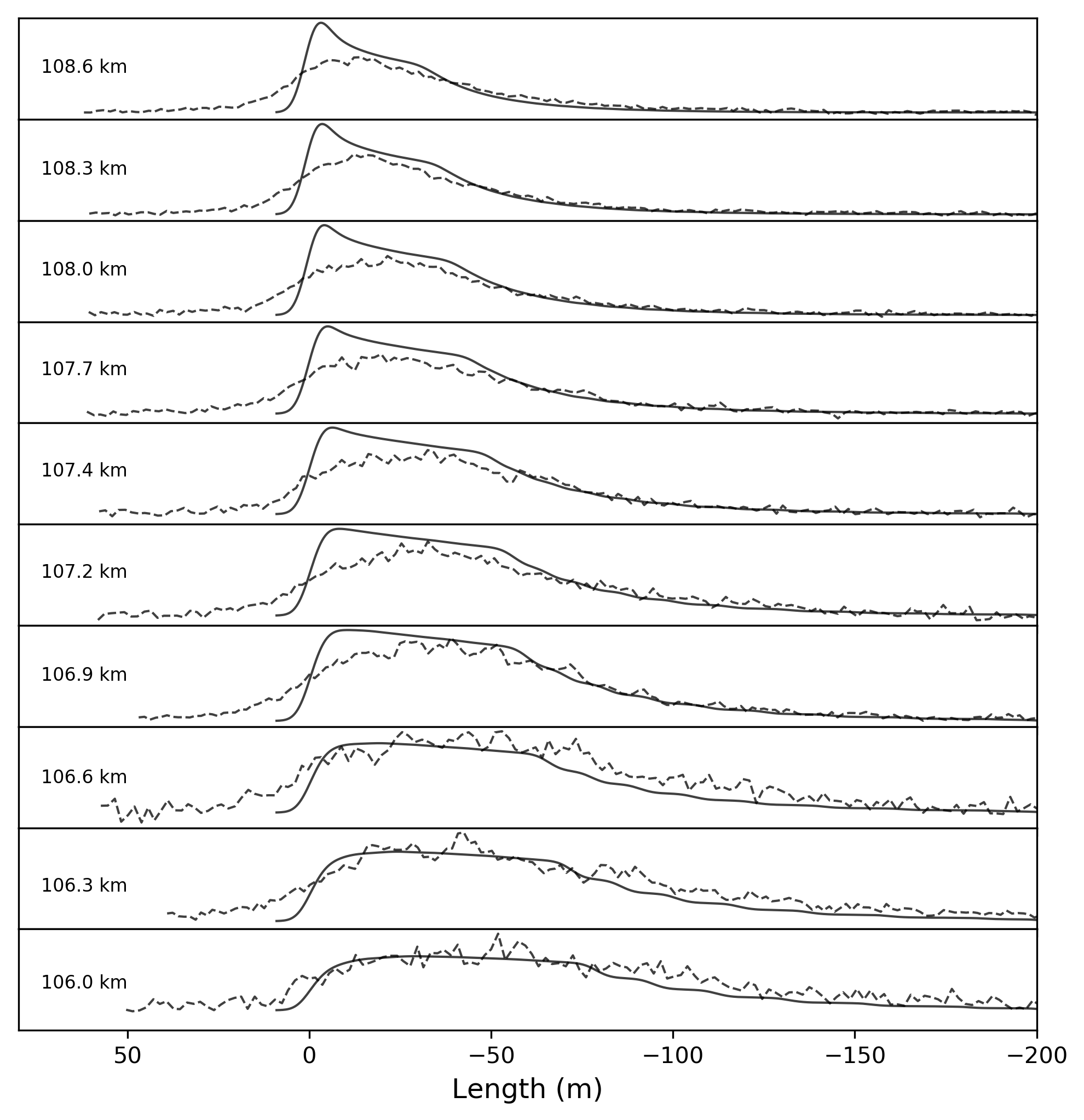}
    \caption{ETA5 2021-05-09 08:34:13}
\end{figure}

\begin{figure}
    \centering
    \includegraphics[width=\linewidth]{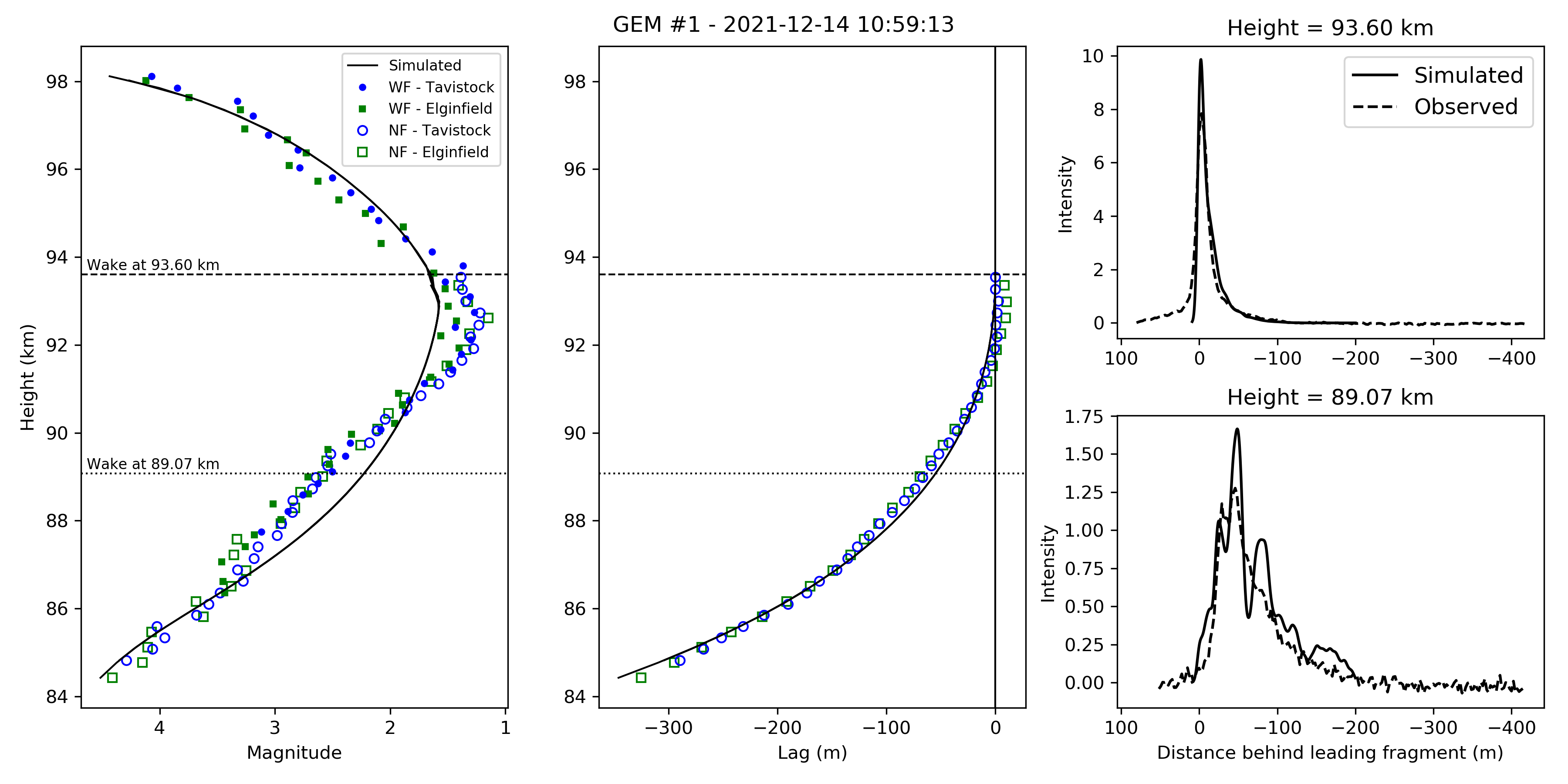}
    \caption{GEM1 2021-12-14 10:59:13}
\end{figure}
\begin{figure}
    \centering
    \includegraphics[width=\linewidth]{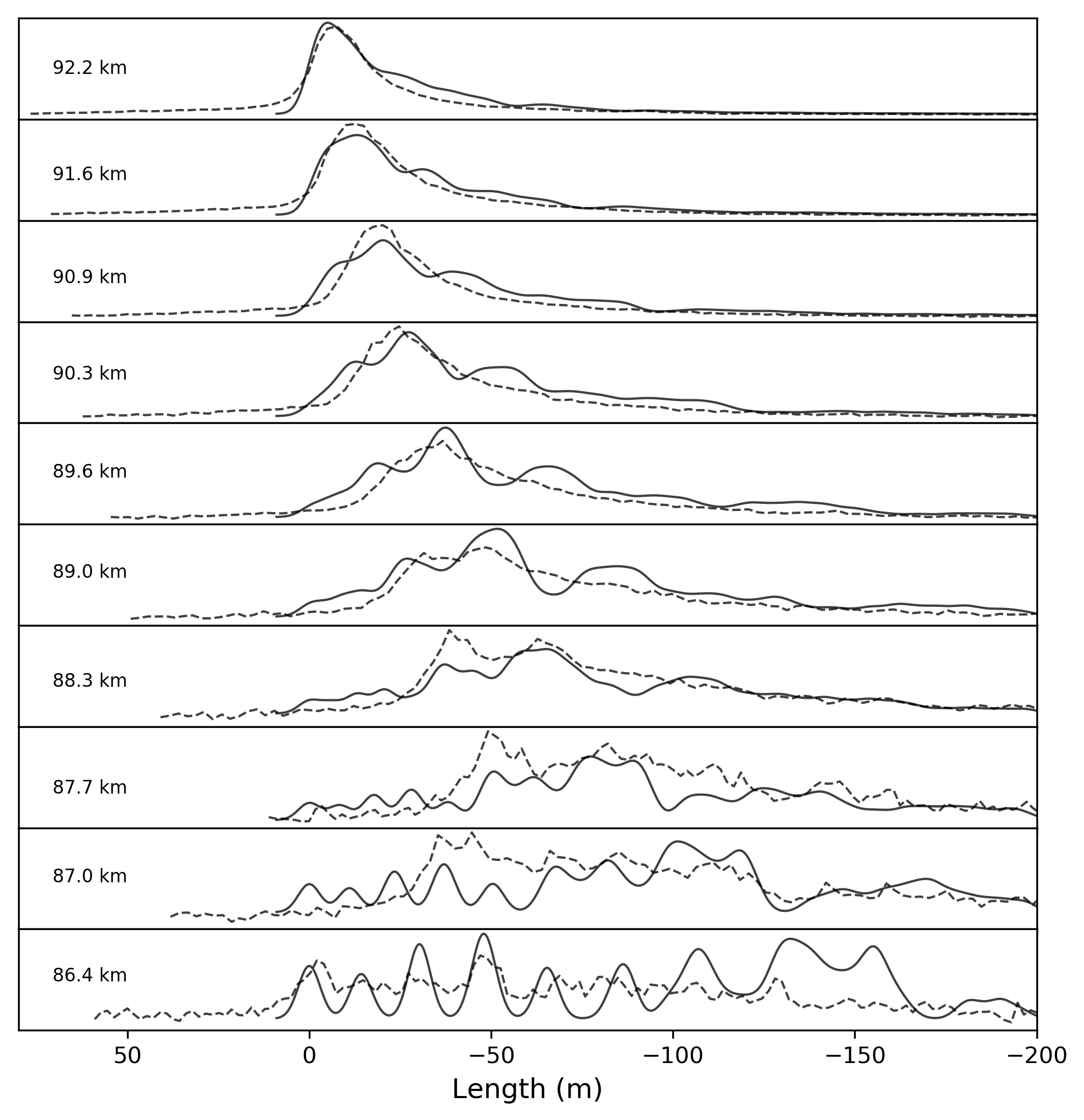}
    \caption{GEM1 2021-12-14 10:59:13}
\end{figure}

\begin{figure}
    \centering
    \includegraphics[width=\linewidth]{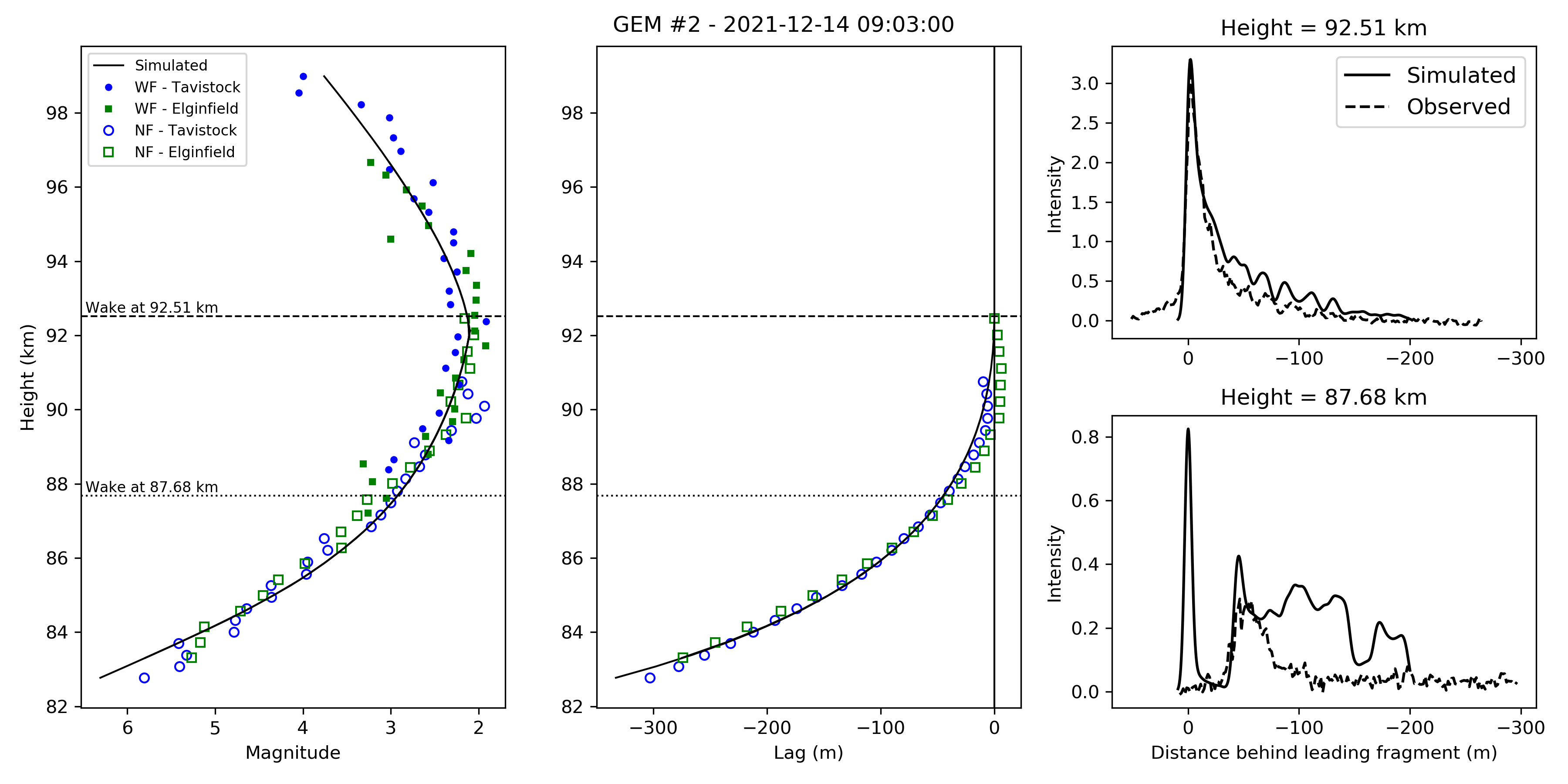}
    \caption{GEM2 2021-12-14 09:03:00}
\end{figure}
\begin{figure}
    \centering
    \includegraphics[width=\linewidth]{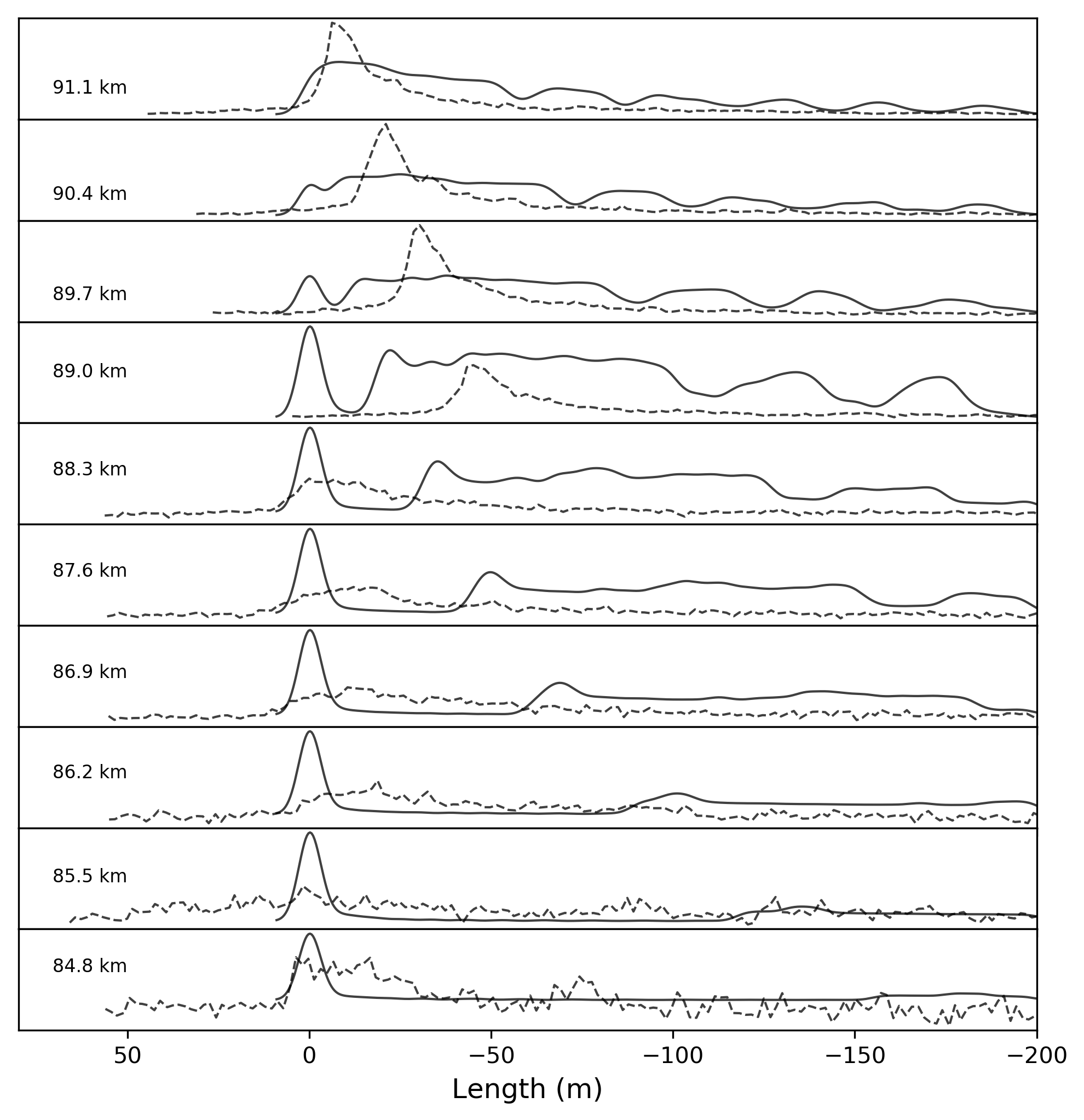}
    \caption{GEM2 2021-12-14 09:03:00}
\end{figure}

\begin{figure}
    \centering
    \includegraphics[width=\linewidth]{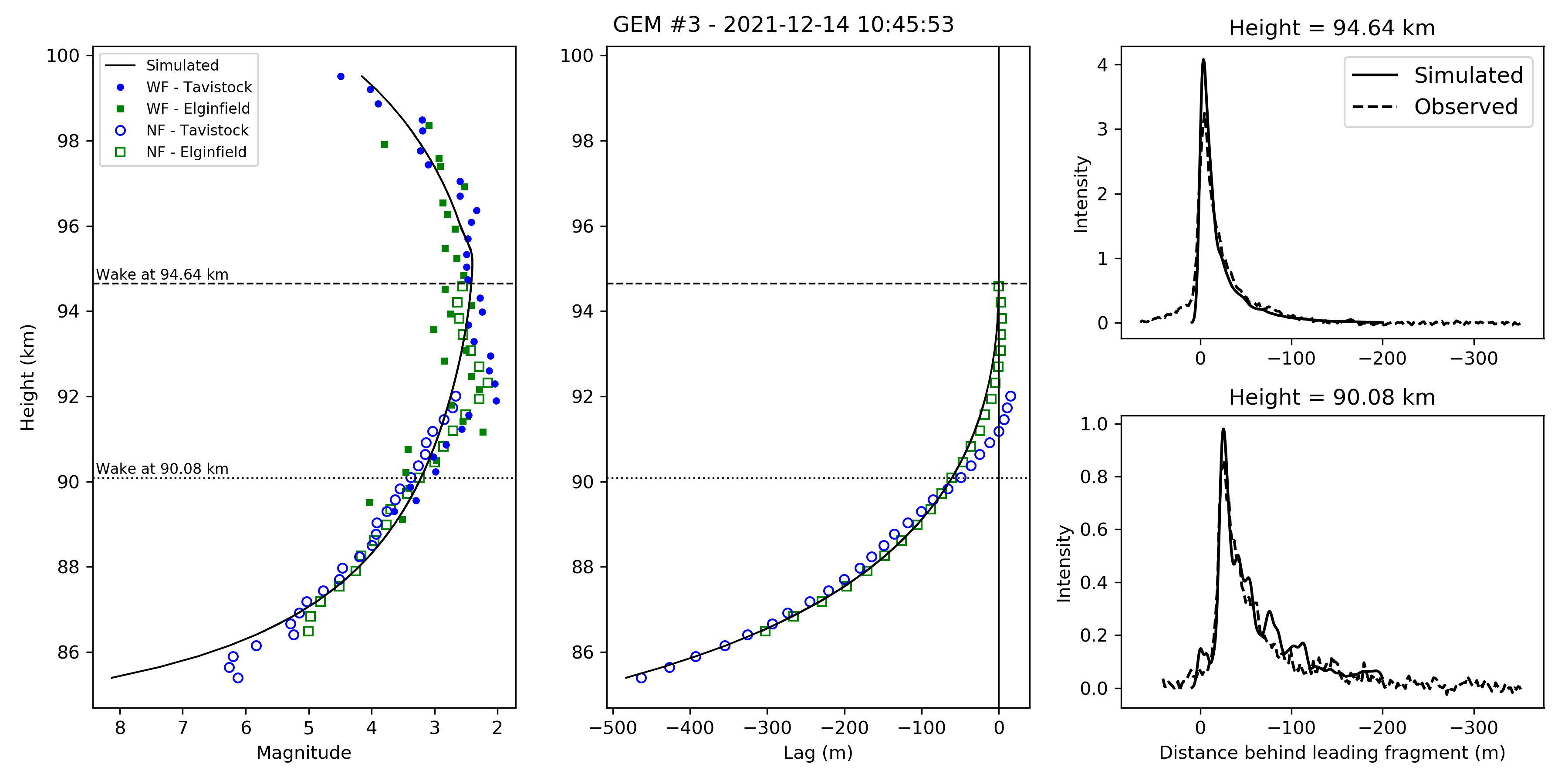}
    \caption{GEM3 2021-12-14 10:45:53}
\end{figure}
\begin{figure}
    \centering
    \includegraphics[width=\linewidth]{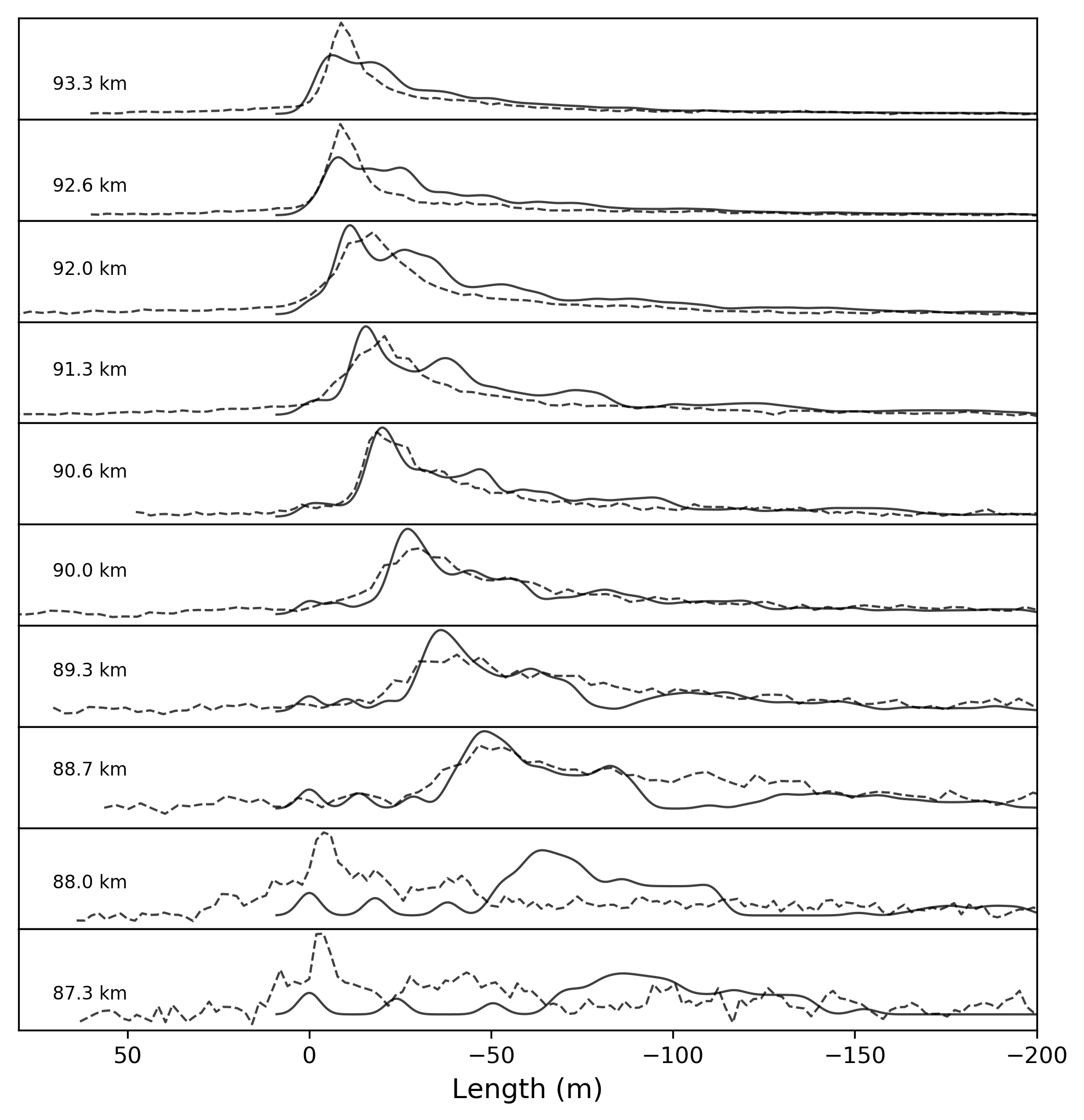}
    \caption{GEM3 2021-12-14 10:45:53}
\end{figure}

\begin{figure}
    \centering
    \includegraphics[width=\linewidth]{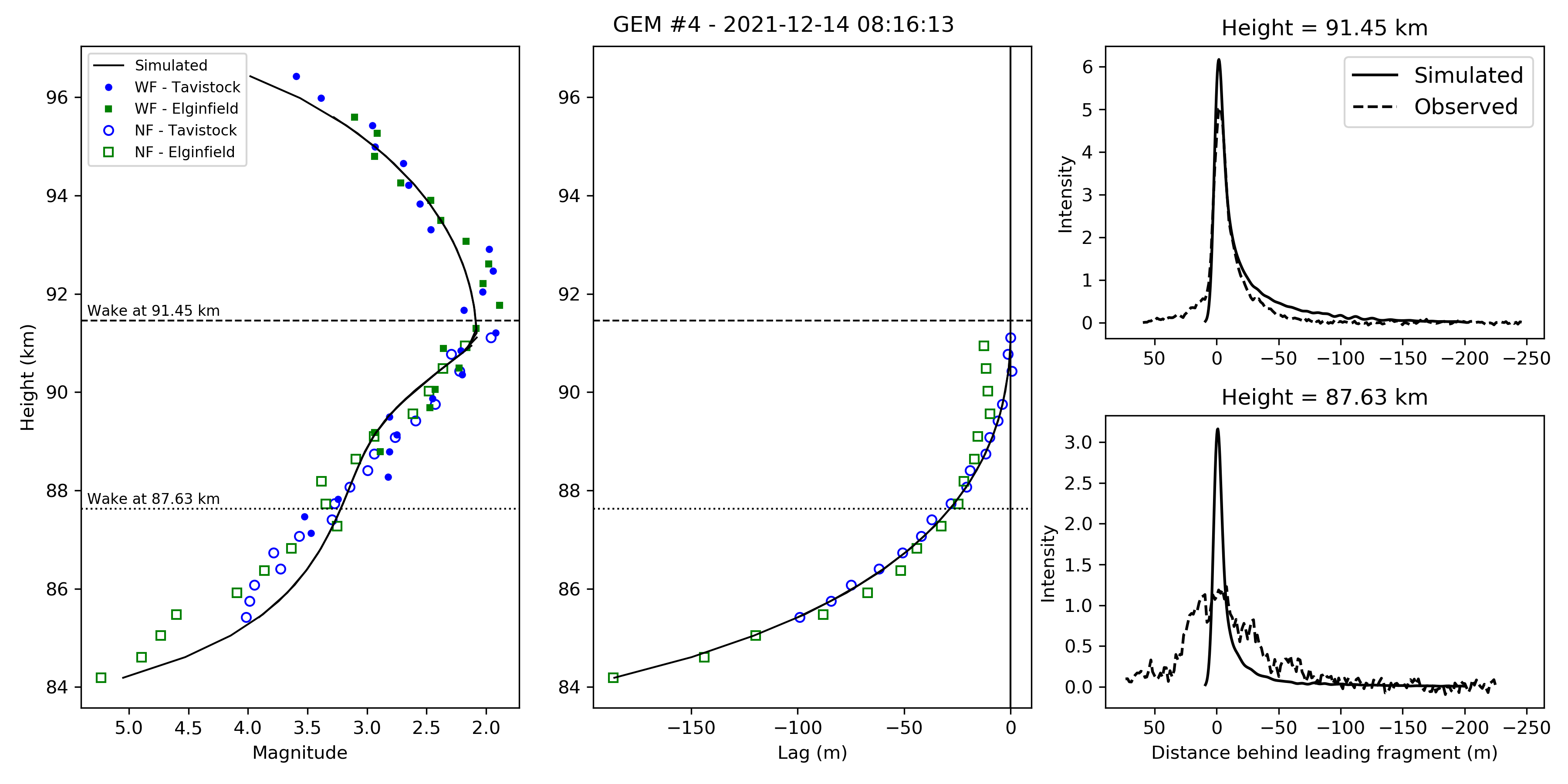}
    \caption{GEM4 2021-12-14 08:16:13}
\end{figure}
\begin{figure}
    \centering
    \includegraphics[width=\linewidth]{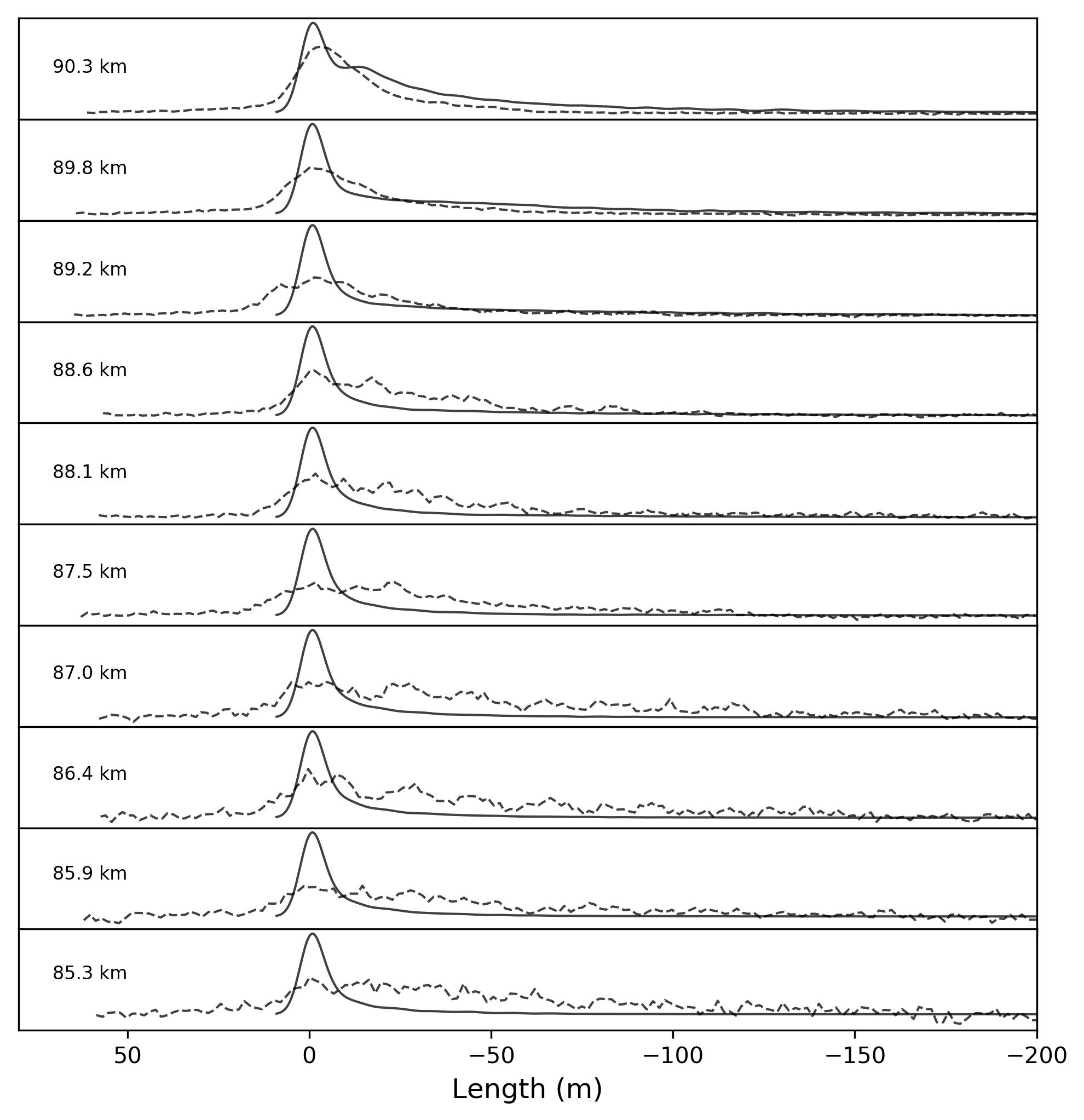}
    \caption{GEM4 2021-12-14 08:16:13}
\end{figure}

\begin{figure}
    \centering
    \includegraphics[width=\linewidth]{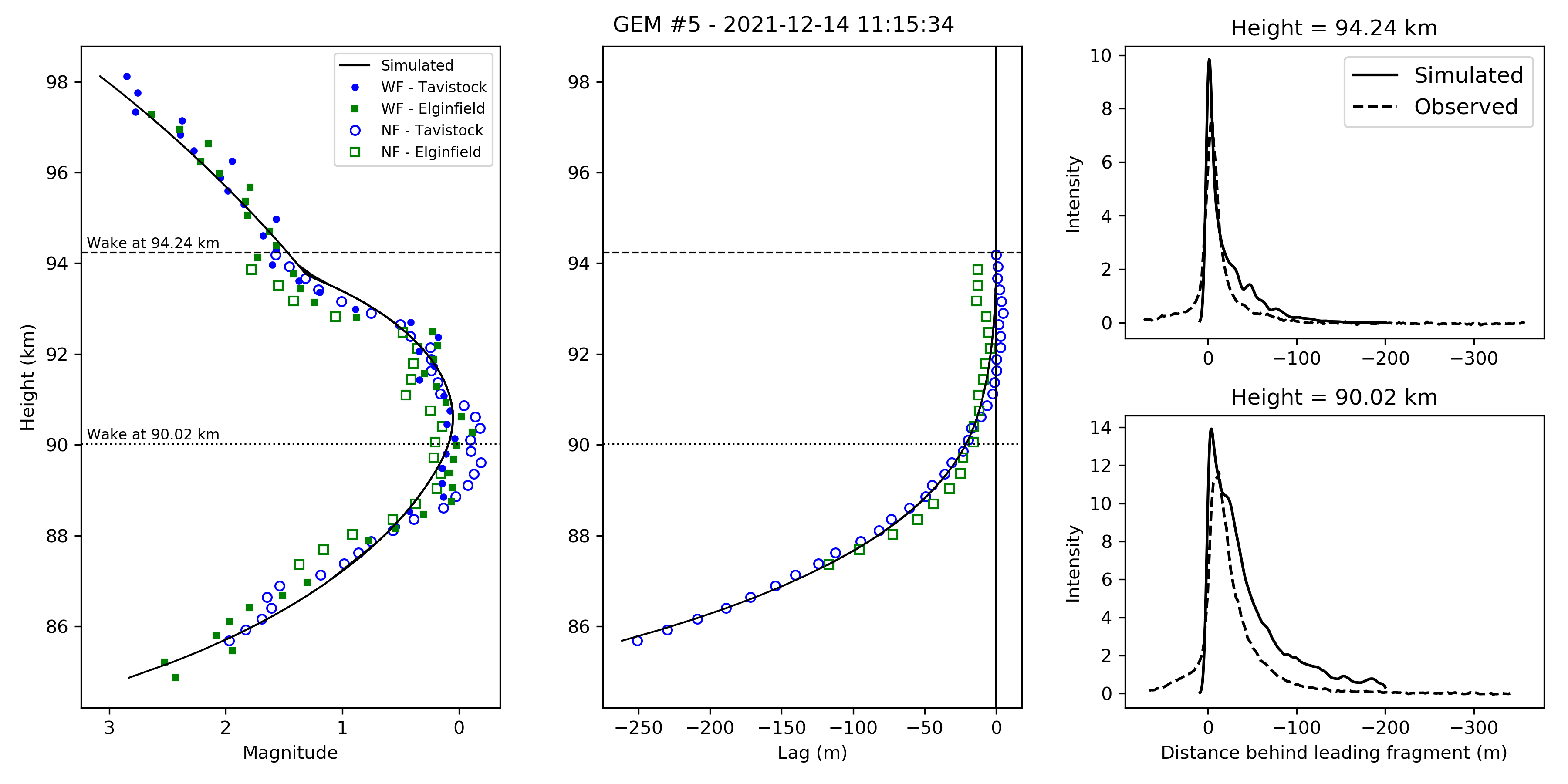}
    \caption{GEM5 2021-12-14 11:15:34}
\end{figure}
\begin{figure}
    \centering
    \includegraphics[width=\linewidth]{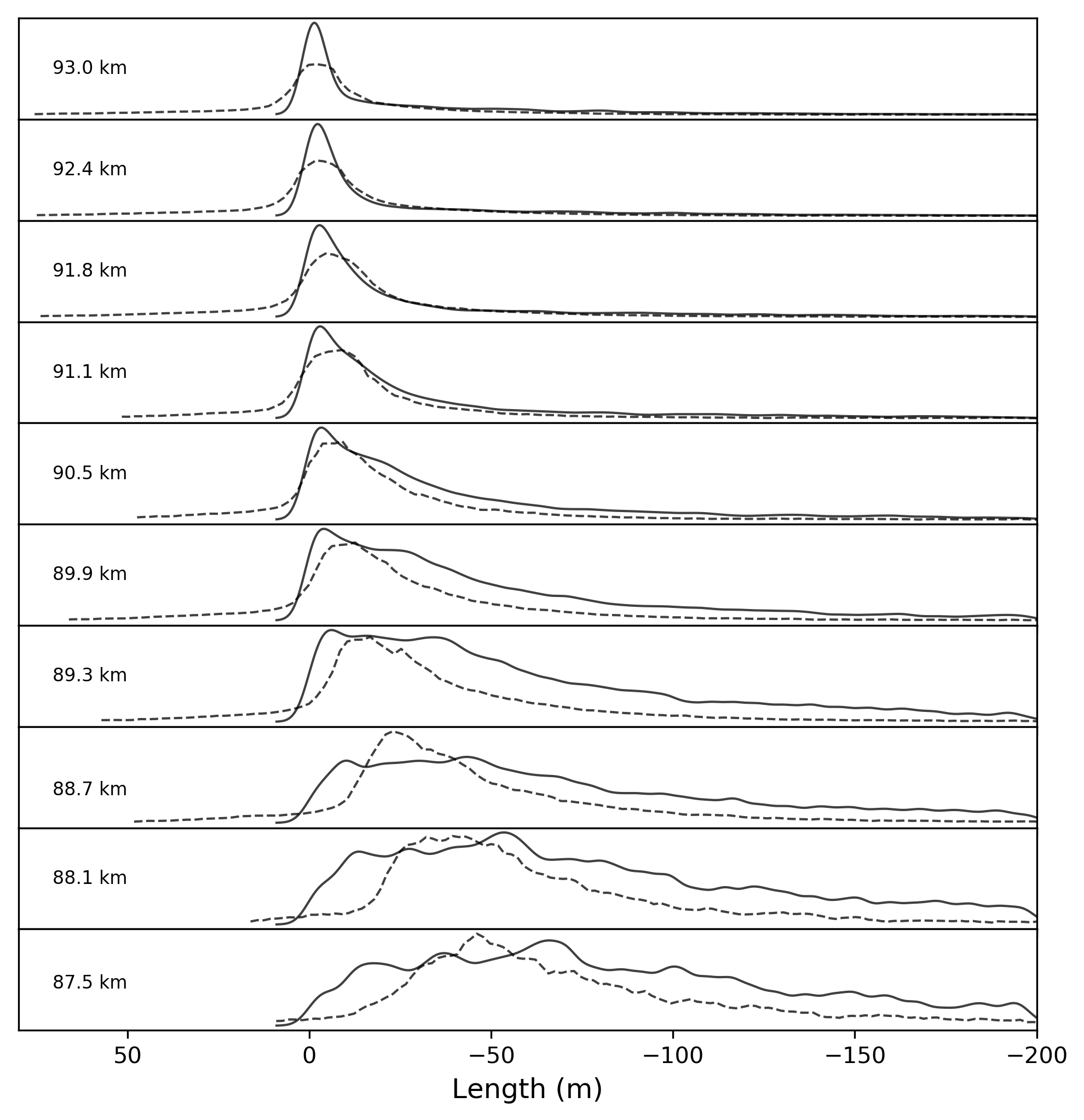}
    \caption{GEM5 2021-12-14 11:15:34}
\end{figure}

\begin{figure}
    \centering
    \includegraphics[width=\linewidth]{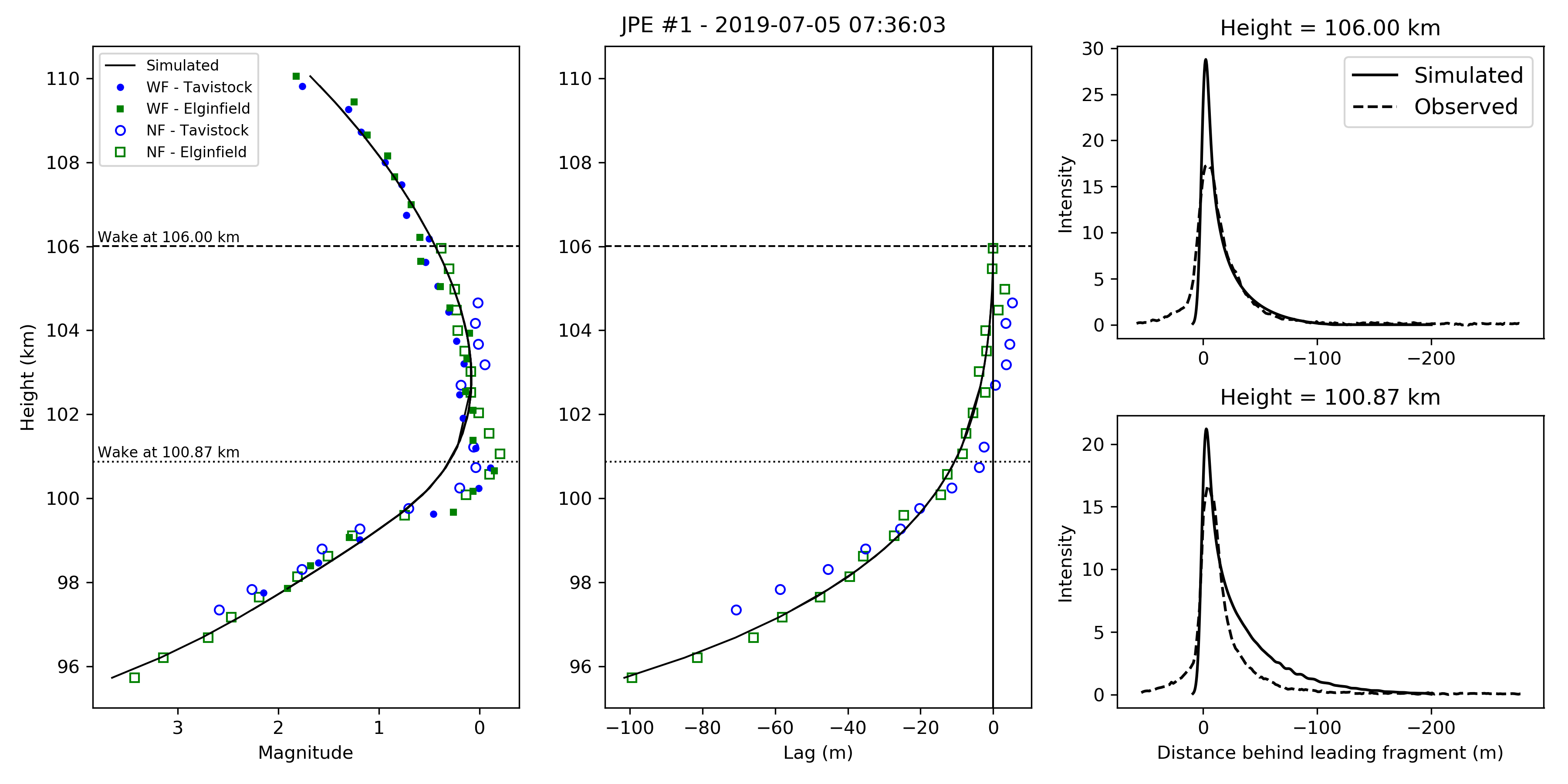}
    \caption{JPE1 2019-07-05 07:36:03}
\end{figure}
\begin{figure}
    \centering
    \includegraphics[width=\linewidth]{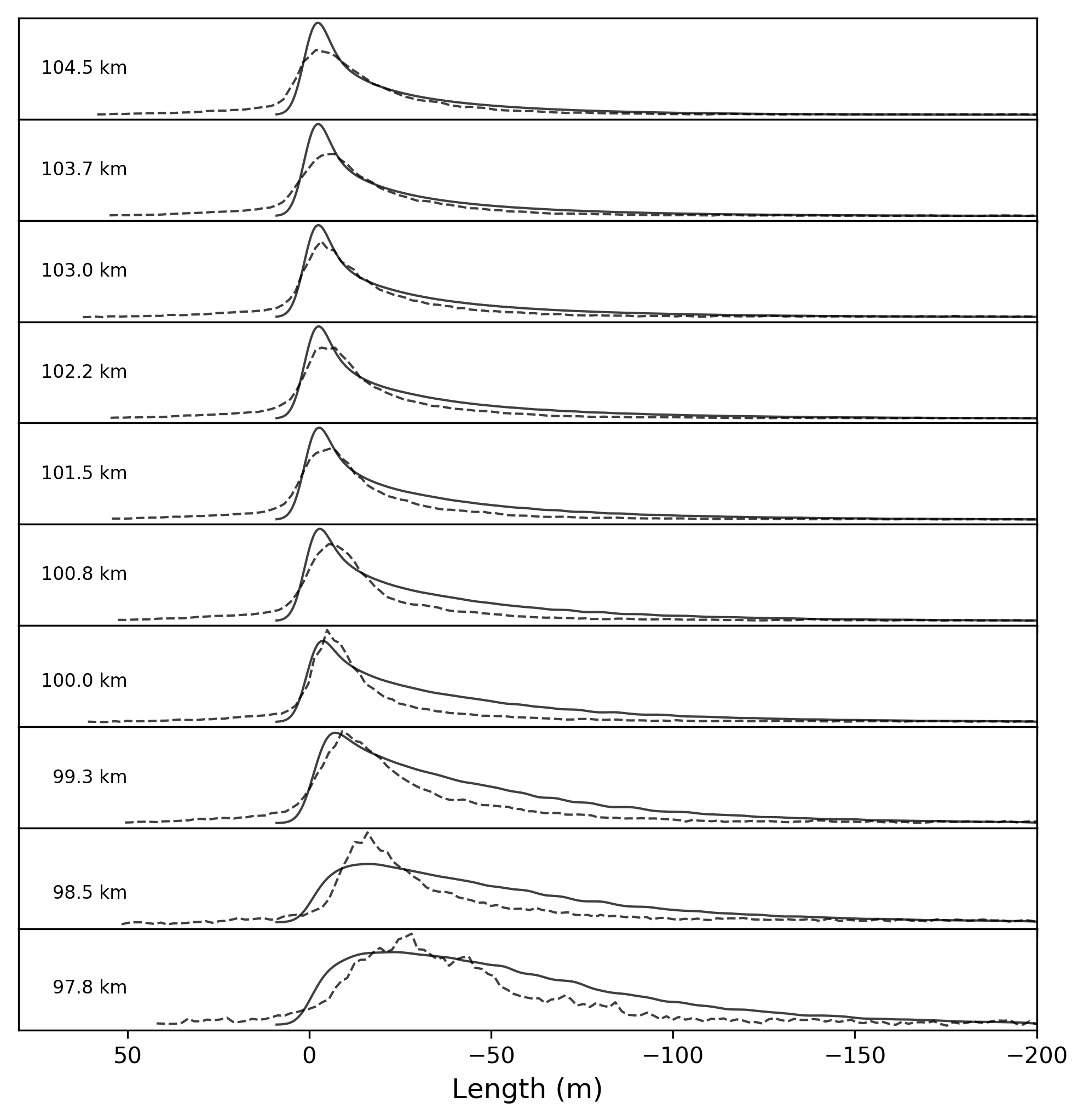}
    \caption{JPE1 2019-07-05 07:36:03}
\end{figure}

\begin{figure}
    \centering
    \includegraphics[width=\linewidth]{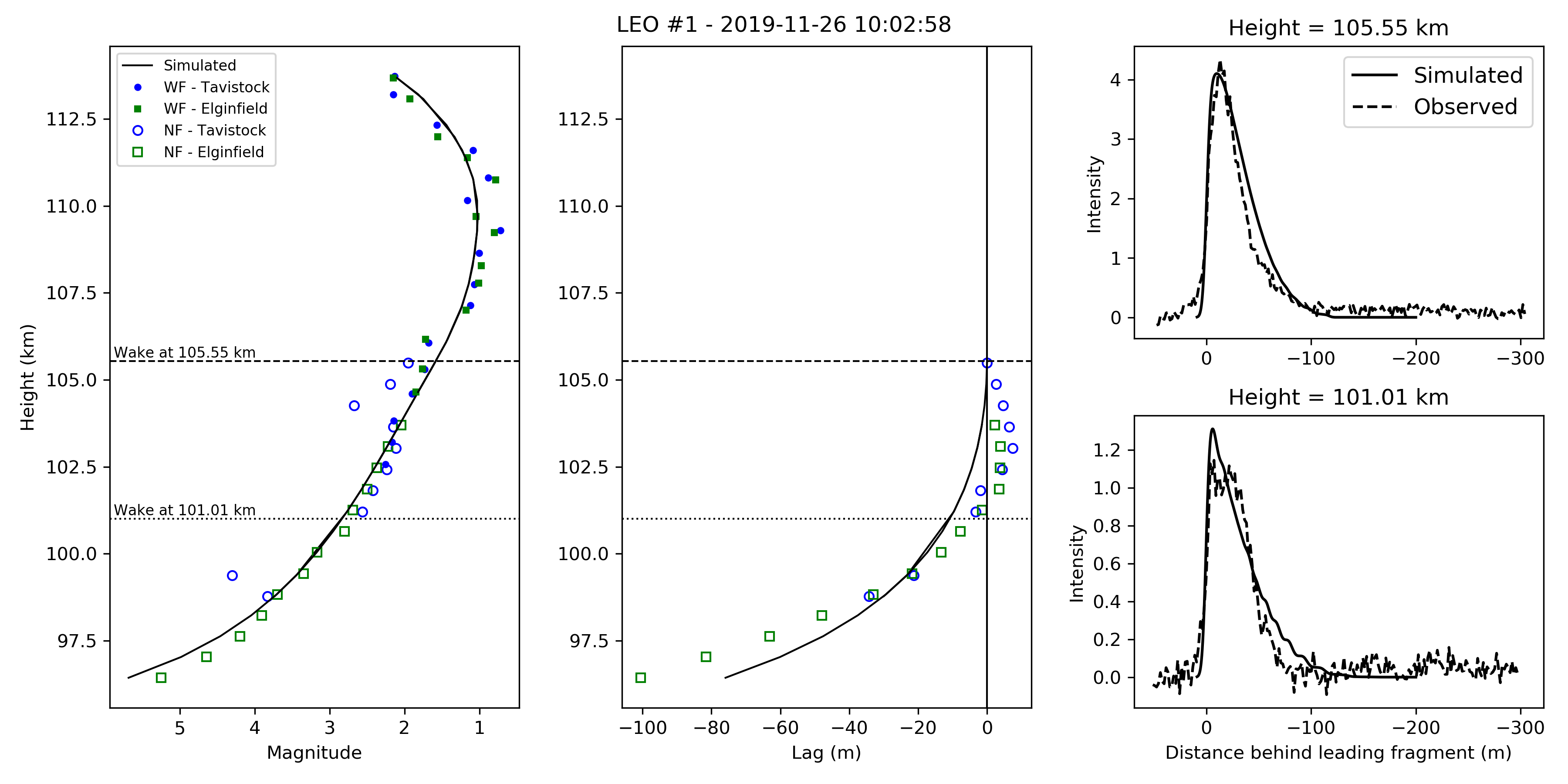}
    \caption{LEO1: 2019-11-26 10:02:58}
\end{figure}
\begin{figure}
    \centering
    \includegraphics[width=\linewidth]{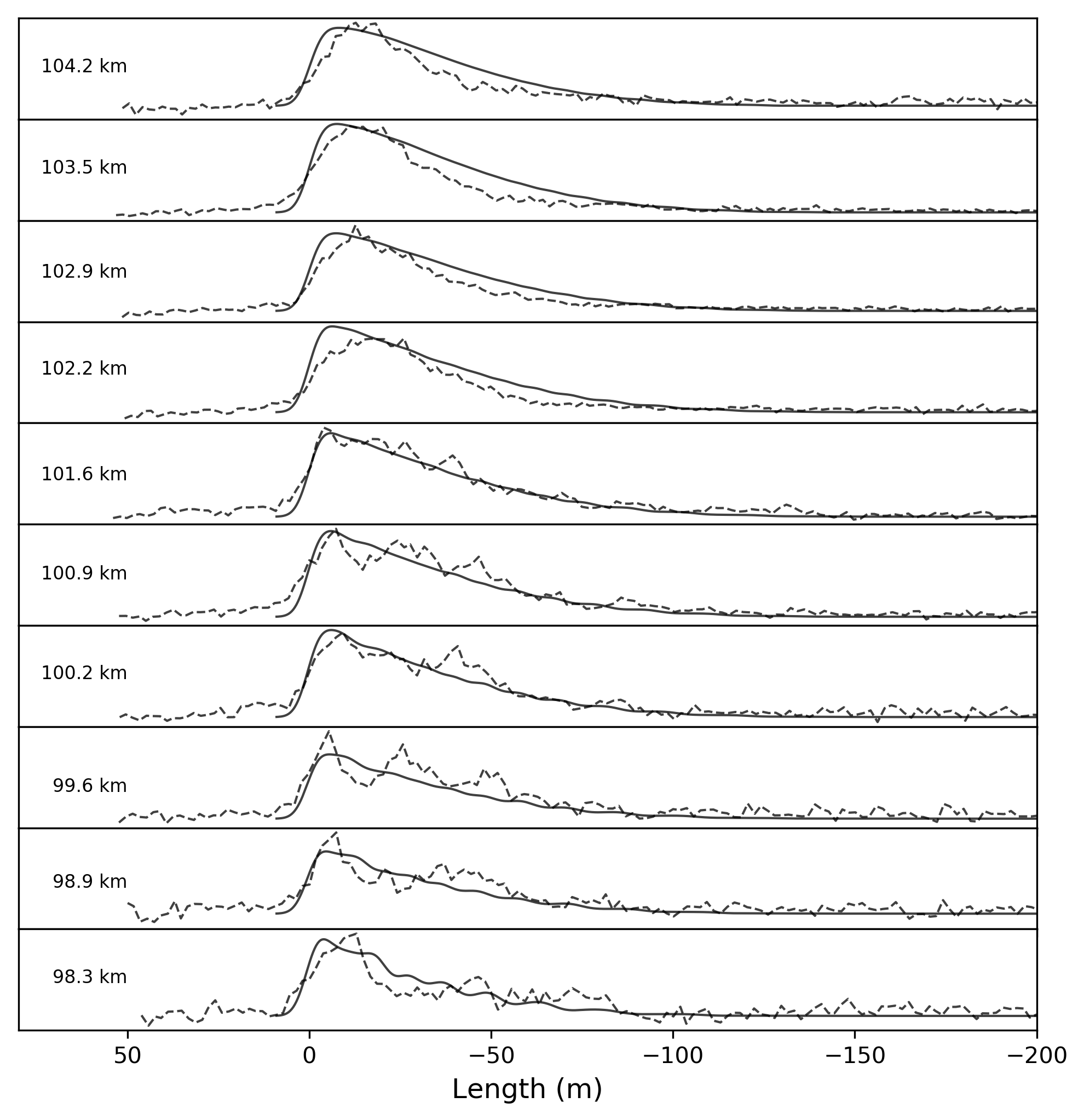}
    \caption{LEO1: 2019-11-26 10:02:58}
\end{figure}

\begin{figure}
    \centering
    \includegraphics[width=\linewidth]{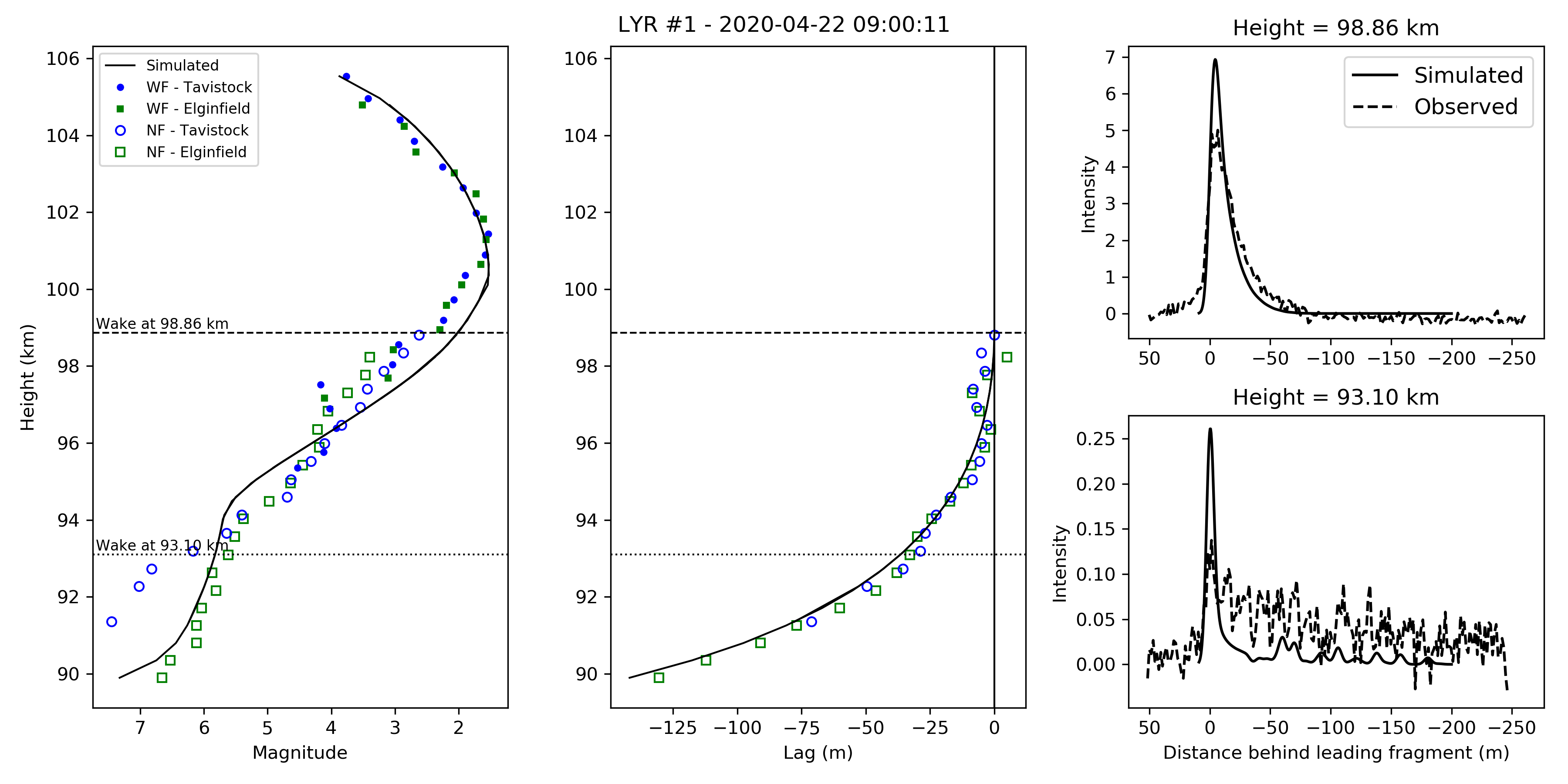}
    \caption{LYR1 2020-04-22 09:00:11}
\end{figure}
\begin{figure}
    \centering
    \includegraphics[width=\linewidth]{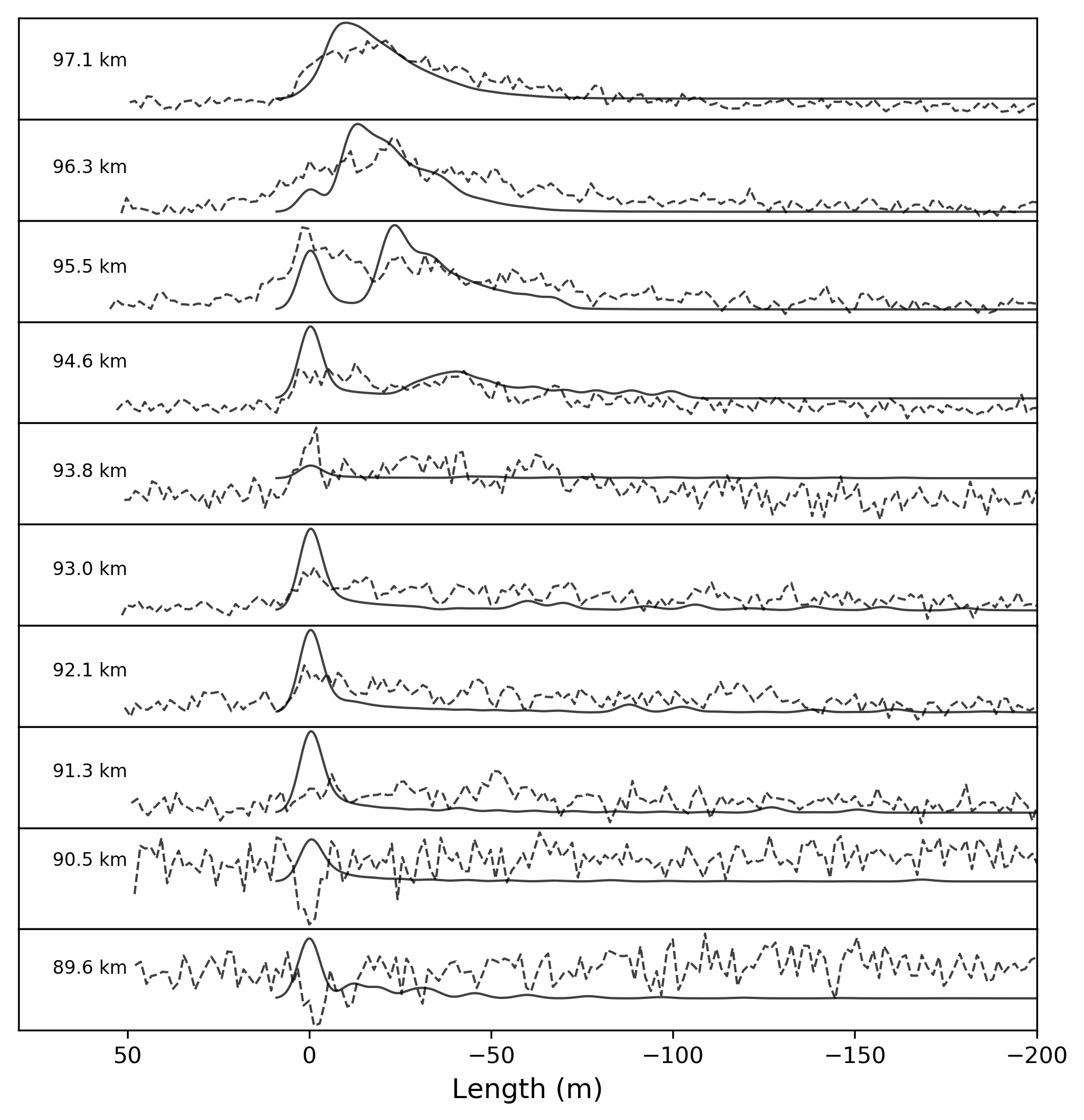}
    \caption{LYR1 2020-04-22 09:00:11}
\end{figure}

\begin{figure}
    \centering
    \includegraphics[width=\linewidth]{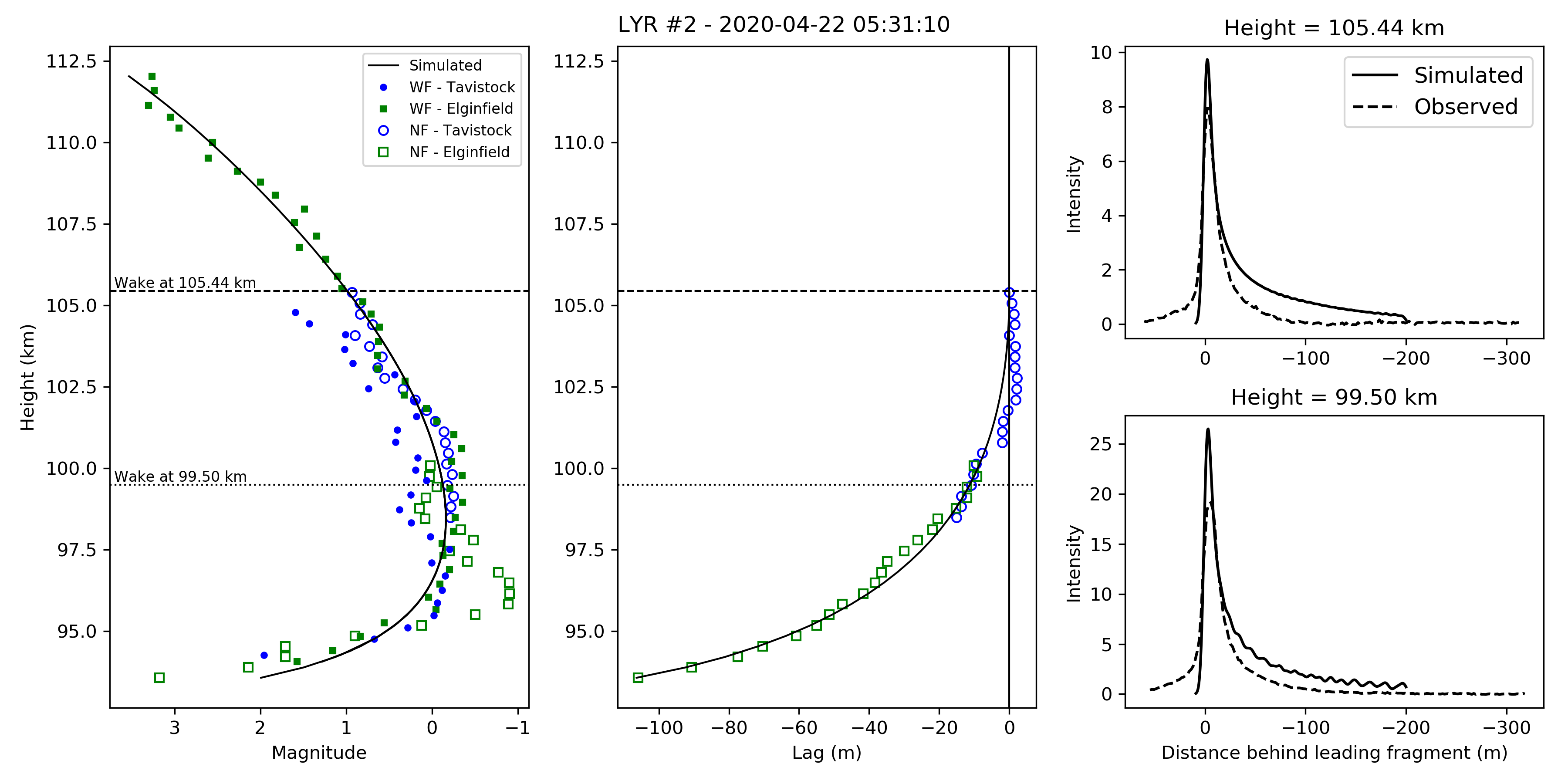}
    \caption{LYR2 2020-04-22 05:31:10}
\end{figure}
\begin{figure}
    \centering
    \includegraphics[width=\linewidth]{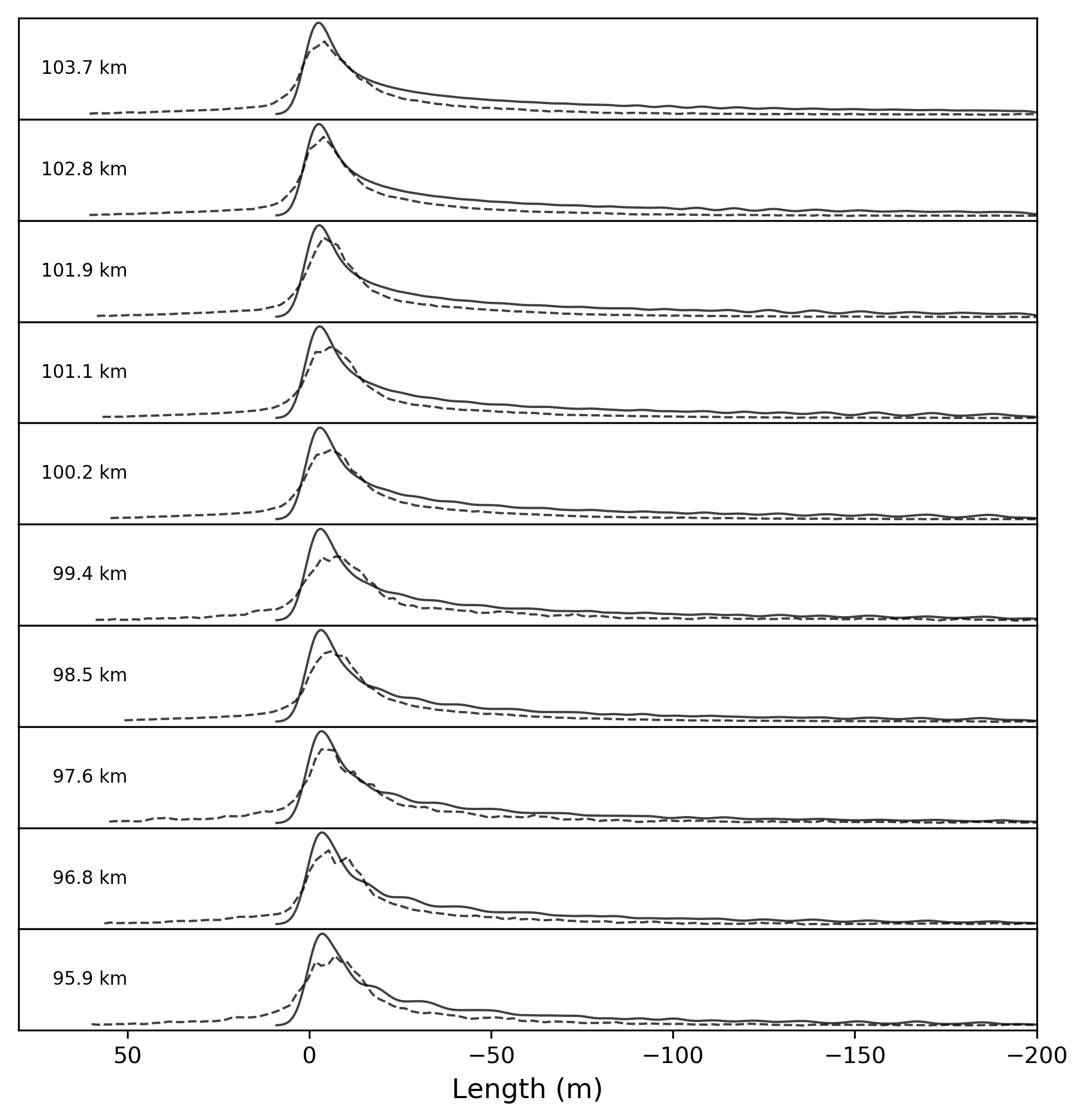}
    \caption{LYR2 2020-04-22 05:31:10}
\end{figure}

\begin{figure}
    \centering
    \includegraphics[width=\linewidth]{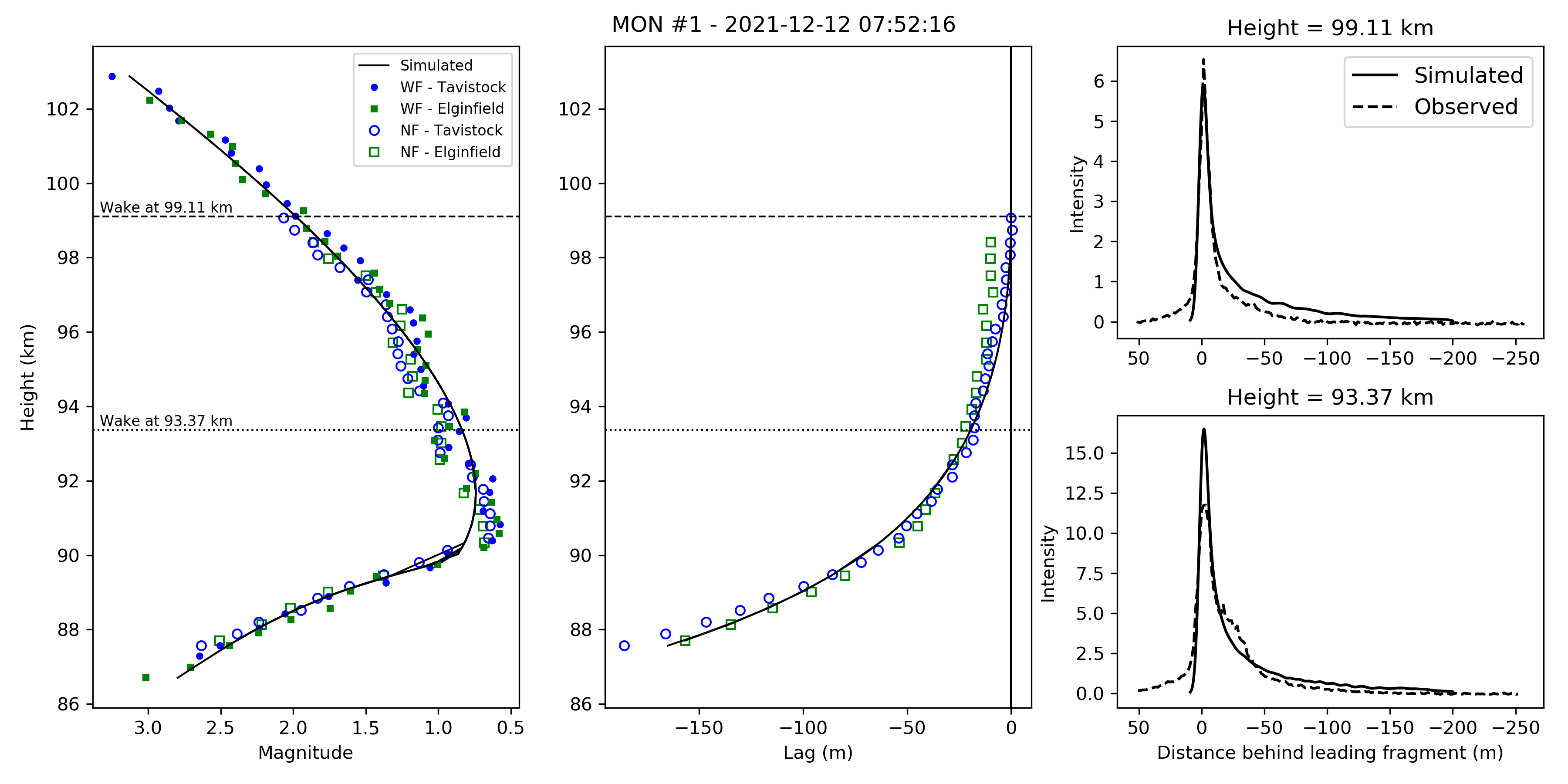}
    \caption{MON1: 2021-12-12 07:52:16}
\end{figure}
\begin{figure}
    \centering
    \includegraphics[width=\linewidth]{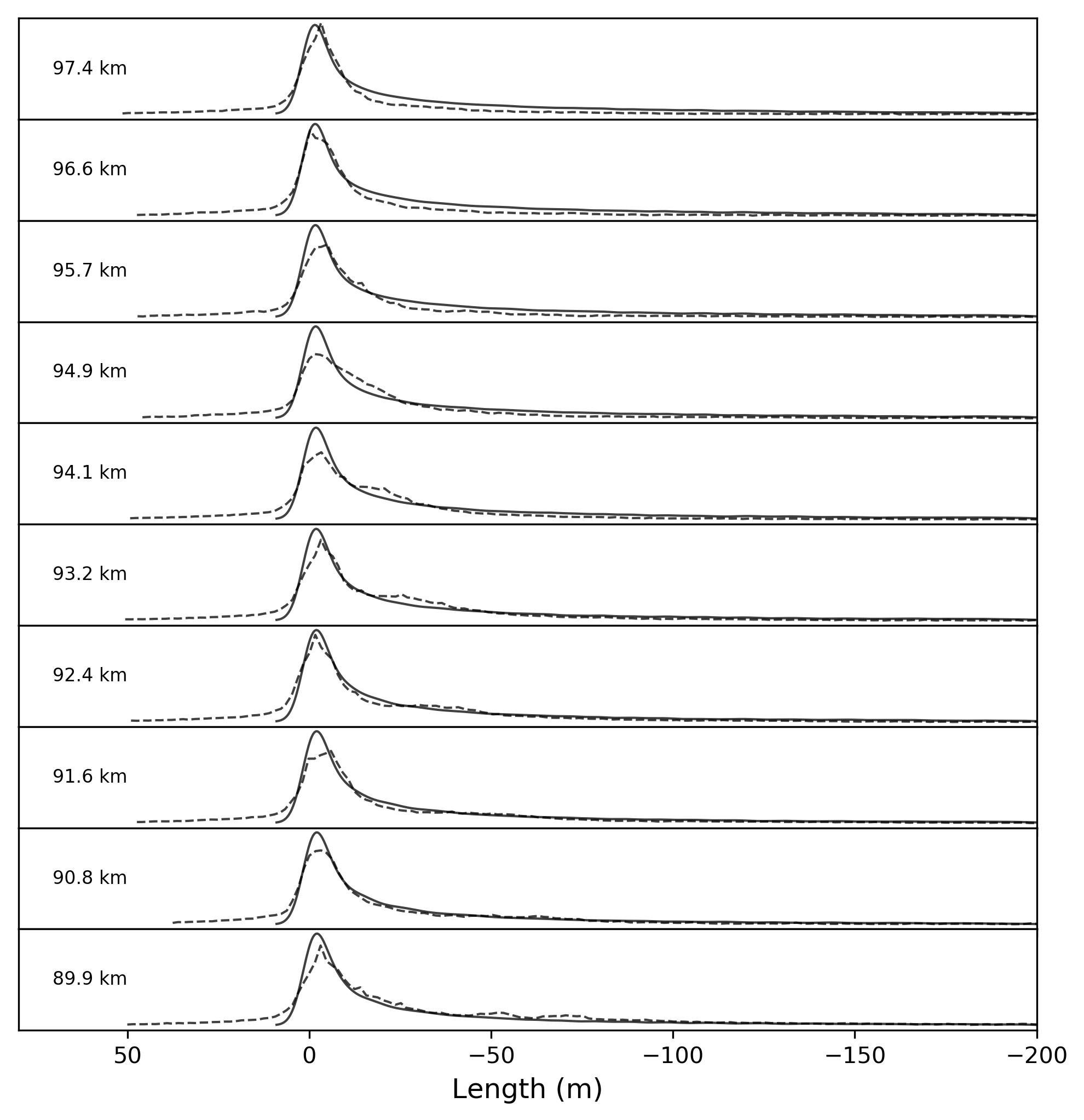}
    \caption{MON1: 2021-12-12 07:52:16}
\end{figure}

\begin{figure}
    \centering
    \includegraphics[width=\linewidth]{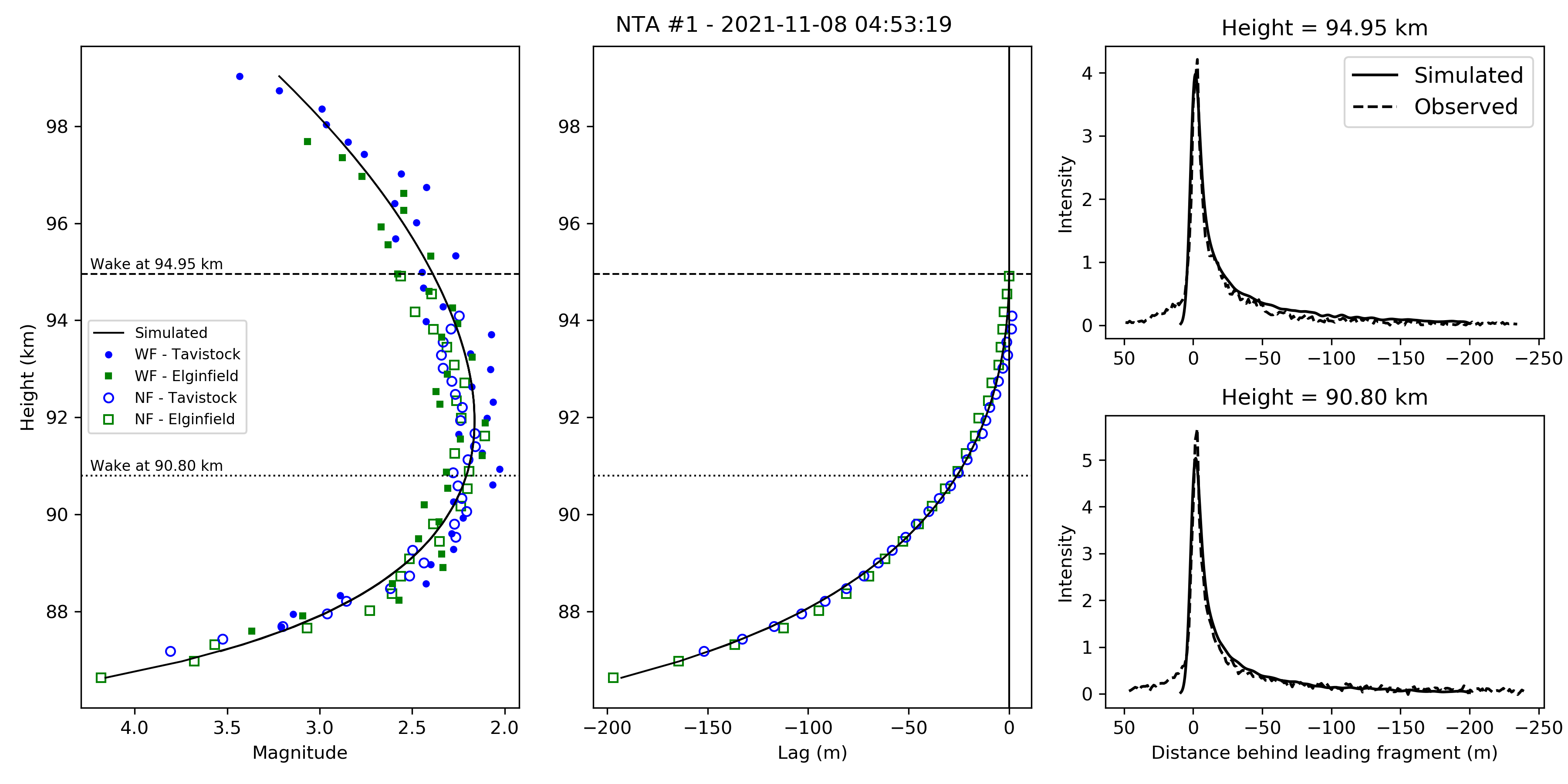}
    \caption{NTA1: 2021-11-08 04:53:19}
\end{figure}
\begin{figure}
    \centering
    \includegraphics[width=\linewidth]{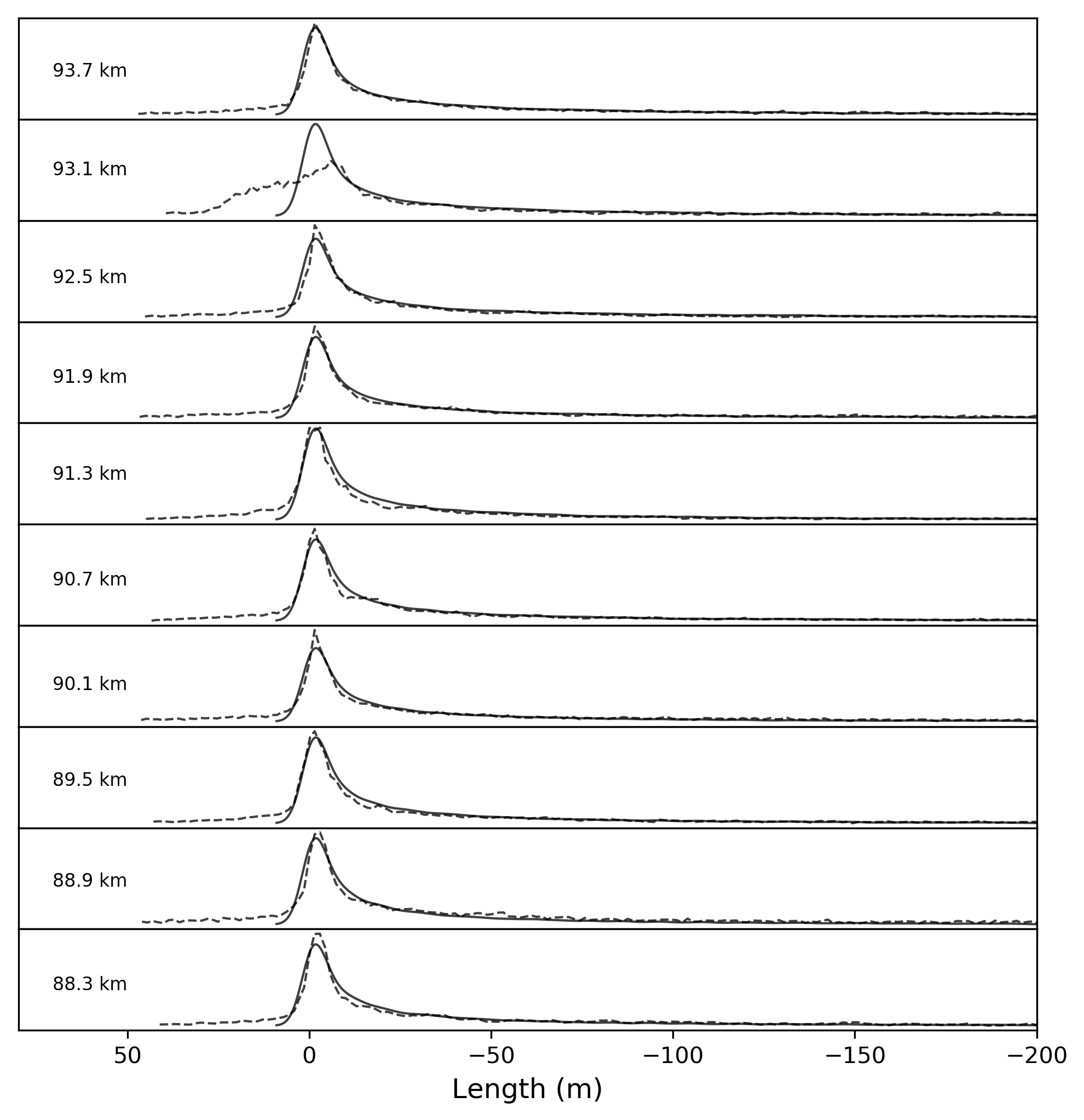}
    \caption{NTA1: 2021-11-08 04:53:19}
\end{figure}

\begin{figure}
    \centering
    \includegraphics[width=\linewidth]{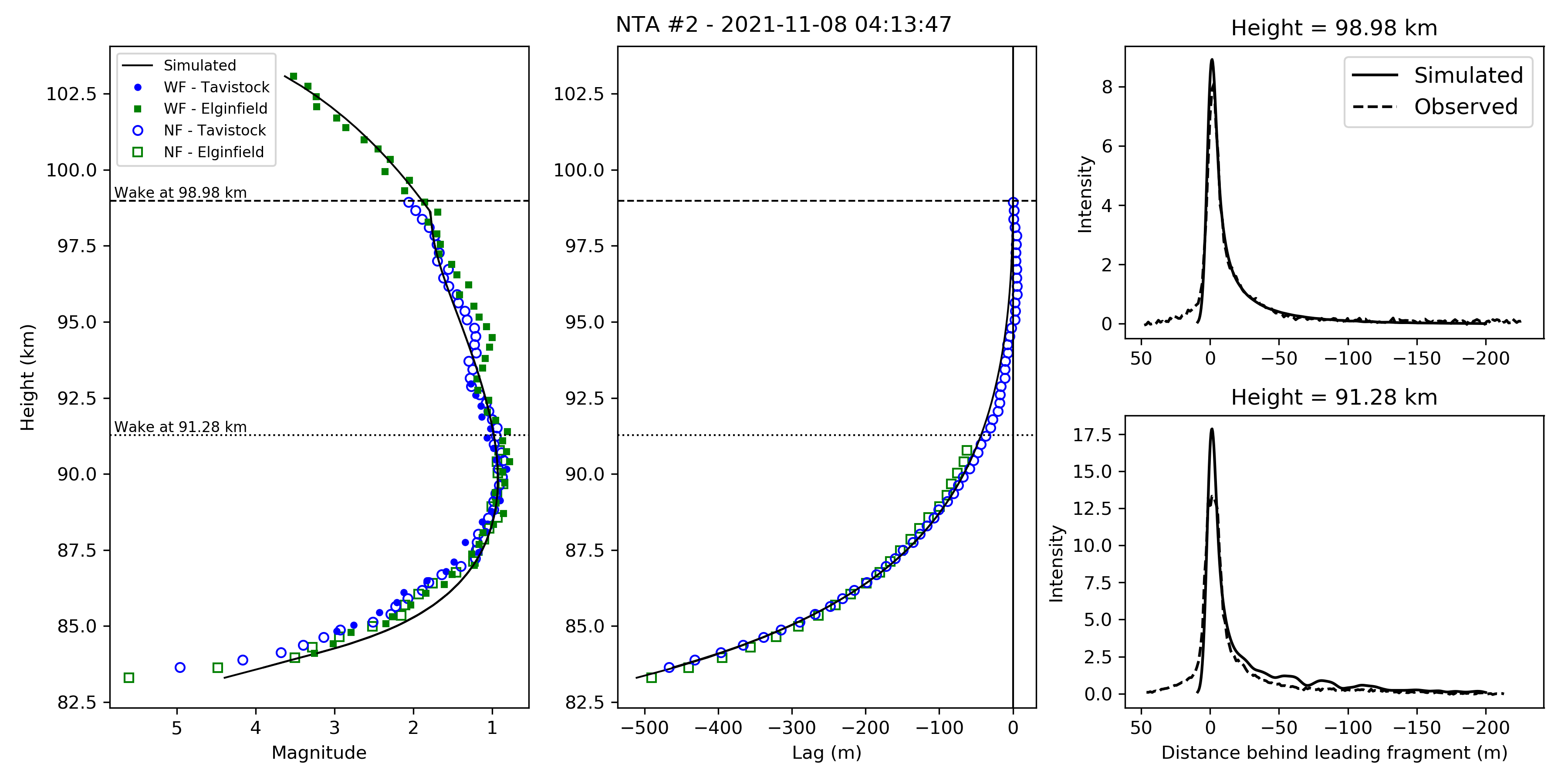}
    \caption{NTA2: 2021-11-08 04:13:48}
\end{figure}
\begin{figure}
    \centering
    \includegraphics[width=\linewidth]{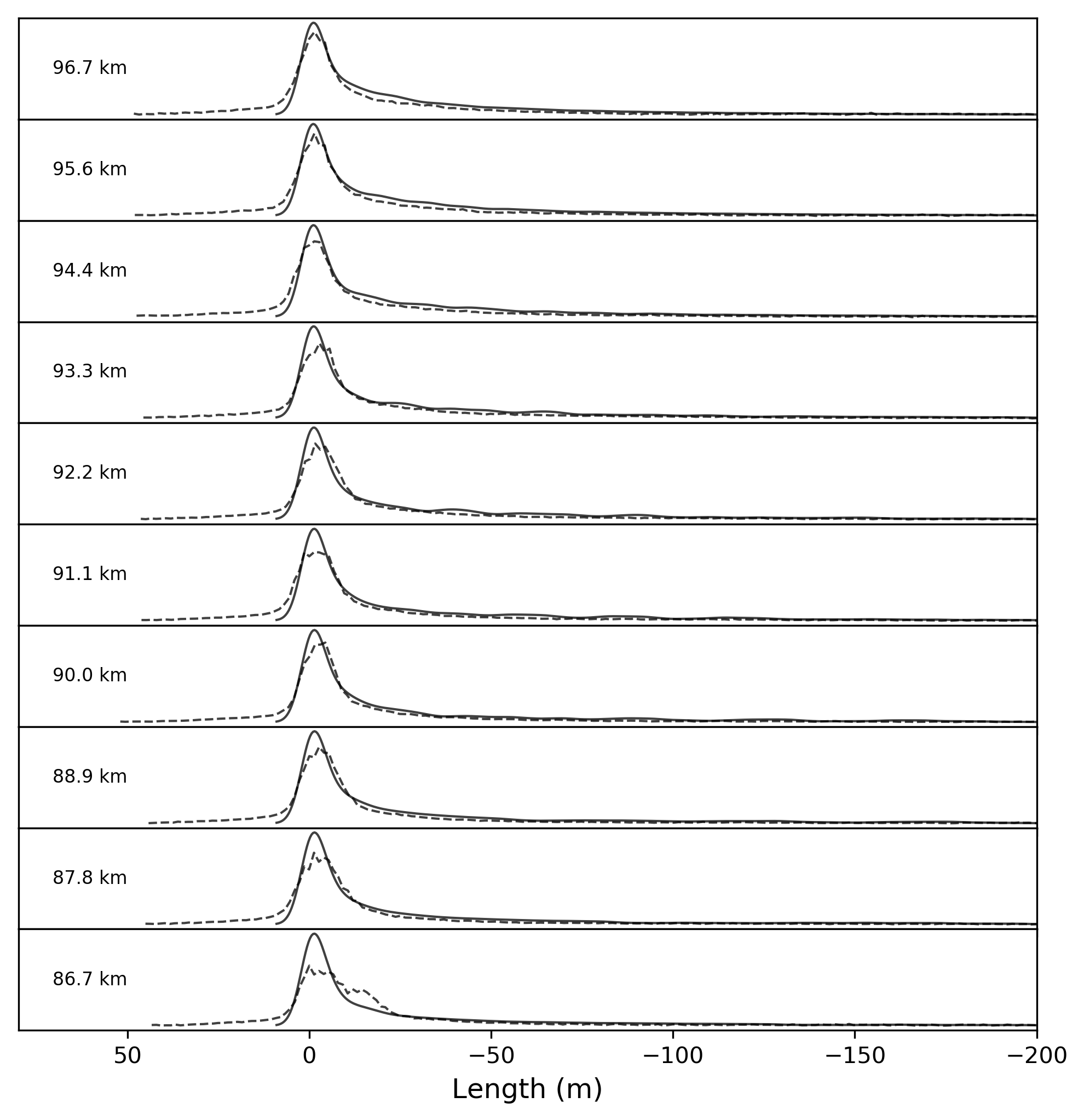}
    \caption{NTA2: 2021-11-08 04:13:48}
\end{figure}

\clearpage

\begin{figure}
    \centering
    \includegraphics[width=\linewidth]{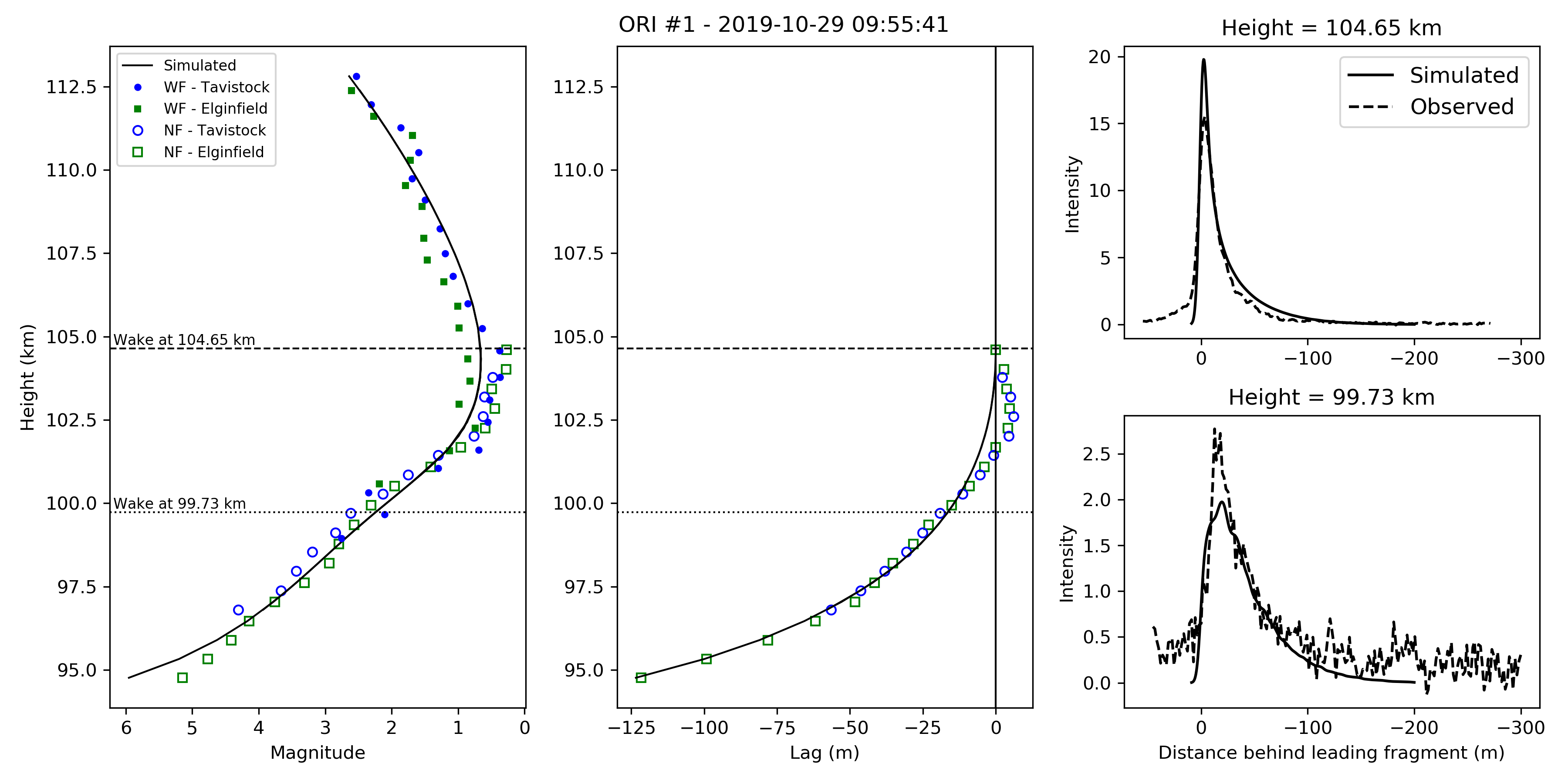}
    \caption{ORI1: 2019-10-29 09:55:41}
\end{figure}
\begin{figure}
    \centering
    \includegraphics[width=\linewidth]{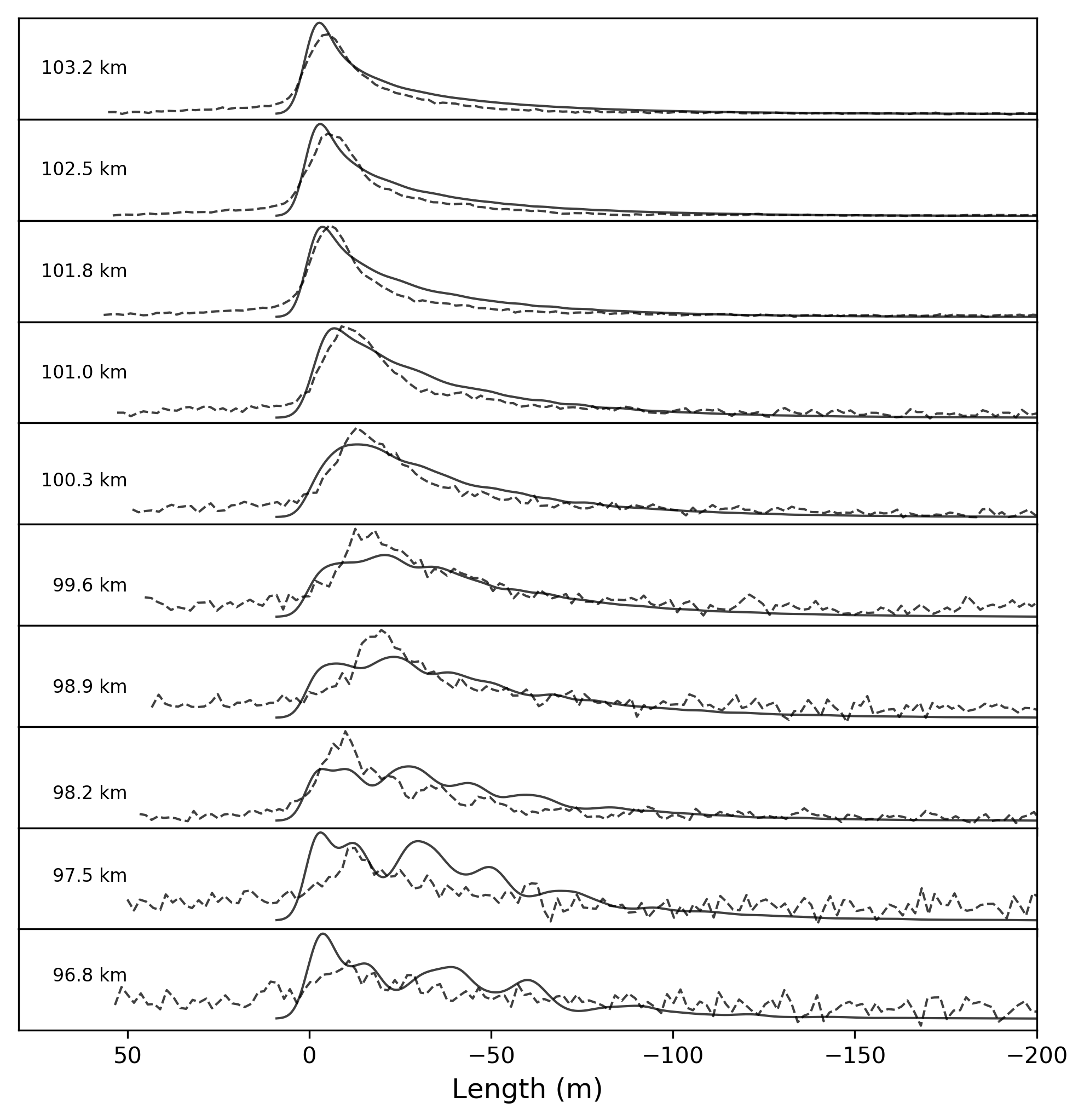}
    \caption{ORI1: 2019-10-29 09:55:41}
\end{figure}

\begin{figure}
    \centering
    \includegraphics[width=\linewidth]{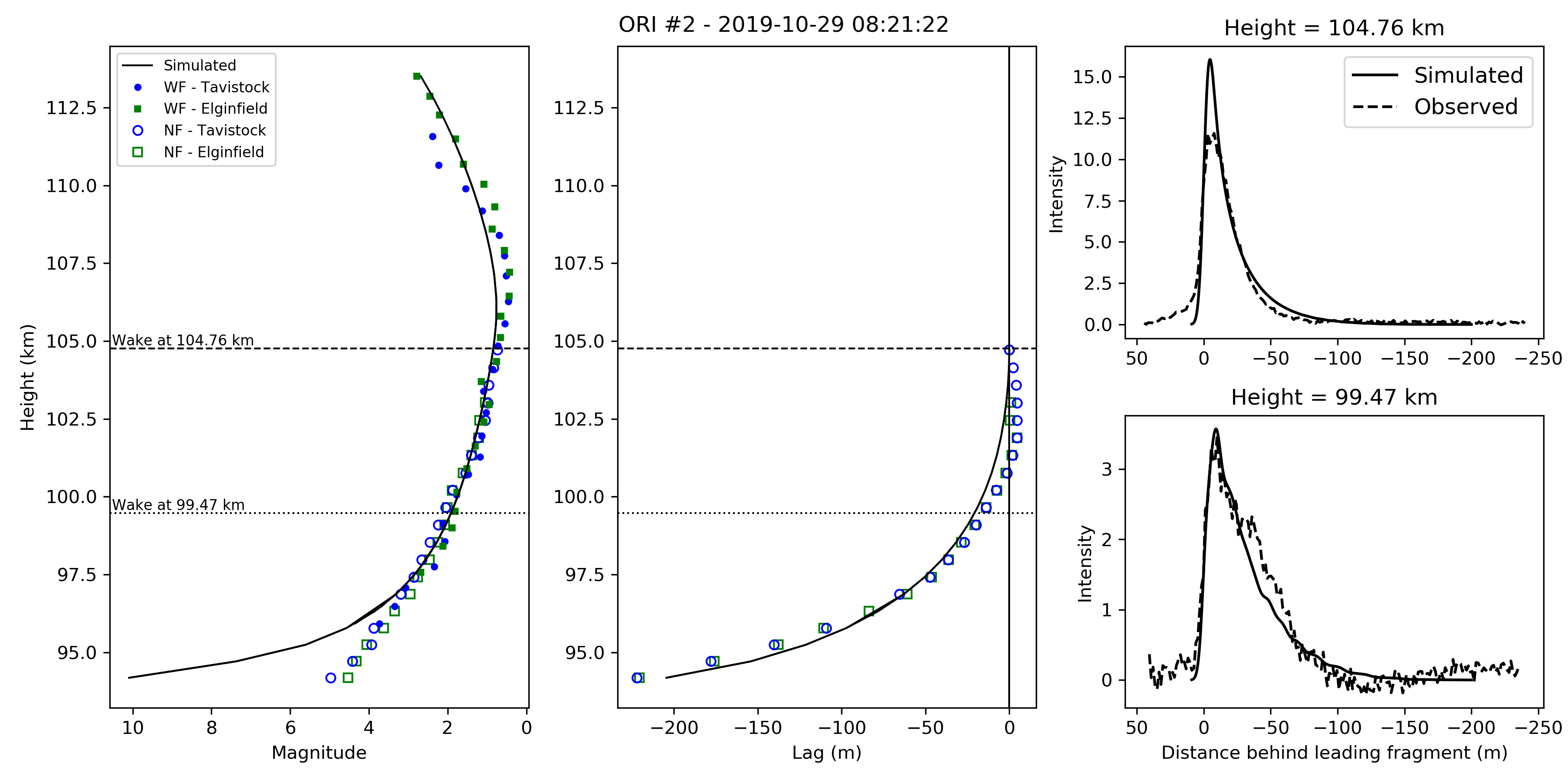}
    \caption{ORI2: 2019-10-29 08:21:22}
\end{figure}
\begin{figure}
    \centering
    \includegraphics[width=\linewidth]{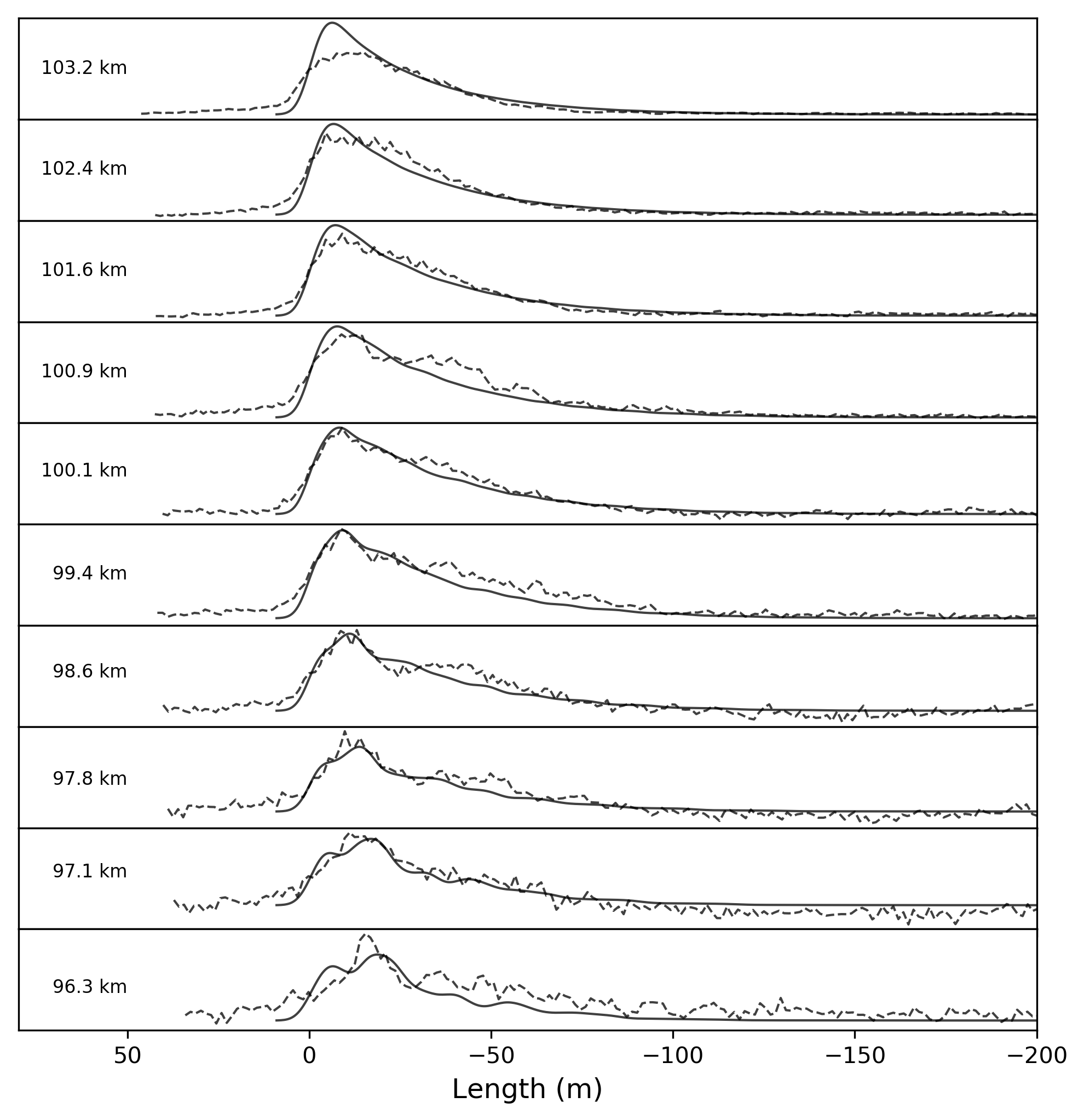}
    \caption{ORI2: 2019-10-29 08:21:22}
\end{figure}

\begin{figure}
    \centering
    \includegraphics[width=\linewidth]{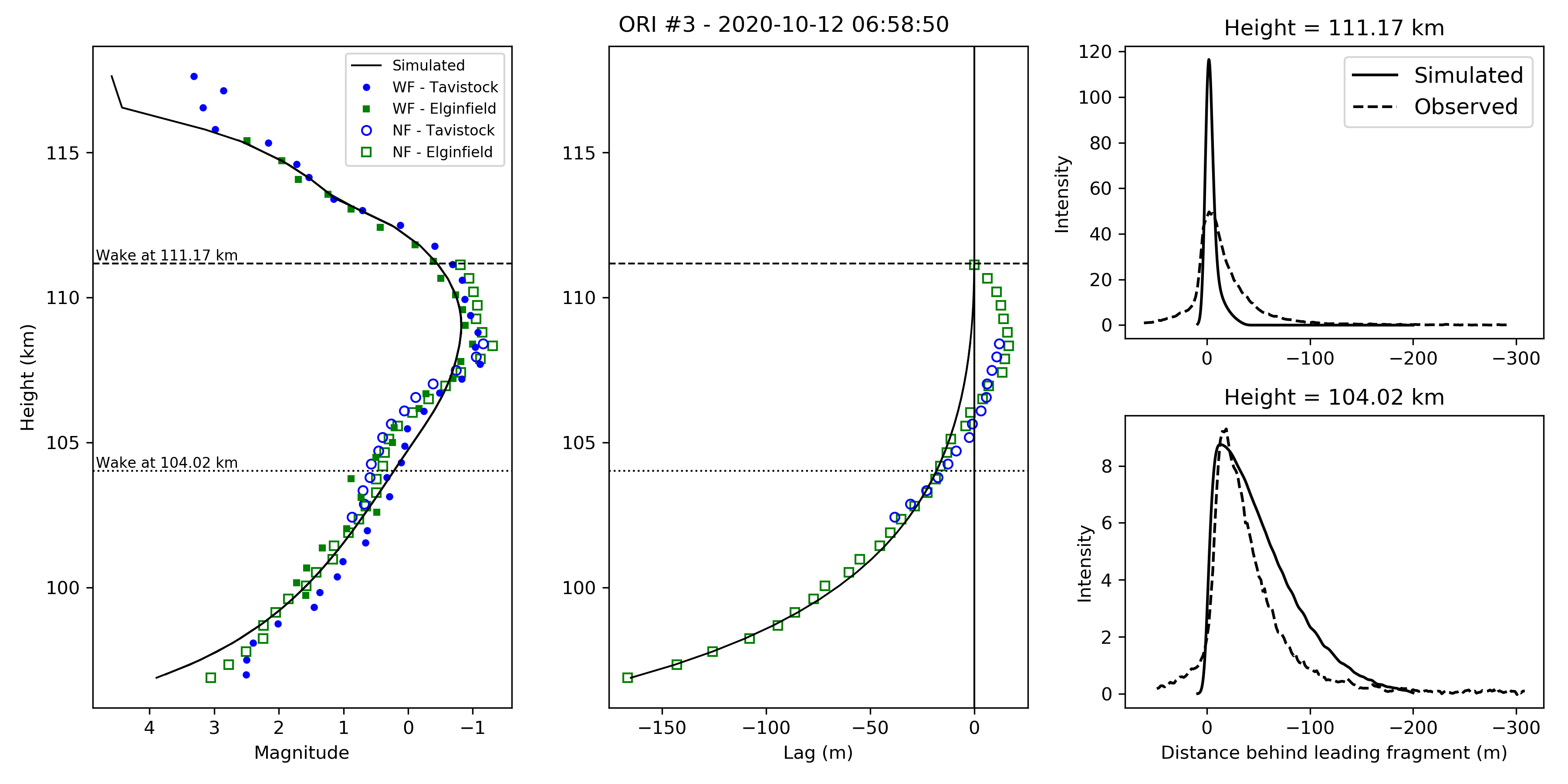}
    \caption{ORI3: 2020-10-12 06:58:50}
\end{figure}
\begin{figure}
    \centering
    \includegraphics[width=\linewidth]{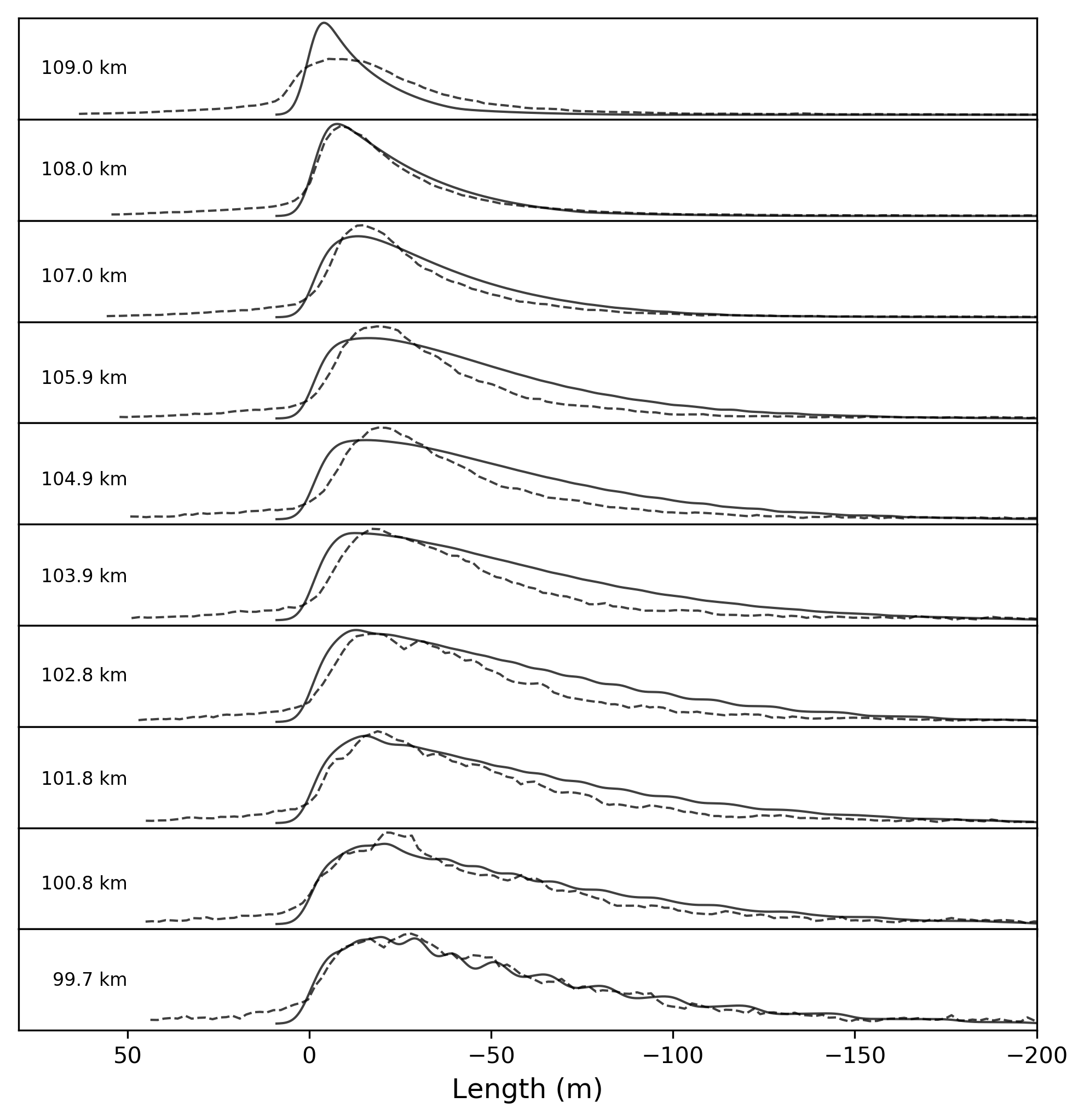}
    \caption{ORI3: 2020-10-12 06:58:50}
\end{figure}

\begin{figure}
    \centering
    \includegraphics[width=\linewidth]{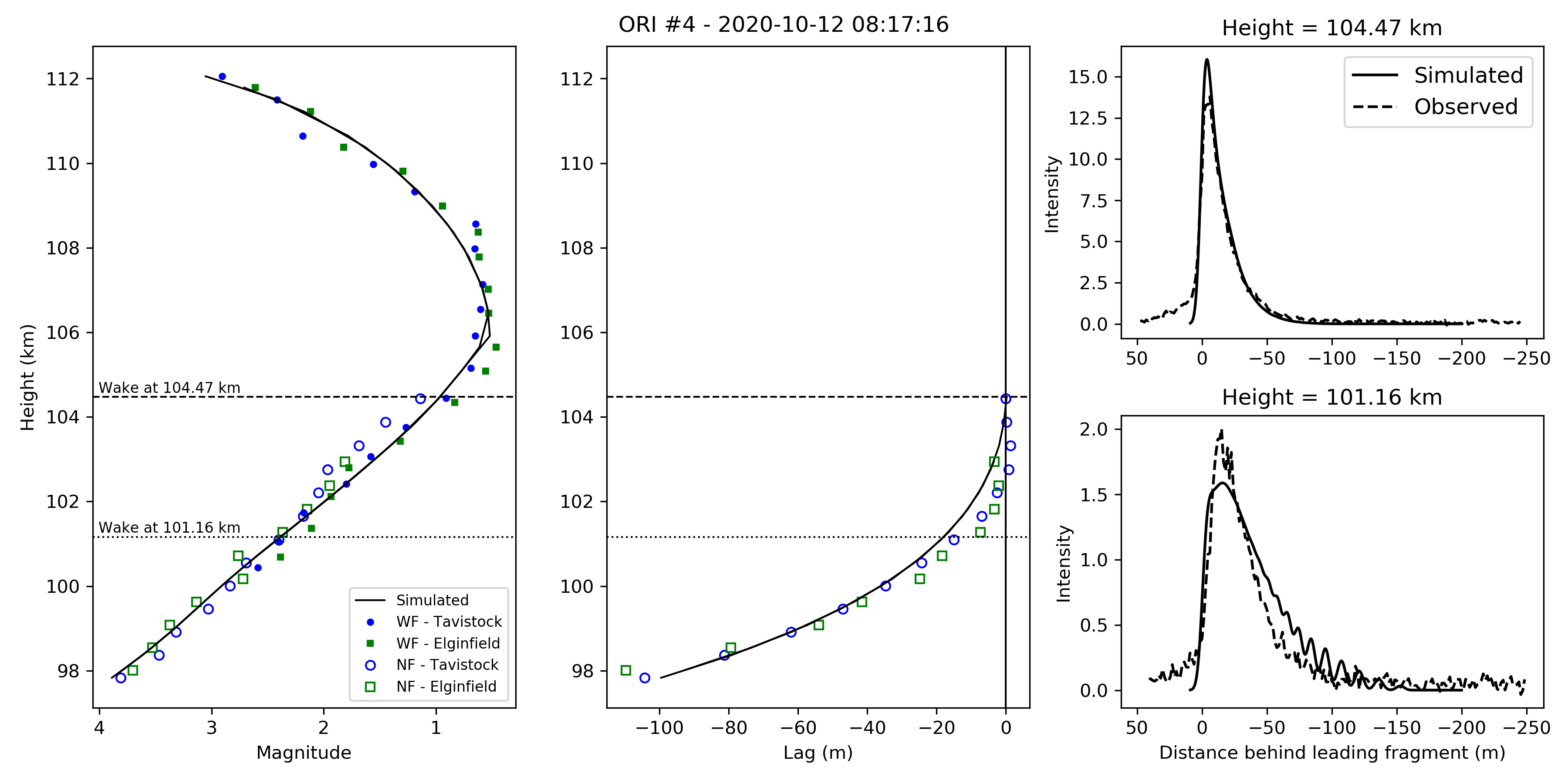}
    \caption{ORI4: 2020-10-12 08:17:16}
\end{figure}
\begin{figure}
    \centering
    \includegraphics[width=\linewidth]{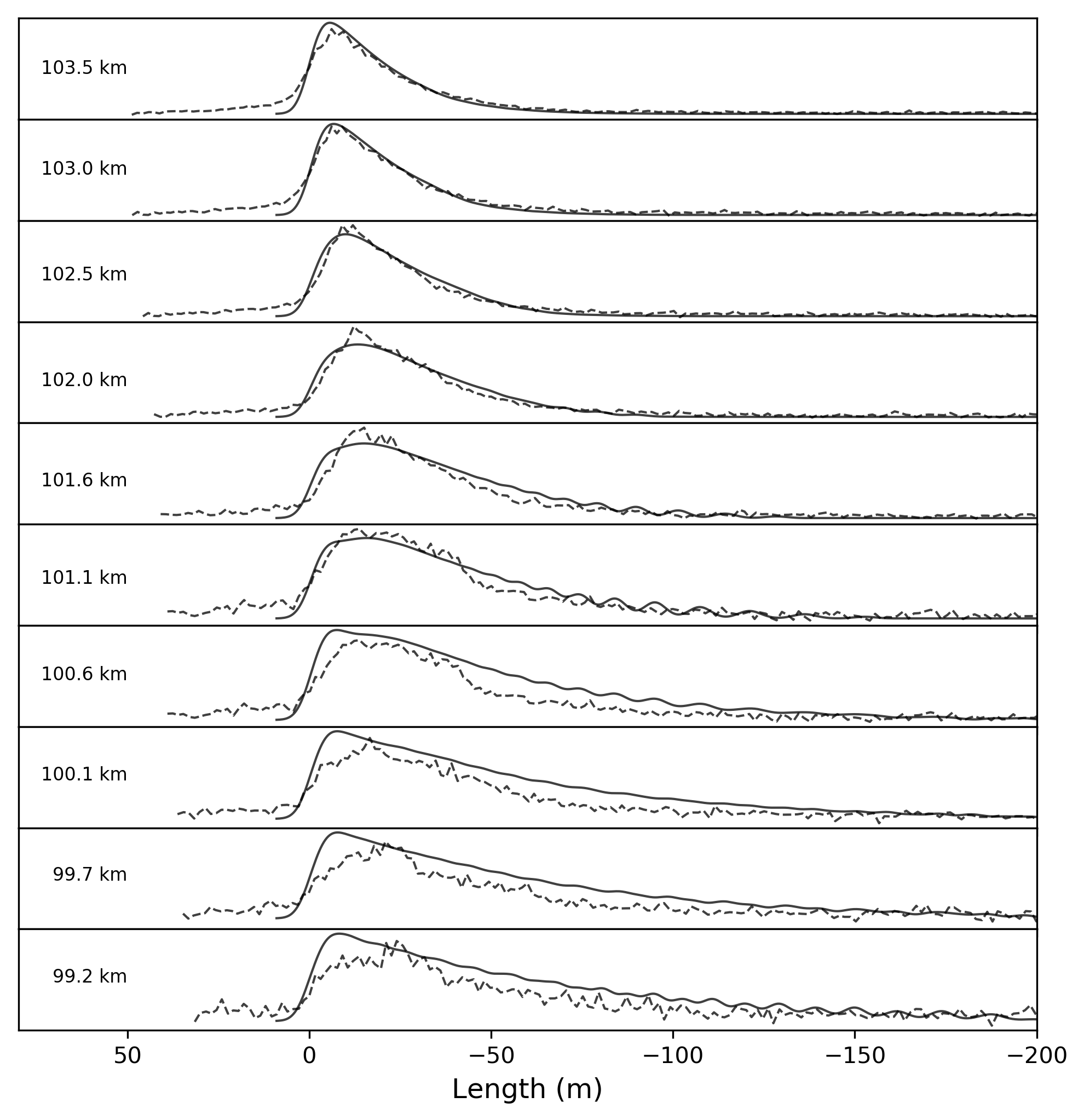}
    \caption{ORI4: 2020-10-12 08:17:16}
\end{figure}

\begin{figure}
    \centering
    \includegraphics[width=\linewidth]{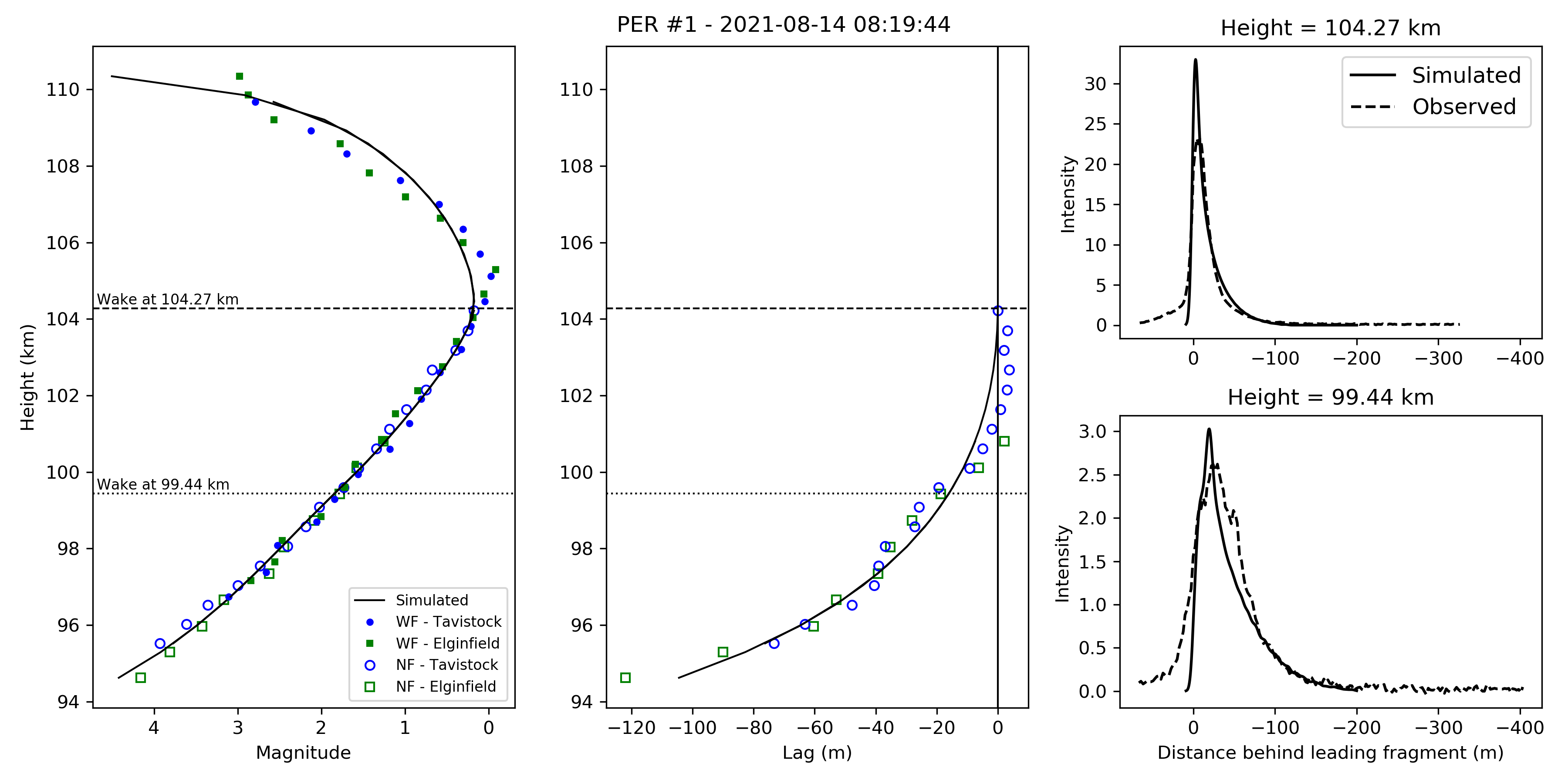}
    \caption{PER1: 2021-08-14 08:19:44}
\end{figure}
\begin{figure}
    \centering
    \includegraphics[width=\linewidth]{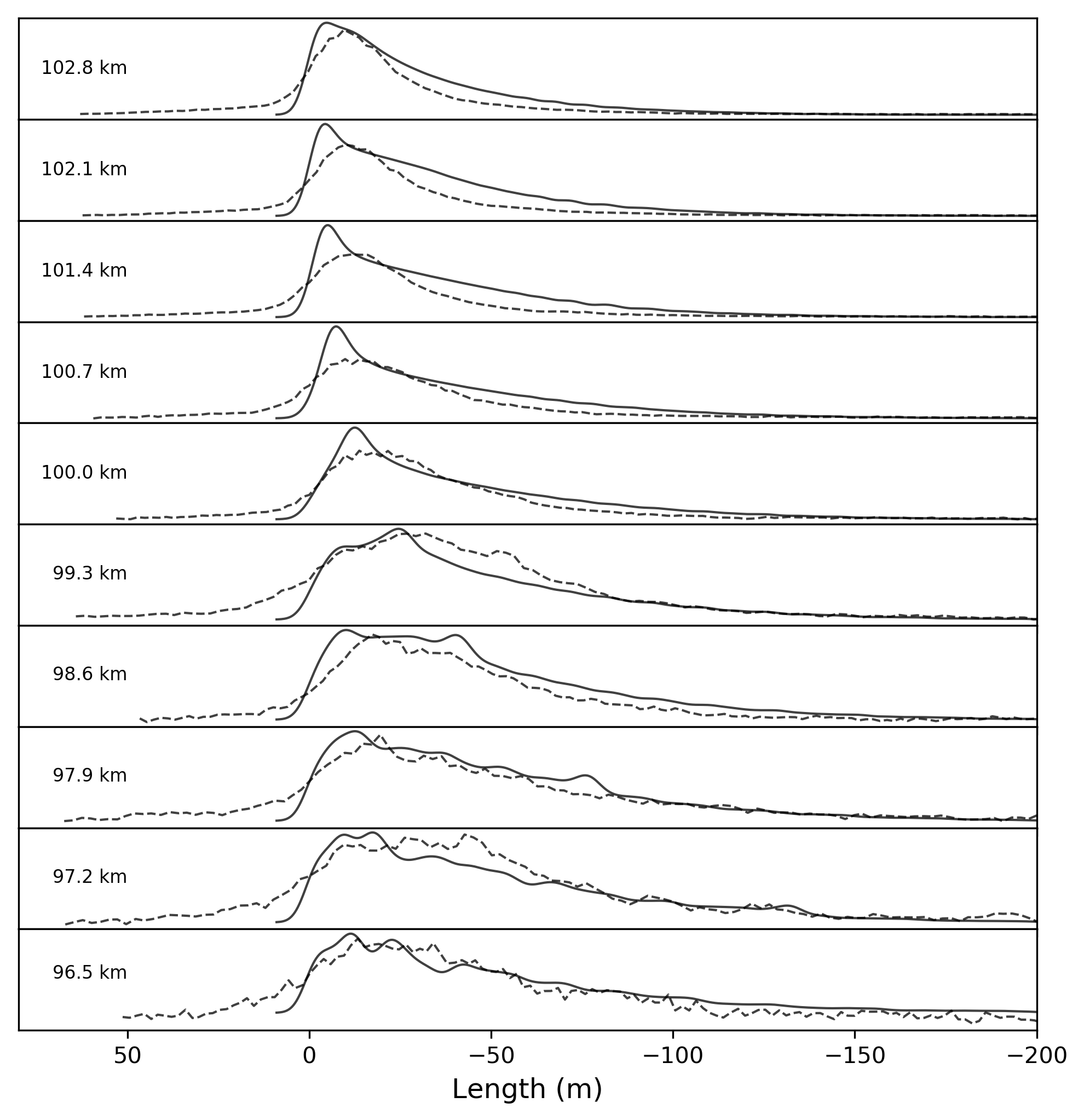}
    \caption{PER1: 2021-08-14 08:19:44}
\end{figure}

\begin{figure}
    \centering
    \includegraphics[width=\linewidth]{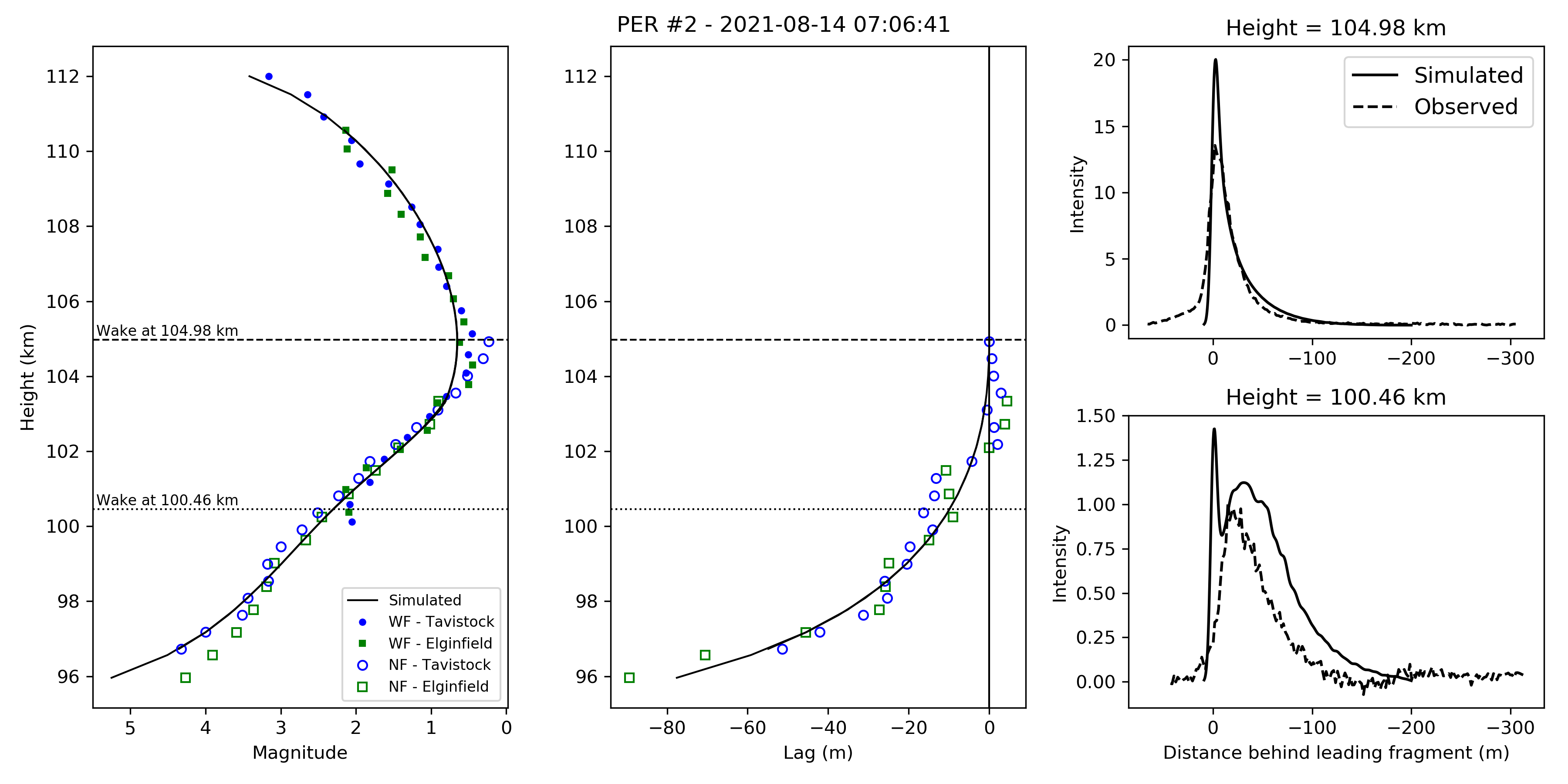}
    \caption{PER2: 2021-08-14 07:06:41}
\end{figure}
\begin{figure}
    \centering
    \includegraphics[width=\linewidth]{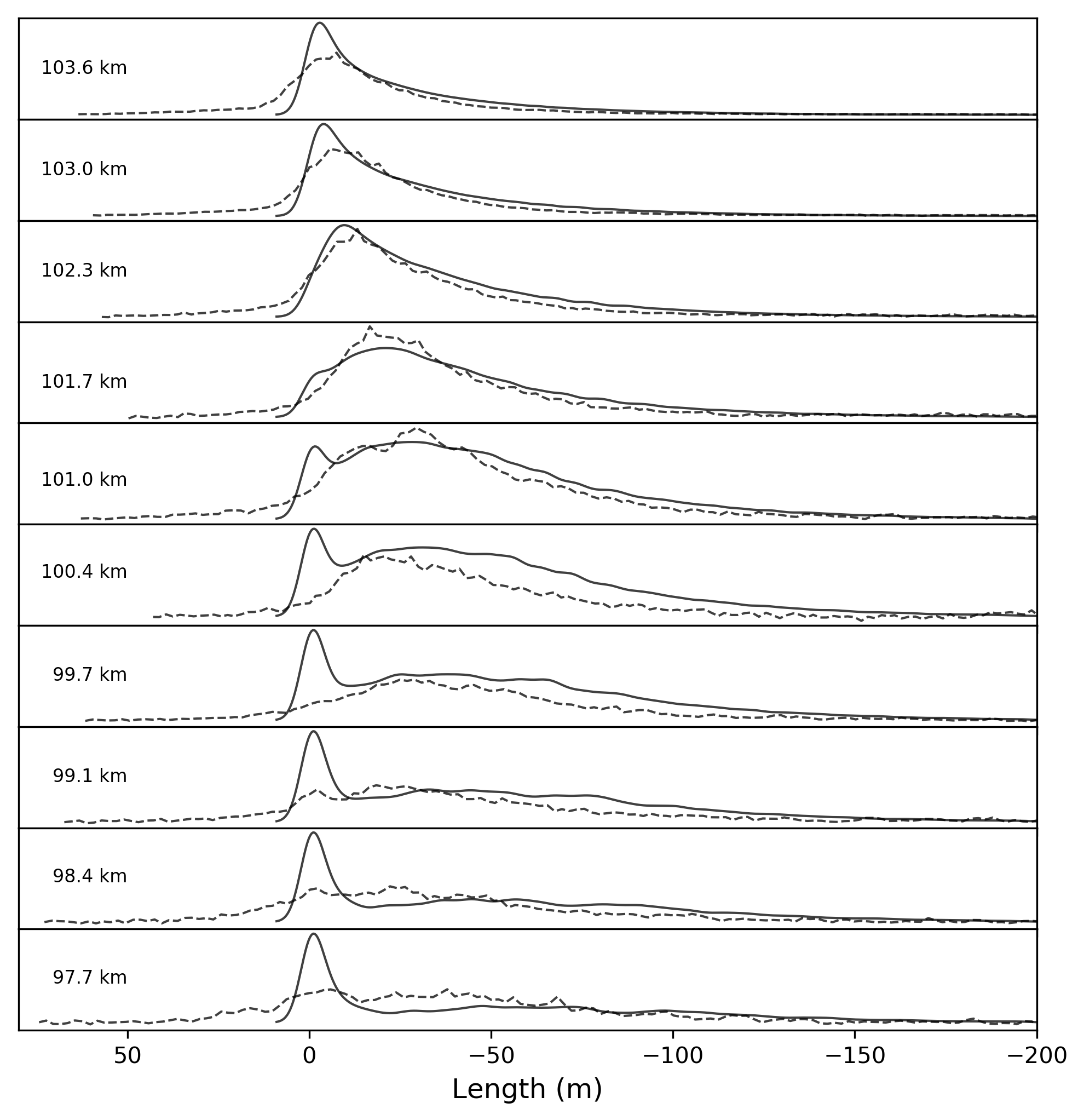}
    \caption{PER2: 2021-08-14 07:06:41}
\end{figure}

\begin{figure}
    \centering
    \includegraphics[width=\linewidth]{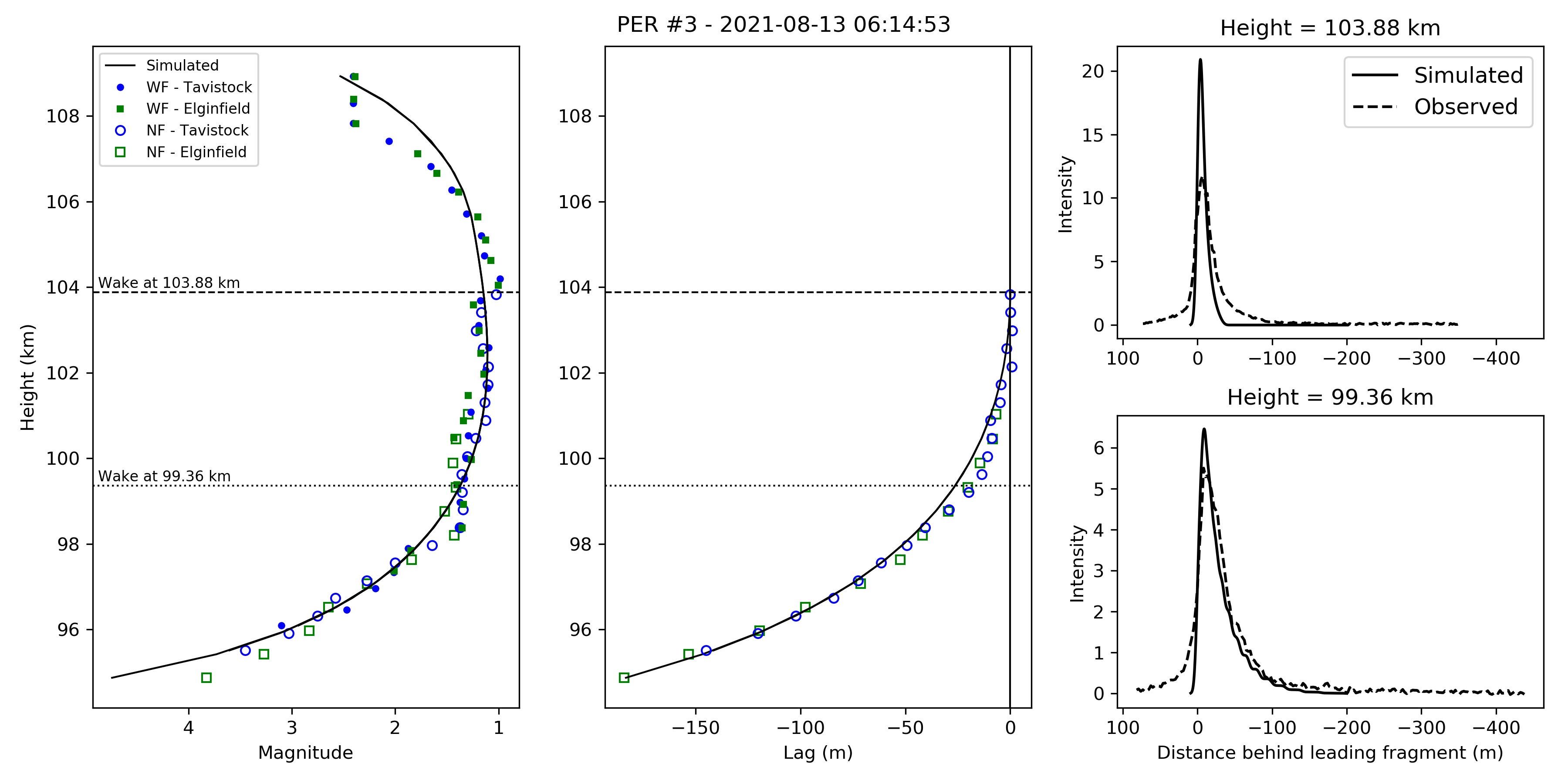}
    \caption{PER3: 2021-08-13 06:14:53}
\end{figure}
\begin{figure}
    \centering
    \includegraphics[width=\linewidth]{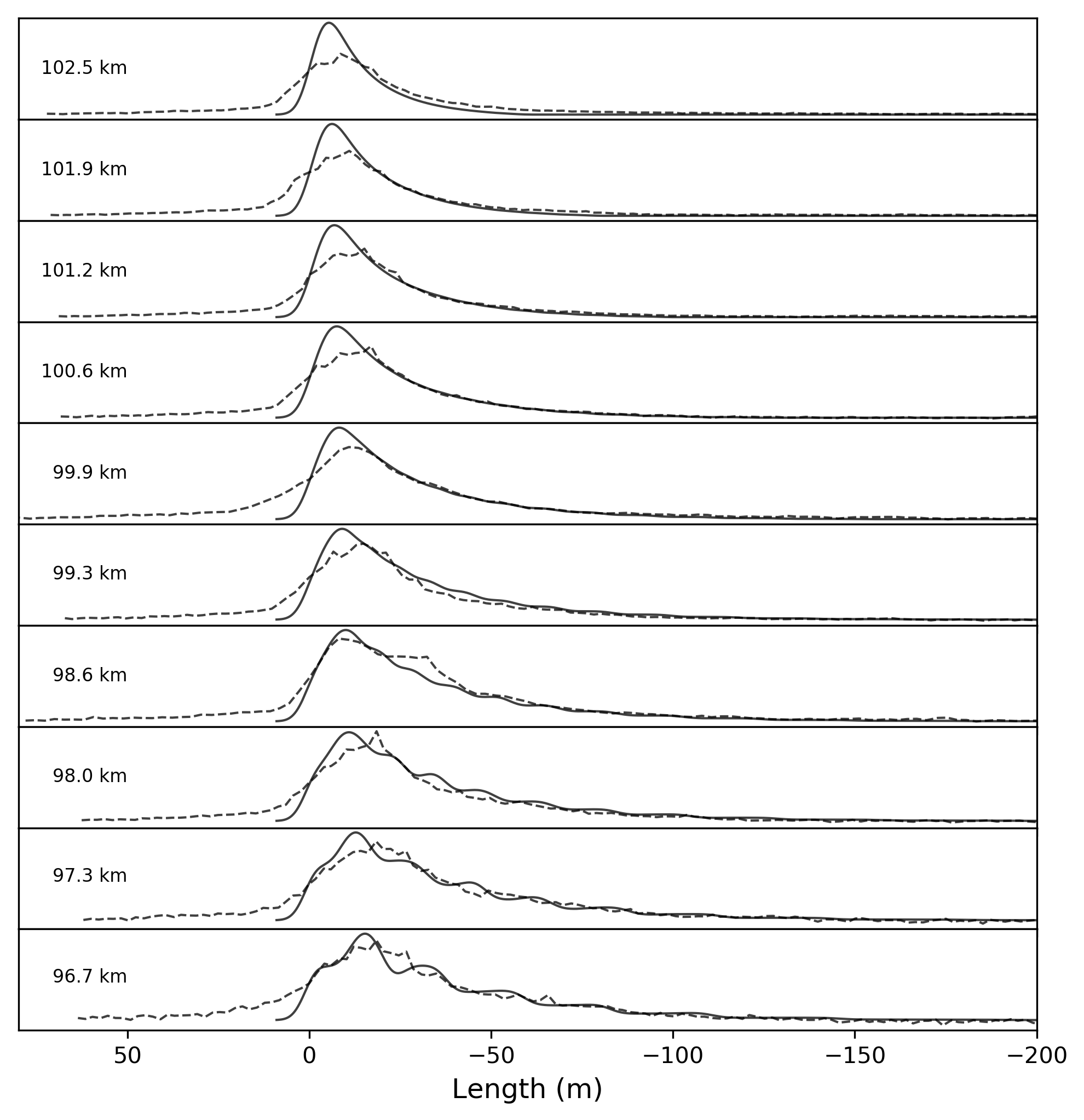}
    \caption{PER3: 2021-08-13 06:14:53}
\end{figure}

\begin{figure}
    \centering
    \includegraphics[width=\linewidth]{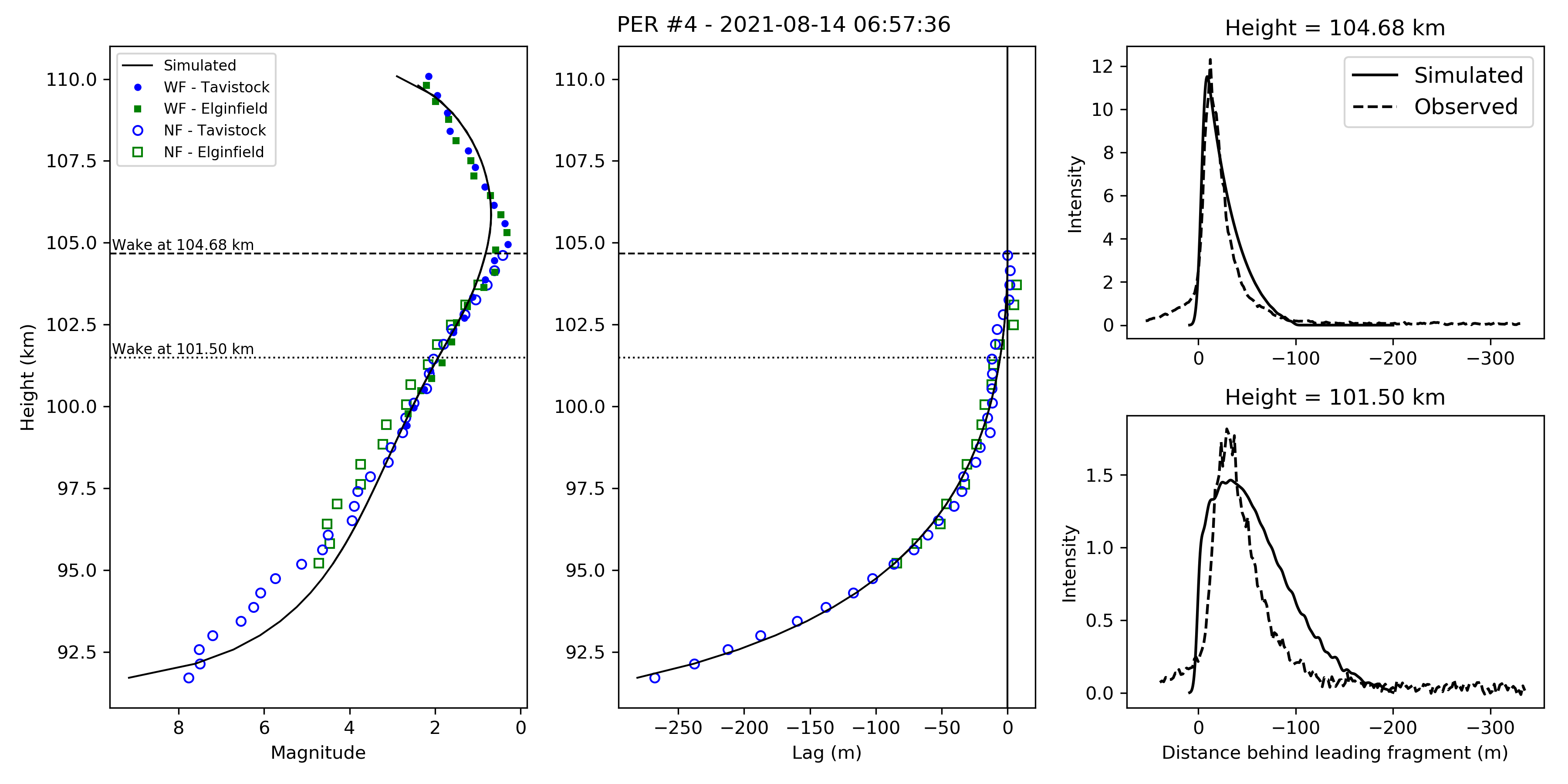}
    \caption{PER4: 2021-08-14 06:57:36}
\end{figure}
\begin{figure}
    \centering
    \includegraphics[width=\linewidth]{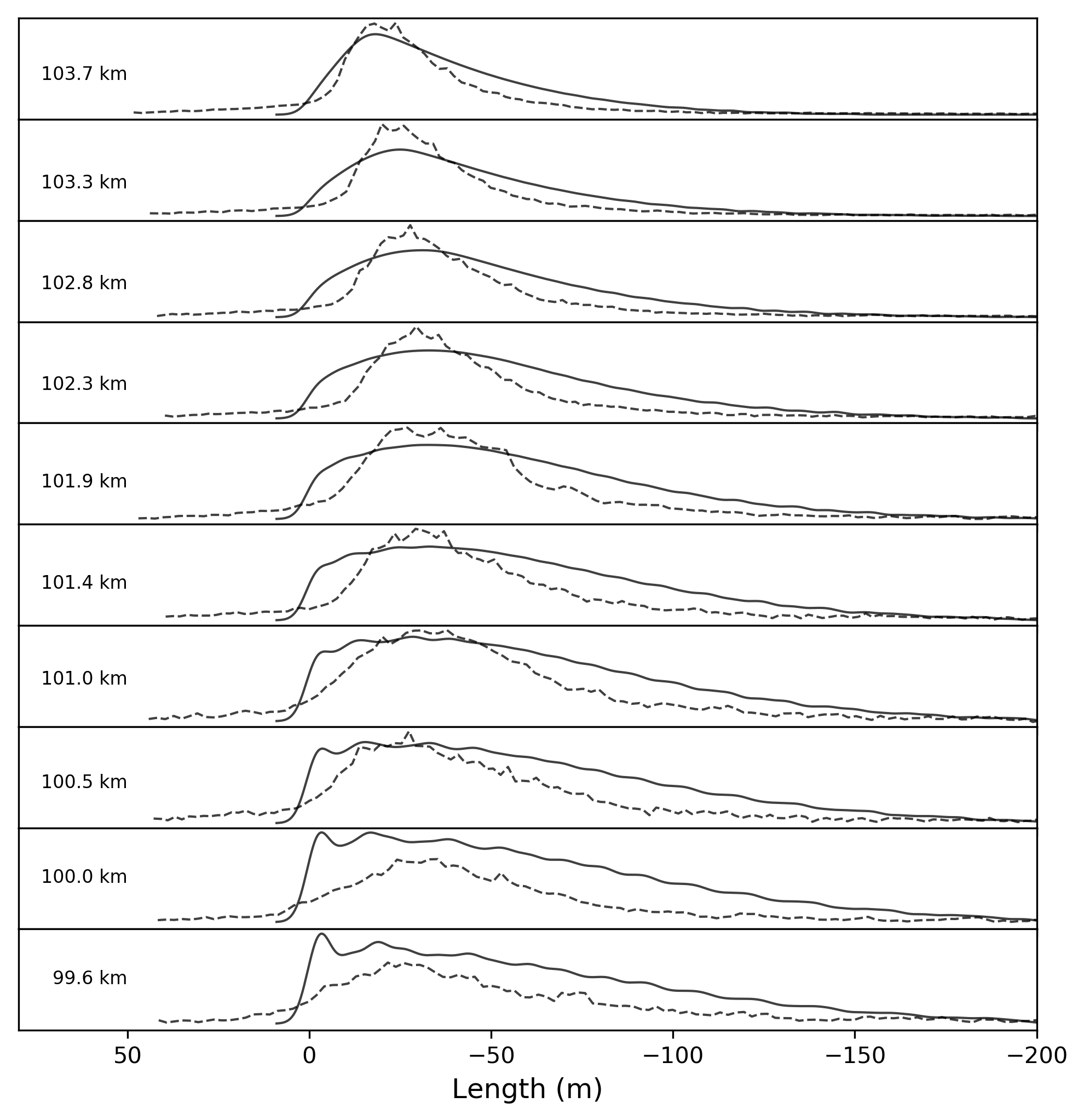}
    \caption{PER4: 2021-08-14 06:57:36}
\end{figure}

\begin{figure}
    \centering
    \includegraphics[width=\linewidth]{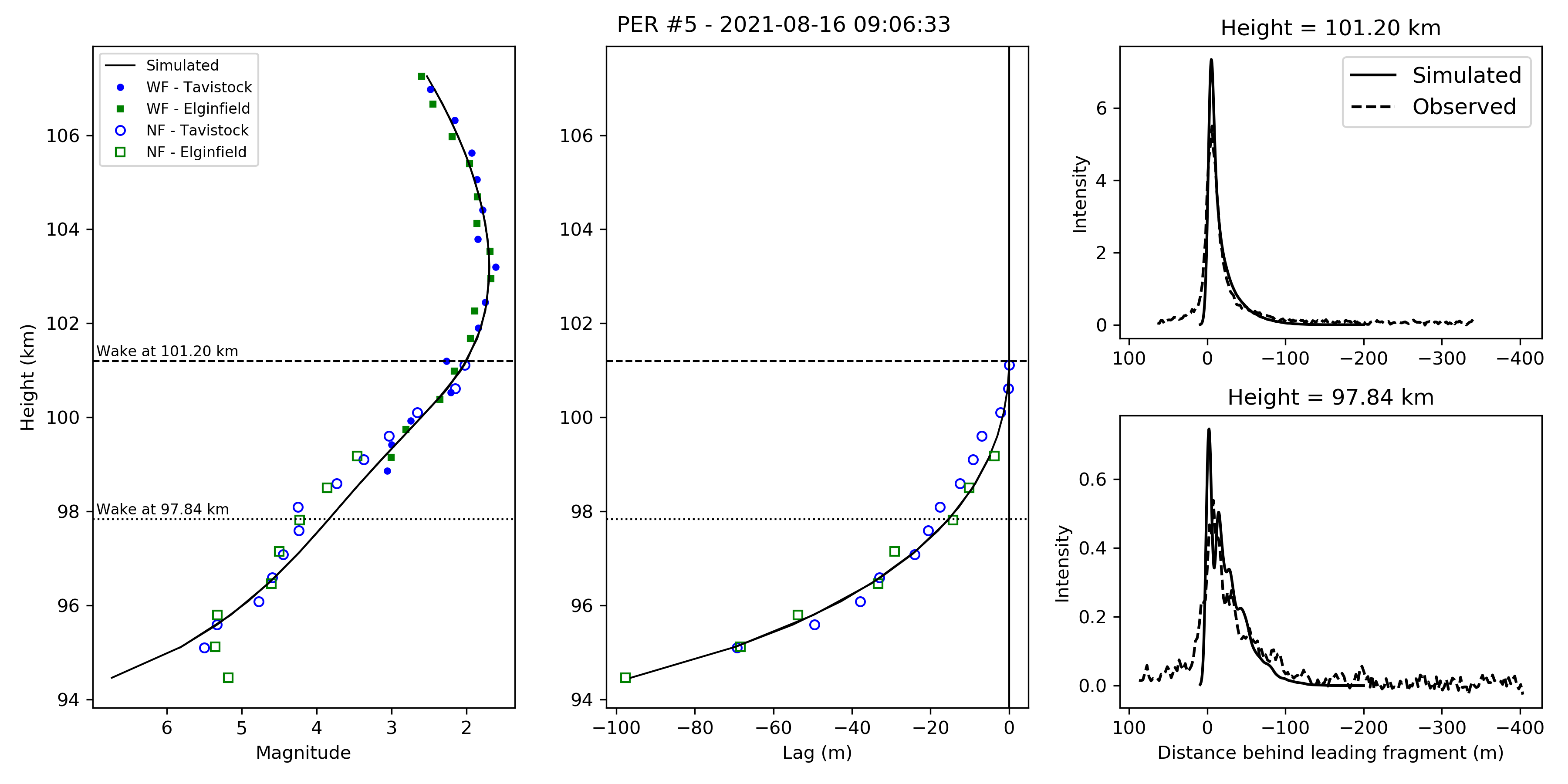}
    \caption{PER5: 2021-08-16 09:06:33}
\end{figure}
\begin{figure}
    \centering
    \includegraphics[width=\linewidth]{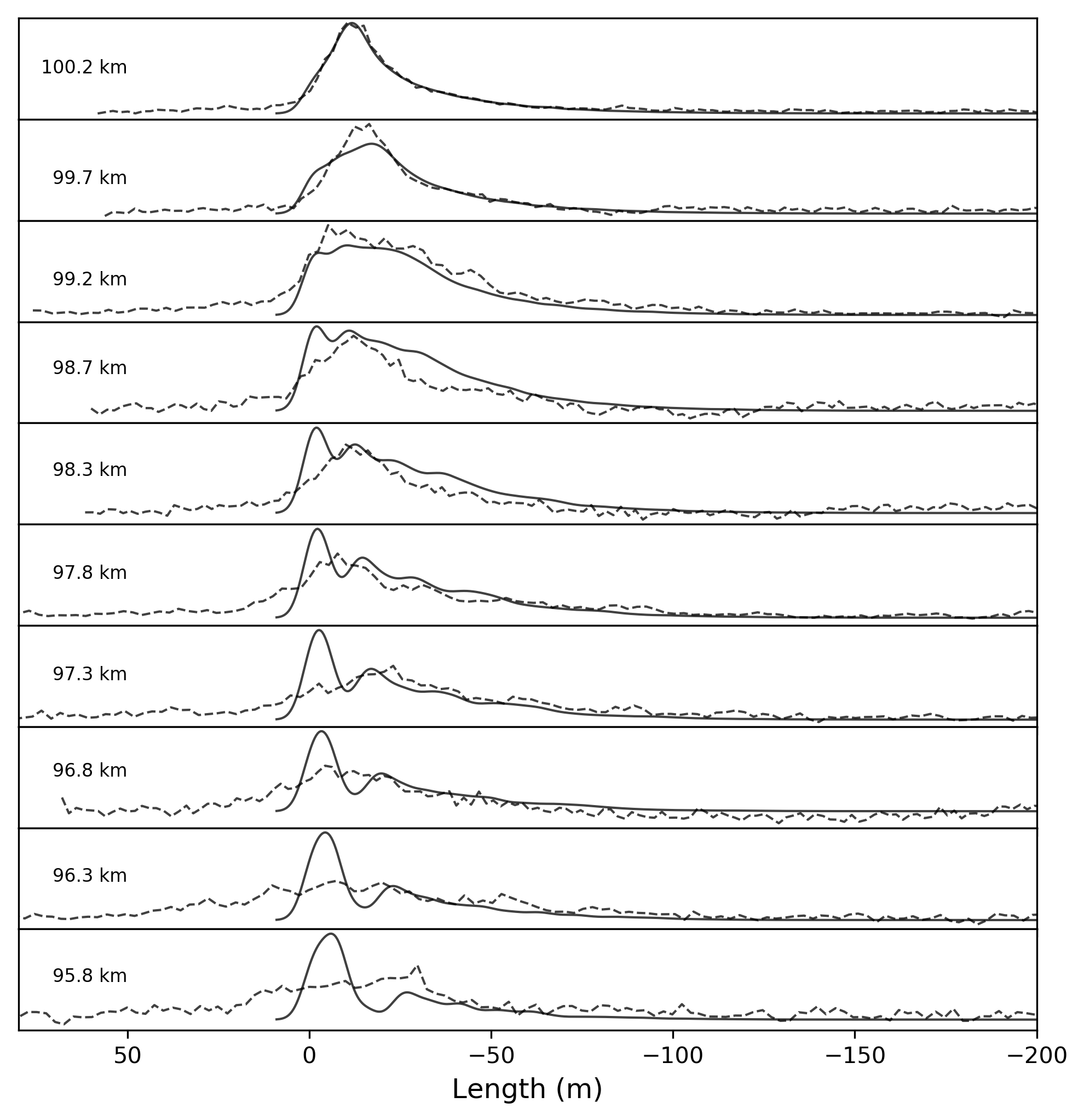}
    \caption{PER5: 2021-08-16 09:06:33}
\end{figure}

\begin{figure}
    \centering
    \includegraphics[width=\linewidth]{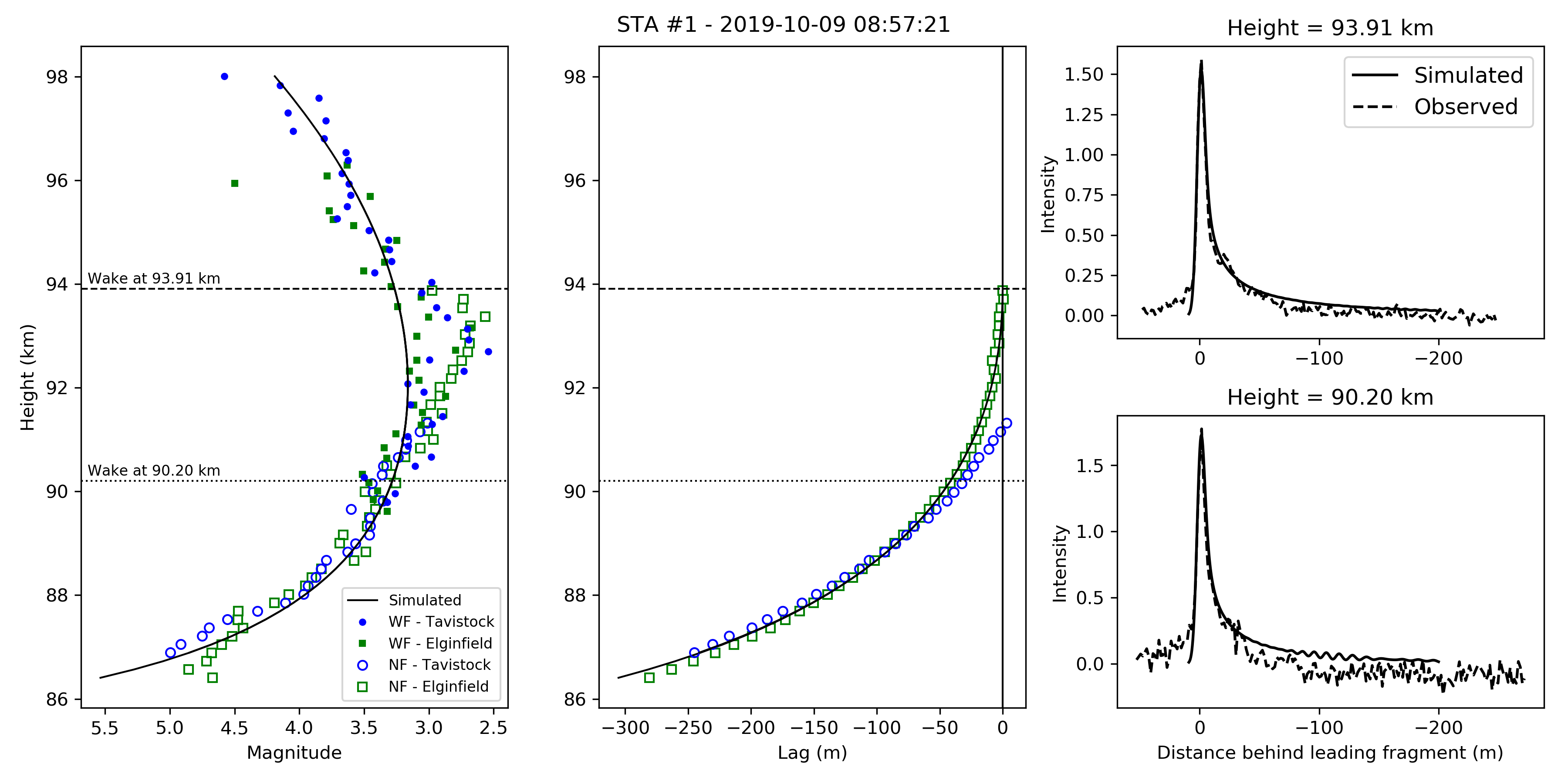}
    \caption{STA1: 2019-10-09 08:57:21}
\end{figure}
\begin{figure}
    \centering
    \includegraphics[width=\linewidth]{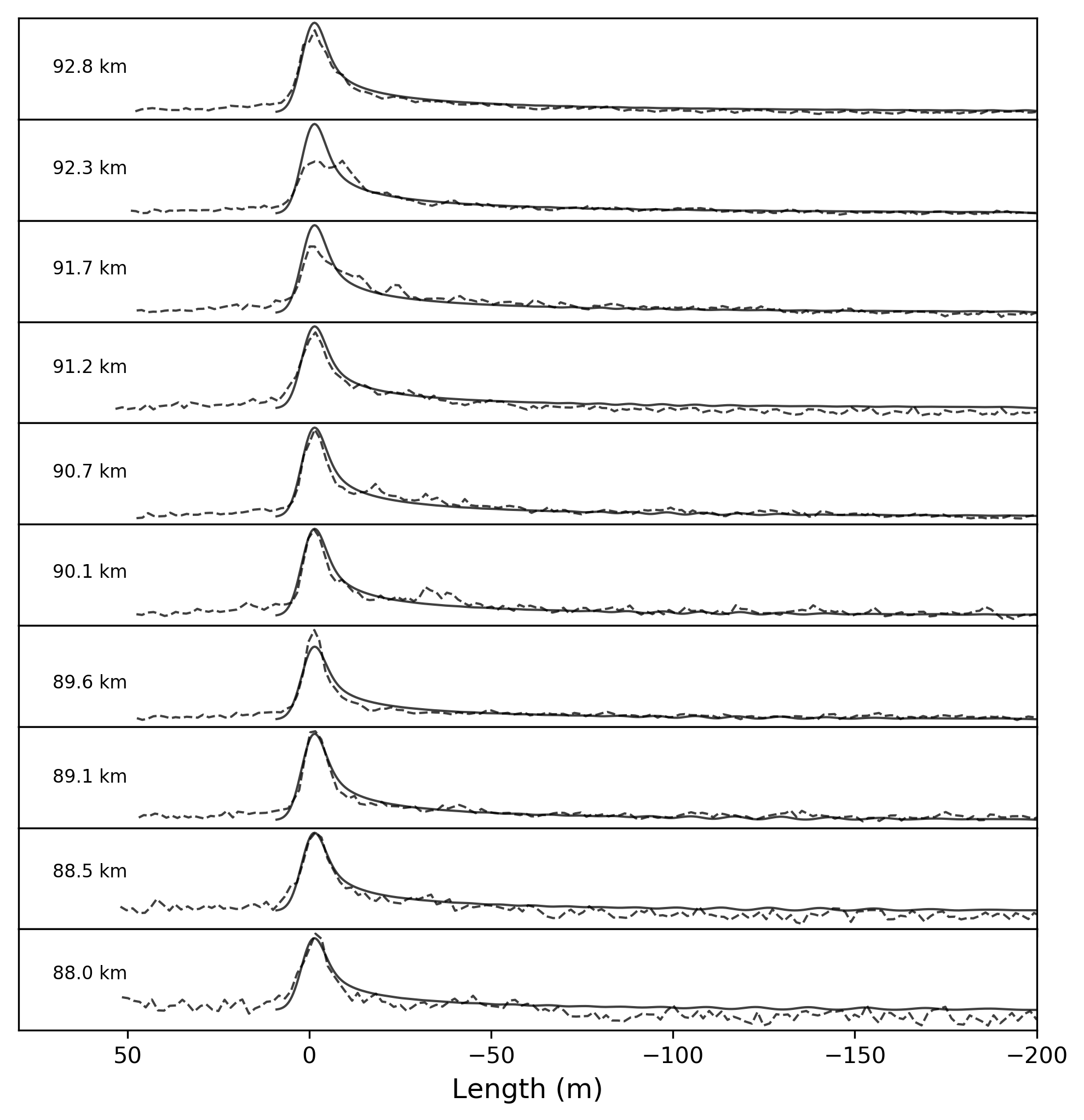}
    \caption{STA1: 2019-10-09 08:57:21}
\end{figure}

\begin{figure}
    \centering
    \includegraphics[width=\linewidth]{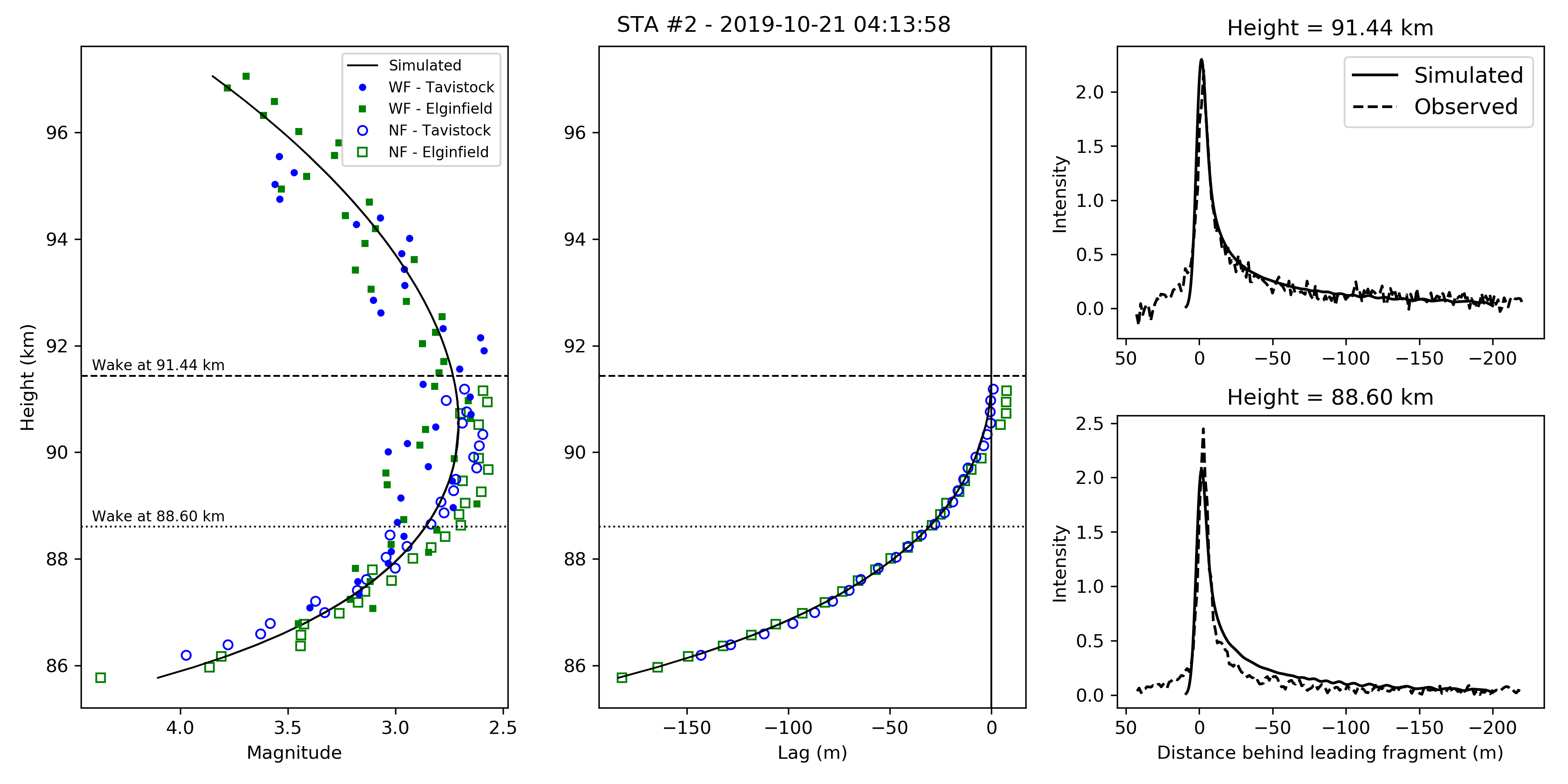}
    \caption{STA2: 2019-10-121 04:13:58}
\end{figure}
\begin{figure}
    \centering
    \includegraphics[width=\linewidth]{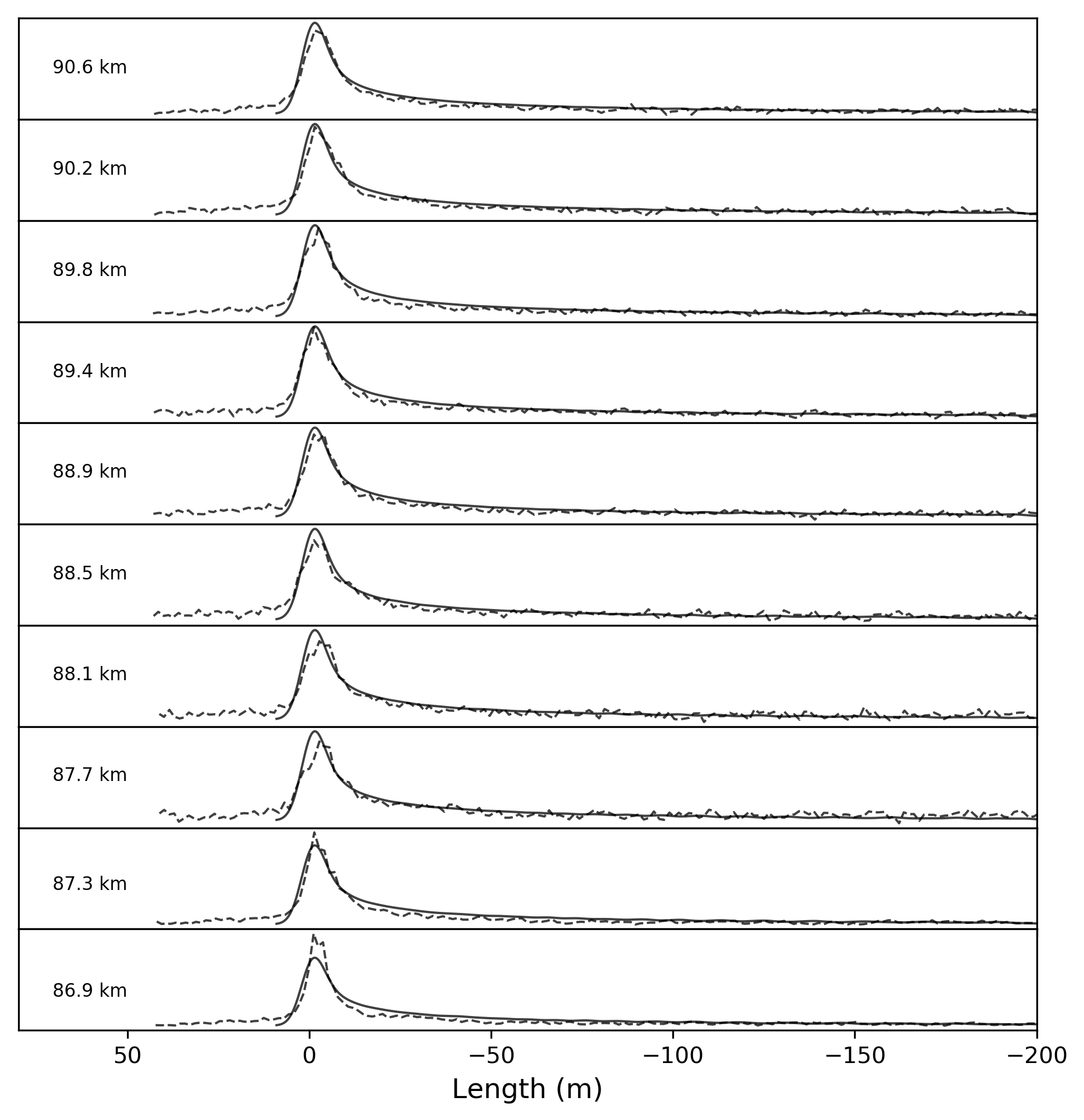}
    \caption{STA2: 2019-10-121 04:13:58}
\end{figure}

\begin{figure}
    \centering
    \includegraphics[width=\linewidth]{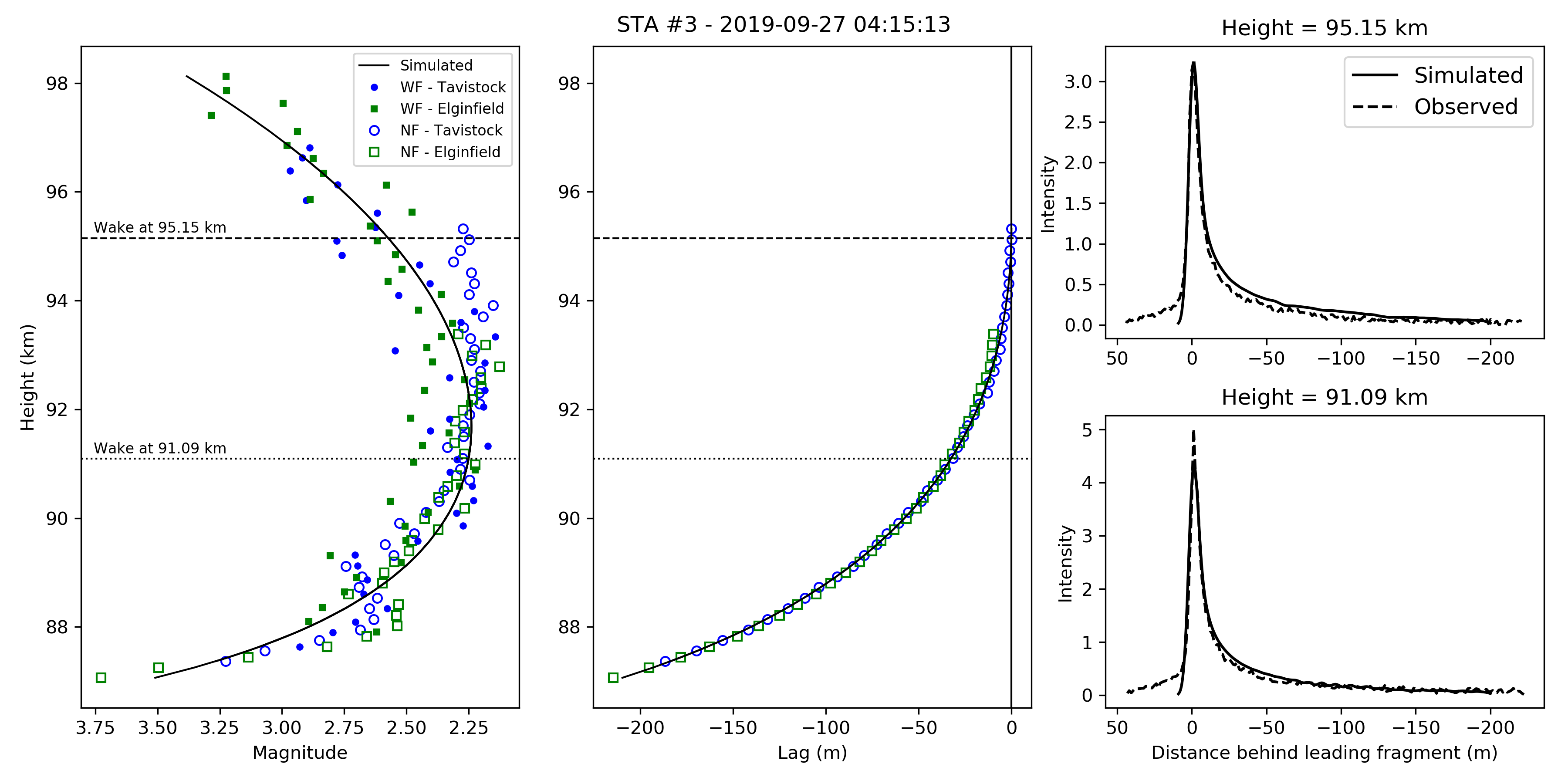}
    \caption{STA3: 2019-09-27 04:15:14}
\end{figure}
\begin{figure}
    \centering
    \includegraphics[width=\linewidth]{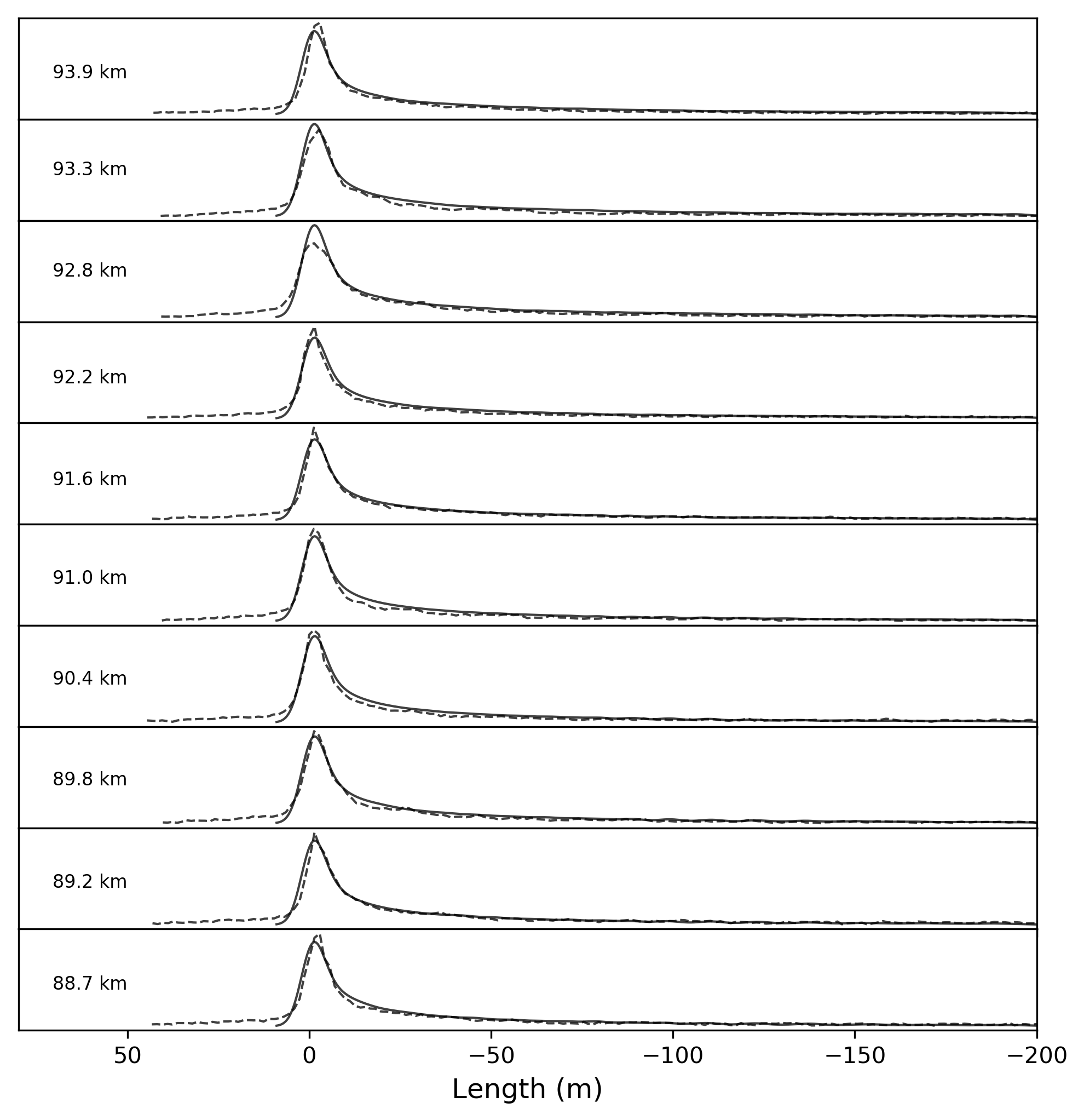}
    \caption{STA3: 2019-09-27 04:15:14}
\end{figure}

\clearpage

\begin{figure}
    \centering
    \includegraphics[width=\linewidth]{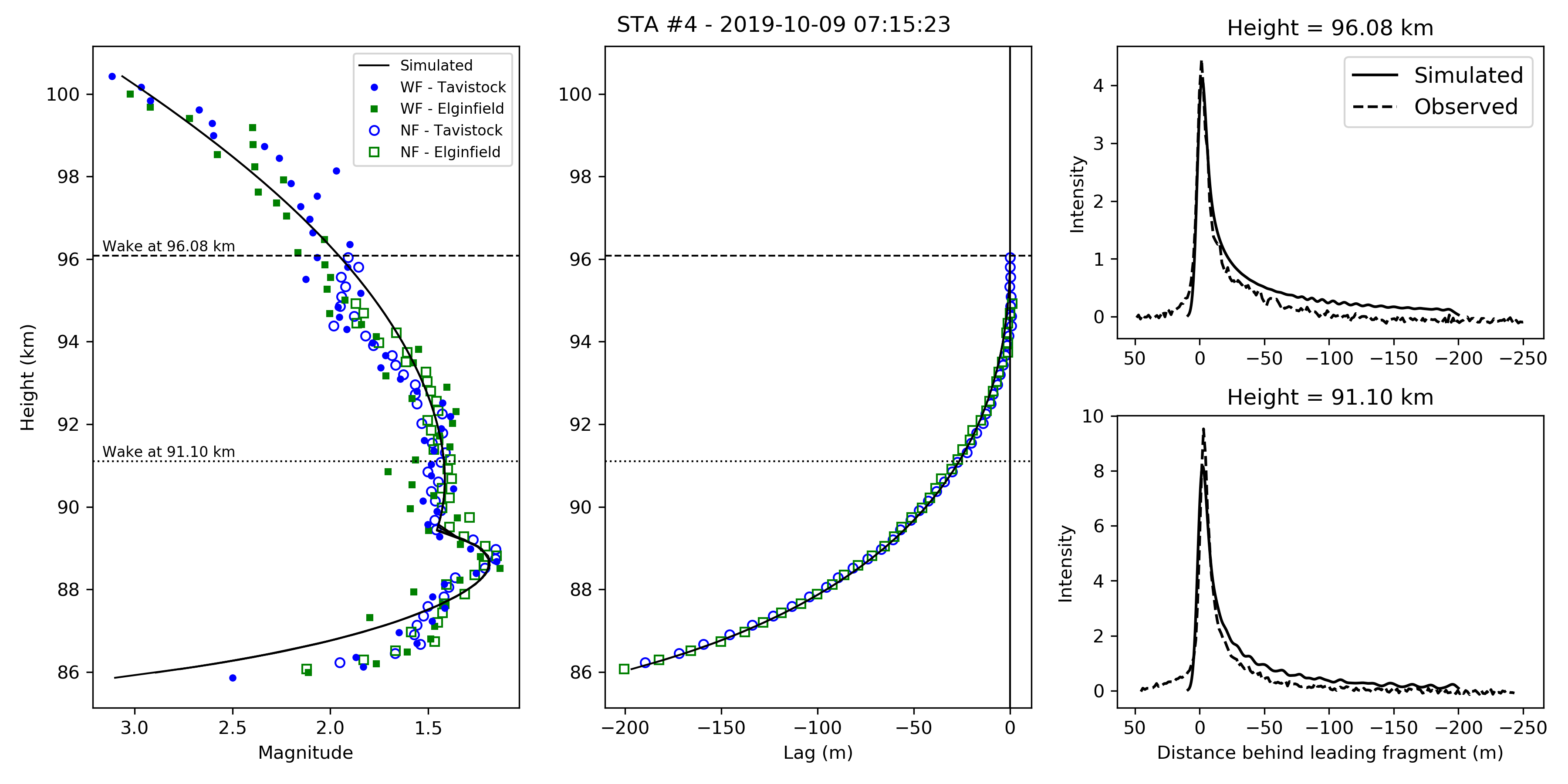}
    \caption{STA4: 2019-10-09 07:15:23}
\end{figure}
\begin{figure}
    \centering
    \includegraphics[width=\linewidth]{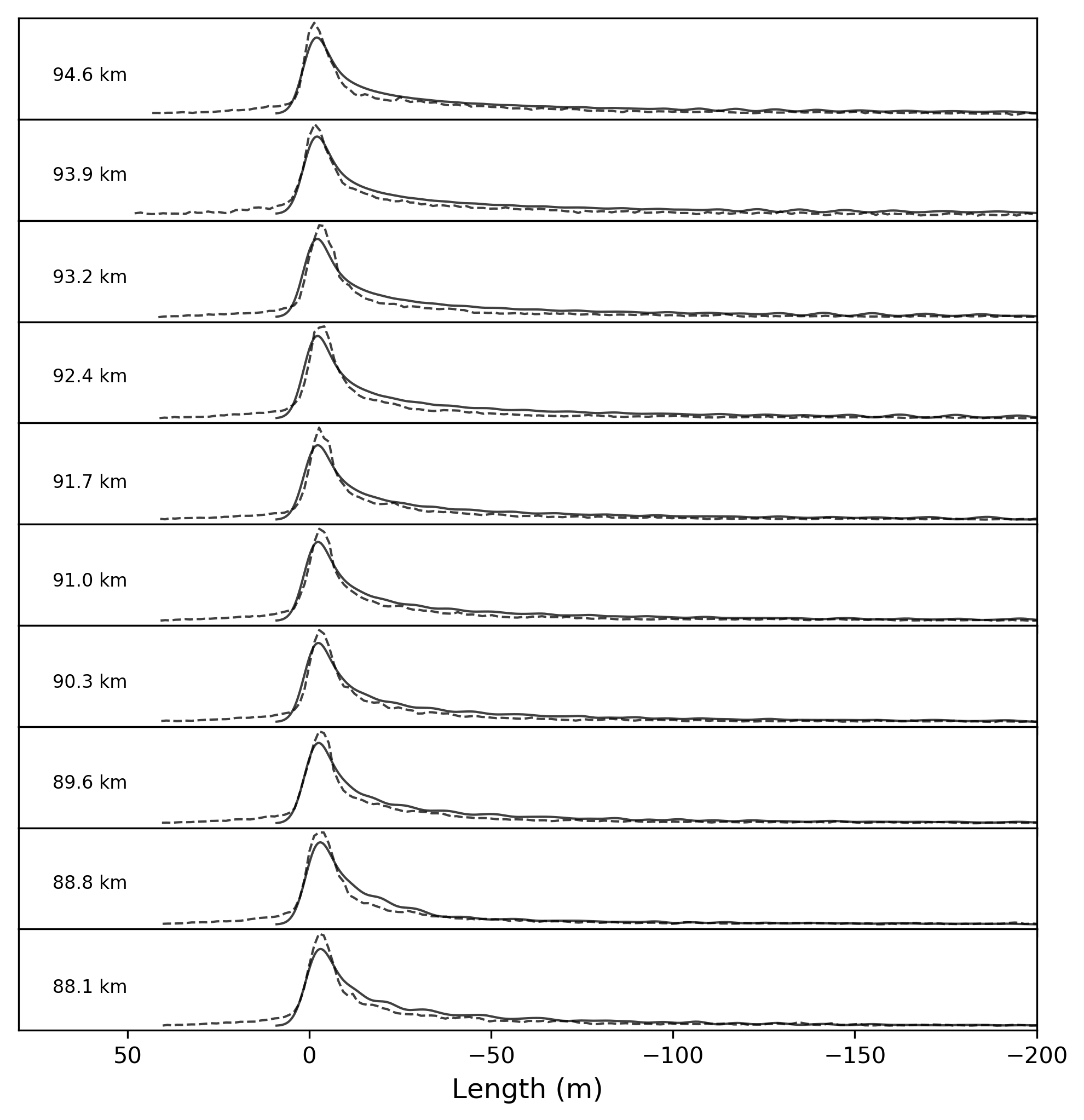}
    \caption{STA4: 2019-10-09 07:15:23}
\end{figure}

\begin{figure}
    \centering
    \includegraphics[width=\linewidth]{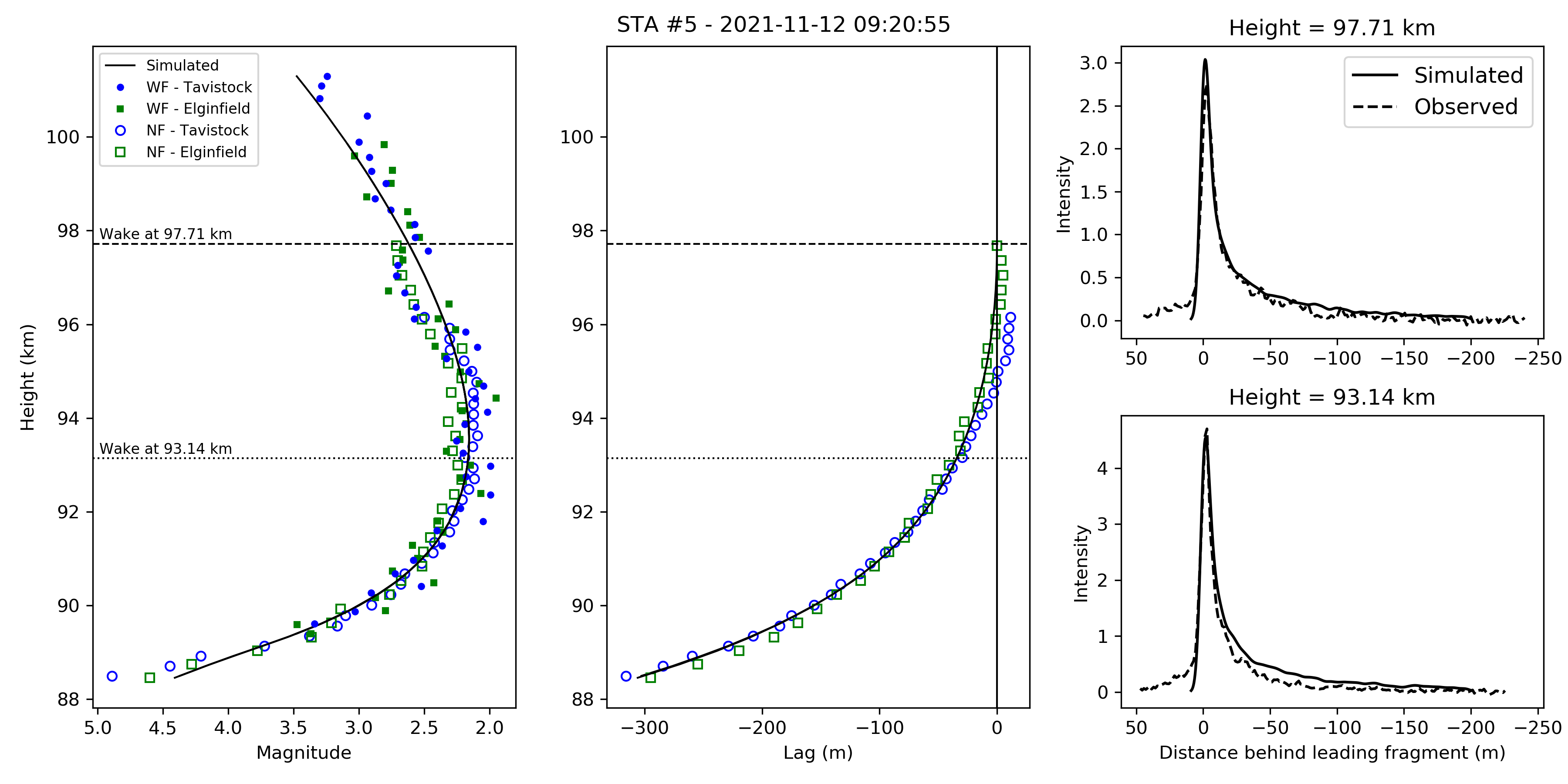}
    \caption{STA5: 2021-11-12 09:20:56}
\end{figure}
\begin{figure}
    \centering
    \includegraphics[width=\linewidth]{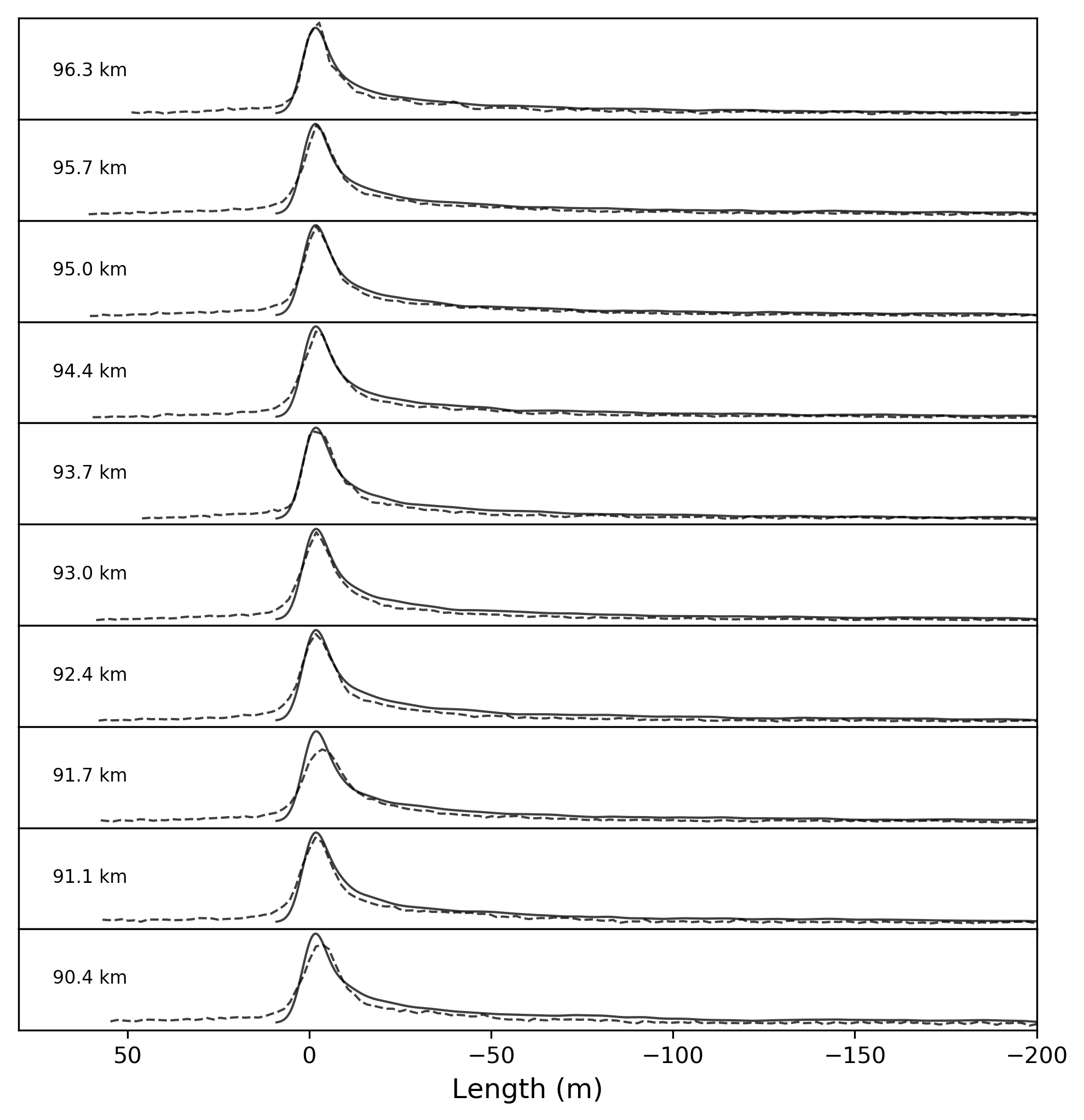}
    \caption{STA5: 2021-11-12 09:20:56}
\end{figure}

\begin{figure}
    \centering
    \includegraphics[width=\linewidth]{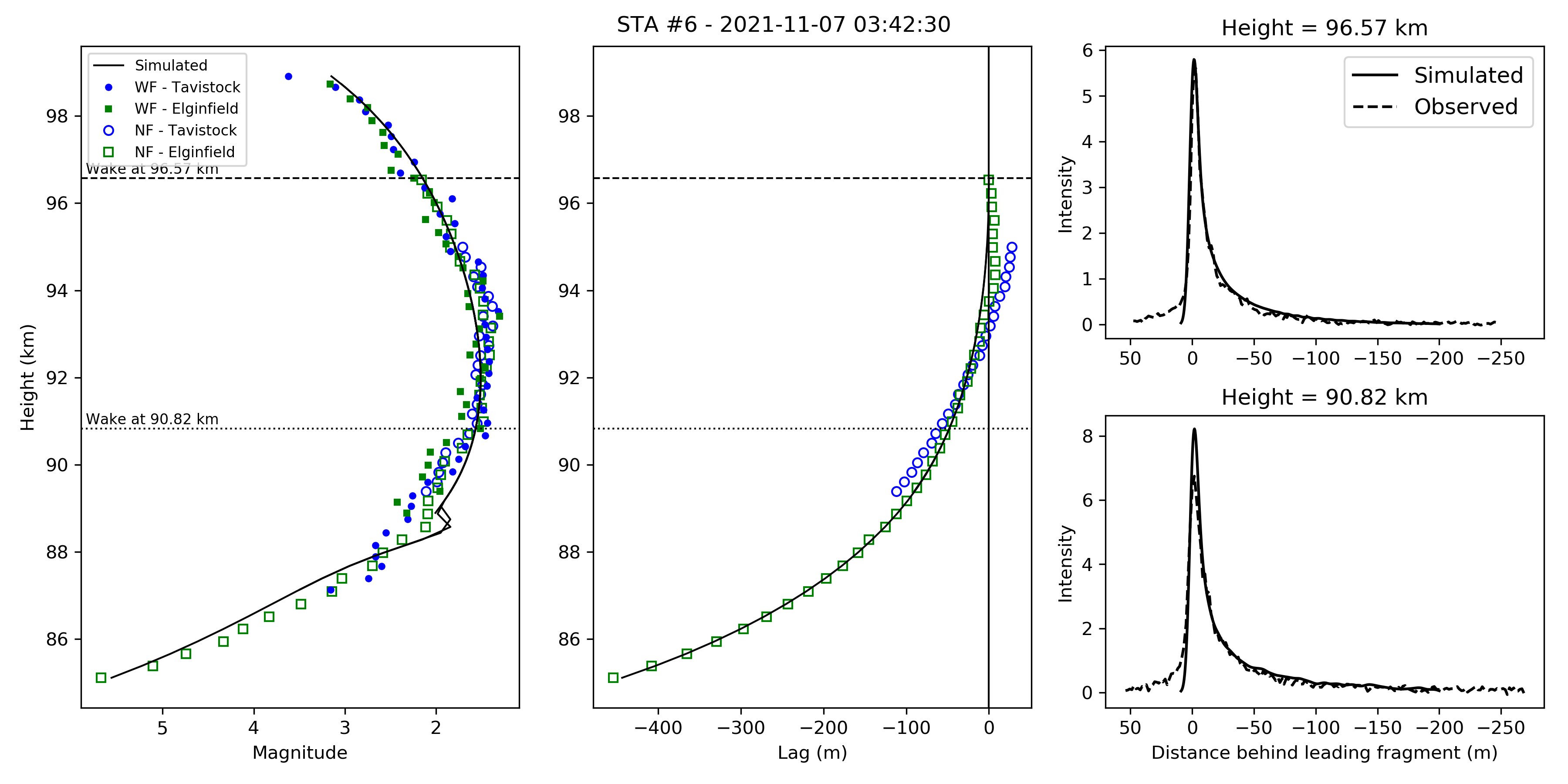}
    \caption{STA6: 2021-11-07 03:42:30}
\end{figure}
\begin{figure}
    \centering
    \includegraphics[width=\linewidth]{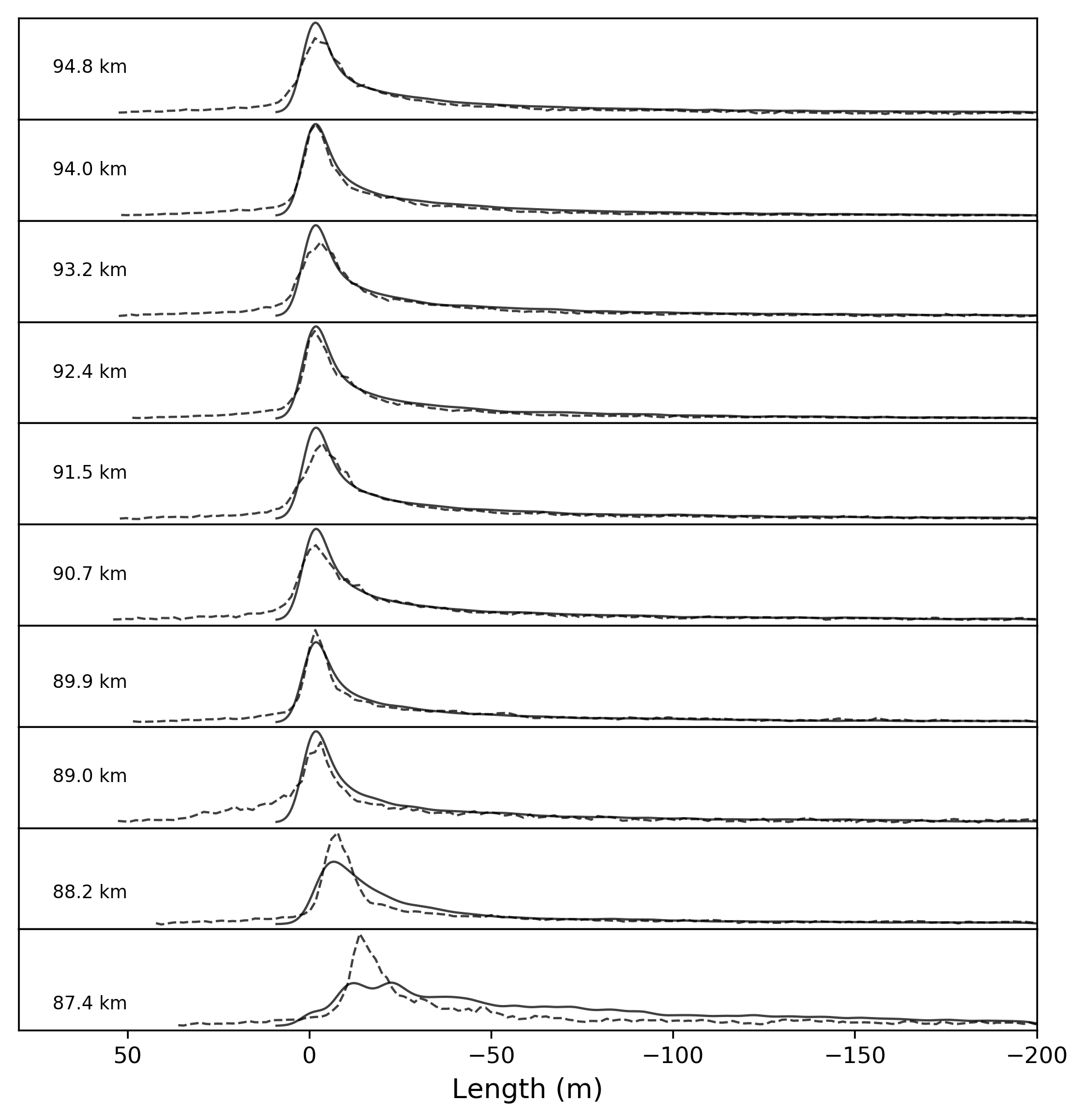}
    \caption{STA6: 2021-11-07 03:42:30}
\end{figure}

\begin{figure}
    \centering
    \includegraphics[width=\linewidth]{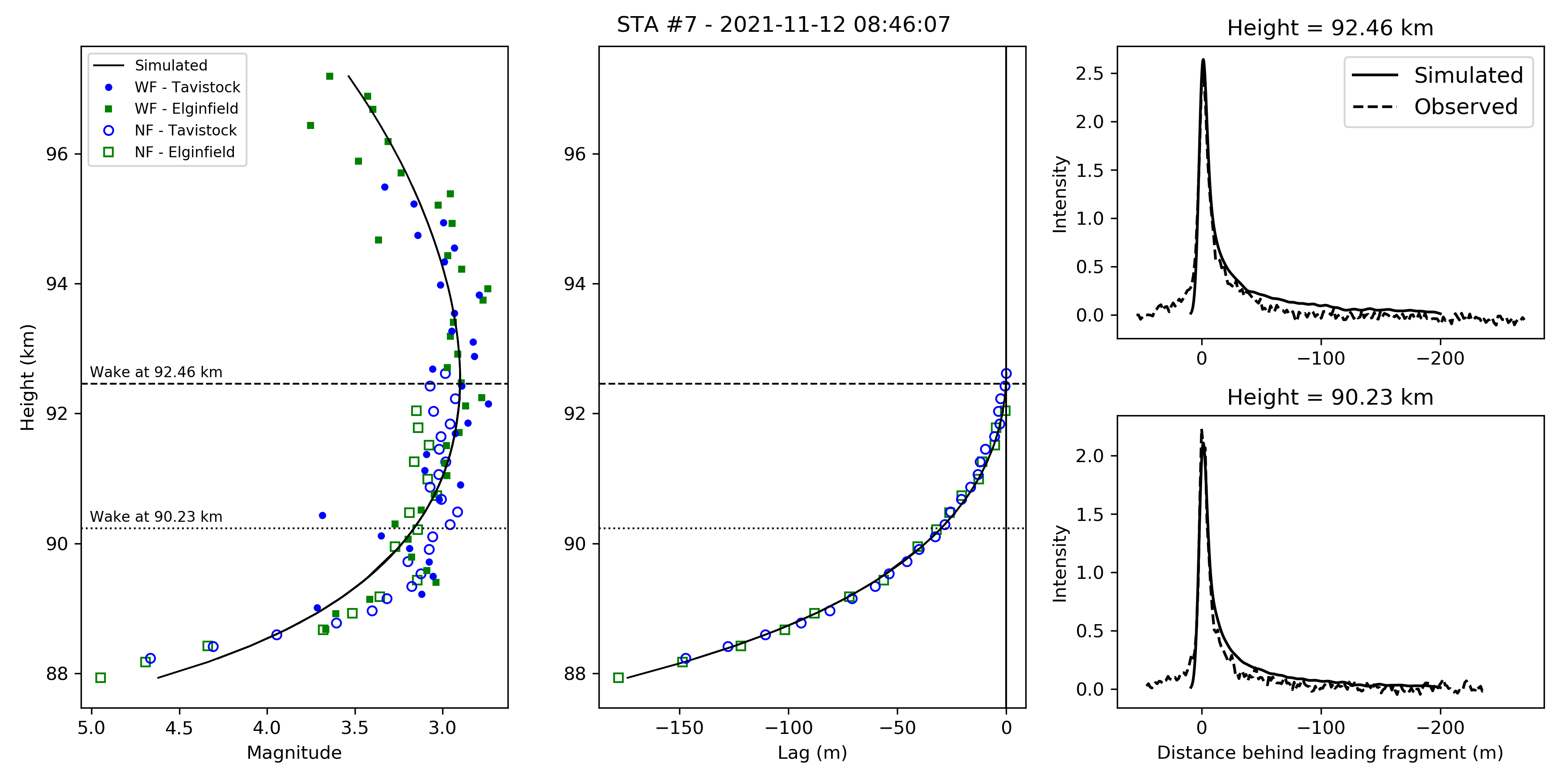}
    \caption{STA7: 2021-11-12 08:46:07}
\end{figure}
\begin{figure}
    \centering
    \includegraphics[width=\linewidth]{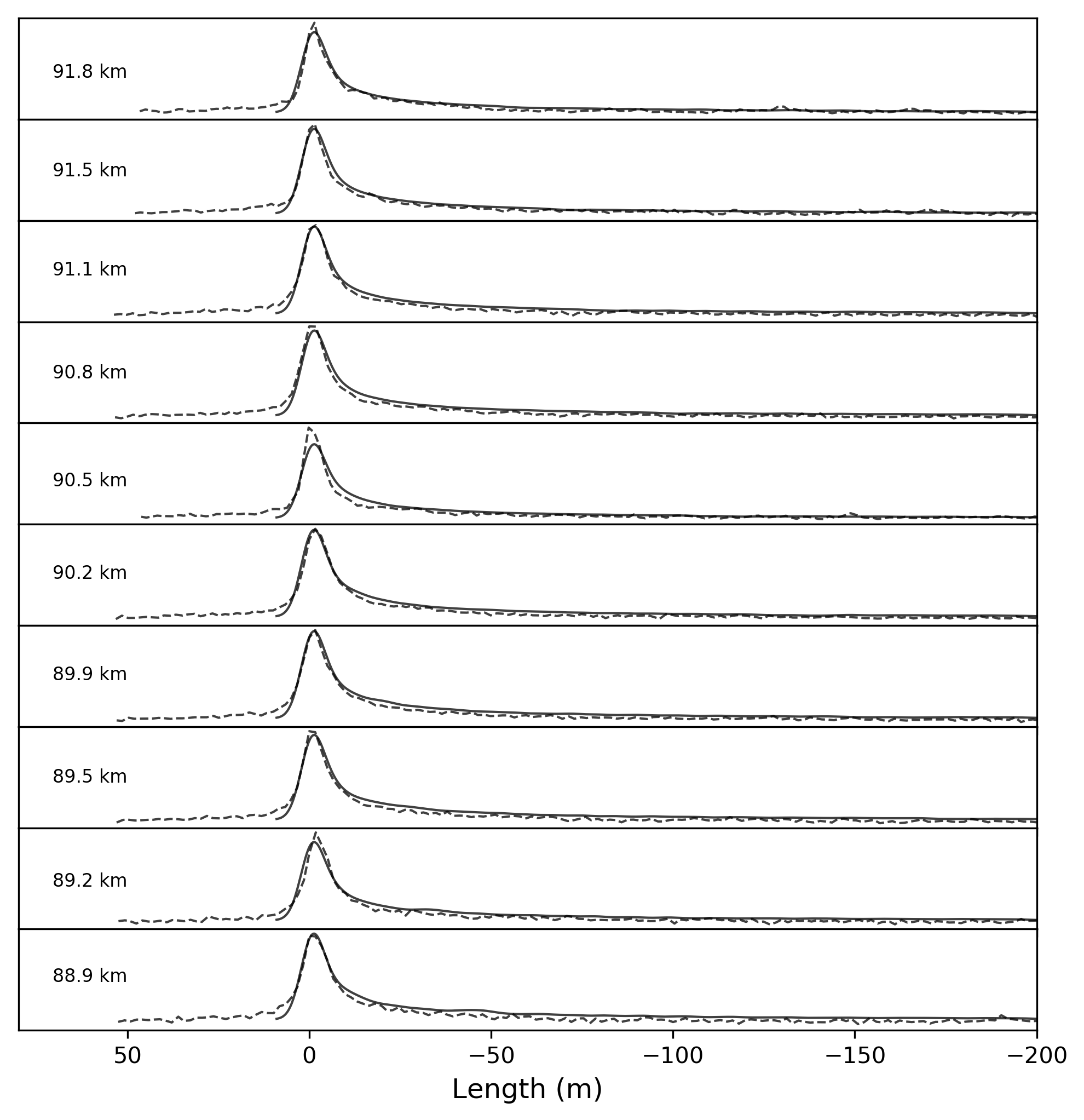}
    \caption{STA7: 2021-11-12 08:46:07}
\end{figure}

\begin{figure}
    \centering
    \includegraphics[width=\linewidth]{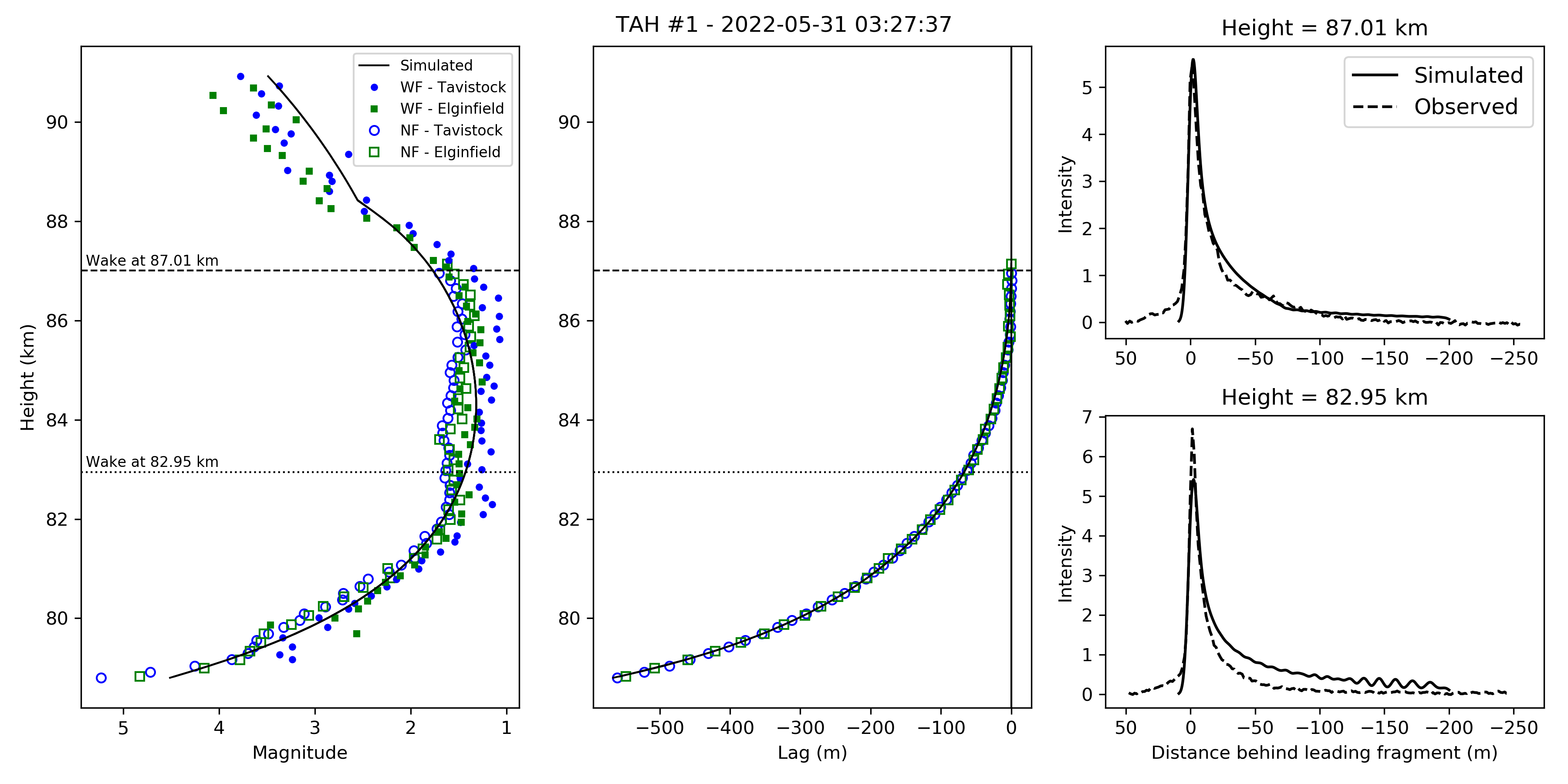}
    \caption{TAH1: 2022-05-31 03:27:37}
\end{figure}
\begin{figure}
    \centering
    \includegraphics[width=\linewidth]{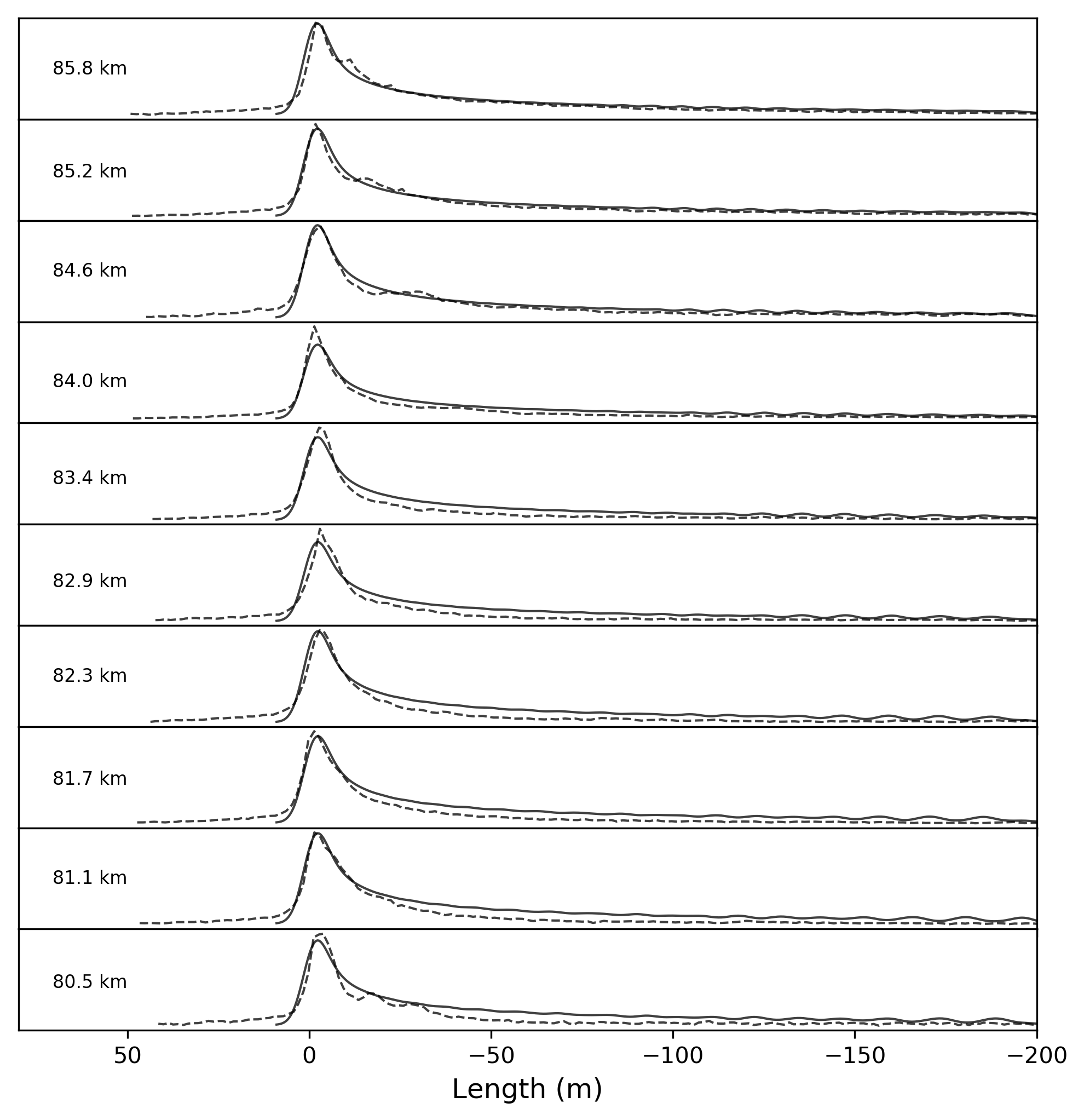}
    \caption{TAH1: 2022-05-31 03:27:37}
\end{figure}

\begin{figure}
    \centering
    \includegraphics[width=\linewidth]{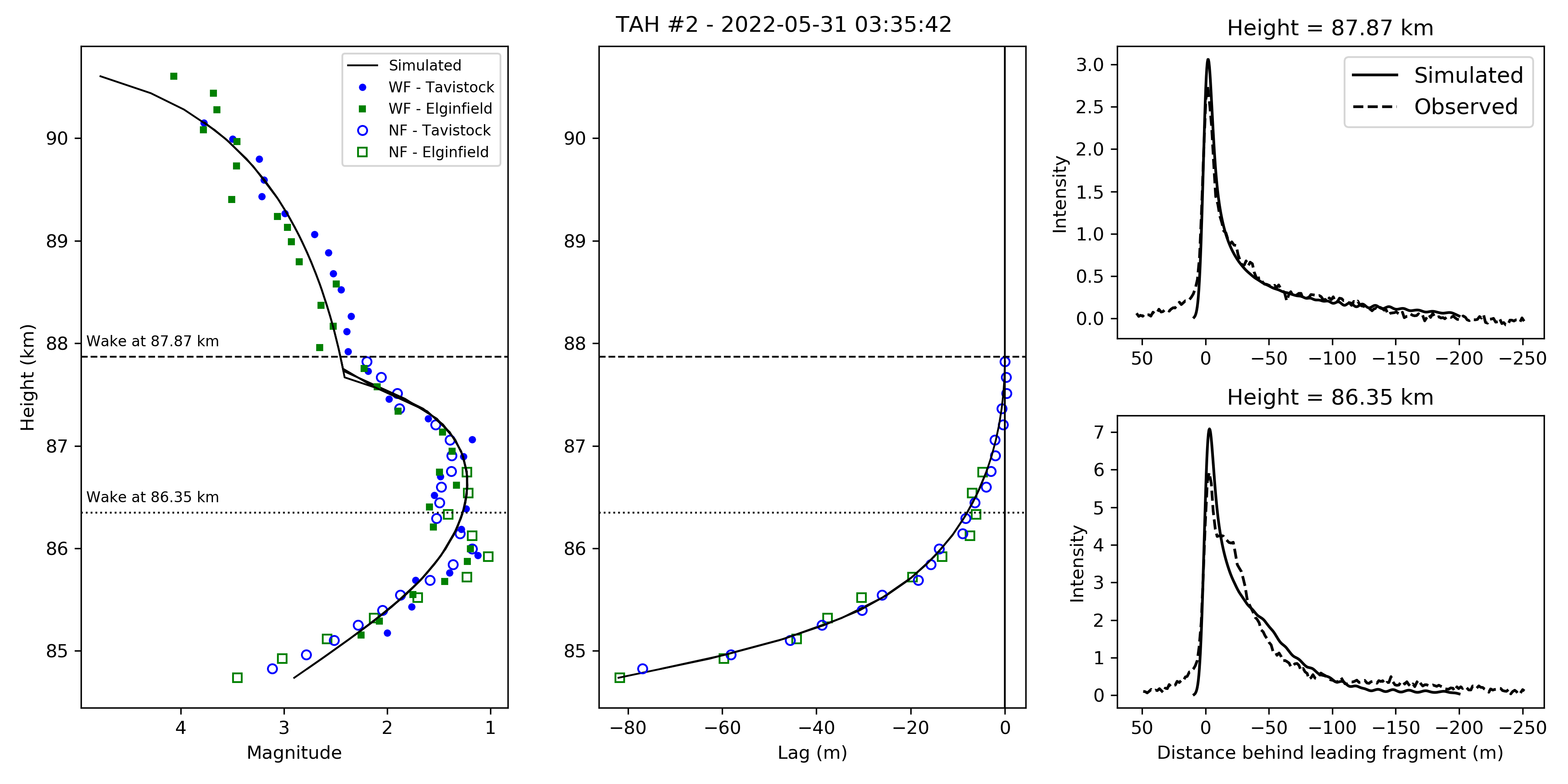}
    \caption{TAH2: 2022-05-31 03:27:37}
\end{figure}
\begin{figure}
    \centering
    \includegraphics[width=\linewidth]{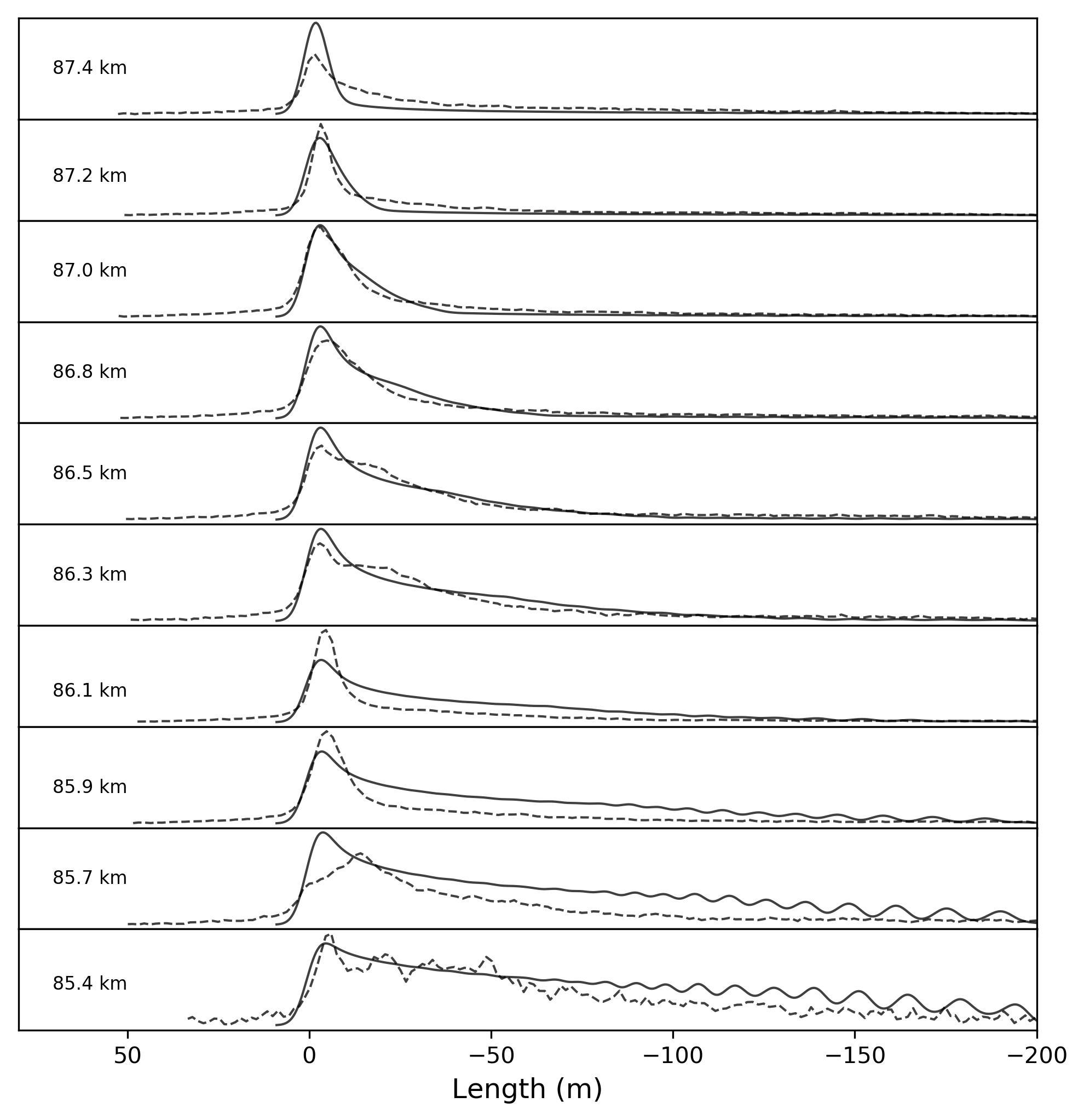}
    \caption{TAH2: 2022-05-31 03:27:37}
\end{figure}

\end{document}